\pdfoutput=1
\documentclass[10pt,twocolumn]{article}

\usepackage[letterpaper,margin=0.75in,columnsep=0.25in]{geometry}
\usepackage[T1]{fontenc}
\usepackage{amsmath,amsthm}
\usepackage{newtxtext,newtxmath}
\usepackage[numbers,sort&compress]{natbib}
\usepackage{graphicx}
\usepackage{booktabs}
\usepackage{xcolor}
\usepackage{array}
\usepackage[shortlabels]{enumitem}
\usepackage[hidelinks]{hyperref}
\hypersetup{pdftitle={Inside the Trap: A Renewal Law for Prefix-Cache Collapse in LLM Agent Serving},pdfauthor={Akshat Valse, Ali Jannesari}}

\setlist{itemsep=1pt,topsep=3pt,parsep=0pt,leftmargin=*}


\newcommand{\fpath}[1]{\nolinkurl{#1}}

\newcommand{\repourl}{\url{https://github.com/AkshatValse/surviving-llm-stampede}}

\newcommand{\lmiss}{\lambda_{\mathrm{miss}}}
\newcommand{\lref}{\lambda_{\mathrm{ref}}}
\newcommand{\ledge}{\lambda_{\mathrm{e}}}
\newcommand{\muinf}{\mu_{\infty}}
\newcommand{\Elb}{E_{\mathrm{lb}}}
\newcommand{\Nlb}{N_{\mathrm{lb}}}
\newcommand{\ncap}{n_{\mathrm{cap}}}
\newcommand{\sps}{\,sessions/s}

\newtheorem{theorem}{Theorem}
\newtheorem{corollary}[theorem]{Corollary}
\newtheorem{proposition}{Proposition}
\newtheorem{lemma}{Lemma}
\newtheorem*{assumption}{Assumption MF}
\theoremstyle{remark}

\newtheorem{nremark}{Remark}

\title{Inside the Trap: A Renewal Law for Prefix-Cache Collapse\\ in LLM Agent Serving}
\author{Akshat Valse\thanks{Corresponding author.} \qquad Ali Jannesari\\[2pt]
Department of Computer Science, Iowa State University, Ames, IA, USA\\[2pt]
\texttt{avalse@iastate.edu} \qquad \texttt{jannesar@iastate.edu}}
\date{}

\begin{document}
\maketitle

\begin{abstract}
Prefix caching lets an LLM agent's follow-up turns reuse cached KV, but a burst can tip an engine into a trap in which every turn misses and the backlog does not drain, at a load it served before. In registered paired tests on two engines, two GPUs and two models (passed at 9 of 10 rates), a 200-session burst left the engine at zero hits to the end in all 30 runs, while every matched no-burst run stayed good and a session cap prevented the trap in all 30 capped runs. We analyse an idealization of the trapped state exactly: for a fluid FIFO server on which every turn misses, we prove that along FIFO orbits its cycles obey a linear renewal recursion. As is classical, the trap drains if and only if arrivals are below the all-miss capacity; above it, a Perron root fixes the dynamics: every trapped state converges to a known self-similar regime in which the gap between turns grows linearly and the turns served are young and short. For our workload the backlog grows at 0.6 to 0.8 of the excess load and sessions complete faster than the all-miss capacity, a youth premium with turn-count and cost-growth factors, so in the model a trap's measured throughput overstates the recovery threshold. Retrospectively, with nothing fitted to trap data, the theory matches 57 stuck runs: the mean served prompt within 3\% in 50 runs, completions above the all-miss capacity in every run, and the pooled gap growth. Blind in outcome, 4 runs trapped by a tool slowdown met 4 of 5 pre-set scores, cut traps persisted, and with a 20-step budget, trapped runs completed below the all-miss capacity, as predicted. Of two exits, refusing arrivals sheds sessions; a session cap restores reuse: in registered tests it rescued 9 of 9 established traps at the same load without shedding, and the engine stayed good after release. We do not prove that the engine converges to the model.
\end{abstract}

\section{Introduction}\label{sec:intro}

\begin{figure*}[t]
\centering
\includegraphics[width=\textwidth]{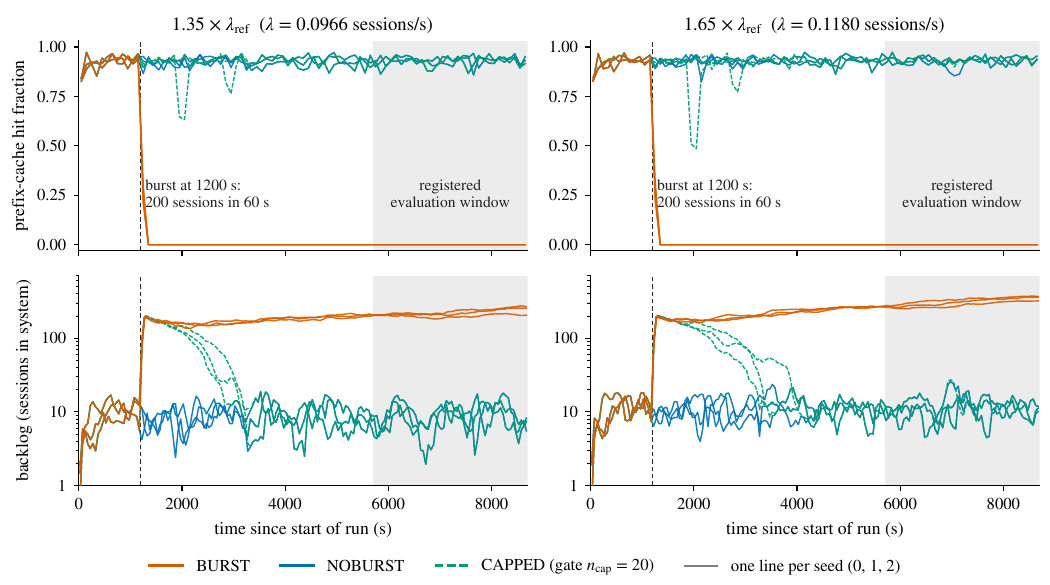}
\caption{The stampede on SGLang: the registered paired test E2 at nominal rates of 1.35~$\lref$ (left) and 1.65~$\lref$ (right),
$\lref=0.0715$\sps, the all-miss session rate measured with the radix cache disabled; one A100, a 300k-token KV pool. Top: hit
fraction (turns with more than half of the prompt cached) in 100\,s windows. Bottom: sessions in the system (log scale), including
those in tool calls. One thin line per seed and arm. Dashed: a burst of 200 sessions at 1200\,s. Shaded: the registered evaluation
window. Without the burst (NOBURST) and with a client-side session cap (CAPPED) the engine stays good; after it (BURST), every turn
misses to the end of the run.}\label{fig:e2}
\end{figure*}

An LLM agent that calls tools sends its serving engine sessions of many turns: each prompt is the previous context plus the
output of the tool just run. Prefix caching keeps the key--value (KV) state of the previous turn, so a follow-up turn whose prefix
is still cached prefills only its new tokens~\citep{zheng2024sglang}. In our workload a hit prefills 551 tokens and a miss about
11{,}100, and a session's prefill work with every turn missing is about 16 times that with every follow-up hitting. The cached KV
belongs to sessions that sit idle in tool calls, and admitting waiting work evicts it. A burst of new sessions can therefore tip
the engine into a state in which every turn misses, the queue grows, the gap between a session's turns grows with it, and idle KV
is evicted before its owner returns. Fig.~\ref{fig:e2} shows this stampede on SGLang: at loads that the engine carried from empty,
a burst of 200 sessions left it at zero hits for the remaining 7500\,s.

\paragraph{The question.} This paper asks what the trapped state \emph{is}. Quasi-static models treat it as the falling branch
of a throughput curve, $\dot n=\lambda-\mu(n)$; read literally in congestion, sessions then complete at the curve's low end and
accumulate at the excess over it. On our engines, where every trapped turn misses, both statements are wrong. Trapped runs of our workload at
loads above the all-miss capacity measured with caching disabled ($\lref$, the unit of every rate below) complete sessions faster
than it, by up to 2.4 times (one from-empty run at 6.5~$\lref$), while over the 57 stuck runs of our paired-test protocol their backlog grows, on
average, at about two thirds of the quasi-static rate. The gap matters to an operator: in model~F below, for our workload, a cut to just below the
throughput measured in an all-miss trap does not clear it.

\paragraph{Contributions.}
\begin{enumerate}[label=(\arabic*),leftmargin=*]
\item \textbf{A renewal law for an idealized trap} (\S\ref{sec:trap}). We idealize the trapped engine as a fluid FIFO server on
which every turn misses and every turn rejoins the tail of the queue (model~F) and prove that along FIFO orbits its cycles obey a
linear renewal recursion (Lemma~\ref{lem:recursion}). Below $\lmiss$ the trap drains, as is classical
(Proposition~\ref{prop:classic}); above it, the recursion's Perron root $z$ carries every trapped state to a self-similar regime
with an exact virtual origin, in which the gap between a session's turns grows at $z-1$, the backlog at $\lambda(1-E[z^{-T}])$ and
completions at $D=\lambda E[z^{-T}]$, with the turn mix tilted to $G_kz^{-k}$ (Theorem~\ref{thm:trap}). We decompose $D/\lmiss$
into a turn-count factor and a cost-growth factor, a youth premium above 1 for our workload at every load we checked
(Theorem~\ref{thm:premium} for $\rho\ge1.04$), and conclude that in model~F with such a premium the throughput measured in a trap overstates the
recovery threshold (Corollary~\ref{cor:trapthr}). The self-similar regime itself is known~\citep{khanin2001steady}, the
recursion is a discrete renewal equation of Lotka--Leslie type~\citep{feller1968introduction} and the premium has a
processor-sharing analogue~\citep{jeanmarie1994transient}; what we add, as far as a targeted search found, is the derivation of the
recursion along FIFO orbits, which carries every trapped state into the regime, and the two factors of the
premium. Corollary~\ref{cor:trapthr} is not new as mathematics; its engine test is.
\item \textbf{The law on the engine} (\S\ref{sec:validation}). With no parameter fitted to any trap measurement, model~F matches,
across 57 stuck runs on SGLang and vLLM, two GPUs and two models (36 of them in the five registered paired-test cells, on 30 distinct arrival
plans: 30 registered runs and 6 earlier exploratory vLLM runs on the same plans), the mean prompt of the turns served (within 3\% in 50 of them), the pooled linear growth of the gap (median ratio of measured to predicted
slope 1.01; per run 0.04--1.81) and completions above the all-miss capacity (in every run). These checks are retrospective. Blind
in outcome, with tolerances set on the earlier runs, the four runs trapped by a tool slowdown alone in two pre-registered cells met
four of the five pre-set scores; a per-run slope check failed in two of them.
A further blind test, MIRROR, with a step-capped
workload for which the theory predicts completions \emph{below} the all-miss capacity, passed at both rates: all six trapped
runs completed 0.50--0.845~$\lref$ over $[6600,8700)$\,s, against a measured 1.06~$\lref$.
\item \textbf{What trapped throughput does not tell} (\S\ref{sec:control}). After a burst, a cut in arrivals to below the trapped
throughput of the same seeds' uncut runs, but above the all-miss capacity, left all three CUT-A runs trapped, and a cut below the
all-miss capacity shrank the three CUT-B backlogs only slowly, all still trapped 6300\,s after the cut, as model~F predicted. Each
arm's registered rule rejected the naive reading, and CUT-A's also the quasi-static one, but not intermediate readings; what singles
out model~F against them is post hoc. Refusing arrivals (SHED) drained a trap by shedding 355--378 sessions. Restoring reuse with a
session cap, which leaves model~F, rescued 9 of 9 established traps at the same offered load without shedding a session, at
a price of 386--569\,s of waiting at the client per paused session and, in a slowdown-only goodput computed post hoc, 45--97\% of the goodput of the hour after the burst, and
the engine stayed good after the cap was removed.
\item \textbf{A registered evidence base} (\S\ref{sec:stampede}, Appendix~\ref{app:stampede}): paired tests with matched no-burst
and capped arms on common random numbers on two engines, two GPUs and two models, a dose response in burst size, registered
negatives, and a first-principles model of the throughput curve with an out-of-sample record (Appendix~\ref{app:entry}).
Table~\ref{tab:ledger-ev} lists the main claims, each with its test, its status and its appendix.
\end{enumerate}

\noindent Model~F is an idealization, and we do not prove that the engine converges to it. Sessions are synthetic (exponential
5\,s tool times, no shared prefixes) and the models are dense and 7--8B in size.

\section{The stampede: registered tests}\label{sec:stampede}

\begin{table*}[t]
\centering\footnotesize
\caption{The registered stampede and recovery tests, all registered and run before Lu et al.'s paper appeared (details,
registrations and timelines in Appendix~\ref{app:stampede}). BURST: 200 sessions within 60\,s at 1200\,s; NOBURST: the same
Poisson stream without them; CAPPED: BURST behind a client FIFO session cap from the start. A rate passes if, in at least two of
three seeds, BURST is trapped (uncached prefill token rate within $\pm15$\% of the all-miss rate and backlog slope above $0.5(\lambda-\lref)$
over $[5700,8700)$\,s), NOBURST good and CAPPED recovered. Pre-registered: the filled registration was committed and pushed to GitHub
before the jobs were submitted (the push log of our private development repository, exported in \fpath{provenance/}); registered:
committed before submission, pushed later. Seeds 0--2 are shared across tests (common random numbers), so counts across tests are
descriptive. Rates are nominal multiples of each cell's $\lref$.}\label{tab:stampede}
\setlength{\tabcolsep}{3pt}
\begin{tabular}{@{}>{\raggedright\arraybackslash}p{2.4cm} >{\raggedright\arraybackslash}p{5.3cm} >{\raggedright\arraybackslash}p{1.9cm} >{\raggedright\arraybackslash}p{7.5cm}@{}}
\toprule
Test & Cell and design & Status & Outcome\\
\midrule
E2 & SGLang 0.5.22, Qwen2.5-7B, A100, 300k pool; 1.35 and 1.65; cap 20 & pre-registered & Pass at both rates: BURST at zero hits 6/6 (trapped 5/6), NOBURST good 6/6, CAPPED recovered 6/6\\
native E2 & SGLang's own pool (955,757 tokens); 2.5 and 4.0; cap 64 & pre-registered & Pass at both rates: 6/6 trapped, 6/6 good, 6/6 recovered\\
vLLM E2 & vLLM 0.32.0, 300k pool; 1.35 and 1.65; cap 20 & registered & Pass at both rates (zero hits 6/6, trapped 4/6); BURST arm repeats earlier exploratory seeds; frozen mechanism-check rule not met (detector alarmed in 2 CAPPED runs)\\
H200 E2 & H200 NVL, 600k pool; 1.5 and 2.0; cap 48 & pre-registered & Pass at both rates: zero hits 6/6 (trapped 5/6)\\
Llama E2 & Llama-3.1-8B, A100, 300k; 1.35 and 1.65; cap 20 & pre-registered & Pass at 1.65, fail at 1.35 (slope rule met by 1 of 3); all six BURST runs at zero hits\\
\addlinespace
BASIN-1, BASIN-2 & 300k, 1.65; bursts of 25 and 100, then 40, 50, 60 & pre-registered & 25 and 40 good 3/3; 60 and 100 stuck 3/3; 50 stuck in 1 of 3; boundary between 40 and 60 in every seed\\
TOOLSLOW & 300k, 1.35, no burst; tool calls 10 or 30\,s longer for 900\,s & pre-registered & Rule not met: +10\,s good 3/3; +30\,s trapped 1/3\\
SHED & 300k, 1.35; arrivals stopped from 1800\,s for 600 or 3600\,s & pre-registered & Pass: 600\,s stuck 2/2 (56--59 sessions refused); 3600\,s good 2/2 (355--378 refused)\\
R1, R4 (rescue) & cap 20 at 1800\,s (300k; 1.35, 1.65) and 64 at 2100\,s (native; 4.0), inside an established trap & pre-registered & Rescued 9/9, no relapse, no session shed; stayed good after release, matching NOBURST\\
\bottomrule
\end{tabular}
\end{table*}

The theory of \S\ref{sec:trap} explains a phenomenon that we tested, independently of Lu et al., under registered rules
(Table~\ref{tab:stampede}). This section summarizes those tests; Appendix~\ref{app:stampede} reports them in full.

\paragraph{Workload and engines.} A session has $T=\min(\mathrm{Geometric}(1/15),60)$ turns ($E[T]=14.76$). Turn $k$ sends a
prompt of $2000+650(k-1)$ tokens, the previous prompt plus 100 decoded tokens and 550 tokens of tool output, as random token ids
(no shared prefixes), and between turns the session waits for an exponential tool time of mean 5\,s. A turn \emph{hits} if more
than half of its prompt is cached. An open-loop client sends Poisson session arrivals to SGLang 0.5.22 (request-level FCFS with its
waiting-prefix refresh, LRU eviction) or vLLM 0.32.0, serving Qwen2.5-7B-Instruct or Llama-3.1-8B-Instruct on one A100 80GB PCIe or
one H200 NVL, with the KV pool capped at 300k tokens (150k or 600k in the pool sweep; 955,757 tokens, SGLang's own, in the native
cell). Before every run the cache is flushed and checked with a canary, and each arm of a test runs in its own job on a fresh
server.

\paragraph{Calibration and the registered rule.} Each cell was calibrated before its test: the closed-population throughput curve
$\mu(n)$ (Stage~C), the all-miss session rate $\lref$ with the radix cache disabled (E0a) and the from-empty edge $\ledge$, the largest
rate at which runs from an empty engine stayed stable for 3000\,s (E0b). In the quasi-static model $\dot n=\lambda-\mu(n)$ the good
state is then metastable in the window $(\lref,\min(\mu^*,\ledge))$ (Appendix~\ref{app:qs}), and a registration placed the test rates
one and two thirds of the way across it and the cap at the smallest $n$ with $\mu(n)\ge0.95\mu^*$, by formula. The paired test E2
runs three arms per seed for 8700\,s, judged over $[5700,8700)$\,s, with the arms of a seed sharing arrival times, turn counts and
tool times, fixed before any run by the client's per-seed plan (Table~\ref{tab:stampede}). Its rule calls a BURST run trapped by tokens, not sessions, because a trapped engine prefills
at the all-miss token rate while, for our workload, completing sessions faster than $\lref$ (\S\ref{sec:validation}). Pre-registered and registered are defined in
Table~\ref{tab:stampede}; neither label says whether a prediction was blind, and no registration is in a third-party registry.

\paragraph{What the tests show.} Over the five paired tests the burst left the engine at zero hits to the end of every one of its
30 BURST runs, 7500\,s after the burst, at rates that the engine carried in the good state for the full 8700\,s of every NOBURST
run; the registered slope rule called 23 of them trapped (post hoc, the plug-in of \S\ref{sec:validation} puts four of the other seven
below the rule's threshold and two within $10^{-4}$\,/s of it). A cap within 5\% of
the measured peak of $\mu(n)$ prevented the trap in all 30 CAPPED runs of the five paired tests, at the price of admission delay at
the client (405--471\,s on average in E2). The registered dose response places the basin boundary between
bursts of 40 and 60 sessions in every seed (50 trapped 1 of 3). A tool slowdown with no burst can also trap the engine: lengthening
every tool call that starts in $[1200,2100)$\,s by 30\,s trapped 1 of 3 runs, so that registered rule (traps in 2 of 3) was not met,
and \S\ref{sec:validation} reports a longer slowdown registered later. On an H200 at the 300k pool, the all-miss capacity rose
2.71-fold and the peak of $\mu(n)$ only 1.63-fold, so the window narrowed to $(1,1.3)\,\lref$, too narrow for the registration rule
to place a test; doubling the pool reopened it to $(1,2.5)\,\lref$ (Appendix~\ref{app:entry} explains the shift). Stopping
arrivals (SHED) and a cap engaged inside an established trap (R1, R4) are the two exits of \S\ref{sec:control}.

\paragraph{The trap persists and outlives its trigger.} Among the runs before the blind tests, no FCFS run trapped by a burst or by a
tool slowdown left the trap on its own. Over 50 such runs, about 100 trapped hours (99--103\,h, depending on where the trap is taken
to start), no 300\,s window after the trap formed reached a follow-up hit fraction of 0.5; the highest was 0.16. Treating the runs as
independent, which shared seeds weaken, a one-sided 95\% bound on a run's probability of escaping within its observed span (at least
6100\,s) is 5.8\%. Stock LPM (SGLang's longest-prefix-match ordering), which leaves request-level FCFS while at most 128 requests wait, escaped in 4 of its 9 trapped runs, and 1 of
the 16 from-empty collapses (150k pool, 1.30~$\lref$) recovered within its 3000\,s run. In 16 stuck runs every trigger session had
left the system 558--6246\,s before the end, and the 727--15{,}453 turns completed after that had no hit in 15 runs (4 hits among
5{,}937 turns in the trapped TOOLSLOW run): the trap outlives its trigger, as Lu et al.\ also observed on
hardware~\citep{lu2026cachetrap}. In the 24 BURST runs of the four A100 paired tests we cannot say so: 3--30 burst sessions were still
present at the end. These counts are post hoc, computed after Lu et al.'s paper appeared (Appendix~\ref{app:dataonly}).

\paragraph{What the record does not show.} The tests use synthetic sessions and dense 7--8B models on one GPU per engine; all but
two use Qwen2.5-7B on the A100. The vLLM registration reached GitHub only after two of its jobs had started, so we call that test
registered, not pre-registered. The 300k H200 cell is a measurement, not a registered test; its closed-curve values for $n\le32$ were
seen by the cell's operating agent and the first author before the model's freeze, and no blind simulator prediction was made for
it. A fresh-seed replication of the
from-empty edge at 300k did not meet its rule (2.5~$\lref$ was stable in 3 of 3 fresh seeds; in realized rates the edge lies at
2.665--2.726~$\lref$). Some predictions for these tests are blind by access only, and the timeline of every registration and
prediction, with its slips, is in Appendix~\ref{app:stampede}.

\section{Model F: a renewal law for the trap}\label{sec:trap}

The quasi-static reading of \S\ref{sec:intro}, whose curve has as its low end Lu et al.'s low-reuse departure rate
$D_{\mathrm{low}}$~\citep{lu2026cachetrap} and, in our earlier model, the all-miss capacity (Appendix~\ref{app:qs}), fails on our
engines (\S\ref{sec:validation}). Here we characterize the trap exactly, within an explicit idealization of the trapped engine.

\paragraph{Model F.} Let a session have $T$ turns, $G_k=P(T\ge k)$ for $k=1,\dots,K$ with $G_K>0$, and let turn $k$ cost $c_k>0$
GPU seconds when it misses. The costs come from an iteration-cost model fitted only to server-log timings
(Appendix~\ref{app:entry}): the miss prefill of turn $k$ plus the all-miss per-turn decode share. Write
\[
\lmiss=\Big(\sum_kc_kG_k\Big)^{-1},\qquad \rho=\lambda/\lmiss ,
\]
the all-miss capacity at the stationary turn mix and the load. Rates in this paper are multiples of $\lref$, the all-miss
session rate measured with the radix cache disabled (E0a, 32 sessions; 0.0715\sps{} for SGLang on the A100); $\lmiss$ is model~F's
value from the cost model, and $\lmiss/\lref=0.96$--$1.035$ across our cells. Model~F is a fluid single-server queue in which
\begin{enumerate}[(F1),leftmargin=*,nosep]
\item every turn misses, so turn $k$ costs $c_k$;
\item the server works at unit rate while anything is pending;
\item requests complete in the order in which they joined (request-level FCFS);
\item a completed turn $k<T$ rejoins the tail of the queue at once as turn $k+1$;
\item new sessions join as turn~1 at rate $\lambda>0$;
\item the state at time 0 is any finite pending mass with positive work.
\end{enumerate}
(F3)--(F4) make model~F a round robin over sessions with visit-dependent quanta $c_k$: a single-station re-entrant
line~\citep{kumar1993reentrant} in overload. A request that joins at time $u$ completes at $f(u)=u+V(u)$, where $V(u)$ is the
pending work; so $V$ is also the gap between two consecutive turns of a session. Call $u_0,\ u_{n+1}=f(u_n)$ a \emph{FIFO orbit}
and $\Delta_n=u_{n+1}-u_n$ its cycles. Let $N(t)$ count the sessions present and $V_{\mathrm{tot}}(t)$ be their total remaining
all-miss work, future turns included, and let $\bar k_c=\sum_kkc_kG_k/\sum_kc_kG_k$ be the cost-weighted mean turn index.

\begin{lemma}[the orbit recursion]\label{lem:recursion}
In model~F, while the queue is busy the remaining work changes at rate $\rho-1$, and along every FIFO orbit the cycles solve the
linear renewal recursion
\begin{equation}\label{eq:renewal}
\Delta_n=\sum_{k=1}^{\min(n,K)}\lambda c_kG_k\,\Delta_{n-k}+h_n ,
\end{equation}
where $h_n=\sum_{k=n+1}^{K}c_kG_k\pi_{k-n}(u_0)/G_{k-n}$, with $\pi_j(u_0)$ the turn-$j$ mass pending at $u_0$, is the work
that the requests pending at $u_0$, and their later turns, bring in generation $n$; so $h_n\ge0$, $h_n=0$ for $n\ge K$ and
$\sum_nh_n=V_{\mathrm{tot}}(u_0)$. An orbit with $u_0<V(0)$ also satisfies, for $n\ge K-1$, the exact identity
$u_{n+1}=V_{\mathrm{tot}}(0)+\sum_k\lambda c_kG_ku_{n+1-k}$.
\end{lemma}

\begin{theorem}[every trapped state converges]\label{thm:trap}
In model~F let $\rho>1$, and let $z>1$ be the Perron root, $\lambda\sum_kc_kG_kz^{-k}=1$. From any initial state:
\begin{enumerate}[(a),leftmargin=*,nosep]
\item as $t\to\infty$ the gap grows linearly, $V(t)/t\to\gamma=z-1$; the backlog grows as $N(t)/t\to\alpha=\lambda(1-E[z^{-T}])$;
sessions complete at rate $D=\lambda E[z^{-T}]$; and the pending requests, and the requests served in each cycle, have the tilted
turn mix $\nu^z_k=G_kz^{-k}/\sum_jG_jz^{-j}$;
\item (affine law with a virtual origin) let $u^*=V_{\mathrm{tot}}(0)/(\rho-1)$ and let $r$ be the largest modulus of the other roots
of $x^K=\sum_k\lambda c_kG_kx^{K-k}$. Every FIFO orbit with $u_0<V(0)$ satisfies $u_n+u^*=Bz^n+O((r+\epsilon)^n)$ for every
$\epsilon>0$, with a constant $B$, so relative to $z^n$ the transient is $O(((r+\epsilon)/z)^n)$. Hence the gap is affine in time with
virtual origin $-u^*$, $V(u_n)=\gamma(u_n+u^*)+O((r+\epsilon)^n)$, and proportional to the remaining work,
$V(u_n)=\frac{\gamma}{\rho-1}V_{\mathrm{tot}}(u_n)+O((r+\epsilon)^n)$; when $r<1$ these errors vanish absolutely. A session that
arrives at $u$ with $T=m$ turns completes at $f^m(u)$, and $f^m(u)+u^*=z^m(u+u^*)(1+o(1))$.
\end{enumerate}
\end{theorem}

\begin{proposition}[the classical parts]\label{prop:classic}
In model~F:
\begin{enumerate}[(a),leftmargin=*,nosep]
\item the queue empties if and only if $\rho<1$, at $V_{\mathrm{tot}}(0)/(1-\rho)$; with arrivals stopped ($\lambda=0$, outside (F5)) it
empties at $V_{\mathrm{tot}}(0)$;
\item if $\rho=1$, $V(t)\to V_{\mathrm{tot}}(0)/\bar k_c$ and $N(t)\to\lambda E[T]V_{\mathrm{tot}}(0)/\bar k_c$;
\item as $\rho\downarrow1$, $\gamma=(\rho-1)/\bar k_c+O((\rho-1)^2)$ and $\alpha/(\lambda-\lmiss)\to E[T]/\bar k_c$.
\end{enumerate}
\end{proposition}

\noindent Part~(a) is classical in substance: at a single work-conserving station the remaining work obeys
$\dot V_{\mathrm{tot}}=\rho-1$ whatever the service order, which gives the fluid busy period, and its stochastic counterpart is
Kendall's branching description of the busy period~\citep{kendall1951some}. Part~(b) is the fluid counterpart of the state-space
collapse of a critically loaded multiclass FIFO feedback queue~\citep{reiman1988multiclass,dai1995multiclass}; we prove it from
Lemma~\ref{lem:recursion} with the discrete renewal theorem~\citep[Ch.~XIII]{feller1968introduction}. Part~(c) is an elementary expansion of
the Perron equation.

\begin{theorem}[youth premium]\label{thm:premium}
In Theorem~\ref{thm:trap}(a), $D/\lmiss=H(z)\,\Gamma(z)$, where
\[
H(z)=\frac{E[T]\,E[z^{-T}]}{\sum_kG_kz^{-k}},\qquad \Gamma(z)=\frac{E_\nu[c]}{E_{\nu^z}[c]},
\]
with $\nu_k\propto G_k$ the stationary and $\nu^z_k\propto G_kz^{-k}$ the trapped turn mix. $H$ depends only on the law of $T$:
$H=1$ for a geometric $T$ on $\{1,2,\dots\}$ ($K=\infty$; the formulas extend for $z>1$), and $H\in(1-q^K,1)$, non-increasing
in $z$, for $T=\min(\mathrm{Geometric}(1-q),K)$ with $K\ge2$. If $c$ is non-decreasing, $\Gamma\ge1$, with equality only for
constant $c$, and $\Gamma$ is non-decreasing in $z$. For that $T$ and a non-decreasing, non-constant $c$ with
$E_\nu[c]/c_1>1/(1-q^K)$, let $z_1$ solve $\Gamma(z_1)=1/(1-q^K)$; then $D>\lmiss$ for every $z\ge z_1$. Near the threshold,
$D/\lmiss-1=(\rho-1)(1-E[T]/\bar k_c)+O((\rho-1)^2)$.
\end{theorem}

\noindent In model~F the trapped server can thus complete sessions faster than its all-miss capacity while its backlog diverges, because
it serves a young backlog of short, cheap turns. The premium is not automatic: truncation lowers $H$, and with
$T=\min(\mathrm{Geometric}(1/15),60)$ and slowly growing costs $c_k=1+10^{-3}(k-1)$ it is negative at every load checked; for a
nearly deterministic $T$, $H$ can be well below 1 (\S\ref{sec:mirror}). For our workload ($c_{60}/c_1=17$--23), $E[T]/\bar k_c=0.60$--$0.62$
in all eight cells, and the premium is positive at every load we checked: proved by Theorem~\ref{thm:premium}'s monotonicity for
$\rho\ge1.04$, shown by a monotone bracket ($H$ falling, $\Gamma$ rising; evaluated in floating point, not interval arithmetic) on
$z\in[1+10^{-12},z_1]$, and only to first order below that (Appendix~\ref{app:trap}).

\begin{corollary}[the drain threshold is $\lmiss$, not the trapped throughput]\label{cor:trapthr}
In model~F at load $\lambda>\lmiss$, let the arrival rate change to $\lambda'$ at a time $t_0$ when the queue is busy. The
queue then empties if and only if $\lambda'<\lmiss$, after the busy period $V_{\mathrm{tot}}(t_0)/(1-\lambda'/\lmiss)$. If
$\lambda'>\lmiss$ the backlog grows as $\alpha(\lambda')t$; at $\lambda'=\lmiss$ the fluid backlog settles at a positive
level. In particular, whenever $D(\lambda)>\lmiss$ (Theorem~\ref{thm:premium}; for our workload at every load we checked), a
cut to any rate in $[\lmiss,D(\lambda))$, below the asymptotic trapped throughput, does not empty model~F's queue.
\end{corollary}

\begin{nremark}[invariance]\label{rem:trapinv}
$z$ solves $\rho\sum_k\omega_kz^{-k}=1$ with the cost-weighted turn mix $\omega_k=c_kG_k/\sum_jc_jG_j$. So $z$, $\gamma$, the
trapped composition, $\alpha/\lmiss$ and $D/\lmiss$ depend on a cell only through $\rho$, $\omega$ and the law of $T$; they are
unchanged by a uniform GPU speed-up at equal $\rho$. This needs proportional cost vectors; our H200 cell model is a uniform speed-up
of the A100's, and the remark is not a general hardware claim.
\end{nremark}

\paragraph{Proof idea.} FIFO and work conservation give $f(u)=V(0)+W(u)$, where $W(u)$ is the work that has joined since
time~0. FIFO maps the requests that join during one cycle onto the requests that complete during the next, so a session
whose first turn joined in $(u_{n-k},u_{n-k+1}]$ rejoins as turn $k$ in $(u_{n-1},u_n]$, and the turn-$k$ mass joining there
is $\lambda G_k\Delta_{n-k}$. Summing its work, $\Delta_n=W(u_n)-W(u_{n-1})$, turns the nonlinear fluid into
Eq.~\eqref{eq:renewal}, whose coefficients are nonnegative and sum to $\rho$ (Lemma~\ref{lem:recursion}). Tilting by $z^{-n}$ and a
finite-memory renewal lemma give Theorem~\ref{thm:trap}(a) and Proposition~\ref{prop:classic}(b); the exact identity gives
Theorem~\ref{thm:trap}(b); Proposition~\ref{prop:classic}(a) is a busy-period argument. Appendix~\ref{app:trap} gives complete
proofs, the stochastic counterpart and a sketch of the processor-sharing variant; \fpath{tests/test_theory_icml.py} checks most
statements numerically (Appendix~\ref{app:trap} lists which); independent AI-agent checks are recorded in \fpath{notes/THEORY_ICML_2026-11.md}, \S10.

\paragraph{What is new.} Eq.~\eqref{eq:renewal} is a discrete renewal equation~\citep{feller1968introduction} of Lotka--Leslie type:
the arrivals $\lambda\Delta_n$ play the role of births, $G_k$ of survival and $\lambda c_kG_k$ of net maternity, and the tilted mix
$G_kz^{-k}$ is Lotka's stable age distribution. Viewed as a steady solution (constant flows from the empty state), the self-similar
regime of Theorem~\ref{thm:trap} is the single-station case of the steady solutions of fluid FIFO re-entrant lines of Khanin et
al.~\citep{khanin2001steady}, which give the Perron equation, the gap slope $z-1$, the $z^{-k}$-tilted mix and the departure flow; they treat
deterministic stages, a closing remark extends their framework to a fixed fraction of the flow leaving at each node (our random
$T$), their normalization gives Remark~\ref{rem:trapinv}, and they leave open the uniqueness of non-steady trajectories. The processor-sharing
version of model~F has the Crump--Mode--Jagers structure of overloaded processor sharing, whose fluid shape converges to an invariant
one~\citep{grishechkin1992processor,jeanmarie1994transient,puha2006overloaded}, and Theorem~\ref{thm:premium} is the visit-indexed
analogue of its departure rate $\lambda E[e^{-\alpha_{\mathrm{PS}}S}]$ ($\alpha_{\mathrm{PS}}$ its growth exponent), which can exceed
$1/E[S]$~\citep{jeanmarie1994transient}. What we add, as far
as a targeted search found (\S\ref{sec:related}), is the derivation of the recursion along FIFO orbits, which carries every solution
from \emph{any} trapped state to the self-similar regime, geometrically in the generation index and with an exact virtual origin,
whereas a burst leaves the engine far from a steady solution; and the turn-count and cost-growth factors of the premium.
Corollary~\ref{cor:trapthr} is not new as mathematics: it is the operational reading of Theorem~\ref{thm:premium} with the classical
drain condition, its processor-sharing counterpart follows from~\citep{jeanmarie1994transient} in the same way, and Lu et al.\ state
that arrivals below their low-reuse departure rate drain the population~\citep{lu2026cachetrap}. What is new about it is its test on
the engine (\S\ref{sec:control}).

\paragraph{Model F is an idealization.} On the engine, (F1) is the measured zero hit fraction of the trap and (F2) its saturated
GPU. (F3) is request-level FCFS admission, up to the $M/E_S$ requests that run concurrently ($M$ the pool in tokens, $E_S$ the
service-weighted KV of a running turn; about 28 at 300k); Proposition~\ref{prop:A} (Appendix~\ref{app:proofs}) bounds the closed system at the stationary mix,
and the open trap exceeds $\lmiss$ only through its tilted mix. (F4) neglects the 5\,s tool time against cycles of 166--516\,s; adding
it changes the fluid's backlog slope, completion rate and mean served prompt over the window by less than 0.02\%. Model~F leaves out
concurrency, preemption (vLLM preempts), and the transitions into and out of the trap, where hits return. We do not prove that the
engine converges to model~F; its test is empirical (\S\ref{sec:validation}).

\paragraph{Why the trap does not heal.} Theorem~\ref{thm:trap}(a) makes the gap between a session's turns grow linearly, while the
trapped engine rewrites its pool at the all-miss token rate, every $R_M=M/\text{churn}\approx26$--78\,s in our cells (churn, the rate
at which new KV is written: uncached prompt plus completion tokens per second). Once the gap is several times $R_M$, as it is in
every BURST run of the first three registered tests (gap$/R_M=4.1$--9.7, Table~\ref{tab:gapR}), an idle session's KV is overwritten
long before it returns, so (F1) is self-consistent, and becomes more so as the trap ages. The pool need not be full: admitted
requests held 55--83\% of it on average in the trapped SGLang runs with 300k and 600k pools (the engine's gauge, which also counts
locked prefixes), and 85--94\% with the native pool and on
vLLM (post hoc, computed after Lu et al.'s paper appeared; Appendix~\ref{app:dataonly}). This is a heuristic consistency argument, not part of the theorem: $R_M$ is a turnover scale, not a pathwise eviction bound. With
Poisson arrivals, Kendall's branching argument puts the probability that the all-miss queue ever empties at 0.35 for $\rho=1.01$ and
$4\times10^{-23}$ at 1.35, for a burst of 200 sessions into an empty engine and the remaining work 3000\,s later taken as realized
and equal to the fluid's (Appendix~\ref{app:trap}); these are not bounds, since hits return first.

\section{The renewal law on the engine}\label{sec:validation}

\begin{figure*}[t]\centering
\includegraphics[width=\textwidth]{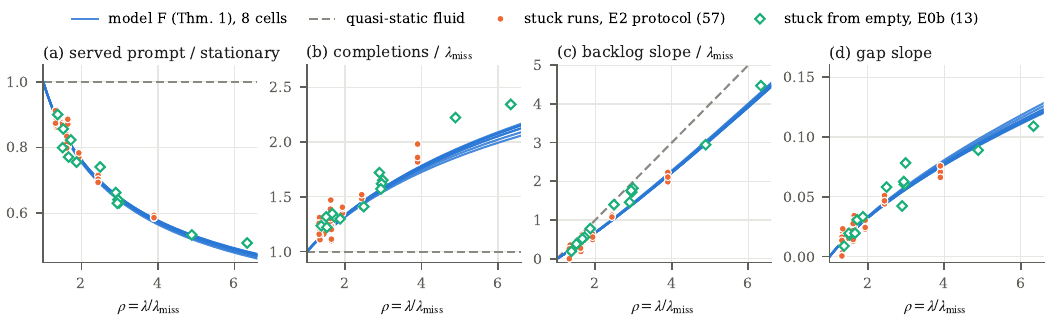}
\caption{Theorem~\ref{thm:trap} against every stuck run of our workload, as a function of the load $\rho=\lambda/\lmiss$. Points: the 57 stuck
runs of the E2 protocol (orange) and the 13 from-empty runs that were stuck over $[1500,3000)$\,s (E0b, hollow). Lines: model~F
for each of the eight cells, nearly coincident (\S\ref{sec:trap}). No parameter is fitted to any trap measurement. (a)~Mean prompt of the turns served, over the
stationary mean: the tilted turn mix. (b)~Session completion rate over $\lmiss$: the youth premium, above 1. (c)~Backlog growth over
$\lmiss$. (d)~Growth rate of the gap between a session's turns, regressed on the turn's completion time, against model~F's
$\gamma/z=1-1/z$. Dashed: the quasi-static fluid read literally in congestion (stationary composition, completion at $\lmiss$,
growth at $\rho-1$). Retrospective: the theory was derived after these runs were read.}\label{fig:trapuniv}
\end{figure*}

\begin{table*}[t]
\centering\footnotesize
\caption{Evidence ledger: the paper's main claims, each with its test, its status and the appendix that reports it. Status:
pre-registered or registered as in Table~\ref{tab:stampede}; blind in outcome, predictions frozen before the runs existed (blind
tests use tolerances set on earlier runs); retrospective, theory derived after the runs were read; post hoc, an analysis chosen
after the outcome. Run counts are runs, not independent trials: seeds 0--2 recur across tests. Commits, tags and clock times are
in Appendix~\ref{app:timeline}.}\label{tab:ledger-ev}
\setlength{\tabcolsep}{3pt}
\begin{tabular}{@{}>{\raggedright\arraybackslash}p{4.0cm} >{\raggedright\arraybackslash}p{3.6cm} >{\raggedright\arraybackslash}p{2.9cm} >{\raggedright\arraybackslash}p{5.85cm} c@{}}
\toprule
Claim & Test, runs & Status & Result & App.\\
\midrule
\multicolumn{5}{@{}l}{\emph{The stampede (\S\ref{sec:stampede}); registered and run before Lu et al.'s paper appeared}}\\
A burst traps the engine at a load it carries & E2 in five cells; 30 BURST, 30 NOBURST runs & pre-registered (vLLM registered) & zero hits to the end in 30/30, NOBURST good 30/30; rule met at 9 of 10 rates (Llama 1.35 failed) & \ref{app:stampede}\\
The basin has a size threshold & BASIN-1, BASIN-2; 15 runs & pre-registered & boundary between 40 and 60 sessions in every seed & \ref{app:stampede}\\
A tool slowdown alone can trap & TOOLSLOW; 6 runs & pre-registered & rule not met: +30\,s trapped 1 of 3, +10\,s none & \ref{app:stampede}\\
The trap persists and outlives its trigger & 50 trapped FCFS runs; 16 with the trigger gone & post hoc (computed after Lu et al.) & no escape in about 100 trapped hours; no hit after the trigger left in 15 of 16 & \ref{app:dataonly}\\
\addlinespace
\multicolumn{5}{@{}l}{\emph{Model~F, retrospective (\S\ref{sec:validation}); nothing fitted to any trap measurement}}\\
The trapped turn mix is $\nu^z$ & 57 E2-protocol stuck runs (36 in paired-test cells: 30 registered, 6 exploratory), 13 from empty & retrospective & served prompt within 3\% in 50 of 57 ($-6.6$ to $+7.7$\%) & \ref{app:trapdata}\\
The gap grows linearly, at $1-1/z$ & same & retrospective & pooled median ratio 1.01; per run 0.04--1.81 & \ref{app:trapdata}\\
Completions exceed $\lmiss$ (youth premium) & same & retrospective & every run above $\lref$ (1.10--2.03; 2.40 from empty); 1.08--1.81 net of the token rate & \ref{app:trapdata}\\
The backlog grows at $\alpha$, not $\lambda-\lmiss$ & same & retrospective & slope over $\alpha$: mean 0.99, median 1.06 (quasi-static: 1.36--1.59) & \ref{app:trapdata}\\
\addlinespace
\multicolumn{5}{@{}l}{\emph{Blind tests, registered on 2026-11-05 after Lu et al.'s paper appeared (\S\ref{sec:validation}, \S\ref{sec:control})}}\\
The trap-state law holds on traps with no burst & TOOLSLOW-W (3 stuck), TOOLSLOW-D15 (1 stuck) & pre-registered; blind in outcome & 4 of 5 pre-set scores pass pooled; the per-run slope check failed in 2 of 4 runs & \ref{app:blind}\\
Which seeds a slowdown traps & TOOLSLOW-D15; 3 runs & pre-registered; two frozen predictors & NO-TRAP, as the simulator said per seed; the first-principles fluid verdict (TRAPS) missed & \ref{app:blind}\\
Model~F's prediction that a near-deterministic turn count reverses the premium & MIRROR; 6 BURST, 3 NOBURST runs & pre-registered; blind in outcome; workload chosen with the model & pass at both rates, RENEWAL 3/3 each; lull 0.50--0.845~$\lref$, below the measured 1.06; rejects only the throughput-curve readings & \ref{app:blind}\\
A cut into $[\lmiss,D)$ does not clear the trap & CUT-A; 3 runs & pre-registered; blind in outcome; rule fixed after the predictions were read & pass: theory-favoured 3/3, stuck 3/3; rejects only the naive and quasi-static readings (a constant rate at the cut fits as well); late slopes post hoc & \ref{app:blind}\\
A cut below $\lmiss$ drains only slowly & CUT-B; 3 runs & as CUT-A & pass: theory-favoured 3/3, backlog falling 3/3, still trapped 3/3 at 8700\,s & \ref{app:blind}\\
\addlinespace
\multicolumn{5}{@{}l}{\emph{The exits (\S\ref{sec:control}); registered before Lu et al.'s paper appeared}}\\
Refusing arrivals drains a trap by shedding & SHED; 4 runs & pre-registered; model~F retrodicted & 600\,s stop stuck 2/2, 3600\,s good 2/2; 355--378 sessions refused & \ref{app:stampede}\\
A cap restores reuse without shedding & R1, R4; 9 runs & pre-registered & rescued 9/9, good after release; 4 of 9 release times outside the frozen $\pm15$\% & \ref{app:stampede}\\
\addlinespace
\multicolumn{5}{@{}l}{\emph{Entry}}\\
The throughput curve from first principles & 8 cells; frozen tags & 5 cells blind (3 in time, 2 by access only, one partly seen) & 0.5-hit crossing off by 1.5 sessions on average (Lu et al.'s curve, read post hoc: 4.1); misses listed & \ref{app:entry}\\
\bottomrule
\end{tabular}
\end{table*}

We test model~F on every trapped run of our workload, in eight cells: SGLang with the 300k pool (FCFS with and without its waiting-prefix
refresh, the BASIN, SHED and TOOLSLOW runs, and stock LPM past its 128-request fallback), SGLang's native pool, vLLM, the H200 at
300k and 600k, Llama-3.1-8B, and the 150k and 600k pools. vLLM, the H200 and Llama take one speed factor, fitted to their own
all-miss token rate, on the cost model of \S\ref{sec:trap}. No parameter is fitted to
any trap measurement. Table~\ref{tab:ledger-ev} collects the evidence.

\paragraph{Retrospective check.} The theory was derived after these runs had been read, so this check is retrospective, not
blind. Over the 57 stuck runs of the E2 protocol (follow-up hit fraction below 0.2 over $[5700,8700)$\,s; 20 cell and rate groups;
36 in the five registered paired-test cells, 30 registered runs and 6 earlier exploratory vLLM runs on the same plans, on 30
distinct arrival plans; Fig.~\ref{fig:trapuniv}, and per cell Table~\ref{tab:trap} in
Appendix~\ref{app:trapdata}):
\begin{itemize}[leftmargin=*,nosep]
\item \emph{Composition.} The mean prompt of the turns served matches the tilted mix $\nu^z$ to within $-6.6$\% and $+7.7$\%,
and to within 3\% in 50 of the 57 runs: the trapped engine serves turns 9--41\% shorter than the stationary mix (the tilted mix predicts 10--42\%).
\item \emph{Gap growth.} The gap between a session's turns grows linearly in time. We regress it on the turn's completion time;
a turn that completes at $t$ joined at about $t/z$, so model~F's slope is $\gamma/z=1-1/z$. Measured over predicted, the ratio has
median 1.01 (mean 0.96). Per run it ranges from 0.04 to 1.81; the low values are the seed-1 runs at 1.35~$\lref$, whose plans
deliver fewer arrivals than nominal (below) and which completed sessions faster than $D$, leaving their backlogs nearly flat.
\item \emph{Youth premium.} Every stuck run completes sessions above $\lref$, at 1.10--2.03~$\lref$ (up to 2.40 in the from-empty
runs), and on average 1.055 times $D$ (0.90--1.21). Part of the excess over $D$ is a congested engine computing prompt tokens faster
than the 32-session reference behind $\lref$, most on the H200 and the native pool (Appendix~\ref{app:trapdata}); on vLLM and Llama
the completion excess is not explained. That speed-up does not make the sign: divided by each run's measured token-rate ratio
(0.95--1.12), completions still exceed $\lref$ by a factor of 1.08--1.81 in every one of the 57 runs
(\fpath{analysis/theory/icml/premium_over_tokrate.py}).
\item \emph{Backlog.} Over model~F's $\alpha$, the backlog slope has mean 0.99 and median 1.06 across the 57 runs (per run 0.00
to 1.62, varying mostly with the plans' window arrivals; $\alpha$ at the nominal rate), where the quasi-static equation, completions
at $\lmiss$, gives 1.36--1.59.
\end{itemize}
These are statements about the runs taken together; per run the gap and backlog statistics scatter widely, mostly with the plan's
arrivals over the window. The 13 from-empty runs stuck over $[1500,3000)$\,s ($\rho=1.4$--6.3) follow the same composition and gap
laws, but their backlog slopes exceed model~F's $\alpha$ by 6--35\% in 11 of the 13 and their completions by up to 15\%, which may
reflect window arrivals 3.8--13.1\% above nominal in these single seed-0 runs; self-similarity (Theorem~\ref{thm:trap}(b)) shows only
where samples are large and censoring small (Appendix~\ref{app:trapdata}).

\paragraph{The registered slope rule, explained post hoc.} Our registered rule calls a BURST run trapped only if its backlog slope
exceeds $0.5(\lambda-\lref)$. At 1.35~$\lref$ Theorem~\ref{thm:trap}(a) puts the true slope at about $0.63(\lambda-\lmiss)$, only
about 25\% above that line, and seed~1's plan delivers 5.8--8.9\% fewer arrivals than nominal over the window at 1.35~$\lref$ in
every cell; 8 of the 9 below-threshold runs (6 of the 7 registered ones) are seed~1. The plug-in $\lambda_{\mathrm{window}}(\text{plan})-D$, computable from the
plan before any run, falls on the measured side of the threshold for 27 of the 30 distinct plans of the registered protocols, against
23 for always predicting ``trapped'' (plans of different cells share seeds, so these are not independent trials); two of the three
exceptions lie within $10^{-4}$\,/s of the threshold, and the third is Llama at 1.35~$\lref$, seed~0, the registered failure. This
account is post hoc and changes no registered verdict.

\paragraph{Blind tests of the trap-state law.} The predictions of Theorem~\ref{thm:trap} for runs that did not yet exist were frozen
under tag \texttt{icml-trap-pred} (pushed at 00:24:47 CST on 2026-11-05), with scoring rules S1--S5 declared beforehand on
composition, gap slope, completions, backlog slope and, per seed, the slope plug-in (tolerances in Appendix~\ref{app:blind}); the
application to a cell and load is refuted if the pooled S1 or S2 fails in a group of at least two trapped runs
(\fpath{notes/PREDICTIONS_icml_trap.md}). The tolerances were set after seeing the committed runs; the runs they score did not exist
at the push, so these tests are blind in outcome only. S2 compares with $\gamma$, not with the completion-time slope $\gamma/z$ used
above; at 1.35~$\lref$ the two differ by 1.3\%, well inside the tolerance. Two pre-registered cells slow the tool calls with no burst; both were designed
after the 900\,s TOOLSLOW arm had trapped 1 of 3 and after reading simulator outputs (Appendix~\ref{app:blind}).
\begin{itemize}[leftmargin=*,nosep]
\item \emph{TOOLSLOW-W}: every tool call that starts in $[1200,3000)$\,s takes 30\,s longer, at 1.35~$\lref$, seeds 0--2, paired
with the committed NOBURST runs of 2026-10-24/25 (same plans, not run alongside); the cell TRAPS if at least 2 of 3 runs are stuck. Outcome: TRAPS. All three runs were stuck (hit
0.000 over $[5700,8700)$\,s) while their paired NOBURST runs stayed good; on the same plans the 900\,s slowdown had trapped 1 of 3. The pooled scores pass: composition $+1.9$\% (limit $\pm5$\%), gap
slope $1.075\gamma$ (1.089 against the unregistered $\gamma/z$), completions $+1.9$\% of $D$ with every run above $\lref$, and backlog slope $+29$\% of $\alpha$ (limit
$\pm35$\%). Per run, composition is within $\pm8$\% in all three, and S5's plug-in lies on the measured side of the threshold in 2 of
3 runs and within 0.01\,/s in 3 of 3. Every model we froze predicted this outcome, so the cell tests the trap-state law on a trap
entered with no burst, not the models against each other; it does not separate the longer gap from the capacity loss and the larger
population that come with the slowdown.
\item \emph{TOOLSLOW-D15}: a 15\,s slowdown over $[1200,2100)$\,s at 1.65~$\lref$, chosen so that our two frozen predictors of entry
disagree: the calibrated simulator, run on the exact client plans, called seed~0 stuck and seeds 1 and~2 good (NO-TRAP), while the
first-principles model's plan-averaged fluid verdict called every seed stuck (TRAPS), and its per-seed layer, with the simulator's
modal calls, gave NO-TRAP a probability of 0.41. Outcome: NO-TRAP, as the simulator said, seed by seed: seed~0 stuck, seeds 1 and~2
at their controls' hit rates. The stuck run passed S1--S4 (composition $+0.1$\%, gap slope $1.02\gamma$, completions $+10$\% of
$D$, backlog slope $-15$\% of $\alpha$); its S5 plug-in missed the 0.01\,/s tolerance by 0.0004\,/s, and with one eligible run the
refutation rule does not apply.
\end{itemize}
Across the four blind runs trapped by a tool slowdown alone, S1--S4 pass in both cells, and S5 is close in 3 of 4 and on the correct
side in 3 of 4. The frozen exploratory measures of both cells, hits and misses, are in Appendix~\ref{app:blind}.

\subsection{A predicted reversal of the youth premium (MIRROR)}\label{sec:mirror}

Theorem~\ref{thm:premium} factors the trapped completion rate as $D=\lmiss H(z)\Gamma(z)$: $\Gamma\ge1$ comes from the growth of
the context, and $H$ from the law of the turn count. $H$ is about 1 for our geometric turn count (0.989 at 1.65~$\lref$), consistent
with every trapped run of our workload at a load above the all-miss capacity completing sessions above it (CUT-B's runs, cut to below it,
completed 0.92--0.97~$\lref$, as model~F predicts). For agents that run to a fixed step budget, a nearly
deterministic $T$ can put $H$ far enough below 1 to outweigh $\Gamma$, and the premium then reverses. MIRROR tests this. It repeats the E2
BURST protocol on the 300k A100 cell, seeds 0--2, at 1.65~$\lref$ (decisive) and 1.35~$\lref$ (secondary), with NOBURST
controls at 1.65~$\lref$ and one change: $T=\min(\mathrm{Geometric}(1/1000),20)$, so about 98\% of sessions run exactly 20 turns. For this
workload model~F's all-miss capacity is 1.046~$\lref$ (1.004 for the standard one), and at 1.65~$\lref$ its asymptotic $D/\lmiss$
is 0.771 ($H=0.699$, $\Gamma=1.103$), against 1.223 for the standard workload. Driven by each seed's exact plan, model~F predicts
three things at 1.65~$\lref$ (Fig.~\ref{fig:mirror} in Appendix~\ref{app:blind}): completions nearly stop for about 40\,min after the burst (0.05--0.09~$\lref$
over $[2400,3600)$\,s), because the burst cohort must pass through 20 FIFO cycles, each longer than the last, before any of it can
finish; the cohort then completes in a wave, its median completion at 5135--5195\,s; and completions then stay near half the
all-miss capacity (0.49--0.60~$\lref$ over $[6600,8700)$\,s, a transient lull; 0.81~$\lref$ asymptotically), while the backlog grows 1.65--1.95 times as fast as the plan's excess load $\lambda_{\mathrm{plan}}-\lmiss$ over $[6000,8700)$\,s (1.4 times asymptotically). Every
throughput-curve reading (the quasi-static model, Lu et al.'s $\dot N=\lambda-X(N)/K$~\citep{lu2026cachetrap}, $K$ their requests per session, read on this workload,
our Eq.~\eqref{eq:fluid} of Appendix~\ref{app:qs}) predicts steady completions near 1.05~$\lref$ and no wave; the standard workload's traps completed
1.05--1.48~$\lref$ over the first window and 1.01--1.35~$\lref$ over the second. The registered rule (tag \texttt{r16-mirror-reg},
pushed at 14:07:35 CST on 2026-11-05, before any job) calls a stuck BURST run RENEWAL if both completion windows, the wave and the
backlog slope fall on model~F's side of frozen thresholds, and CURVE if both windows are at least 0.85~$\lref$; the cell passes with
2 of 3 RENEWAL and fails with 2 of 3 CURVE (Appendix~\ref{app:blind}). The predictions are blind in outcome only:
the workload was chosen after model~F's and our fluid model's predictions for it and for a few alternatives had been computed, the timing tolerance comes
from the standard workload's committed runs, no simulator prediction was made, and the test came after Lu et al.'s paper. Outcome:
PASS at both rates, RENEWAL in 3 of 3 seeds each: at 1.65~$\lref$, completions of 0.06--0.09~$\lref$, waves at 4945--5089\,s,
then 0.50--0.61~$\lref$, below the measured capacity of 1.06~$\lref$. MIRROR measures that lull, not the asymptotic $D$, and rejects only the throughput-curve readings; a
secondary check, the calibrated wave time, missed in 1 of 6 runs.

\section{What trapped throughput does not tell}\label{sec:control}

\begin{figure*}[t]
\centering
\includegraphics[width=\textwidth]{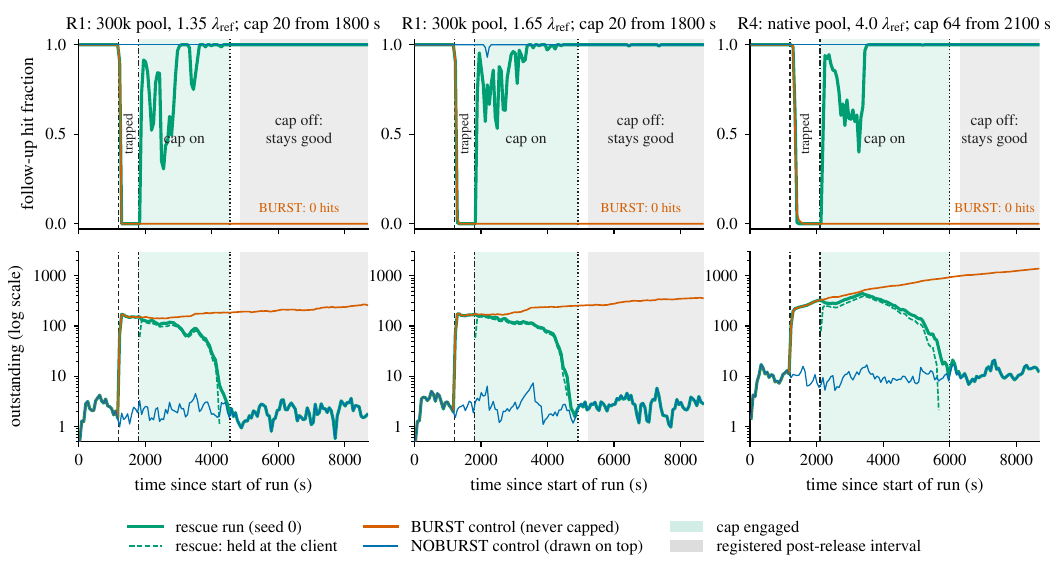}
\caption{Trap, rescue, release and persistence: the registered rescue tests R1 and R4, seed 0 of each cell. Each rescue run
(green) replays the arrival plan of a paired test's BURST run until a client-side session cap is engaged at $t_{\mathrm{on}}$
(dash-dotted); the cap is removed at
$t_{\mathrm{rel}}$ (dotted) once both client queues have been empty for 300\,s. The same-seed BURST (vermillion) and NOBURST (blue)
runs share its arrivals. Dashed: the burst of 200 sessions at 1200\,s. Top: follow-up hit fraction in 60\,s bins. Bottom:
outstanding work, requests in the engine plus sessions held at the client (log scale); dashed green: sessions held at the
client. Seeds 1 and 2 behave alike (Table~\ref{tab:rescue}).}\label{fig:rescue}
\end{figure*}

An operator who sees a trapped engine complete sessions at some rate may expect a cut to just below that rate to clear it.
Corollary~\ref{cor:trapthr} says that in model~F, for our workload, it does not: the trap clears only below $\lmiss$, after a busy period that diverges
as the cut approaches $\lmiss$. In the all-miss fluid at 1.65~$\lref$ (300k), a cut to $1.18\lmiss$, below $D=1.22\lmiss$, makes the
backlog dip by 0.3\% and grow again, and a cut to $0.9\lmiss$ drains it only after about 15 hours (Appendix~\ref{app:trap}). That
leaves two exits: starve the trap, which is slow and, done by refusing arrivals, sheds sessions; or leave model~F, by making turns hit
again, which needs the sessions that compete for the pool to fall below the all-miss onset $n_b$ (Appendix~\ref{app:entry}) without
lowering the offered load. Lu et al.\ reverse the outcome in simulation by lowering post-surge arrivals~\citep{lu2026cachetrap}; their
systems keep some reuse in the trap, so Corollary~\ref{cor:trapthr} does not describe them (Appendix~\ref{app:lu}).

\paragraph{Cutting arrivals: CUT-A and CUT-B (blind in outcome).} CUT tests the corollary's setting on the engine: the 1.65~$\lref$ BURST
protocol, then from 2400\,s arrivals cut to 1.15~$\lref$ (CUT-A, $\rho'=1.145$, inside $(\lmiss,D(\lambda))$) or to 0.85~$\lref$
(CUT-B, $\rho'=0.846$), seeds 0--2. Model~F's predictions, driven by each seed's exact client plan, were frozen under tag
\texttt{icml-cut-pred}: all six runs stay stuck over $[5700,8700)$\,s; CUT-A, although its rate is below the trapped throughput of
the same seed's uncut run, does not clear, its backlog dipping by less than 15\% on 300\,s bins and growing again in seeds 0 and 2 (seed~1's plan
has few post-cut arrivals); CUT-B's backlog shrinks slowly and would empty only at about 27{,}100\,s, the busy period of
Proposition~\ref{prop:classic}(a). The decision rule, registered after the predictions were pushed and before any CUT job (tag
\texttt{r15-cut-reg}), compares each seed's backlog slope $S$ over $[3000,8700)$\,s with model~F's frozen slope and with two
constant-throughput comparators: a naive one, in which the trap drains at the trapped throughput of the same seed's uncut control
(1.32--1.41~$\lref$), and, for CUT-A, a quasi-static one, in which it completes sessions at $\lref$. A seed is theory-favoured if
model~F's slope is strictly the closest; an arm passes if at least 2 of 3 seeds are theory-favoured and the sign clause of Appendix~\ref{app:blind} holds, and fails if a
comparator is favoured, or the run leaves the trap, in 2 of 3. The rule was fixed after the per-seed predictions had been read, and it
differs from the draft rule simulated in the prediction note in five ways: three stricter, one stricter for two seeds and looser for
the third, and one looser, which in the note's simulation raised CUT-A's probability of passing from 0.82 to 0.99
(Appendix~\ref{app:blind}).

Both arms passed. With arrivals cut to a realized 1.12--1.20~$\lref$, all three CUT-A runs stayed stuck to the end at zero hits, and
all three seeds were theory-favoured: slopes of $+0.0040$, $-0.0022$ and $+0.0029$\,/s against model~F's $+0.0039$, $-0.0026$ and
$+0.0028$, the naive comparator's $-0.0094$ to $-0.0204$ and the quasi-static one's $+0.0086$ to $+0.0146$. After the cut to a realized
0.81--0.88~$\lref$, the CUT-B backlog fell, slowly, at $-0.0085$ to $-0.0126$\,/s against model~F's $-0.0095$ to $-0.0131$ and the
naive $-0.033$ to $-0.043$; all three seeds were theory-favoured, and all three runs were still stuck at 8700\,s, with 140, 77 and 89
sessions present against model~F's 139, 79 and 94.

The passes are narrower than the corollary. Both comparators assume a constant throughput, so each rule rejects only the
constant-throughput readings it was built against: the naive one in both arms and, in CUT-A, the quasi-static one. Intermediate
readings fit as well: over $[3000,8700)$\,s constant completions at the nominal cut rate fit CUT-A as well as model~F, and the
quasi-static reading, not a registered comparator for CUT-B, fits its slopes about as well (had it been one, CUT-B would still have
passed, 2 of 3). What singles out model~F against them is post hoc: CUT-A's late-window slopes ($+0.0071$, $-0.0027$ and $+0.0131$\,/s
against model~F's frozen $+0.0069$, $-0.0036$ and $+0.0129$, while constant completions at the trapped throughput, at $\lref$, at their
midpoint or at the nominal cut rate each missed by about 0.004\,/s or more) and CUT-B's completions below the all-miss capacity as the trapped mix aged (0.918--0.974~$\lref$ against model~F's
0.918--0.979, although at the frozen tolerance the all-miss capacity also lies just inside). Persistence discriminates only in part:
every constant reading also leaves CUT-A stuck to 8700\,s, so there it rules out only a drain at the completion rate seen just before
the cut (1.21--1.70~$\lref$ over $[1800,2400)$\,s), which would have reached the onset near 6300\,s in seeds 0 and~1, the operator-facing
content of Corollary~\ref{cor:trapthr}; in CUT-B the naive reading, or a drain at the pre-cut completion rate, would have reached the
onset by 4900--7600\,s. The test has limits: the cut comes 1200\,s after the burst, so it probes the young trap's transient and the start
of its regrowth, not the asymptotic rate; CUT-B cannot show recovery within the run; CUT-A's runs shared one node and CUT-B's ran on
two; and ``did not clear'' means within 8700\,s.

\paragraph{Refusing arrivals sheds sessions (SHED).} The registered SHED test at 1.35~$\lref$ on the 300k pool stopped arrivals
entirely from 1800\,s, 600\,s after a burst. A 600\,s stop refused 56--59 sessions and left both runs stuck; a 3600\,s stop refused
355--378 and left both good, hits returning 2300--2600\,s into the stop (post hoc). In model~F the fluid backlog falls to the onset $n_b=32.7$
after 2{,}020\,s of the stop and the queue empties after 3{,}180\,s, the remaining work $V_{\mathrm{tot}}(1800)$
(Proposition~\ref{prop:classic}(a)); the quasi-static model's frozen minimum stop was 1{,}795\,s, early by about a quarter, and
model~F's time to the onset is early too, by about 15\%. Model~F's numbers are a retrodiction, computed after the SHED runs had ended,
and the test does not discriminate: any drain time between 600 and 3600\,s fits its two stop lengths.

\paragraph{Restoring reuse with a session cap.} A FIFO gate that admits at most $\ncap$ sessions, each holding its slot until its
last turn ends, makes the admitted sessions a closed system of $\min(n,\ncap)$ sessions (Proposition~\ref{prop:cap}, Appendix~\ref{app:qs}). In the
quasi-static model the gate then empties if and only if $n_1<\ncap<n_2$, the stable and unstable fixed points of
$\dot n=\lambda-\mu(n)$, and once it is empty and removed the population stays in the good state's basin
(Corollary~\ref{cor:rescue}, Appendix~\ref{app:qs}). In model-F terms the cap keeps the admitted sessions below the onset, so their turns hit again, and the
excess waits at the client instead of in the engine's queue, where every turn would miss. The registered rescue tests R1 (300k pool,
cap 20, engaged at 1800\,s, 600\,s after the burst, at 1.35 and 1.65~$\lref$) and R4 (native pool, cap 64, at 2100\,s, at
4.0~$\lref$) engaged the registered CAPPED caps inside established traps: follow-up hit fraction 0.000--0.002 over the preceding
300\,s, with 139--324 requests or sessions outstanding. All 9 runs were rescued without shedding a session (Fig.~\ref{fig:rescue};
Table~\ref{tab:rescue}). Hits returned within three minutes, the backlog drained, and the cap was released at 3980--6140\,s, once the
client queues had been empty for 300\,s. With the cap removed, every run stayed good at the same arrival process for the remaining
2560--4720\,s, matching its NOBURST control in follow-up hit fraction, outstanding work and turn latency, while the same-seed uncapped
BURST run stayed at zero hits. The price is waiting at the client: 386--569\,s per paused session on average, 38--78 (300k) and 222--230
(native) session-hours per run. In a slowdown-only form of Lu et al.'s SLO goodput (their
inter-token-latency condition omitted, our own unloaded reference; post hoc, Appendix~\ref{app:dataonly}), the burst cost all 30
registered BURST runs 99.2--100\% of the next hour's goodput against the same-seed NOBURST run, and the rescue 45--97\% (depending
on the reference), but the rescue restored the evaluation window's goodput to within 1\% of NOBURST.

\paragraph{What the rescue does and does not show.} Persistence here means that the drained state sustains itself once the cap no
longer binds: the release rule waits until the gate is empty, so the tests do not cover a release while the cap still binds, and
the cap is engaged at a fixed time, not on an online signal. The frozen first-principles model predicted the rescue in every cell,
but 4 of its 9 release times fell outside its $\pm15$\% tolerance (R4 released 19--22\% late; Appendix~\ref{app:rescuetests} and Table~\ref{tab:frozensum}). The tests show a registered, replicated rescue of an \emph{established} trap on
the engine at the same offered load, without shedding, with the control removed and the good state persisting after it.

\section{Related work}\label{sec:related}

\paragraph{Concurrent work.} Concurrently with and independently of this work, Lu et al.~\citep{lu2026cachetrap} studied the same
collapse on prefill--decode disaggregated SGLang. Our stampede and recovery experiments (\S\ref{sec:stampede}) were run before their
paper appeared, and the theory and blind tests after it (Appendix~\ref{app:reuse}); Appendix~\ref{app:lu} relates the two.

\paragraph{Thrashing, metastability and overloaded queues.} The thrashing collapse and the working-set remedy are
Denning's~\citep{denning1968thrashing,denning1968working}; metastability from a unimodal throughput curve and bistability under
finite sources were analysed for random-access channels and multiprogrammed
systems~\citep{kleinrock1975aloha,carleial1975bistable,courtois1975decomposability,nelson1987stochastic}. Metastable failures were
named by Bronson et al.~\citep{bronson2021metastable}, studied in production incidents and modelled by Huang et
al.~\citep{huang2022metastable}, whose trigger-and-control design our paired test follows, and analysed with queueing models and
Markov chains~\citep{isaacs2025analyzing,habibi2024msf,alvaro2025formal}; Dong poses metastability in queues as an open
problem~\citep{dong2022metastability}. As far as we found, none of these gives the bad regime's growth law or the composition of
its population; service systems whose service slows in congestion show related bistability~\citep{chan2014speedup,dong2015slowdowns}.
Overloaded processor sharing and the steady solutions of fluid FIFO re-entrant lines~\citep{khanin2001steady}, the work closest to
model~F, are discussed in \S\ref{sec:trap}; overloaded multiclass FIFO queues with abandonment and no re-entry have fluid
limits~\citep{jennings2013fluid}, and overloaded networks have bottleneck fluid approximations~\citep{chen1991discrete}. Re-entrant
lines~\citep{kumar1993reentrant} and their fluid models~\citep{dai1996stability,bramson1996convergence} are studied mostly for
stability, and for convergence to equilibria at load at most one. FIFO queues with feedback are classical in the stable regime~\citep{takacs1963feedback,vandenberg1989deterministic};
round robin with feedback to the tail of the queue, which model~F is at the level of sessions, was analysed in equilibrium for
time-shared systems~\citep{kleinrock1967timeshared,coffman1968feedback}; overloaded polling systems have a random fluid limit with a
branching structure~\citep{remerova2014random}; and in heavy traffic multiclass FIFO feedback queues show state-space
collapse~\citep{reiman1988multiclass,dai1995multiclass}.

\paragraph{LLM serving and KV management for agents.} Ao et al.\ give fluid-guided scheduling with memory
constraints~\citep{ao2025wait} and show that, in a decode-only model without prefix reuse, the eviction-free fixed point can be
unstable~\citep{ao2026service}; Dai et al.\ prove throughput optimality of work-conserving policies without cache
state~\citep{dai2025throughput}; Nie et al.\ give stability conditions under KV constraints for one-shot
requests~\citep{nie2026queueing}; Dexter et al.\ analyse FCFS and longest-prefix-match scheduling under prefix
reuse~\citep{dexter2025prefix}; and Yao et al.\ derive a mean-field LRU hit ratio for multi-turn sessions with exogenous gaps between
turns~\citep{yao2026multiturn}, which our entry model extends with hit-dependent running reservations (Appendix~\ref{app:entry}).
Agent-serving systems keep, offload or evict the KV that sessions leave idle in tool calls, order requests to preserve reuse, and bound
the number of active agents: PEEK~\citep{xie2026peek}, Continuum~\citep{li2025continuum}, Sutradhara~\citep{biswas2026sutradhara},
CONCUR~\citep{chen2026concur}, ThunderAgent~\citep{kang2026thunderagent} and its Dynamo port~\citep{dhanani2026dynamo},
EfficientAgent~\citep{shao2026efficientagent}, AgentKV~\citep{anon2026agentkv}, MISA-T~\citep{hong2026misat},
MARS~\citep{wang2026mars}, VAMP~\citep{kim2026vamp} and KAIROS~\citep{yuan2026kairos}, among others (Appendix~\ref{app:lu}). Several
report hit-rate collapse as concurrency grows~\citep{chen2026concur,yuan2026agentic,hong2026misat,kim2026vamp}, describe the loop
between eviction, cold prefill and queueing for closed or bounded populations, or assert that a short overload can leave an instance
persistently inefficient~\citep{yuan2026kairos}. We claim none of these mechanisms; a session cap is classical load
control~\citep{denning1976optimal,cherkasova2002session}.

\section{Limitations}\label{sec:limits}

\paragraph{The theory.} Model~F is an idealization: it assumes that every turn misses, that requests complete in FIFO order and
that a finished turn rejoins at once. The engine runs about 28 requests concurrently at 300k (14--90 across our pools), vLLM preempts, and hits return as the backlog
falls toward the onset, so model~F describes the trap, not the transitions into and out of it. We do not prove that the engine
converges to model~F, and the theorem's limits are reached only geometrically: in the registered window, about 33--37 generations
after the burst, a transient of order 10\% remains for our workload (MIRROR's late window, model~F's post-wave lull, lies 20--40\% below its asymptote), so we compare with the exact fluid as well as with the limits. The premium is
proved positive for our workload at $\rho\ge1.04$; below that it rests on a floating-point bracket and a first-order expansion.

\paragraph{The validation.} The check on the 57 stuck runs is retrospective: the theory was derived after they were read, and its
scoring tolerances were set on them. Only the runs submitted after tag \texttt{icml-trap-pred} test it blind, and only in outcome:
the four runs trapped by a tool slowdown (three TOOLSLOW-W, one TOOLSLOW-D15), the six CUT runs and MIRROR's runs, all on the A100
with the 300k pool. In the CUT runs each arm's registered rule rejected the naive reading, and CUT-A's also the quasi-static one;
what singles out model~F against intermediate readings is post hoc; MIRROR's rule rejects only the throughput-curve readings.
The from-empty runs are single runs per rate, and at 300k the
seed-0 bracket did not replicate on fresh seeds (\S\ref{sec:stampede}). In an exploratory replication on two plans chosen because
their outcomes were known, three runs each reproduced both originals, one collapse and one stable run, which suggests that the spread
across seeds at the edge comes from the plans rather than from service noise (six runs, on two GPUs of one node). The trapped windows
are at most 7500\,s long, and ``does not escape'' means within that horizon.

\paragraph{The experiments.} Sessions are synthetic (random tokens, exponential 5\,s tool times, geometric turn counts, fixed tool
outputs and output lengths, no shared prefixes), clients never retry or abandon, and real agent traces have tool times far shorter
and more dispersed~\citep{zhu2026tracelab}. Engines are colocated, with one GPU each, dense 7--8B models and pools of 150k to 956k tokens; disaggregated
serving, offloading, larger models and parallelism are untested. Seeds are shared across tests. The rescue engages a fixed cap at a
fixed time, the arrival stop ran at one rate, and the tool slowdowns at two. Appendix~\ref{app:threats} gives the full list.

\section{Conclusion}\label{sec:conclusion}

A burst can tip a prefix-cached agent server into a trap in which its cache stops paying; Lu et al., who posted first, and we
observed this independently (in their disaggregated system the hit rate falls to about 0.3, in ours every turn misses). Idealized as
model~F, the trap has an exact law: a linear renewal recursion along FIFO orbits carries every trapped state to the known
self-similar regime~\citep{khanin2001steady}, fixed by a Perron root, in which the gap between turns grows linearly, the backlog grows
at a load-dependent multiple of the excess load and the turns served are young and short; for our workload the multiple is 0.6 to 0.8 at our loads
and sessions complete faster than the all-miss capacity; with a 20-step budget they complete slower, and a blind test on the engine found completions below that capacity in the transient after a burst. Checked
retrospectively on 57 stuck runs on two engines, two GPUs and two models, with no parameter fitted to any trap measurement, these laws
hold for the runs taken together, and, with predictions frozen before the runs, on four runs in two cells trapped by a tool slowdown
alone, in four of the five pre-set scores (a per-run slope check failed in two). The practical reading is
Corollary~\ref{cor:trapthr}: in model~F, for our workload, a cut to just below the throughput measured in an all-miss trap does not clear it, and a cut
below the all-miss capacity drains the trap only slowly and by turning sessions away. On the engine, a cut between the two left all three CUT-A traps in place
to the end of the run, and a cut below the all-miss capacity shrank the three CUT-B backlogs only slowly, all still trapped at 8700\,s; each arm's registered rule rejected the naive reading, and what singles out
model~F against intermediate readings is post hoc. The other exit, a session cap that restores reuse, rescued all 9 established traps
of our registered tests at the same load without shedding, and the engine stayed good after the cap was removed.

\section*{Code and data availability}
{\raggedright The code (simulator, load generator, analysis, the first-principles model and model~F, the Slurm job scripts), the
experiment configurations, every registration file, the frozen predictions with their tags, all real-engine data with the job logs,
the research log, the scripts behind every figure and derived table, and the source of every number in the paper
(\fpath{paper/NUMBERS.md}) are in our repository, \repourl{}. Its history is a redacted copy of the authors' development repository,
whose commit hashes this paper cites; \fpath{provenance/} maps them and holds the timing evidence behind our registration claims
(Appendix~\ref{app:prov}).\par}

\section*{Use of AI tools}
Claude and OpenAI Codex assisted with experiments, code, analysis, theoretical development, and writing. The authors directed the work and take full responsibility for its content.

\section*{Acknowledgments}
The experiments ran on the Nova cluster at Iowa State University. The research reported in this paper is partially supported by the HPC@ISU equipment at Iowa State University, some of which has been purchased through funding provided by NSF under MRI grants number 1726447 and MRI2018594.

\bibliographystyle{plainnat}
\bibliography{refs}

\clearpage
\appendix
\section{Model F: proofs and remarks}\label{app:trap}

\paragraph{Model F.} Fix $K\in\mathbb N$, costs $c_1,\dots,c_K>0$ (all-miss GPU seconds of turn $k$), and
$1=G_1\ge G_2\ge\dots\ge G_K>0$ with $G_{K+1}=0$ ($G_k=P(T\ge k)$). Write
\begin{gather*}
p_k=G_k-G_{k+1},\qquad r_k=G_{k+1}/G_k,\\
a_k=\lambda c_kG_k,\qquad \rho=\textstyle\sum_ka_k=\lambda/\lmiss,
\end{gather*}
with $\lmiss=1/\sum_kc_kG_k$. A session whose next turn is $k$ has remaining work $R_k=\sum_{j\ge k}c_jG_j/G_k$, so that
$R_1=1/\lmiss$ and $R_k=c_k+r_kR_{k+1}$. The model assumes:
\begin{enumerate}[(F1),leftmargin=*,nosep]
\item every turn misses, so turn $k$ costs $c_k$;
\item the server does one GPU second of work per second while work is pending;
\item requests complete in the order in which they joined;
\item a completed turn $k<T$ rejoins at once as turn $k+1$, so a fraction $r_k$ of the class-$k$ mass completing at any time
rejoins as class $k+1$;
\item new sessions join as class-1 mass at rate $\lambda$;
\item at time 0 a pending mass $m^0_k\ge0$ of each class is served first, with $V_0=\sum_kc_km^0_k>0$, through absolutely
continuous non-decreasing class-completion functions $C^0_k$ on $[0,V_0]$ with $C^0_k(0)=0$, $\sum_kc_kC^0_k(t)=t$ and
$C^0_k(V_0)=m^0_k$.
\end{enumerate}
A \emph{solution} consists of non-decreasing absolutely continuous $J_k$, $C_k$ (joins and completions after time 0) on
$[0,\tau)$ such that
\begin{itemize}[leftmargin=*,nosep]
\item $J_1(t)=\lambda t$ and $J_{k+1}=r_kC_k$;
\item $C_k=C^0_k$ on $[0,V_0]$, and $C_k(f(u))=m^0_k+J_k(u)$ for $u\ge0$ (FIFO);
\item $\sum_kc_kC_k(t)=t$ for $t<\tau$ (F2);
\item the pending work $V(u)=\sum_kc_k\pi_k(u)$ is positive on $[0,\tau)$;
\item either $\tau=\infty$ or $V(t)\to0$ as $t\uparrow\tau$.
\end{itemize}
Here $f(u)=u+V(u)$ and $\pi_k=m^0_k+J_k-C_k$. The last condition makes $\tau\le\infty$ the time at which the queue first
empties, rather than any earlier cut-off. With zero think time every
session present has one pending request, so the number in the system is $N=\sum_k\pi_k$. The remaining work is
$V_{\mathrm{tot}}=\sum_k\pi_kR_k$, and the departures are $\mathrm{Dep}(t)=\sum_km^0_k+\lambda t-N(t)$.

On the engine, (F1) is the measured zero hit fraction of the trap and (F2) its saturated GPU. (F3) is request-level FCFS
admission (Proposition~\ref{prop:A}), up to the $M/E_S\approx28$ requests that run concurrently. For (F4), the 5\,s tool
time is short against cycles of 166--516\,s, and including it changes the window statistics of the fluid by less than
$10^{-3}$ (\fpath{analysis/theory/icml/support.py}). The costs are $c_k=t^m_k+\delta$: the calibrated miss prefill of
Appendix~\ref{app:entry} plus the all-miss per-turn decode share $\delta$, chosen so that $\sum_k(G_k/E[T])c_k=1/\nu_b$. Then
$\lmiss$ above equals the model's $\mu_b$ (Proposition~\ref{prop:nb}(a)).

\begin{lemma}[Perron root]\label{lem:perron}
$\varphi(x)=\sum_ka_kx^{-k}$ decreases strictly from $\infty$ to $0$ on $(0,\infty)$, and $\varphi(1)=\rho$. Hence
$\varphi(z)=1$ has a unique root $z>0$, with $z>1$ if and only if $\rho>1$.
\end{lemma}

\begin{lemma}[well-posedness, work, orbit recursion]\label{lem:orbit}
Under (F1)--(F6):
\begin{enumerate}[(i),leftmargin=*,nosep]
\item model F has exactly one solution;
\item while the server is busy, $f(u)=V_0+W(u)$ with $W=\sum_kc_kJ_k$, and $f$ is continuous and strictly increasing;
\item while busy, $V_{\mathrm{tot}}(t)=V_{\mathrm{tot}}(0)+(\rho-1)t$.
\end{enumerate}
\begin{enumerate}[(i),leftmargin=*,nosep,resume]
\item For $u_0\in[0,V_0)$, let $u_{n+1}=f(u_n)$ and $\Delta_n=u_{n+1}-u_n$. Then
\begin{gather*}
\Delta_n=\sum_{k=1}^{\min(n,K)}a_k\Delta_{n-k}+h_n(u_0),\\
h_n(u_0)=\sum_{k=n+1}^{K}c_k\frac{G_k}{G_{k-n}}\pi_{k-n}(u_0),
\end{gather*}
so that $h_0=V(u_0)$ and $h_n=0$ for $n\ge K$. Each $h_n$ is continuous in $u_0$, and $\sum_nh_n(u_0)=V_{\mathrm{tot}}(u_0)$.
\item For $n\ge1$, $\pi_k(u_n)=\lambda G_k\Delta_{n-k}$ if $k\le n$, and $\pi_k(u_n)=(G_k/G_{k-n})\pi_{k-n}(u_0)$ otherwise.
The requests completed in $(u_n,u_{n+1}]$ are those pending at $u_n$.
\end{enumerate}
\end{lemma}

\begin{proof}
\emph{(ii).} If the server is busy on $[0,f(u)]$, then by (F2) it has completed $s$ units of work by each time $s$ in that
interval. By (F3), the request joining at $u$ completes once the work joined before it, $V_0+W(u)$, is done. So
$f(u)=V_0+W(u)$, which is continuous. Since $W'\ge\lambda c_1>0$ almost everywhere, $f$ is strictly increasing, and
$f(0)=V_0$.

\emph{(i).} Put $b_0=0$, $b_1=V_0$ and $b_{n+1}=f(b_n)$. On $[0,b_1]$ the completions are $C^0$, which give $J$ on
$[0,b_1]$. Suppose $J$ is known on $[0,b_n]$. Then:
\begin{itemize}[leftmargin=*,nosep]
\item $f=V_0+W$ is known on $[b_{n-1},b_n]$ and maps it onto $[b_n,b_{n+1}]$;
\item FIFO gives $C_k(f(u))=m^0_k+J_k(u)$ on $[b_n,b_{n+1}]$;
\item re-entry extends $J_{k+1}=r_kC_k$ to $[b_n,b_{n+1}]$.
\end{itemize}
Each step is forced by FIFO, re-entry, the arrivals and (F2); this proves uniqueness. For existence, two facts show that the
construction is a solution. First, $\sum_kc_kC_k(f(u))=V_0+W(u)=f(u)$, so the server works at unit rate and the constructed
$f$ equals $u+V(u)$. Second, $V(f(v))=W(f(v))-W(v)\ge\lambda c_1V(v)>0$, so $b_{n+1}>b_n$. Absolute continuity follows from
$C_k'=J_k'(u)/f'(u)\le J_k'(u)/(\lambda c_1)$. Moreover $V\ge\eta V_{\mathrm{tot}}$ with $\eta=\min_kc_k/R_k>0$, because both
are linear in $\pi\ge0$. By (iii), which needs only the unit work rate:
\begin{itemize}[leftmargin=*,nosep]
\item if $\rho\ge1$, then $V\ge\eta V_{\mathrm{tot}}(0)>0$ throughout, $b_{n+1}-b_n\ge\eta V_{\mathrm{tot}}(0)$, and
$b_n\to\infty$;
\item if $\rho<1$, then $b_{n+1}-b_n=V(b_n)$ lies between $\eta V_{\mathrm{tot}}(b_n)$ and $V_{\mathrm{tot}}(b_n)$, where
$V_{\mathrm{tot}}(b_n)=V_{\mathrm{tot}}(0)-(1-\rho)b_n$; so $b_n$ increases to $V_{\mathrm{tot}}(0)/(1-\rho)$, and
$V(t)\le V_{\mathrm{tot}}(t)\to0$ there.
\end{itemize}

\emph{(iii).} With $j_k=J_k'$ and $\dot C_k=C_k'$, we have $\sum_kc_k\dot C_k=1$, $j_1=\lambda$ and $j_{k+1}=r_k\dot C_k$. Hence
\begin{align*}
V_{\mathrm{tot}}'&=\textstyle\sum_kR_k(j_k-\dot C_k)=\lambda R_1+\sum_k\dot C_k(r_kR_{k+1}-R_k)\\
&=\textstyle\lambda R_1-\sum_kc_k\dot C_k=\rho-1.
\end{align*}

\emph{(iv), (v).} Let $\Delta J_k(n)=J_k(u_n)-J_k(u_{n-1})$. By (ii), $\Delta_n=W(u_n)-W(u_{n-1})=\sum_kc_k\Delta J_k(n)$ for
$n\ge1$, and $\Delta_0=V(u_0)=\sum_kc_k\pi_k(u_0)$. For class 1, $\Delta J_1(n)=\lambda\Delta_{n-1}$. For $k\ge2$ and
$n\ge2$, FIFO gives $C_{k-1}(u_n)=m^0_{k-1}+J_{k-1}(u_{n-1})$ and the same at $u_{n-1}$, so
$\Delta J_k(n)=r_{k-1}\Delta J_{k-1}(n-1)$. For $n=1$, $C_{k-1}(u_1)-C_{k-1}(u_0)=\pi_{k-1}(u_0)$. Unrolling with
$\prod_{j=k-n}^{k-1}r_j=G_k/G_{k-n}$ gives (v) and then (iv).

The pending mass at $u_n$ is the joins in $(u_{n-1},u_n]$, because the completions by $u_n=f(u_{n-1})$ are the initial mass
plus the joins by $u_{n-1}$. For the sum, substitute $j=k-n$:
\[
\sum_nh_n=\sum_j\pi_j\sum_{n\ge0}c_{j+n}G_{j+n}/G_j=\sum_j\pi_jR_j=V_{\mathrm{tot}}.
\]
\end{proof}

\begin{lemma}[finite-memory renewal]\label{lem:renewal}
Let $A(s)=\sum_{k=1}^Kb_ks^k$ with $b_k\ge0$, $b_1>0$ and $A(1)=1$. Let
$y_n=\sum_{k=1}^{\min(n,K)}b_ky_{n-k}+g_n$, with $g_n\ge0$, $g_0>0$ and $g_n=0$ for $n\ge K$. Then $y_n=L+\epsilon_n$, where
$L=\sum_ng_n/\sum_kkb_k>0$ and $|\epsilon_n|\le C_r\vartheta^n$ for some $\vartheta<1$ that depends only on $(b_k)$, with
$C_r$ linear in $(g_n)$.
\end{lemma}

\begin{proof}
As formal power series, $\sum_ny_ns^n=g(s)/(1-A(s))$, where $g(s)=\sum_{n<K}g_ns^n$. The root $s=1$ of $1-A$ is simple, since
$A'(1)>0$. If $|s|<1$, then $|A(s)|\le A(|s|)<1$. If $|s|=1$ and $A(s)=1=\sum_kb_k$, then $s^k=1$ whenever $b_k>0$, in
particular $s=1$. So every other root lies outside the closed unit disk, at modulus at least $\vartheta_0^{-1}>1$. Fix
$\vartheta\in(\vartheta_0,1)$. Partial fractions give $g(s)/(1-A(s))=L/(1-s)+Q(s)$. Here $L=g(1)/A'(1)$ is the coefficient
of $1/(1-s)$, that is, minus the residue at $s=1$, and $Q$, a rational function plus a polynomial, is analytic for
$|s|<\vartheta_0^{-1}$. Cauchy's estimates on $|s|=\vartheta^{-1}$ finish the proof. This is the finite-memory case of the
discrete renewal theorem~\citep[Ch.~XIII]{feller1968introduction}.
\end{proof}

\begin{proof}[Proof of Lemma~\ref{lem:recursion}]
The rate of the remaining work is Lemma~\ref{lem:orbit}(iii), and the recursion with its properties is Lemma~\ref{lem:orbit}(iv):
by (F6) the state at any time $u_0$ is an admissible initial state, so it suffices to take $u_0\in[0,V_0)$.
For the exact identity, which an independent referee pointed out, take $u\ge b_{K-1}$. All class-$k$ descendants of the initial
requests have then joined, and the arrivals' class-$k$ requests joined by $u$ are those whose first request joined by
$f^{-(k-1)}(u)$. Hence
\[
f(u)=V_{\mathrm{tot}}(0)+\sum_ka_kf^{-(k-1)}(u).
\]
An orbit with $u_0\in[0,V_0)$ has $u_n\in[b_n,b_{n+1})$, so $u_n\ge b_{K-1}$ for $n\ge K-1$, and then
$u_{n+1}=V_{\mathrm{tot}}(0)+\sum_ka_ku_{n+1-k}$.
\end{proof}

\begin{proof}[Proof of Theorem~\ref{thm:trap}]
\emph{(a).} Apply Lemma~\ref{lem:renewal} to $y_n=\Delta_nz^{-n}$, $b_k=a_kz^{-k}$ and $g_n=h_nz^{-n}$. It gives
$\Delta_n=z^n(L(u_0)+\epsilon_n)$ with $L$ continuous and positive on $[0,V_0]$ (since $h_0=V(u_0)>0$) and
$|\epsilon_n|\le C\vartheta^n$ uniformly. At $t=u_n(u_0)$,
\begin{gather*}
t=u_0+\textstyle\sum_{j<n}\Delta_j=\dfrac{Lz^n}{z-1}\bigl(1+o(1)\bigr),\\
V(t)=\Delta_n=Lz^n\bigl(1+o(1)\bigr),
\end{gather*}
so $V(t)/t\to z-1$ uniformly. The relative error is $O(n\max(\vartheta,1/z)^n)$. As $t\to\infty$, the generation index $n$
of $t$ tends to infinity because every $b_n$ is finite. By Lemma~\ref{lem:orbit}(v), $N(t)=\lambda\sum_kG_k\Delta_{n-k}$, and the
identity $(z-1)\sum_kG_kz^{-k}=1-E[z^{-T}]$ gives $N(t)/t\to\lambda(1-E[z^{-T}])$. The departure limit follows from
$\mathrm{Dep}=\sum_km_k^0+\lambda t-N$. By Lemma~\ref{lem:orbit}(v), the composition limit holds for the pending requests, and for
the requests completed over the next cycle, which are the same. Over a cycle $(t,f(t)]$, any per-turn weight $P_k$ is processed at
the average rate $\sum_kP_k\pi_k/\sum_kc_k\pi_k$.

\emph{(b).} $f(u)/u=1+V(u)/u\to z$, and $f^m(u)/u\to z^m$ by induction on $m$. A session that arrives at $u$ joins as turn~1 at
$u$ and, with zero think time and FIFO, completes turn $m$ at $f^m(u)$. By Lemma~\ref{lem:recursion}, for $n\ge K-1$ the orbit
solves $u_{n+1}=V_{\mathrm{tot}}(0)+\sum_ka_ku_{n+1-k}$. The constant $V_{\mathrm{tot}}(0)/(1-\rho)=-u^*$ is a particular solution,
so $w_n=u_n+u^*$ solves the homogeneous recursion, and $w_n=Bz^n+O((r+\epsilon)^n)$, with $r$ the largest modulus among the other
roots. Since $V_{\mathrm{tot}}(u_n)=(\rho-1)w_n$ (Lemma~\ref{lem:orbit}(iii)) and $V(u_n)=w_{n+1}-w_n$,
\[
V(u_n)=\frac{z-1}{\rho-1}V_{\mathrm{tot}}(u_n)+O((r+\epsilon)^n),
\]
that is, $V(u_n)=\gamma\,(u_n+u^*)+O((r+\epsilon)^n)$, and $w_{n+m}/w_n\to z^m$ gives the centred form
$f^m(u)+u^*=z^m(u+u^*)(1+o(1))$. In all eight cells, over $\rho\in[1.01,6.6]$, $r=0.946$--$0.966<1$, so the deviation vanishes
absolutely. The transient decays by $\vartheta=r/z=0.84$--$0.95$ per generation, and the registered window opens 33--37 generations
after the burst. So we compare with the exact fluid as well as with the limits.
\end{proof}

\begin{proof}[Proof of Proposition~\ref{prop:classic}]
\emph{(a).} Let $\rho<1$ and take $u_0=0$. Summing the recursion, whose terms are all nonnegative, gives
$\sum_{n\le N_0}\Delta_n\le V_{\mathrm{tot}}(0)+\rho\sum_{n\le N_0}\Delta_n$, and in the limit
$\sum_n\Delta_n=V_{\mathrm{tot}}(0)/(1-\rho)$. Thus $b_n\uparrow\tau=V_{\mathrm{tot}}(0)/(1-\rho)$. For every orbit start
$u_0\in[0,V_0)$, each $\Delta_n(u_0)$ is positive, since $\Delta_n\ge a_1\Delta_{n-1}$. Every $t<\tau$ lies on such an
orbit, so the server is busy on $[0,\tau)$, and $V(b_n)=\Delta_n\to0$. (Equivalently, in the exact identity of
Lemma~\ref{lem:recursion} the constant $V_{\mathrm{tot}}(0)/(1-\rho)$ is a particular solution, which gives $u_n\to\tau$ at once.) If
$\rho\ge1$, the queue never empties: $V\ge\eta V_{\mathrm{tot}}(0)>0$ throughout, by the proof of Lemma~\ref{lem:orbit}(i).

Without arrivals ($\lambda=0$, used for SHED; outside (F5)), the orbit argument does not apply. A solution still exists: the
construction of Lemma~\ref{lem:orbit}(i) runs up to the first zero of $V$, and each $C_k$ is $1/c_k$-Lipschitz, because
$\sum_kc_kC_k(t)=t$ and every $C_k$ is non-decreasing. There, $V_{\mathrm{tot}}'=-1$ and $V\ge\eta V_{\mathrm{tot}}$ give
$\tau=V_{\mathrm{tot}}(0)$ directly.

\emph{(b), $\rho=1$.} Lemma~\ref{lem:renewal} with $b_k=a_k$ gives $\Delta_n(u_0)\to V_{\mathrm{tot}}(u_0)/\sum_kka_k$,
uniformly in $u_0\in[0,V_0]$. Here $V_{\mathrm{tot}}(u_0)=V_{\mathrm{tot}}(0)$ and $\sum_kka_k=\bar k_c$. Every
$t\ge V_0$ equals $f^n(u_0)$ for a unique $u_0\in[0,V_0)$ and $n\ge1$. So
$V(t)\to V_{\mathrm{tot}}(0)/\bar k_c$, and by Lemma~\ref{lem:orbit}(v), $N\to\lambda E[T]V_{\mathrm{tot}}(0)/\bar k_c$.

\emph{(c).} $F(x,\rho)=\rho\sum_k\omega_kx^{-k}-1$, with $\omega_k=c_kG_k/\sum_jc_jG_j$, satisfies $F(1,1)=0$ and
$\partial_xF(1,1)=-\bar k_c$. So $z'(1)=1/\bar k_c$. Then $E[z^{-T}]=1-(z-1)E[T]+O((z-1)^2)$, so
$\alpha=\lmiss(E[T]/\bar k_c)(\rho-1)+O((\rho-1)^2)$.
\end{proof}

\begin{proof}[Proof of Theorem~\ref{thm:premium}]
\emph{The product.} Since $\lambda=1/\sum_kc_kG_kz^{-k}$, we have
$D/\lmiss=E[z^{-T}]\sum_kc_kG_k/\sum_kc_kG_kz^{-k}$. Multiplying and dividing by $\sum_kG_kz^{-k}$ and by
$E[T]=\sum_kG_k$ gives $H\Gamma$.

\emph{The factor $\Gamma$.} The ratio $\nu^z_k/\nu_k$ is proportional to $z^{-k}$, which is strictly decreasing, so $\nu^z$ is
below $\nu$ in the likelihood-ratio order. For non-decreasing $c$ this gives $E_{\nu^z}[c]\le E_\nu[c]$. By Abel summation,
\[
E_\nu[c]-E_{\nu^z}[c]=\sum_{k<K}\bigl(F_{\nu^z}(k)-F_\nu(k)\bigr)(c_{k+1}-c_k),
\]
in which every difference of distribution functions is positive. So equality holds only for constant $c$. The same argument
applied to $z'>z$ shows that $\Gamma$ is non-decreasing in $z$.

\emph{The factor $H$.} Since $\sum_kG_kz^{-k}=(1-E[z^{-T}])/(z-1)$, $H\ge1$ if and only if $E[z^{-T}](1+E[T](z-1))\ge1$.
For a geometric $T$ with mean $m$, $E[z^{-T}]=1/(1+m(z-1))$. For $T=\min(\mathrm{Geometric}(1-q),K)$ we have
$E[z^{-T}]=(1-q)\sum_kG_kz^{-k}+qG_Kz^{-K}$ and $(1-q)E[T]=1-q^K$. Hence
\[
H=1-q^K\Bigl(1-E[T]\Big/\sum_kG_kz^{K-k}\Bigr).
\]
For $z>1$ and $K\ge2$, the sum in the denominator exceeds $E[T]$ and increases with $z$. So $H$ lies in $(1-q^K,1)$ and
is non-increasing in~$z$.

\emph{Near the threshold.} The expansion follows from Proposition~\ref{prop:classic}(c), with
$\alpha=\lmiss(E[T]/\bar k_c)(\rho-1)+O((\rho-1)^2)$, and from $D=\lambda-\alpha$. When $\bar k_c=E[T]$, the sign of the
premium is decided at second order.

\emph{Monotonicity.} $\Gamma$ is strictly increasing for non-constant non-decreasing $c$ (by the Abel argument), and it
tends to $E_\nu[c]/c_1$ as $z\to\infty$. If that limit exceeds $1/(1-q^K)$, there is a unique $z_1$ with
$\Gamma(z_1)=1/(1-q^K)$. For $z\ge z_1$, $H\Gamma>(1-q^K)\Gamma(z_1)=1$. For $z\in[1+10^{-12},z_1]$ in our cells, the
monotonicity of $H$ and $\Gamma$ bounds the premium below on each bracket $[z_i,z_{i+1}]$ by $H(z_{i+1})\Gamma(z_i)-1$. That
bound is positive on a geometric grid of 53 brackets in all eight cells (\fpath{analysis/theory/icml/support.py},
evaluated in floating point).
\end{proof}

\begin{proof}[Proof of Corollary~\ref{cor:trapthr}]
The state at the switching time $t_0$ is an admissible initial condition (F6), and $V(t_0)>0$. The claim then follows
from Proposition~\ref{prop:classic}(a)--(b) and Theorem~\ref{thm:trap}(a) applied with rate $\lambda'$. If $\lambda'\in(\lmiss,D(\lambda))$, then $\rho'>1$, so
by Lemma~\ref{lem:orbit}(iii) $V_{\mathrm{tot}}(t)\ge V_{\mathrm{tot}}(t_0)$, and $V\ge\eta V_{\mathrm{tot}}(t_0)>0$:
the queue never empties, and it grows by Theorem~\ref{thm:trap}(a). At $\rho'=1$ the same bound holds, and the backlog tends
to $\lmiss E[T]V_{\mathrm{tot}}(t_0)/\bar k_c$.
\end{proof}

\noindent\emph{Proof of Remark~\ref{rem:trapinv}.}

Divide Lemma~\ref{lem:perron}'s equation by $\sum_jc_jG_j$. The other quantities are explicit functions of $z$, $G$ and
$\omega$. Under $c\mapsto c/s$, $\omega$ is unchanged and $\lmiss\mapsto s\lmiss$. \qed

\paragraph{Remarks.}
\begin{itemize}[leftmargin=*,nosep]
\item \emph{Processor sharing.} The layer RR of Appendix~\ref{app:entry} advances every session at $1/V(t)$ turns per second. Use
the clock $x=\int_0^t ds/V(s)$ and the continuous renewal theorem; the kernel is a step function with compact support, hence
non-lattice. This yields the Euler--Lotka equation $\lambda\sum_kc_kG_k\int_{k-1}^ke^{-\alpha_{\mathrm{PS}}a}\,da=1$ for the growth exponent
$\alpha_{\mathrm{PS}}$, the backlog growth $\lambda(1-E[e^{-\alpha_{\mathrm{PS}}T}])$, departures $\lambda E[e^{-\alpha_{\mathrm{PS}}T}]$
and the cycle slope $\alpha_{\mathrm{PS}}$. In overloaded processor sharing with service times $S$ the departures are
$\lambda E[e^{-\alpha_{\mathrm{PS}}S}]$ ($\alpha_{\mathrm{PS}}$ its own growth exponent), a rate that can exceed $1/E[S]$: the
youth premium of Theorem~\ref{thm:premium} is the visit-indexed analogue of that property. This is the Crump--Mode--Jagers
structure of overloaded processor sharing~\citep{grishechkin1992processor,jeanmarie1994transient,puha2006overloaded}, with
reproduction rate $\lambda c_{\lfloor a\rfloor+1}$ at age~$a$; we do not write out the details. Near the threshold, the
backlog growth over $\lambda-\lmiss$ tends to $E[T]/(\bar k_c-\tfrac12)$, a half-turn offset from model~F. On the 300k cell,
$\alpha/(\lambda-\lmiss)$ is 0.647 and 0.666 under processor sharing at 1.35 and 1.65~$\lref$, against 0.634 and 0.653 for
model~F.
\item \emph{Stochastic version.} Take Poisson arrivals and any work-conserving order. The all-miss busy period started from
realized remaining works $w_i$ is the total work of a Galton--Watson tree, finite exactly when its progeny is. It ends with
probability $\exp(-\lambda(1-\xi)\sum_iw_i)$, where $\mathcal W_T=\sum_{k\le T}c_k$ is a session's total all-miss work and $\xi$
is the least root of $\xi=E[e^{-\lambda\mathcal W_T(1-\xi)}]$; the workload ever falls to a level $w_b$ with probability
$\exp(-\lambda(1-\xi)(\sum_iw_i-w_b))$~\citep{kendall1951some}. Take $w$ from the fluid of a burst of 200 sessions into an
empty engine on the 300k cell, 3000\,s after the burst, with the remaining work taken as realized and equal to the fluid's
(\fpath{support.py}, \texttt{kendall\_numbers}). Emptying then has probability 0.35 at $\rho=1.01$, 0.12 at 1.02,
$4\times10^{-3}$ at 1.05, $10^{-5}$ at 1.10, $4\times10^{-23}$ at 1.35 and $2\times10^{-54}$ at 1.65. A realized $w$
fluctuates around the fluid value, and if only the class composition is known, with the turn counts unrealized, the probability is
the product of per-session transforms, which by Jensen's inequality is larger. Hits return before the queue empties, so these
numbers are not bounds on a spontaneous exit; they indicate that it matters only within a few percent of $\lmiss$. The branching description is classical and claimed as such.
\item \emph{Steady solutions of FIFO re-entrant lines.} Model~F is a single-station re-entrant line~\citep{kumar1993reentrant}
in overload. Khanin et al.~\citep{khanin2001steady} study steady (constant-flow) solutions of fluid FIFO re-entrant lines from the
empty state, with stage-dependent service times. At an overloaded station the steady flows satisfy
$x_s=x_{s-1}/\sum_jx_{j-1}v_j$; at a single station with $n$ fixed stages this gives $x_s=a^s$ with $\sum_jv_ja^j=1$, which is
Lemma~\ref{lem:perron}'s equation with $a=1/z$ and $T\equiv n$, and the steady flows give Theorem~\ref{thm:trap}(a)'s gap slope,
composition and departure rate; a closing remark extends their uniqueness theorem to a fixed fraction of the flow leaving at each node, which in
a fluid is our random $T$, and their normalization gives Remark~\ref{rem:trapinv}. They prove existence, uniqueness for
Kelly-type lines and non-uniqueness examples for two stations, and leave the uniqueness of general (non-steady) trajectories open.
Relative to theirs, our contribution is the renewal recursion along FIFO orbits from an arbitrary initial state, hence the
convergence of every trapped trajectory to the steady regime with an exact virtual origin and rate, and the turn-count and
cost-growth factors of Theorem~\ref{thm:premium}; Corollary~\ref{cor:trapthr} is their operational reading with the classical drain
condition. The critical case,
Proposition~\ref{prop:classic}(b), is the fluid counterpart of heavy-traffic state-space collapse in multiclass FIFO feedback
queues~\citep{reiman1988multiclass,dai1995multiclass}. Overloaded multiclass FIFO queues with abandonment and no re-entry have fluid
limits~\citep{jennings2013fluid}; fluid metastability of queues with load-dependent service is posed by
Dong~\citep{dong2022metastability}; and within-request KV growth under continuous batching is analysed by Ao et
al.~\citep{ao2026service}, a different instability.
\item \emph{Think time.} With a 5\,s think time (the workload's mean tool time) the exact fluid's statistics over the registered
window agree with zero think time to within 0.02\% (backlog slope, completion rate, mean served prompt); the mean cycle
is 5\,s shorter and its slope the same.
\item \emph{How fast the limits are reached.} The transient shrinks by $\vartheta=r/z=0.84$--$0.95$ per generation over
$\rho\in[1.01,6.6]$. At 300k the registered window starts about 33 (1.65~$\lref$) or 37 (1.35~$\lref$) generations after the
burst, so a transient of order $\vartheta^{35}\approx0.1$ remains; the asymptotic values and the exact fluid over the window
differ by about 1\% for the composition and by 4--12\% for the backlog slope. No general rate is claimed: a nearly periodic kernel
$\lambda c_kG_k$ converges slowly.
\end{itemize}

\paragraph{Corollary~\ref{cor:trapthr} illustrated.} All-miss fluid of the 300k cell at 1.65~$\lref$, where
$D=1.223\lmiss$; arrivals cut at 5000\,s, when the backlog is 251 sessions (\fpath{notes/THEORY_ICML_2026-11.md}, \S6.7; reproduced
with independent code by a referee):
\begin{center}\footnotesize
\begin{tabular}{@{}l l@{}}
\toprule
cut to & backlog after the cut\\
\midrule
$1.18\lmiss$ & dips to 250.7 ($-0.3$\%), then grows to 445 by 30{,}000\,s\\
$1.10\lmiss$ & dips to 246 ($-2$\%), then grows to 352\\
$1.05\lmiss$ & dips to 241 ($-4$\%), then grows to 295\\
$1.02\lmiss$ & dips to 237.6 ($-5.5$\%), then grows to 261\\
$1.00\lmiss$ & undershoots to 234.6 ($-6.7$\%), settles at 238.7 (Prop.~\ref{prop:classic}(b))\\
$0.90\lmiss$ & drains slowly; empty at 59{,}655\,s, the busy period $V_{\mathrm{tot}}/(1-\rho')$\\
\bottomrule
\end{tabular}
\end{center}
In the fluid, an operator who sets arrivals between $\lmiss$ and the trapped throughput sees at most a brief dip, then renewed
growth; on one arrival path the backlog can keep falling slowly for the whole run (CUT-A seed~1).

\paragraph{Numerical checks and data.} The tests in \fpath{tests/test_theory_icml.py} (20) cover:
\begin{itemize}[leftmargin=*,nosep]
\item the root and its sign;
\item the decomposition, $H$ and $\Gamma$;
\item the absence of a premium for a geometric $T$ with constant costs;
\item the near-threshold coefficients;
\item $\sum_nh_n=V_{\mathrm{tot}}$;
\item the critical limit;
\item the orbit recursion on the simulated fluid (residual below $2\times10^{-3}$, from discretization);
\item convergence of the fluid to the asymptotics;
\item the busy period.
\end{itemize}
They do not test the exact identity of Lemma~\ref{lem:recursion} or the affine law of Theorem~\ref{thm:trap}(b),
Remark~\ref{rem:trapinv}, Corollary~\ref{cor:trapthr} beyond the subcritical busy period, or the monotonicity in $z$ and the
statement about $z_1$ in Theorem~\ref{thm:premium}; the bracket of \fpath{support.py} is a computation, not a test. The theory
note \fpath{notes/THEORY_ICML_2026-11.md}, whose \S10 records the independent checks, numbers its statements as in an earlier
draft: its Lemmas 1 and 2 are Lemmas~\ref{lem:perron} and~\ref{lem:orbit} here (its Lemma 2(iv) is
Lemma~\ref{lem:recursion}); its Theorem 1(a), (b), (c), (d) with (f), and (e) are Proposition~\ref{prop:classic}(a),
Proposition~\ref{prop:classic}(b), Theorem~\ref{thm:trap}(a), Theorem~\ref{thm:trap}(b) and Proposition~\ref{prop:classic}(c);
and its Corollary 4 is Remark~\ref{rem:trapinv} (Corollary~\ref{cor:rescue} here is the quasi-static cap rescue).
\fpath{analysis/theory/icml/run_check.py} produces \fpath{results/theory/icml/trap_check.md}
(Appendix~\ref{app:trapdata}). That file covers all 57 stuck runs of the E2 protocol: SGLang 300k with and without the waiting-prefix refresh, BASIN, SHED, TOOLSLOW, LPM beyond its
fallback, the native pool, vLLM, the H200 with 600k, and Llama-3.1-8B. It also covers the 13 unstable E0b runs, including the
pool sweep and the 300k H200. \fpath{analysis/theory/icml/support.py} computes every summary number quoted in the main text
into \fpath{results/theory/icml/support.json}. With arrivals stopped at 1800\,s, the fluid reaches $n_b$ after 2{,}020\,s and
empties after 3{,}180\,s, consistent with the registered SHED outcomes (600\,s stuck, 3600\,s good). This is a retrodiction,
and it does not discriminate: any drain time between 600 and 3{,}600\,s would fit.

\paragraph{Frozen predictions.} Tag \texttt{icml-trap-pred} (\texttt{6b47a6c}) froze
\fpath{configs/icml_trap_predict.yaml} before any later run existed. For eight cells it lists the asymptotic quantities at
loads from 1.1 to 6.5~$\lref$, the E2-protocol fluid and stochastic slope-test power, and per-seed plug-ins. The scoring
rules S1--S5 are stated in \fpath{notes/PREDICTIONS_icml_trap.md}.

\section{Model F against the engine: full tables}\label{app:trapdata}

\begin{table}[htbp]
\centering\footnotesize
\caption{Theorem~\ref{thm:trap} against the stuck runs of the registered E2-protocol cells (follow-up hit fraction below 0.2
over $[5700,8700)$\,s; 36 of the 57 stuck runs, the others below), ranges over seeds (vLLM: registered
and exploratory runs, which share plans). $\rho=\lambda/\lmiss$ with the model's $\lmiss$. $\bar L$: mean prompt of the turns
served, measured, against the tilted-mix prediction (stationary mean about 10{,}500 tokens). Gap slope: trend of the median gap
between a session's turns against the turn's completion time, against $1-1/z$. $D$: session completion rate in units of $\lref$.
Backlog: slope over model~F's $\alpha$, which the law puts at 1, with the quasi-static reading's $(\lambda-\lmiss)/\alpha$ in
parentheses (\fpath{analysis/theory/icml/backlog_over_alpha.py}). Retrospective; no parameter fitted to any trap
measurement; vLLM, the H200 and Llama take one speed factor fitted to their all-miss token rate.}\label{tab:trap}
\setlength{\tabcolsep}{1.6pt}
\resizebox{\columnwidth}{!}{%
\begin{tabular}{@{}l c c c c c@{}}
\toprule
cell, rate & $\rho$ & $\bar L$ meas.\ / pred. & gap slope / $1-1/z$ & $D/\lref$ / pred. & backlog$/\alpha$ (q.-s.)\\
\midrule
300k 1.35 & 1.34 & 9125--9556 / 9324 & .004--.020 / .013 & 1.13--1.28 / 1.13 & 0.23--1.42 (1.58)\\
300k 1.65 & 1.64 & 8417--8830 / 8606 & .015--.027 / .022 & 1.25--1.35 / 1.23 & 0.92--1.20 (1.53)\\
native 2.5 & 2.44 & 7264--7484 / 7374 & .044--.051 / .043 & 1.44--1.56 / 1.49 & 1.09--1.17 (1.46)\\
native 4.0 & 3.90 & 6127--6213 / 6115 & .066--.076 / .075 & 1.87--2.03 / 1.81 & 0.93--1.05 (1.36)\\
vLLM 1.35 & 1.32 & 9097--9547 / 9393 & .000--.016 / .012 & 1.18--1.28 / 1.14 & 0.00--1.26 (1.59)\\
vLLM 1.65 & 1.62 & 8104--8778 / 8676 & .013--.028 / .022 & 1.30--1.42 / 1.24 & 0.65--1.28 (1.55)\\
H200 1.5 & 1.45 & 9015--9177 / 9062 & .015--.022 / .016 & 1.20--1.27 / 1.20 & 0.83--1.32 (1.57)\\
H200 2.0 & 1.93 & 8058--8230 / 8082 & .023--.031 / .031 & 1.40--1.45 / 1.36 & 0.80--1.02 (1.51)\\
Llama 1.35 & 1.33 & 9075--9512 / 9356 & .003--.016 / .012 & 1.22--1.33 / 1.14 & 0.00--1.23 (1.57)\\
Llama 1.65 & 1.63 & 8459--8757 / 8631 & .013--.021 / .022 & 1.34--1.49 / 1.23 & 0.46--0.99 (1.53)\\
\bottomrule
\end{tabular}}
\end{table}

\paragraph{Inputs and definitions.} Each cell's costs come from its frozen cell model (\fpath{analysis/theory/cells/*.yaml}): the
calibrated miss prefill of turn $k$ plus the all-miss decode share, chosen so that $1/\sum_kc_kG_k$ equals the first-principles
model's $\mu_b$. vLLM, the H200 and Llama take one speed factor fitted to their own E0a token rate, as in the paper's earlier
predictions. A run is stuck if its follow-up hit fraction over its window is below 0.2: $[5700,8700)$\,s for E2-protocol runs,
$[1500,3000)$\,s for unstable from-empty (E0b) runs. The composition statistic is the mean prompt of the turns completed in the
window; the gap slope is the least-squares slope, over the window, of the 300\,s medians of the gap between a turn's completion
and the same session's previous one; completions and backlog slope are over the window. Source:
\fpath{results/theory/icml/trap_check.md} and \fpath{.json}, produced by \fpath{analysis/theory/icml/run_check.py}; summary
numbers by \fpath{support.py} into \fpath{support.json}. An AI referee recomputed six statistics for 22 of the runs from the raw
files with independent code and matched them to five or more digits. The gap slope is regressed on the turn's completion time,
so its model-F constant is $\gamma/z=1-1/z$ (\fpath{analysis/theory/icml/gap_slope_completion_time.py}); the theory note's tables
compare it with $\gamma$, which is 1--13\% larger at these loads. The backlog column of the table below, as in
Table~\ref{tab:trap}, is the measured slope over model~F's $\alpha$, with the quasi-static $(\lambda-\lmiss)/\alpha$ in parentheses
(\fpath{analysis/theory/icml/backlog_over_alpha.py}, \fpath{results/theory/icml/backlog_over_alpha.json}).

\paragraph{From-empty collapses.} The 13 from-empty runs that were stuck over
$[1500,3000)$\,s are younger traps:
\begin{center}\footnotesize
\setlength{\tabcolsep}{2.5pt}
\resizebox{\columnwidth}{!}{%
\begin{tabular}{@{}l c c c c c@{}}
\toprule
cell, rate & $\rho$ & $\bar L$ meas.\ / pred. & gap slope / $1-1/z$ & $D/\lref$ / pred. & backlog$/\alpha$ (q.-s.)\\
\midrule
SGLang 300k 2.5 & 2.49 & 7{,}755 / 7{,}274 & 0.058 / 0.045 & 1.42 / 1.46 & 1.35 (1.44)\\
SGLang 300k 3.0 & 2.99 & 6{,}613 / 6{,}754 & 0.078 / 0.057 & 1.66 / 1.57 & 1.28 (1.40)\\
H200 300k 1.4 & 1.38 & 9{,}422 / 9{,}233 & 0.009 / 0.014 & 1.26 / 1.16 & 0.80 (1.57)\\
H200 300k 1.55 & 1.53 & 8{,}965 / 8{,}866 & 0.018 / 0.019 & 1.24 / 1.21 & 1.15 (1.55)\\
H200 300k 1.75 & 1.72 & 8{,}613 / 8{,}444 & 0.031 / 0.025 & 1.34 / 1.27 & 1.18 (1.52)\\
H200 600k 3.0 & 2.90 & 6{,}939 / 6{,}868 & 0.042 / 0.054 & 1.78 / 1.61 & 1.09 (1.42)\\
Llama 300k 3.0 & 2.96 & 6{,}714 / 6{,}754 & 0.060 / 0.057 & 1.64 / 1.57 & 1.23 (1.39)\\
native 6.5 & 6.33 & 5{,}332 / 5{,}064 & 0.109 / 0.116 & 2.40 / 2.19 & 1.06 (1.27)\\
150k pool 1.45 & 1.51 & 8{,}369 / 8{,}879 & 0.020 / 0.019 & 1.27 / 1.13 & 1.14 (1.52)\\
150k pool 1.6 & 1.66 & 8{,}071 / 8{,}528 & 0.020 / 0.024 & 1.30 / 1.18 & 1.17 (1.50)\\
150k pool 1.8 & 1.87 & 7{,}909 / 8{,}123 & 0.033 / 0.030 & 1.25 / 1.23 & 1.32 (1.48)\\
600k pool 5.0 & 4.89 & 5{,}591 / 5{,}581 & 0.089 / 0.093 & 2.27 / 1.97 & 0.99 (1.31)\\
vLLM 300k 3.0 & 2.94 & 6{,}598 / 6{,}826 & 0.063 / 0.055 & 1.60 / 1.60 & 1.28 (1.41)\\
\bottomrule
\end{tabular}}
\end{center}
The composition is within $-5.7$\% and $+6.6$\% and the gap-slope ratio between 0.63 and 1.37 (mean 1.03). Completions and
backlog slopes both run above the asymptotic values in 9 of the 13 runs. In model~F the two add up to the arrival rate, so this
may reflect window arrival rates above nominal (each is a single seed-0 run, with window arrivals 3.8--13.1\% above nominal) rather than the transient of a young trap.

\paragraph{E2-protocol runs beyond Table~\ref{tab:trap}.} The other stuck runs of the E2 protocol behave like the registered ones:
the SGLang 300k grid with the waiting-prefix refresh disabled (r0: composition 8{,}406--9{,}532 against 8{,}606 and 9{,}324), the
BASIN runs with bursts of 50, 60 and 100 sessions (8{,}372--9{,}272 against 8{,}606), SHED's 600\,s stop, TOOLSLOW's trapped run and
stock LPM past its 128-request fallback. Across all 57 runs the engine completes 0.90--1.21 times $D$, 1.055 on average; seven are
below $D$ (the late-forming BASIN traps of 50 and 60 sessions, seed~1, at 0.90 and 0.91; native 2.5~$\lref$ seed~1; H200 1.5~$\lref$
seed~1; three LPM runs). The measured token ratios are 0.95--1.12 against 1.01--1.05 from the composition alone; the excess is largest on the H200 and
the native pool, while the completion excess on vLLM and Llama is not explained by the token rate (the paper's checks of its
congestion assumptions, Appendix~\ref{app:proofs}).

\paragraph{The slope rule and the plans.} The client's constant-rate plan fixes each seed's base arrival times before any run, and
reproduces the measured window arrival rates exactly. Seed~1's plan delivers 0.8--8.9\% fewer arrivals than nominal over the window
in every cell and rate (5.8\% at 300k, 7.9\% on vLLM and 8.9\% on Llama at 1.35~$\lref$). The plug-in $\lambda_{\mathrm{window}}
(\text{plan})-D$ lies within 0.01\,/s of the measured backlog slope in 49 of the 57 runs and within 0.002\,/s in 18, with median
residual $-0.0022$\,/s because the engine completes slightly more than $D$. On the 36 runs of the five registered paired-test cells (30 registered runs and 6 earlier exploratory vLLM runs; 30 distinct
plans, the 12 vLLM runs being six plans run twice; plans of different cells share seeds) it falls on the measured side of the registered threshold in 33 runs and for 27
plans, against 23 for always predicting ``trapped''. The exceptions are H200 1.5~$\lref$ seed~1 (plug-in 0.04849 against a threshold
of 0.04847), Llama 1.65~$\lref$ seed~1 (above it by 0.00005) and Llama 1.35~$\lref$ seed~0 (0.0205 against a measured 0.0052).

\paragraph{Self-similarity.} By Theorem~\ref{thm:trap}(b), among completed sessions with similar $T$ that arrive in
$[2500,7000)$\,s, the sojourn should grow linearly in the arrival time with slope about $E[z^T-1]$. Measured over predicted
slope, per seed: 0.86--1.19 in every turn-count bin on the H200 at 600k (160--400 sessions per bin), 0.85--0.96 for short sessions on
the native pool at 2.5~$\lref$, 0.2--1.6 (noisy, 40--120 sessions per bin) on SGLang 300k, vLLM and Llama, and 0.1--0.8 for long
sessions at high $z$, where right-censoring at 8700\,s keeps only the faster ones. This check is descriptive, not a test.

\section{The blind tests in detail}\label{app:blind}

This appendix gives the blind tests of \S\ref{sec:validation} and \S\ref{sec:control} in full, as scored: the frozen
predictions, the registered rules and their history, every outcome, and the frozen exploratory measures, hits and misses.
Table~\ref{tab:frozensum15} lists them with the other frozen predictions, and Table~\ref{tab:timeline-r15} gives their time
record. Section numbers in the text below refer to the main text.

\paragraph{Frozen predictions and scoring rules.} The predictions of Theorem~\ref{thm:trap} for runs that did not yet exist were frozen under tag
\texttt{icml-trap-pred} (commit \texttt{6b47a6c}, pushed at 00:24:47 CST on 2026-11-05 by GitHub's activity log), with scoring rules
S1--S5 declared beforehand: per-run composition within $\pm8$\% (pooled $\pm5$\%); pooled gap slope within $[0.6,1.4]\gamma$; completion
above $\lref$ in every trapped run at realized load $\ge1.1\lref$ (pooled within $\pm15$\% of $D$); pooled backlog slope within $\pm35$\%
of $\alpha$ or of the E2 fluid slope; and, per seed, the plug-in's side of the registered threshold and
$|$slope$-$plug-in$|\le0.01$\,/s. Theorem~1's application to a cell and load is refuted if the pooled S1 or S2 fails in a group of at
least two trapped runs (\fpath{notes/PREDICTIONS_icml_trap.md}). The tolerances were set after seeing the committed runs; the runs
they score did not exist at the push. S2 compares with $\gamma$, not with the completion-time slope $\gamma/z$; at 1.35~$\lref$ the
two differ by 1.3\%, well inside the tolerance.
\begin{itemize}[leftmargin=*,nosep]
\item \emph{TOOLSLOW-W} (pre-registered: tag \texttt{r15-reg}, commit \texttt{e05a262}, pushed at 03:18:36 CST, before the first job
was submitted at 03:21:38; designed after the 30\,s TOOLSLOW arm, whose slowdown ended at 2100\,s, trapped 1 of 3, and after
reading simulator outputs, so it is blind in outcome only): no burst; every tool call that starts in
$[1200,3000)$\,s takes 30\,s longer, at 1.35~$\lref$, seeds 0--2, paired with the committed NOBURST runs of 2026-10-24/25 (same plans, not run alongside). The
registered rule calls the
cell TRAPS if at least 2 of the 3 runs are stuck, NO-TRAP if at least 2 are good, and MIXED otherwise; its trapped runs, if any, are
scored by S1--S5. Outcome: TRAPS. All three runs were stuck (hit 0.000 over $[5700,8700)$\,s; backlog slopes 0.0192, 0.0184
and 0.0232\,/s against a threshold of 0.0125\,/s), while their paired NOBURST runs stayed good (hit 0.930--0.934); on the same
plans, the slowdown that ended at 2100\,s had trapped 1 of 3. All three runs are eligible for S1--S5, and the pooled scores pass:
composition $+1.9$\% (limit $\pm5$\%), gap slope $1.075\gamma$ (1.089 against the unregistered $\gamma/z$), completions $+1.9$\% of
$D$ with every run above $\lref$, and backlog slope $+29$\% of $\alpha$ (limit $\pm35$\%); Theorem~1's application to this cell is
not refuted. Per run, composition is within $\pm8$\% in all three (seed~1 at $+7.9$\%), and S5's plug-in lies on the measured side
of the threshold in 2 of 3 runs and within 0.01\,/s in 3 of 3. Every model we froze (the calibrated simulator in both modes and the
first-principles model) predicted this outcome, so the cell tests the trap-state law on a trap entered with no burst, not the models
against each other. Like TOOLSLOW's, its outcome shows that a temporary increase in tool latency, with no burst, can trigger the
trap; it does not separate the longer gap from the capacity loss and the larger population that come with it. Of the frozen
exploratory measures, the first 150\,s window below a follow-up hit of 0.5 fell at 1950, 1950 and 2100\,s, inside its predicted
band, and the sessions alive at 2100 and 3000\,s met their reference accuracy; the outstanding requests at 3300\,s missed it by 8 and
6 in seeds 0 and~2, and pred2's queue-free plan walk undercounted the sessions alive at 3000\,s by 7--18.
\item \emph{TOOLSLOW-D15} (pre-registered with TOOLSLOW-W under tag \texttt{r15-reg}; jobs 17256873--75 submitted at 03:57:14
CST; designed, like TOOLSLOW-W, after the 900\,s arm and after reading simulator outputs, so it is blind in outcome only):
no burst; every tool call that starts in
$[1200,2100)$\,s takes 15\,s longer, at 1.65~$\lref$, seeds 0--2, paired with the committed NOBURST runs of 2026-10-24/25 (same plans, not run alongside), under
the same three-way
rule; its stuck runs, if any, are scored by S1--S5. It was chosen so that our two frozen predictors disagree: the calibrated
simulator, run on the exact client plans, called seed~0 stuck and seeds 1 and~2 good (NO-TRAP) in both modes, while the
first-principles model's plan-averaged fluid verdict called every seed stuck (TRAPS); the model's per-seed layer made the
simulator's modal calls, but with $P(\text{NO-TRAP})=0.41$. Outcome: NO-TRAP, as the simulator said, seed by seed. Seed~0 was stuck
(hit 0.000; backlog slope 0.0258 against a threshold of 0.02325\,/s), and seeds 1 and~2 recovered to their controls' hit rates
(0.934 and 0.924, against 0.934 and 0.921). The fluid verdict was wrong, and the per-seed layer gave the outcome a low probability
(log score $-0.89$); over the six TOOLSLOW-W and TOOLSLOW-D15 runs the simulator was right in 6, the per-seed modal calls in 6
and the fluid verdict in 4. The stuck run passed S1--S4 (composition $+0.1$\%, gap slope $1.02\gamma$, completion $+10$\% of $D$,
backlog slope $-15$\% of $\alpha$); its S5 plug-in missed the 0.01\,/s tolerance by 0.0004\,/s, and with one eligible run the
refutation rule does not apply. Across the four blind runs trapped by a tool slowdown alone, S1--S4 pass in both cells, and S5 is close
in 3 of 4 and on the correct side in 3 of 4. Of the frozen exploratory measures, the sessions alive at 2100\,s were off by 4 in
seed~1 (reference accuracy 1), the outstanding requests at 3300\,s by 7 in seed~0 (reference 5), and seed~0's first 150\,s window
below a follow-up hit of 0.5 came one bin early; pred2's plan walk missed seed~0's sessions alive at 3000\,s by 51.
\item \emph{CUT} (a blind test of model~F's trajectory under an arrival cut, the setting of Corollary~\ref{cor:trapthr}): the
1.65~$\lref$ BURST protocol, then from 2400\,s arrivals cut to 1.15~$\lref$ (CUT-A, $\rho'=1.145$, inside $(\lmiss,D(\lambda))$) or
to 0.85~$\lref$ (CUT-B, $\rho'=0.846$), seeds 0--2. Model~F's predictions, driven by each seed's exact client plan, were frozen under
tag \texttt{icml-cut-pred} (commit \texttt{2f90267}, pushed at 02:25:43 CST on 2026-11-05; \fpath{notes/PREDICTIONS_icml_cut.md}).
Model~F predicts that all six runs stay stuck over $[5700,8700)$\,s; that CUT-A, although its rate is below the trapped throughput
of the same seed's uncut run, does not clear, its backlog dipping by less than 15\% on 300\,s bins and growing again in seeds 0 and 2
(predicted slopes $S$ of the backlog over $[3000,8700)$\,s of $+0.0039$ and $+0.0028$\,/s; seed~1's plan has few post-cut arrivals,
and its predicted slope is $-0.0026$\,/s); and that CUT-B's backlog shrinks slowly (predicted slopes $-0.0095$ to $-0.0131$\,/s) and would empty
only at about 27{,}100\,s, the busy period of Proposition~\ref{prop:classic}(a), far beyond the run.

The decision rule was registered after the predictions were pushed and before any CUT job (commit \texttt{11b0e4e}, tag
\texttt{r15-cut-reg}, pushed at 03:54:49 CST; jobs 17256867--72 submitted at 03:57). Per seed, the measured slope $S$ over
$[3000,8700)$\,s is compared with model~F's frozen slope and with two comparators: a naive one,
$\lambda'_{\mathrm{plan}}-D_{\mathrm{trap}}$, where $D_{\mathrm{trap}}$ is the trapped throughput of the same seed's uncut 1.65~$\lref$
control over $[3000,8700)$\,s, fixed beforehand at 0.0956, 0.1005 and 0.0946\,sessions/s (1.34, 1.41 and 1.32~$\lref$), the reading
in which a cut below the measured throughput drains the trap; and, for CUT-A, the quasi-static $\lambda'_{\mathrm{plan}}-\lref$, in
which the trap completes sessions at the all-miss capacity. From the frozen plans the naive comparator is about $-0.009$ to
$-0.020$\,/s for CUT-A and $-0.033$ to $-0.043$\,/s for CUT-B, and the quasi-static one $+0.009$ to $+0.015$\,/s for CUT-A, so model~F's
values lie well between them. A seed is theory-favoured if model~F's slope is strictly the closest to $S$, and a run that leaves the
trap counts against the theory. CUT-A passes if at least 2 of 3 seeds are theory-favoured and the slope over $[5700,8700)$\,s is
positive in at least 2 of 3 (model~F predicts that clause from seeds 0 and 2); CUT-B passes if at least 2 of 3 seeds are
theory-favoured and $S<0$ in at least 2 of 3. Either arm fails if a comparator is favoured, or the run leaves the trap, in at least
2 of 3 seeds; otherwise it is mixed. The
rule was fixed after the per-seed predictions had been read, and it differs from the draft rule simulated in the prediction note in
five ways: it adds, for CUT-A only, the quasi-static comparator (stricter; the prediction note had put it 0.010--0.011\,/s from
model~F's slope there and within 0.002\,/s in CUT-B, where it was not added); where the note counted a run that leaves the trap as not
theory-favoured, the rule also counts it toward a fail (stricter); it takes $D_{\mathrm{trap}}$ from the frozen controls instead of
each run's own completion rate over $[1800,2400)$\,s (stricter for seeds 0 and~1, looser for seed~2); it adds $S<0$ in at least 2 of
3 to CUT-B (stricter); and it moves CUT-A's sign clause from $[3000,8700)$ to $[5700,8700)$\,s, as the note recommended (looser: in
the note's simulation this raised CUT-A's probability of passing from 0.82 to 0.99). The test has limits: the cut comes 1200\,s
after the burst, so it probes the young trap's transient and the start of its regrowth, not the asymptotic rate, and CUT-B cannot
show recovery within the run.

\emph{CUT-A outcome: pass.} With arrivals cut to a nominal 1.15~$\lref$ (realized 1.12--1.20~$\lref$ over $[3000,8700)$\,s; the
plans are thinned Poisson streams), below the trapped throughput of the same seeds' uncut runs (1.32--1.41~$\lref$), all three runs
stayed stuck to the end of the run, at zero hits over $[5700,8700)$\,s. Over $[3000,8700)$\,s the backlog's slope was $+0.0040$,
$-0.0022$ and $+0.0029$\,/s, against model~F's frozen $+0.0039$, $-0.0026$ and $+0.0028$, the naive comparator's $-0.0094$, $-0.0204$
and $-0.0099$ and the quasi-static one's $+0.0146$, $+0.0086$ and $+0.0132$, so all three seeds were theory-favoured; over
$[5700,8700)$\,s it was positive in seeds 0 and~2 and negative in seed~1, as model~F predicted for that plan. The draft rule gives
the same verdict. The secondary predictions frozen with model~F (tolerances declared in the prediction note, not part of the registered
rule) held as well: completions over $[5700,8700)$\,s within 0.001\,/s of model~F's per-seed values, the served prompt within 0.9\%, a
dip below the $[2100,2400)$\,s level of at most 11.0\% on 300\,s bins (frozen bound 15\%; on 10\,s bins, against the backlog at the
cut, up to 19.1\%), and more than 150 sessions present over the last 300\,s. Model~F's backlog levels ran 5--10 sessions high (186
against 176 over $[2100,2400)$\,s in seed~0).

The pass is narrower than Corollary~\ref{cor:trapthr}. Both comparators assume a constant throughput, so the registered rule
rejects only the two constant-throughput readings it was built against, in which the trap drains at the throughput it showed or completes sessions at
the all-miss capacity: over $[3000,8700)$\,s, completions equal to the nominal cut rate fit as well as model~F, and both match the
measured slopes to within the engine's run-to-run jitter (0.0001--0.0005\,/s, from the identical pre-cut parts of runs with the same
plan). One result singles out model~F: in the late window, scored post hoc against slopes frozen with the predictions, the
backlog's slope was $+0.0071$, $-0.0027$ and $+0.0131$\,/s against model~F's $+0.0069$, $-0.0036$ and $+0.0129$, while constant
completions at $D_{\mathrm{trap}}$, at $\lref$, at their midpoint or at the nominal cut rate each missed by about 0.004\,/s or more
(the smallest, 0.0040, at the nominal cut rate in seed~0). The completion rate there supports it, also post hoc: it varied with the
plan as model~F said (1.05, 1.18 and 1.10~$\lref$ against 1.04, 1.20 and 1.09), and no single constant rate comes within 0.004\,/s
of all three seeds, although the best one (0.080\,/s) stays within the frozen tolerance of 0.006\,/s. All three traps also
persisted to 8700\,s, which each constant reading above also predicts: the naive one still leaves about 50--130 sessions at the end,
far above the onset $n_b=32.7$. Persistence rules out only a drain at the completion rate seen just before the cut, 1.70, 1.65 and
1.21~$\lref$ over $[1800,2400)$\,s, which would have reached the onset near 6300\,s in seeds 0 and~1; that is the operator-facing
content of Corollary~\ref{cor:trapthr}, and on the engine it held within the run. CUT-A's three runs shared one node, and ``did not
clear'' means within 8700\,s.

\emph{CUT-B outcome: pass.} After the cut to 0.85~$\lref$ (realized 0.81--0.88~$\lref$ over $[3000,8700)$\,s) the backlog
fell, slowly, at $-0.0085$, $-0.0126$ and $-0.0124$\,/s over $[3000,8700)$\,s, against model~F's frozen $-0.0095$, $-0.0131$
and $-0.0123$ and the naive comparator's $-0.033$, $-0.043$ and $-0.034$, so all three seeds were theory-favoured and all three
slopes were negative. All three runs were still stuck at 8700\,s (hit 0.000 over $[5700,8700)$\,s; 140, 77 and 89 sessions present,
against model~F's 139, 79 and 94). Here persistence discriminates: from the backlog over $[3000,3300)$\,s, the naive reading would
have reached the hit-recovery onset $n_b=32.7$ at 5656--7245\,s, and a drain at the completion rate seen just before the cut (1.68,
1.65 and 1.23~$\lref$ over $[1800,2400)$\,s) at 4877--7598\,s; model~F's fluid empties only at about 27{,}100\,s, the busy period of
Proposition~\ref{prop:classic}(a). A reading in which the trap completes sessions at the all-miss capacity, not a registered comparator for
this arm, fits the slopes about as well: had it been one, seed~0 would have favoured it (error 0.00036 against model~F's
0.00099\,/s), and CUT-B would still have passed, 2 of 3. Model~F's distinctive prediction for CUT-B, completions below that capacity as the trapped
mix ages, held after the fact: 0.918, 0.932 and 0.974~$\lref$ over $[5700,8700)$\,s against model~F's 0.918, 0.955 and 0.979, with
served prompts of 10.9--11.5k tokens against 8.4--8.8k in the uncut runs, although at the frozen tolerance of 0.006\,/s the
all-miss capacity also lies (just) inside. The frozen secondary predictions held too: completions within 0.002\,/s, the served
prompt within 1\%, and a negative trend of the median gap in all three seeds. CUT-B's runs ran on two nodes, and none left the trap within 8700\,s.

Together, on the engine a cut below the trapped throughput but above the all-miss capacity left every run trapped, and a cut below
the all-miss capacity shrank the backlog only slowly, every run still trapped at 8700\,s: the pattern of Proposition~\ref{prop:classic}(a) and Corollary~\ref{cor:trapthr}. Each
arm's registered rule rejected the naive reading, and CUT-A's also the quasi-static one; what singles out model~F against
intermediate readings is post hoc: CUT-A's late-window slopes and CUT-B's completions below the all-miss capacity.
\end{itemize}

\begin{figure*}[tbp]\centering
\includegraphics[width=\textwidth]{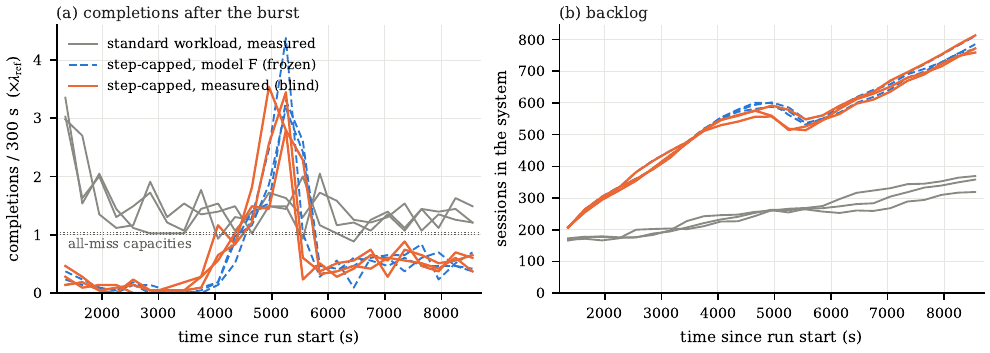}
\caption{MIRROR at 1.65~$\lref$: the standard workload's three committed traps (grey), model~F's frozen predictions for the
step-capped workload (blue, dashed) and the MIRROR runs (orange). (a)~Session completions per 300\,s bin, in $\lref$; dotted:
the model's all-miss capacities of the two workloads (1.004 and 1.046~$\lref$). (b)~Sessions in the system, 300\,s means.
Orange: the decisive cell's three BURST runs (jobs 17264055--57, seeds 0--2), all RENEWAL; the dashed lines were frozen under
tag \texttt{r16-mirror-reg} before any MIRROR job. Grey: jobs 17093390, 17093396 and 17093402
(\fpath{figures/r16/fig_mirror.py}).}\label{fig:mirror}
\end{figure*}

\paragraph{MIRROR.} The test of \S\ref{sec:mirror} runs the E2 BURST protocol on cell \texttt{sglang-fcfs-r1-mirror-k20} (SGLang
0.5.22, FCFS with the waiting-prefix refresh, 300k pool, Qwen2.5-7B-Instruct, A100 80GB PCIe) with
\fpath{configs/agentload_e2_mirror_k20.yaml}, which is \fpath{configs/agentload_e2_burst.yaml} with the mean turn count changed from
15 to 1000 and the maximum from 60 to 20, and nothing else but its description text; rates are multiples of the standard workload's registered
$\lref=0.071534$\sps. The arms run one per job, seeds 0--2, for 8700\,s, with the burst of 200 sessions in $[1200,1260)$\,s: BURST at
1.65~$\lref$ (the decisive cell), BURST at 1.35~$\lref$ (secondary) and NOBURST at 1.65~$\lref$ (controls), and one E0a run of the
workload measures its all-miss capacity, which only the secondary check W10 uses. The step cap $K=20$ was chosen by computation only, comparing $K=15$,
20 and 25 and mean turn counts of 30, 40 and 60 in model~F and our fluid model: at $K=20$ the model's all-miss capacity matches the standard
workload's (1.046 against 1.004~$\lref$) and the good state's basin at 1.65~$\lref$ is wide (stable point 14.3 sessions, unstable
point 41.8), so the NOBURST controls should stay good; no engine run of the workload existed. Asymptotically, model~F gives
$\rho=1.577$, $D/\lmiss=0.771$ ($H=0.699$, $\Gamma=1.103$; the standard workload 1.223, with $H=0.989$ and $\Gamma=1.236$) and
$\alpha/(\lambda-\lmiss)=1.398$ (standard 0.65) at 1.65~$\lref$, and $\rho=1.291$, $D/\lmiss=0.870$ ($H=0.825$, $\Gamma=1.055$;
standard 1.126) and $\alpha/(\lambda-\lmiss)=1.447$ at 1.35~$\lref$ (\fpath{results/theory/r16/mirror_predictions.json}). Its frozen
per-seed predictions, from the plan-driven FIFO on each seed's exact client plan (rates in $\lref$; wave: the median completion time
of the 200 burst sessions; slopes over $[6000,8700)$\,s in sessions/s, model~F against the quasi-static $\lambda_{\mathrm{plan}}-\lmiss$):
\begin{center}\footnotesize
\resizebox{\columnwidth}{!}{%
\begin{tabular}{@{}c c c c c c@{}}
\toprule
rate, seed & $[2400,3600)$ & wave (s) & burst IQR (s) & $[6600,8700)$ & slope F / QS\\
\midrule
1.65, 0 & 0.047 & 5195 & 4803--5392 & 0.546 & 0.0789 / 0.0478\\
1.65, 1 & 0.070 & 5135 & 4876--5299 & 0.599 & 0.0857 / 0.0441\\
1.65, 2 & 0.093 & 5157 & 4788--5376 & 0.486 & 0.0930 / 0.0518\\
1.35, 0 & 0.000 & 4883 & 4497--5022 & 0.599 & 0.0565 / 0.0274\\
1.35, 1 & 0.058 & 4802 & 4500--4973 & 0.686 & 0.0534 / 0.0192\\
1.35, 2 & 0.070 & 4870 & 4546--5068 & 0.726 & 0.0603 / 0.0278\\
\bottomrule
\end{tabular}}
\end{center}
In the quasi-static reading both completion windows are at 1.046~$\lref$.

The rules (\fpath{analysis/r16_mirror.py}; registration \fpath{configs/agentload_prereg_2026-11-05_mirror_k20.yaml}, commit
\texttt{bc81ee7}, tag \texttt{r16-mirror-reg}, pushed at 14:07:35 CST on 2026-11-05, before any job;
\fpath{notes/PREDICTIONS_r16_mirror.md}) score each BURST run: M1, completions over $[2400,3600)$\,s at most 0.30~$\lref$; M2, the
wave within $0.15(F-1200\,\mathrm{s})$ of model~F's time $F$; M3, completions over $[6600,8700)$\,s at most 0.85~$\lref$; M4, the
least-squares slope of the 10\,s backlog over $[6000,8700)$\,s closer to model~F's than to the quasi-static slope. A seed is RENEWAL
if its run is stuck (the frozen endpoints of the basin tests) and M1--M4 hold, CURVE if both completion windows are at least
0.85~$\lref$, NOT-TRAPPED if the run is not stuck, and MIXED otherwise; the cell passes with RENEWAL in 2 of 3 seeds, fails with CURVE
in 2 of 3 and is NO-TRAP, leaving model~F untested, with NOT-TRAPPED in 2 of 3. A run counts only if it is the registered cell's,
completed, bound to the registration and driven by the frozen plan. Registered secondary checks, reported but not in the verdict:
W5, the wave within $0.05(F_{\mathrm{cal}}-1200\,\mathrm{s})$ of the calibrated time $F_{\mathrm{cal}}=1200+0.9223(F-1200)$\,s; W6,
the wave earlier at 1.35 than at 1.65~$\lref$ in each seed (model~F: by 287--334\,s); W7, the mean prompt of the turns completed over
$[5700,8700)$\,s within 5\% of model~F's; W8, the late completions within 0.25~$\lref$ of model~F's; W9, every NOBURST control good;
W10, the workload's measured all-miss capacity, and whether the late completions lie below it. The prediction note committed
with the registration quoted the asymptotic 1.4 for the backlog's growth in the late window and 270--330\,s for W6's shift; both
were corrected in an erratum (commit \texttt{d8c49b7} on \texttt{r16/mirror}, pushed at 16:19:36, during the runs; that no MIRROR run
output had been read by then rests on our own record), and M4 scores the per-seed slopes of the table above and W6 only the order of the waves, so neither error enters a
rule. The registration's push time is GitHub's record of the commit's push, which created branch \texttt{r16/mirror} at
\texttt{bc81ee7}; GitHub keeps no push time for the tag object. It came 41.5\,min before the first MIRROR job, a cache-off smoke
test, was submitted and 43.3\,min before the nine counted jobs; an erratum to the prediction note's writing time (\texttt{18ac5e6})
reached GitHub 17\,s after the smoke test was submitted and before the counted jobs (Table~\ref{tab:timeline-r15}). The tolerances come from committed
data only: on the six committed standard-workload BURST runs, model~F's plan-driven FIFO times the burst sessions' completions about
8\% late (measured over model elapsed time 0.9125--0.9327 per run, mean 0.9223), which M2's 15\% covers and W5's factor removes; M1,
M3 and the CURVE threshold lie between model~F's values and the quasi-static ones, fixed from those values and from the standard
workload's measured traps. Not blind: the workload was chosen after model~F's and our fluid model's predictions for it and for a few alternatives had been
computed; the timing factor comes from the standard workload's committed runs; no simulator prediction was made for MIRROR; and the
test came after Lu et al.'s paper. Outcome: PASS in both cells, RENEWAL in 3 of 3 seeds each (the registered scorer at the tag's code, reading the
predictions from the tag; \fpath{results/agentload_analysis_r16_mirror/20261105-173136/}). The nine runs (jobs 17264055--60 and
17264062--64, submitted at 14:50:52--53 CST and ended at 17:20:42--17:21:03) all count: each is the registered cell's, finished
at 8700\,s, is bound to the registration and was driven by its frozen plan (exact in sessions, turn counts and arrival times; tool
times within $10^{-13}$\,s). Measured, with model~F's frozen value in parentheses (rates in $\lref$; slopes in sessions/s, model~F's
and the quasi-static one):
\begin{center}\footnotesize
\resizebox{\columnwidth}{!}{%
\begin{tabular}{@{}c c c c c@{}}
\toprule
rate, seed & $[2400,3600)$ & wave (s) & $[6600,8700)$ & slope (F / QS)\\
\midrule
1.65, 0 & 0.070 (0.047) & 4996 (5195) & 0.592 (0.546) & 0.0767 (0.0789 / 0.0478)\\
1.65, 1 & 0.058 (0.070) & 4945 (5135) & 0.606 (0.599) & 0.0837 (0.0857 / 0.0441)\\
1.65, 2 & 0.093 (0.093) & 5089 (5157) & 0.499 (0.486) & 0.0935 (0.0930 / 0.0518)\\
1.35, 0 & 0.105 (0.000) & 4759 (4883) & 0.619 (0.599) & 0.0557 (0.0565 / 0.0274)\\
1.35, 1 & 0.093 (0.058) & 4673 (4802) & 0.739 (0.686) & 0.0504 (0.0534 / 0.0192)\\
1.35, 2 & 0.163 (0.070) & 4576 (4870) & 0.845 (0.726) & 0.0496 (0.0603 / 0.0278)\\
\bottomrule
\end{tabular}}
\end{center}
Every seed was called RENEWAL. All six BURST runs were stuck (hit fraction 0.000 over $[5700,8700)$\,s), and in each all 200
burst sessions completed by 8700\,s. The rules held with room in every run but one. The blackout windows held 5--14 completions, against M1's ceiling of 25.75;
the waves came 69--295\,s before model~F's times (1.7--8.0\% of the elapsed time, against M2's 15\%) and peaked at 2.8--3.5~$\lref$
per 300\,s bin at 1.65~$\lref$ (model~F 3.2--4.4); the backlog grew 1.61--1.90 times as fast as the plan's excess load at
1.65~$\lref$ (model~F 1.65--1.95), and every slope was nearer model~F's than the quasi-static one by at least 0.011\,/s. At
1.35~$\lref$, seed~2's 127 late completions were 0.7 below M3's ceiling of 127.7 (0.845 against 0.85): one more would have made that
seed MIXED, and the cell would still have passed, 2 of 3. In the decisive cell, changing any seed's call would have taken at
least 18 more blackout completions (M1) or 37 more late ones (M3), a wave at least 399\,s further from model~F's time (M2), or a
slope at least 0.013\,/s nearer the quasi-static one (M4). Secondary checks: W5 held in 5 of 6 runs and failed at 1.65~$\lref$,
seed~2 (5089\,s against $F_{\mathrm{cal}}=4850\pm182$\,s); on this workload model~F's timing ran 2--8\% slow
($(\text{wave}-1200)/(F-1200)=0.92$--0.98; 2--5\% in five of the six runs), so in those five the factor 0.9223, which removes 8\%, over-corrected. W6 held in 3 of 3 seeds, with shifts of
237, 271 and 513\,s against model~F's 312, 334 and 287\,s. W7 held in 6 of 6 (served prompt $+0.2$ to $+2.6$\% of model~F's) and W8
in 6 of 6 (late completions 0.007--0.120~$\lref$ above model~F's). W9: the three NOBURST controls stayed good, with hit fraction
0.949--0.950, the ceiling $1-1/E[T]$ of this workload, completions of 1.66--1.82~$\lref$ in both windows and backlog slopes within
0.001\,/s of zero. W10: the E0a of this workload (one run, seed~0, closed population 32, cache off) measured an all-miss rate of
1.062~$\lref$ turns-based, against the model's $\lmiss=1.046~\lref$; the late completions were 47--57\% of it at 1.65~$\lref$
and 58--80\% at 1.35~$\lref$. Counted in sessions, the E0a's rate depends on the window (0.99--1.17~$\lref$), because synchronized
cohorts complete in waves; every late value lies below the lowest of these. The registered E0a job (17264061) failed its own
pre-run time check, which needed 7080\,s where 5369\,s of the registered 01:30:00 limit were left, so no run started; it was
resubmitted as 17264093 (14:53:08) with \texttt{-{}-time 02:15:00} and nothing else changed, and only W10 uses it.

What the pass shows. The registered alternatives are throughput-curve readings, which predict neither the blackout nor the wave, so
the pass rejects them but does not separate model~F from other FIFO or age-structured models with the same turn count, and no
alternative wave time was registered. M4 is not an independent fourth check: the plan
fixes the arrivals, so the backlog slope is the plan's arrival rate less a weighted completion rate over $[6000,8700)$\,s, nearly
what M3 measures over $[6600,8700)$\,s; the pass rests on three observations, the blackout, the wave and the lull. The late window measures
model~F's post-wave lull (frozen at 0.49--0.73~$\lref$), not its asymptotic $D$ (0.81 and 0.91~$\lref$ at 1.65 and 1.35~$\lref$); at 1.35~$\lref$ a second
completion wave enters the window, and seed~2 completed 1.01~$\lref$ over $[7500,8700)$\,s (model~F 0.87). So the test confirms the plan-driven prediction of the
reversal; it does not measure the asymptotic premium. The NOBURST
controls ran at 1.65~$\lref$ only, by design, so the inference that the burst caused the 1.35~$\lref$ traps rests on them and, post hoc, on
the 1.35~$\lref$ runs' follow-up hit fraction of 1.00 before the burst. The three decisive runs ran concurrently on one node (separate GPUs);
the NOBURST runs use the same base plans as the BURST runs and the same seeds recur at both rates, so the runs are not independent
trials. The test covers one engine, GPU type, model, pool (300k), synthetic workload family and burst, two rates and three seeds
per cell.

\section{Persistence, the trigger, pool occupancy and goodput}\label{app:dataonly}

Four analyses (D1, D2, D5 and the pool occupancy), computed on 2026-11-04/05 after Lu et al.'s paper appeared, use only runs
committed before 2026-11-02; a fifth, D4, a post-hoc
reading of Lu et al.'s model on our cells, is in Appendix~\ref{app:lu}. Each was computed by one agent and then recomputed by a second agent
with its own code from the raw files; where they differ we report the checked value and say so. They are post hoc. Scripts and
outputs: \fpath{analysis/r15_dataonly/} and \fpath{results/r15_dataonly/} (\fpath{PROVENANCE.md} there).

\paragraph{Persistence (D5).} Rolling 300\,s windows of the follow-up hit fraction; a trap starts at the first window after the
trigger with hit fraction at most 0.2, and an escape is any later window with at least 0.5 before the run ends or an intervention
begins (a rescue cap, a SHED stop). In the 50 trapped FCFS runs committed before the blind tests of 2026-11-05 (49 after a burst, registered and exploratory,
and the trapped run of the 900\,s TOOLSLOW arm; the TOOLSLOW-W and CUT runs are not included) there was no escape; the highest window after onset reached 0.164. Trapped time totals 99--103\,h, depending on whether the
trap is taken to start at the beginning or the end of its first window (the first computation gave 104.5\,h, which the recomputation
did not reproduce), and the shortest trapped span is 6100--6400\,s, so a one-sided 95\% upper bound on a run's escape probability
within such a span is $1-0.05^{1/50}=5.8$\%. Stock LPM escaped in 4 of its 9 trapped runs (several barely above 0.5), and LPM with its
fallback disabled in its one trapped run. Of the 16 from-empty runs that collapsed, one recovered within its 3000\,s run (150k pool,
1.30~$\lref$, seed~0).

\paragraph{The trap outlives its trigger (D1).} In 16 stuck runs committed before those tests, every trigger session (the burst's, or for TOOLSLOW the sessions
with a slowed tool call) had left the system 558--6246\,s before the end: the 600k H200 paired test (6 of 6), BASIN bursts of 50 (1),
60 (3) and 100 (3), SHED's 600\,s stop (2) and TOOLSLOW's trapped run (1). The 727--15{,}453 turns completed after the last trigger
session left had no hit in 15 of these runs; in the TOOLSLOW run 4 of 5{,}937 hit, all within 715\,s after its last slowed session left. In
the 24 BURST runs of the four A100 paired tests (SGLang 300k and native, vLLM, Llama) 3--30 burst sessions were still present at the
end (at most 4.1\% of the sessions present; the first computation's ``5.8\%'' could not be reproduced), so for those runs we do not
claim that the trap outlived its trigger.

\paragraph{Pool occupancy in the trap.} From the engines' 5\,s metrics, the KV held by admitted requests (SGLang's used, non-evictable
tokens, which also count locked prefixes; vLLM's cache usage) averaged over the window was 63--79\% of the pool in the six registered
SGLang 300k BURST runs, 55--83\% across all 34 stuck SGLang FCFS runs with 300k and 600k pools, 85--94\% with the native pool and 92--94\%
on vLLM, with follow-up hit fraction zero throughout. (The first computation's ``63--79\% in registered traps'' holds only for the
SGLang 300k runs.)

\paragraph{SLO goodput (D2).} Lu et al.\ define goodput as sessions completed per hour with slowdown at most 4 and at least 95\% of
decode steps meeting an inter-token latency target~\citep{lu2026cachetrap}. We have no per-token latencies, so we omit the second
condition from both arms. That makes each goodput an upper bound but does not bound the loss in either direction, except where no
BURST session meets the slowdown condition; there (26 of the 30 runs in the first hour, all 30 in the evaluation window) the loss is
100\% under either definition, provided the NOBURST goodput stays positive. Slowdown is the session's elapsed time from its planned
arrival, over an unloaded reference built from the NOBURST runs (first computation: per-turn minimum latencies; recomputation: a
least-squares fit of latency on new-suffix and decode tokens; both reported). Against the same-seed NOBURST run, base sessions that
arrived in the hour after the burst lost 99.2--100\% of goodput in all 30 registered BURST runs (no session met the target in 26), and
those arriving in the evaluation window lost 100\% in all 30, for reference latencies up to twice the unloaded ones. In the evaluation
window the CAPPED arms were within 0--1.7\% and the rescue runs within $-0.3$ to $+0.9$\% of NOBURST. In the first hour after the
burst the cap cost 4--57\% (CAPPED) and the rescue 45--97\%; these two ranges depend on the reference, and a reference twice as large
lowers them by a quarter to a third. Lu et al.\ report about 88\% from one run of a different system, which is not comparable in
detail.

\section{The stampede experiments}\label{app:stampede}

This appendix reports the experiments of \S\ref{sec:stampede} in full: the workload, engines and harness, the calibration of each
cell, the quasi-static model that placed the registered rates and caps, the registered tests and their results, the first-principles
model of entry with its frozen predictions, the registrations, the timeline of every registration and prediction, and the
provenance of the data. It condenses the experimental part of an earlier, longer draft of this work, which was withdrawn from arXiv before
announcement and never published (its sources are in the repository history); what was cut is listed in
\fpath{notes/RESTRUCTURE_2026-11-05.md}.

\subsection{Dates and reuse}\label{app:reuse}
The experiments of \S\ref{sec:stampede} (the closed curves, calibrations and from-empty runs, the five registered paired tests,
BASIN-1 and BASIN-2, TOOLSLOW, SHED and the rescue tests R1 and R4) were designed and run, and the tests among them registered,
between 21 October and 1 November 2026, before Lu et al.'s paper appeared on 2 November 2026~\citep{lu2026cachetrap}, and they are reported in full here.
The blind tests TOOLSLOW-W, TOOLSLOW-D15, CUT and MIRROR, and the exploratory REPLICATE-EDGE runs, were registered on
5 November 2026 and run from then on, after Lu et al.'s
paper had appeared; so were the theory of \S\ref{sec:trap} and its frozen predictions. Table~\ref{tab:ledger} lists every
registration with its commit and push time, and Tables~\ref{tab:timeline}, \ref{tab:timeline-vllm}, \ref{tab:timeline-r11}
and~\ref{tab:timeline-r15} give the order of events.

\subsection{Workload, engines and harness}\label{app:harness}

\paragraph{Workload and notation.}\label{sec:notation}
A session has $T\sim\min(\mathrm{Geometric}(1/15),60)$ turns, so $E[T]=14.761$. Turn $k$ sends a prompt of $L_k=L_0+(k-1)(\Delta+d)$ tokens: a first prompt of $L_0=2000$ tokens, $\Delta=550$ tokens of tool output and $d=100$ decoded tokens per turn. Between turns the session waits for a tool for a time $Z$, independent across turns, exponential with mean 5\,s in the workload of record. A turn \emph{hits} if more than half of its prompt is served from cache; on the real engine a hit almost always reuses everything but the 550 new tokens and the last decoded token, so it prefills 551 tokens (all but 95 of the 185,613 hits in the registered paired test E2 of \S\ref{app:e2tests}). First turns always miss, so in the long run the hit fraction is at most $1-1/E[T]=0.932$; finite windows can lie slightly above it (up to 0.94 in our runs). With every turn missing, a session's prefill work is $W(T)=2000\,T+325\,T(T-1)$ tokens with $E[W]=154{,}529$; with every follow-up hitting it is $2000+551(T-1)$, about 9,600 in expectation.

The engine has a KV pool of $M$ tokens. In the \emph{closed} system $n$ sessions are present and a finished session is replaced at once. The thrashing curve $\mu(n)$ is the long-run session completion rate (sessions/s); $\mu^*=\max_n\mu(n)$ is attained at $n^*$, and $\muinf=\lim_{n\to\infty}\mu(n)$ is the \emph{congested service rate}, a long-run rate of the closed system (an open system in congestion can complete sessions faster than $\muinf$ for a long time; Theorem~\ref{thm:premium},
\S\ref{sec:validation}). The all-miss capacity $\lmiss$ is the throughput of the saturated engine with prefix caching disabled. In the \emph{open} system sessions arrive as a Poisson process of rate $\lambda$, and $N(t)$ counts sessions in the system (waiting, running or in a tool call). A session rate is estimated either as sessions completed per second or, with much lower variance because session work is heavy-tailed in $T$, as turns per second divided by $E[T]$ (``turns-based'').

\paragraph{Engines.}
LLM engines batch per model iteration: running requests decode one token per iteration and waiting requests join between iterations~\citep{yu2022orca}; KV is stored in pages~\citep{kwon2023vllm}. SGLang keeps cached prefixes in a radix tree, evicts least-recently-used (LRU) leaves, and can order waiting requests by longest prefix match (LPM)~\citep{zheng2024sglang}. In SGLang 0.5.22 the default policy is FCFS and the default eviction rule LRU; LPM stops sorting its queue when more than 128 requests wait, a fallback that PEEK also documents~\citep{xie2026peek}; a waiting request does not lock its cached prefix, which is locked when the request is admitted; and under FCFS, including LPM above that threshold, an LRU refresh touches the prefixes of the first 128 waiting requests once every 32 scheduling passes. vLLM 0.32.0 schedules FCFS by default, and a freed cached block keeps its hash until it is reallocated. The paragraph `Engine sources' of Appendix~\ref{app:proofs} cites the source lines at the release tags. Our synthesis of these rules is that in both engines a session's KV is protected only while one of its requests runs.

\paragraph{Real-engine harness.}\label{sec:harness}
All real-engine runs used one GPU per job on the Nova SLURM cluster at Iowa State University, an NVIDIA A100 80GB PCIe except in the two H200 cells (one H200 NVL; \S\ref{sec:h200}), with SGLang 0.5.22 (FlashInfer attention~\citep{ye2025flashinfer}, or on the H200 FlashAttention~3~\citep{shah2024flashattention3}, SGLang's default there; chunked prefill of 8192 tokens, at most 256 running requests, LRU eviction) serving Qwen2.5-7B-Instruct~\citep{qwen2024qwen25} or, in one A100 cell, Llama-3.1-8B-Instruct~\citep{grattafiori2024llama3} from a mirror whose weight shards are byte-identical to Meta's. The KV pool was capped at 300,000 tokens, at 150,000 or 600,000 in the pool sweep and at 600,000 in the second H200 cell (\S\ref{sec:frozen}), or, in the native-pool runs, sized by SGLang at 955,757 tokens. Decoding was greedy, 100 tokens per turn (Table~\ref{tab:workload} lists every setting). The vLLM 0.32.0 runs used the same model, GPU, client and workload with a 300,000-token pool (\S\ref{sec:vllm}).

An open-loop client on the same node sends each turn as a completion request whose prompt is a list of token ids: random content from a seeded generator, extended each turn by the token ids the server returned plus 550 new tool tokens, so a hit covers exactly the previous prompt and output; during a tool call the client sleeps. Cached tokens come from the server's usage report, which the engine's counters and log confirmed in the closed-population runs (Stage~C, \S\ref{sec:stages}). Before every run the cache is flushed and checked with a canary: a 1024-token request that reused 1023 tokens before the flush (1008 on vLLM, which caches whole 16-token blocks) must reuse at most 15 after; it reused none. In the cache-off calibration runs (E0a, \S\ref{sec:stages}) the check only confirms that the engine is idle. Each arm of a registered test ran in its own job on a fresh server.

The BURST arm adds 200 fresh sessions within 60\,s at $t_b=1200$\,s to the Poisson stream; NOBURST is the same stream without them; CAPPED is BURST plus a client-side FIFO gate admitting at most $\ncap$ sessions, each holding its slot until its last turn ends (the gate of Proposition~\ref{prop:cap}); the rescue tests of \S\ref{sec:rescue} engage such a gate only after the trap is established. The burst equals 12 to 38 minutes of base arrivals at the registered rates, and its 2000-token first prompts alone (400k tokens) exceed the 300k-token pool; we took its size from our earlier simulator, where bursts of 100 or more trapped every work-conserving policy, so the test asks whether a burst large enough to cross the basin boundary traps the engine, not how likely such a burst is (\S\ref{sec:basin}). The arms of one seed share arrival times, turn counts and tool times (common random numbers), so BURST differs from NOBURST by the burst and CAPPED from BURST by the gate, up to GPU timing nondeterminism: each arm ran on a fresh server in its own job, and the arms of one seed did not always share a node. Seeds 0, 1 and 2 are reused across rates, pools, policy cells, engines and tests: for a given seed the 200 burst sessions are identical everywhere, the arrival times at a lower rate are a subset of those at a higher rate, and the $i$-th session to arrive has the same turn count, tool times and client-generated tokens everywhere, so runs that share a seed are not independent across these.

\begin{table}[htbp]
\centering\footnotesize
\caption{Workload and engines of the real-engine runs.}\label{tab:workload}
\begin{tabular}{@{}l l@{}}
\toprule
First prompt $L_0$ & 2000 tokens\\
Tool output per turn $\Delta$ & 550 tokens\\
Output per turn $d$ & 100 tokens, greedy, \texttt{ignore\_eos}\\
Turns $T$ & $\min(\mathrm{Geometric}(1/15),60)$, $E[T]=14.761$\\
Tool time $Z$ & exponential, mean 5\,s\\
Prompt at turn $k$ & $2000+650(k-1)$ tokens\\
Output feedback & server-returned token ids\\
\midrule
Engine & SGLang 0.5.22, FlashInfer; FA3 on the H200\\
Scheduling & FCFS, waiting-prefix refresh on\\
Batching & chunked prefill 8192, $\le256$ running\\
KV pool & 300,000 tokens; native 955,757\\
 & pool sweep 150,000, 600,000; H200 also 600,000\\
Context limit & 49,152 tokens\\
Model & Qwen2.5-7B-Instruct, rev.\ a09a3545\\
 & one cell: Llama-3.1-8B-Instruct, rev.\ d10aef7\\
 & (NousResearch mirror, shards identical to Meta's)\\
GPU & A100 80GB PCIe, 81,920\,MiB, driver 580.178.04\\
 & two cells: H200 NVL, 143,771\,MiB, same driver\\
\midrule
vLLM & 0.32.0, FCFS (default), prefix caching on\\
 & chunked prefill 8192, $\le256$ running\\
 & 300,000 tokens (18,750 blocks of 16)\\
\bottomrule
\end{tabular}
\end{table}

Table~\ref{tab:workload} lists the workload and engine settings of the real-engine runs.

\subsection{Calibration and the from-empty edge}\label{app:calib}

\paragraph{Staged protocol.}\label{sec:stages}
Each stage's inputs came from the previous one (Table~\ref{tab:stages} gives the job IDs). \emph{Stage~C} measured $\mu(n)$ in closed populations, with all $n$ sessions starting at $t=0$, one run per $n$ and an 1800\,s window after 900\,s of warmup. \emph{E0a} measured the all-miss session rate with the radix cache disabled; its turns-based value $\lref$ is the unit of every arrival rate, and numeric multiples of a rate in this paper are multiples of the $\lref$ measured with the same engine, GPU and model at 300k (0.0715\sps{} on SGLang and the A100, 0.0657 on vLLM, 0.194 on the H200, 0.0596 with Llama-3.1-8B), except that the native-pool cell uses its own (0.0718); the 150k and 600k cells thus use the 300k value of their GPU. Model~F's $\lmiss$ (\S\ref{sec:trap}) is instead the cost model's value, within 4\% of $\lref$ ($\lmiss/\lref=0.96$--$1.035$). \emph{E0b} ran Poisson arrivals from an empty engine for 3000\,s per rate, with arrival seed 0 at every rate; a run is stable if its final queue is below 50 and its queue rises by at most 10 over the last 1500\,s. The registration was then filled and committed, and \emph{E2}, the registered paired test, ran the three arms; the native pool and vLLM repeated the sequence, and the later registered tests follow the paired-test protocol.

\begin{table}[htbp]
\centering\footnotesize\renewcommand{\arraystretch}{0.9}
\caption{Real-engine stages (SLURM job IDs on Nova). Submit, start, end and node of every job: \fpath{provenance/sacct_cited_jobs_2026-11-01.txt} (local time; \fpath{_utc.txt}, UTC; see the text).}\label{tab:stages}
\begin{tabular}{@{}l l p{3.6cm}@{}}
\toprule
Stage & Jobs & What it measures\\
\midrule
Stage 0 & 17060165 & smoke test\\
Stage C & 17062103 & closed $\mu(n)$, 300k pool, FCFS, $n\in\{16,20,24,32\}$ (node gpu-1)\\
 & 17062104 & FCFS, $n\in\{48,96,300\}$ (gpu-3)\\
 & 17062105 & LPM, $n\in\{48,300\}$ (gpu-2)\\
E0a & 17072634--37 & $\lmiss$, cache off\\
E0b & 17084750--55 & from-empty stability\\
E2 & 17093387--404 & registered paired test, 2 rates $\times$ 3 seeds $\times$ 3 arms\\
native C & 17096856 & $\mu(n)$, native pool, $n\in\{24,48,64,80\}$ (gpu-2)\\
 & 17096857 & $n\in\{96,128,300\}$ (gpu-1)\\
native E0a & 17096858 & $\lmiss$, native pool\\
grid & 17097086--103 & BURST arm, 3 policy cells\\
vLLM & 17097972--74 & install, GPU check, smoke\\
native E0b & 17100938--39 & from-empty stability, native pool\\
vLLM E0a & 17100940 & $\lmiss$, vLLM, cache off\\
native E2 & 17102173--90 & registered paired test, native pool, 2 rates $\times$ 3 seeds $\times$ 3 arms\\
vLLM BURST & 17102318--23 & BURST arm, vLLM, exploratory\\
vLLM C & 17162959 & $\mu(n)$, vLLM, 300k pool, $n\in\{16,20,24\}$ (gpu-3)\\
 & 17162960 & $n\in\{28,32,40\}$ (gpu-3)\\
 & 17162961 & $n\in\{48,96,300\}$ (gpu-2)\\
vLLM E0b & 17162962--63 & from-empty stability, vLLM\\
vLLM E2 & 17166121--38 & registered paired test, vLLM, 2 rates $\times$ 3 seeds $\times$ 3 arms\\
\midrule
\multicolumn{3}{@{}l}{\emph{Rounds 11 to 14 (Table~\ref{tab:frozensum})}}\\
H200 & 17166807 & smoke, H200 NVL, 300k pool\\
 & 17166810--13 & E0a and its 64-session check\\
 & 17166818--22 & Stage C, $n=16$--300 (nine runs, with the next row)\\
 & 17166828, 30, 32, 34 & Stage C, the rest of the nine runs\\
 & 17168377--82 & E0b\\
 & 17169668--71 & E0b reseeds, added after 17168377--82 were read (post hoc)\\
 & 17169612--14 & Stage C, $n=36$, 44, 52\\
 & 17169672--80, 82--87 & Stage C, E0b, 600k pool\\
 & 17171020--37 & E2, 600k pool\\
Llama & 17166808 & smoke, Llama-3.1-8B\\
 & 17166814--16 & Stage C\\
 & 17166823, 29, 31, 33 & E0a and its 64-session check\\
 & 17169565--70 & E0b\\
 & 17170409--26 & E2\\
pools & 17168864--81 & Stage C, E0b; 150k, 600k\\
BASIN-1 & 17168500--05 & bursts of 100, then 25\\
E0F & 17168506--07 & from-empty, seeds 3--5\\
LPM-INF & 17168743--45 & BURST, no LPM fallback\\
BASIN-2 & 17169893--901 & bursts of 40, 60, then 50\\
TOOLSLOW & 17169902--07 & $\delta=10$, then 30\,s\\
SHED & 17169908--11 & $\tau=600$, then 3600\,s\\
rescue & 17168965 & smoke test of the gate (never decisive)\\
 & 17168966--71 & R1, 300k pool, 2 rates $\times$ 3 seeds\\
 & 17168972--74 & R4, native pool, 3 seeds\\
\bottomrule
\end{tabular}
\end{table}

\noindent The accounting export \fpath{provenance/sacct_cited_jobs_2026-11-01.txt} was made after the last cited job had ended; its times are Slurm's local time, CDT until 02:00 on 2026-11-01 and CST after it, and \fpath{provenance/sacct_cited_jobs_2026-11-01_utc.txt} gives the same records in UTC. The earlier export \fpath{provenance/sacct_cited_jobs_2026-10-30.txt} covers every job through the vLLM E0b, and the data directories of the vLLM paired test and of the BASIN-1, fresh-seed, rescue, pool-sweep, BASIN-2, TOOLSLOW and SHED jobs hold their own copies (\fpath{sacct_17166121-17166138.txt}, \fpath{sacct_17168500-17168507.txt}, \fpath{sacct_1716896x.txt}, \fpath{sacct_17168864-17168881.txt}, \fpath{sacct_17169893-17169901.txt}, \fpath{sacct_17169902-17169911.txt}).

\paragraph{Stage 0.} The smoke test (job 17060165) verified that every engine flag exists; that cached tokens are reported (a repeated 64-token prompt had 63 tokens cached); that the server returns output token ids; that contexts up to 40,350 tokens are accepted; that the KV pool is exactly 300,000 tokens; that client and server agree on prompt length (0 mismatches over 3191 requests); and that the flush empties the cache (canary 1023 cached tokens before, 0 after).

\paragraph{Stage C and E0a.} Closed population; all sessions start at $t=0$ and a finished session is replaced at once; 900\,s warmup and an 1800\,s measured window; one run per $n$ (seed 0); flush and canary check between runs. The native-pool and vLLM closed curves used the same protocol, with the populations of Table~\ref{tab:stages}. E0a disables the radix cache and runs 32 sessions with seeds 0, 1 and 2, plus 64 sessions with seed 0 as a saturation check. Its per-run values over the window were (sessions/s, turns-based sessions/s, uncached tokens/s): seed 0, 0.0567, 0.0643, 11,076; seed 1, 0.0817, 0.0773, 11,342; seed 2, 0.0767, 0.0730, 11,312; 64 sessions, 0.0544, 0.0676, 11,431. The spread comes from in-flight composition: the mean turn index of the requests in the window was 15.9, 13.2 and 14.1 for seeds 0, 1 and 2. vLLM E0a ran the same protocol with vLLM's prefix caching off: seed 0, 0.0539, 0.0603, 10,208; seed 1, 0.0728, 0.0705, 10,487; seed 2, 0.0700, 0.0662, 10,352; 64 sessions, 0.0467, 0.0631, 10,531.

\paragraph{E0b.} Poisson session arrivals from an empty cache and queue at 1.25, 1.5, 1.75, 2.0, 2.5 and 3.0 $\lref$; one 3000\,s run per rate; stable if the final queue is below 50 and the fitted rise of the queue over the last 1500\,s is at most 10. All runs used arrival seed 0, so they share one random-number stream and the arrivals at each rate are a subset of those at the next higher rate. The six runs received 302, 359, 416, 472, 585 and 697 sessions, 8.3--12.6\% above $\lambda T$ and about 2.1 Poisson standard deviations above it in every run. Native E0b used the same rule at 1.5, 2.5, 3.5, 4.5, 5.5 and 6.5 times the native $\lref$ (361, 587, 811, 1034, 1257 and 1478 sessions, 5.6--11.7\% above $\lambda T$), and vLLM E0b the multiples of the 300k runs of vLLM's $\lref$ (279, 331, 383, 435, 539 and 642 sessions, 8.6--13.3\% above $\lambda T$; stable through 2.0, final queue 81.6 and 247.1 at 2.5 and 3.0).

\begin{figure*}[t]
\centering
\includegraphics[width=0.92\textwidth]{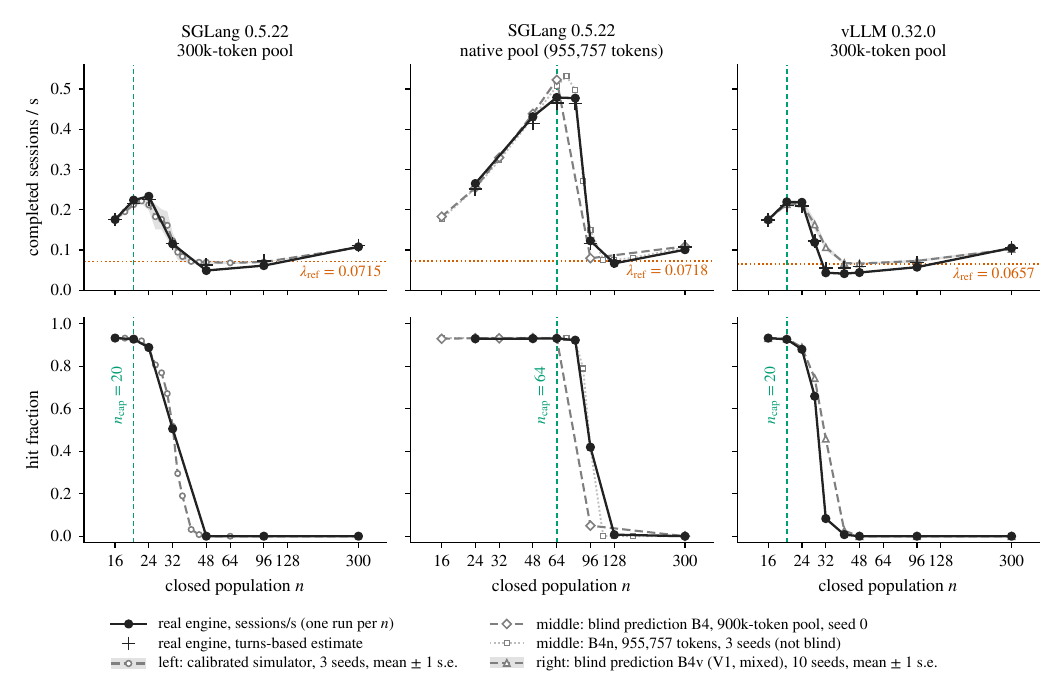}
\caption{Closed-population thrashing curve, real engine (black, filled; one run per $n$) against the simulator (gray, open, dashed). Top: $\mu(n)$ in completed sessions/s (plus signs: turns-based estimate); bottom: hit fraction. Left: SGLang, 300k-token pool; calibrated simulator, mean of 3 seeds $\pm$1 s.e. Middle: SGLang, native pool (955,757 tokens); blind prediction B4 at 900k tokens (seed 0); light gray, dotted: B4n at the real pool (3-seed mean), which is not blind. Right: vLLM 0.32.0, 300k-token pool; blind prediction B4v (variant V1: token costs scaled to vLLM's all-miss rate; mixed batching), mean of 10 seeds $\pm$1 s.e. (\S\ref{sec:vllm}). Dotted: each cell's measured $\lref$, the turns-based E0a estimate of $\lmiss$. Green dashed: the registered caps, $n=20$, 64 and 20.}\label{fig:curve}
\end{figure*}

\paragraph{Closed curves and from-empty runs.}
With the 300k pool (Fig.~\ref{fig:curve}, left), $\mu(n)$ rises from 0.176\sps{} at $n=16$ to 0.224 at 20 and 0.233 at 24, with hit fractions 0.93, 0.93 and 0.89. The run at $n=24$, which sets $\mu^*$, contains a degraded episode inside its window (hit fraction 0.66 over $[900,1200)$\,s, 0.46 in its worst 100\,s); over $[1200,2700)$\,s alone $\mu(24)=0.257$ against $\mu(20)=0.231$, so the near-tie that sets the cap of 20 owes part of its closeness to this episode. The run at $n=32$ mixes two regimes: degraded from about 450 to 2400\,s, then recovered to 0.80, for a hit fraction of 0.51. At $n=48$ none of the 20.0M prompt tokens in the window was served from cache: evictable tokens survive about 6.3\,s before eviction (estimated as mean evictable KV over the eviction rate) while a session returns after 30.6\,s on average (2.9\,s against 66\,s at $n=96$). The hit fraction crosses 0.5 at $n=32.2$, about 3.7 sessions above the capacity ratio $\Nlb\approx28.5$ and just below the all-miss onset $n_b=32.7$ of the first-principles model, which predicts this curve without using its points (Appendix~\ref{app:entry}). Token-level hit rates have been reported to collapse as concurrent agents grow on SGLang~\citep{chen2026concur} and on vLLM~\citep{yuan2026agentic,hong2026misat}, and, on prefill--decode disaggregated SGLang, once the working set reaches about twice the cache~\citep{lu2026cachetrap}; what we add is $\mu(n)$ in completed sessions per second, the quantity the open-system model uses, at four pool sizes on the A100, two on an H200 and, in \S\ref{sec:vllm}, on a second engine. Beyond the peak the curve is not monotone: it falls to $\mu(48)=0.049$\sps{} and rises again, $\mu(96)=0.061<\mu(300)=0.108$\sps. We attribute both to the synchronized start, which leaves old, long turns in the window at $n=48$ (served prompts 14\% above the stationary mean) and young sessions with short prompts at $n=300$ (71\% of the stationary length in flight); no longer run has tested this; the all-miss level implied by token throughput is roughly flat at 0.072--0.081\sps, so the collapse from the peak is about threefold. Stock LPM keeps 39\% of turns hitting at $n=48$ (rising from 0.28 to 0.44 through the window); at $n=300$ its queue of 269 exceeds the 128-request threshold above which, by the source, SGLang's LPM stops re-sorting its queue, and no turn hits; whether the fallback causes the zero was not tested (at $n=300$ the pool may be oversubscribed under any ordering).

With the native pool (Fig.~\ref{fig:curve}, middle), SGLang sized the KV pool at 955,757 tokens, 3.19 times the 300k pool. $\mu$ peaks at 0.479\sps{} at $n=64$ (0.477 at 80) with 92--93\% of turns hitting, then falls to 0.123 at $n=96$ (hit fraction 0.42), 0.067 at 128 (0.007) and 0.101 at 300. The hit fraction crosses 0.5 between $n=80$ (0.92) and 96 (0.42), near 93 by linear interpolation, against just above 32 at 300k; the peak is at 64--80 against 20--24. With one run per $n$, the cliff moves out roughly threefold. The all-miss capacity is unchanged within its standard error: $0.0718\pm0.0037$\sps{} turns-based against $0.0715\pm0.0038$, and $11{,}282\pm51$ against $11{,}243\pm84$ all-miss tokens/s. The 300k-token pool limit did not create the cliff; it moved it. The ratio $\mu^*/\lref$ is 6.67 with the native pool and 3.26 at 300k.\footnote{These are the registered values: the peak in completed sessions over the turns-based $\lref$. Turns-based over turns-based gives 6.48 and 3.15.}

\paragraph{The all-miss capacity and the from-empty edge.}\label{sec:edge}

E0a gives $\lref=0.0715\pm0.0038$\sps{} turns-based (standard error over three seeds), our estimate of $\lmiss$, $0.0717\pm0.0076$ in completed sessions, and $11{,}243\pm84$ all-miss prefill tokens/s. The token rate varies by $\pm1.2\%$ across seeds and the session rate by $\pm17\%$: the engine is bound by prefill tokens, and the sessions completed in 1800\,s are a small sample of a heavy-tailed quantity. The token rate divided by $E[W]$ gives 0.0728\sps, within 2\% of the turns-based value.

\begin{figure}[t]
\centering
\includegraphics[width=\columnwidth]{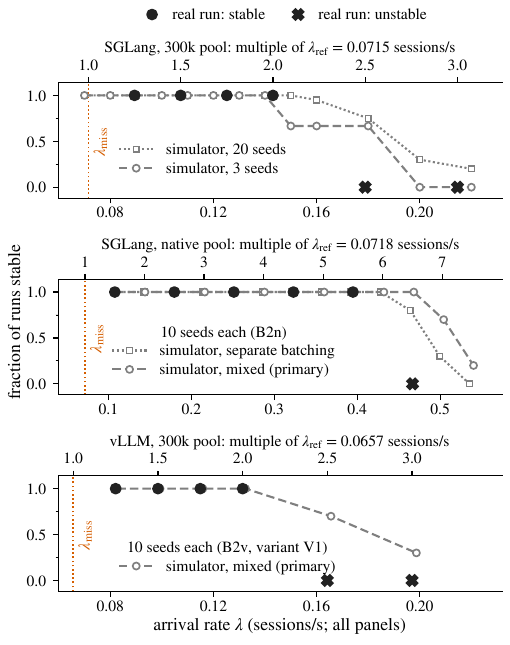}
\caption{From-empty stability. Real engine: one 3000\,s run per rate from an empty cache and queue (1 = stable, 0 = unstable), plotted at its nominal rate; all runs used arrival seed 0, whose realized rates were 6--13\% higher (the fresh-seed runs of \S\ref{sec:edge} are not plotted). Gray: blind simulator predictions, the fraction of simulated runs stable at multiples of the simulator's own $\lmiss$. Top, SGLang with the 300k pool: B2 (3 seeds dashed, 20 dotted). Middle, SGLang with the native pool: B2n (10 seeds; mixed batching, the primary mode, dashed; separate, dotted). Bottom, vLLM 0.32.0 with the 300k pool: B2v (variant V1, costs scaled to vLLM; mixed batching, 10 seeds). Top axes: multiples of each cell's $\lref$ (the simulators' own $\lmiss$ are 0.0722 at 300k, 0.0720 and 0.0714 native, and 0.0662 for B2v).}\label{fig:edge}
\end{figure}

From an empty cache and queue (Fig.~\ref{fig:edge}, top) the engine stayed stable at 1.25, 1.5, 1.75 and 2.0 $\lref$, ending each run with an empty queue and with 91--93\% of turns hitting over the run, and was unstable at 2.5 $\lref$ (final queue 127, rising by 146 over the last 1500\,s) and 3.0 (289 and 196). On this grid, which used the single arrival seed 0, the edge lies between 2.0 and 2.5 $\lref$ in nominal rates, well below $\mu^*/\lref=3.26$, so the from-empty edge, not the peak, bounds the window. The registered window was $(1,\,2.0)\,\lref=(0.0715,\,0.143)$\sps. Each rate is one 3000\,s run; a longer horizon could find a lower edge, and at 2.0 $\lref$ the hit fraction fell from 0.934 in $[300,1500)$\,s to 0.903 in the second half of the run. All six runs used seed 0, so they share one random-number stream: their arrival sequences are nested, and each arrival count lies about 2.1 standard deviations above its Poisson mean, 8--13\% above the nominal rate (302, 359, 416, 472, 585 and 697 sessions; the runs labelled 2.0 and 2.5 $\lref$ received 2.20 and 2.73 $\lref$). This makes the stable side of the bracket conservative and its unstable side weaker than the labels suggest; in realized arrivals the bracket is still below $\mu^*/\lref$.

An exploratory replication with fresh seeds 3, 4 and 5 at 2.0 and 2.5 $\lref$ (jobs 17168506--07; \S\ref{sec:frozen}) found all six runs stable, each ending with an empty queue and with 92--93\% of turns hitting. Its frozen rule asked for 2.0 $\lref$ to be stable in all three seeds and 2.5 $\lref$ unstable in at least two, so the seed-0 bracket is not replicated: near 2.5 $\lref$ the nominal edge depends on the seed (2.5 $\lref$ was stable in 3 of 4 seeds). The rule's fallback restates the bracket in realized arrival rates. Over the twelve from-empty runs at this pool (seeds 0 and 3--5), every run that received at most 2.665 $\lref$ (0.191\sps) was stable and the two that received 2.726 and 3.248 $\lref$ were unstable, so stability is monotone in the realized rate and the edge lies between 2.665 and 2.726 $\lref$, still below $\mu^*/\lref$. The registered window was derived from the nominal seed-0 bracket and stands as frozen; the registered points, 1.35 and 1.65 $\lref$, lie inside the from-empty stable range on either reading. The calibrated simulator had predicted all six fresh-seed runs stable, blind in time. The heuristic H2 of Appendix~\ref{app:entry} was pessimistic: it puts the per-seed probability of instability at 0.15 at 2.0 and 0.69 at 2.5 $\lref$, so three stable runs at 2.5 had probability 0.03 under it (\S\ref{sec:frozen} states which of its values are blind).

With the native pool (Fig.~\ref{fig:edge}, middle; native E0b, rates in multiples of the native $\lref=0.0718$\sps) the engine stayed stable at 1.5, 2.5, 3.5, 4.5 and 5.5 $\lref$, ending each run with an empty queue and with 93\% of turns hitting over the run, and was unstable at 6.5 $\lref$: its final queue was 505, rising by 477 over the last 1500\,s, and its hit fraction fell from 0.84 in $[300,1500)$\,s to 0.00 in the second half. In nominal rates the edge lies between 5.5 and 6.5 $\lref$ (0.395 and 0.467\sps), so the native window is $(1,\,5.5)\,\lref$. Unlike at 300k, this does not show the edge below the peak ($\mu^*/\lref=6.67$, or 6.48 turns-based): these runs also used seed 0 (361 to 1478 sessions, 5.6--11.7\% above nominal), and the 6.5 run received 0.493\sps{} (6.86 $\lref$), above the measured peak, so overload alone can explain its instability. The stable 5.5 run received 0.419\sps{} (5.84 $\lref$). Its upper edge is 2.75 times higher than at 300k in units of $\lref$ (5.5 against 2.0), and the window is $4.5\,\lref$ wide against $1.0$. Again each rate is a single 3000\,s run.

\subsection{The quasi-static model, the window and the cap}\label{app:qs}

\paragraph{The thrashing curve for agent sessions.}\label{sec:theory-curve}
Let $\Elb$ be the time-averaged KV footprint of a session over its lifetime, a length-biased mean in which long sessions and late turns weigh more (during the tool call after turn $k$ a session holds $L_k+d$ tokens). By the $H=\lambda G$ relation~\citep{brumelle1971relation,heyman1980relation}, the mean KV demand of $n$ present sessions is $n\Elb$. For our workload $\Elb$ is about 10.5k tokens when tool calls dominate the lifetime, so a 300k pool holds $\Nlb=M/\Elb\approx28.5$ sessions on average (weighting every turn equally, the last included, gives 10.6k and 28.4). While $n\Elb$ is well below $M$, idle KV survives the tool call and $\mu(n)$ rises with $n$ through batching. Beyond it, admitting a waiting turn evicts idle KV, which belongs to sessions in tool calls or already waiting in the queue; they miss when their next turn is admitted. The peak $n^*$ lies below $\Nlb$ mainly because idle KV is evicted at a finite LRU age, so hits fall before mean demand reaches the pool, and to a lesser degree because demand fluctuates around its mean. The capacity ratio is the collapse predictor of AgentKV~\citep{anon2026agentkv} and the basis of MISA-T's session cap~\citep{hong2026misat}. Appendix~\ref{app:entry} derives the curve from first principles; its all-miss onset at 300k, $n_b=32.7$, is the sessions in tool calls plus one pool of running sessions (Proposition~\ref{prop:nb} of \S\ref{sec:fp-results}), so the measured peak, the capacity ratio and the onset are ordered $n^*=24<\Nlb<n_b$.

\paragraph{The open system as a quasi-static approximation.}\label{sec:fluid}
Replacing the departure rate of the open system by its closed-system value at the current population, the classical decomposition approximation~\citep{courtois1975decomposability,kleinrock1975aloha}, gives
\begin{equation}\label{eq:fluid}
\dot n=\lambda-\mu(n).
\end{equation}
Extend $\mu$ from integer $n$ to all $n\ge0$ by linear interpolation, so that it is continuous and piecewise linear with $\mu(0)=0$ and Eq.~\eqref{eq:fluid} has unique solutions, and assume that the stationary curve is \emph{unimodal}: strictly increasing on $[0,n^*]$ to $\mu^*$, non-increasing on $[n^*,\infty)$ with limit $\muinf<\mu^*$, and strictly decreasing wherever $\mu>\muinf$. This allows a tail on which $\mu=\muinf$, as on the predicted curves of Appendix~\ref{app:entry}; the statements below are for $\lambda\ne\muinf$. For $0<\lambda<\mu^*$ let $n_1<n^*$ solve $\mu(n_1)=\lambda$. If $\lambda<\muinf$, $n_1$ is the only fixed point of Eq.~\eqref{eq:fluid}, and it attracts every trajectory. If $\muinf<\lambda<\mu^*$, there is one more, $n_2>n^*$, which is unstable, and from any $n>n_2$, $n\to\infty$: the good state is metastable. If $\lambda>\mu^*$ there is no good state. A solution from a congested state $x_0>n^*$ returns to $n_1$ if and only if $\mu>\lambda$ on $(n_1,x_0]$, the endpoint $n_1$ excluded because $\mu(n_1)=\lambda$; since $\mu>\lambda$ on $(n_1,n^*]$ and $\mu\ge\mu(x_0)$ on $[n^*,x_0]$, this holds if and only if $\lambda<\mu(x_0)$. As $x_0$ grows, $\mu(x_0)$ falls to $\muinf$, so $\muinf$ is the \emph{recovery threshold} from deep congestion. After a finite burst $x_0$ is finite, so recovery above $\muinf$ is possible; the threshold for a given burst depends on its size. The end of this subsection proves these statements and treats a curve without the monotone tail, such as our measured one.

In a finite system fluctuations can carry $N$ past $n_2$ without a burst, and an experiment sees only escapes within its horizon. We therefore use a measured \emph{from-empty edge} $\ledge$, the largest rate at which runs started from an empty engine stay stable over a stated horizon, and define the window
\begin{equation}\label{eq:window}
\mathcal W=\big(\muinf,\ \min(\mu^*,\ledge)\big).
\end{equation}
Inside $\mathcal W$ the good state exists from empty, and a burst that crosses the basin boundary leaves the system in a regime whose service rate is below the arrival rate. That regime is transient (the backlog grows), not a second stationary state, so we say metastable rather than bistable; two stable states, the bistability of ALOHA-type channels~\citep{carleial1975bistable} and the stochastic cusp of multiprogramming models~\citep{nelson1987stochastic}, arise here only for the admitted population of a capped engine ($c$ sessions at $\mu(c)$ while the gate grows; Corollary~\ref{cor:rescue}) or, when the closed curve folds (dispersed tool times, not our workload), in the closed system itself, where stable means stable under a relaxation of the hit probability at fixed $n$ (not shown here).

Eq.~\eqref{eq:fluid} assumes that the internal state relaxes quickly compared with changes in $n$, and it is not quantitative in congestion, where a trapped engine serves a young backlog of short turns; we use it for the window, as a long-run statement for the recovery threshold, and for the capped system's drain time, not for burst sizes (\S\ref{sec:fp-weak}) or backlog growth (Theorem~\ref{thm:trap}, \S\ref{sec:validation}).

Without a cap, deep congestion of the closed system is bounded by the all-miss capacity: under request-level FCFS and assumptions (A1)--(A3) of Appendix~\ref{app:proofs} (one of which is that the congested engine computes prompt tokens no faster than with caching disabled), no online eviction rule, including queued-prefix protection and FCFS-order TTL pins, lets $\mu(n)$ exceed $\lmiss$ as $n\to\infty$, and the rate of hit turns tends to 0 (Proposition~\ref{prop:A}). It covers SGLang's stock LPM through its 128-request fallback, but not a pure resident-first order; Appendix~\ref{app:proofs} states and proves it and discusses its scope. Under request-level FCFS and (A1)--(A3), then, $\muinf\le\lmiss$, with equality if the congested engine also computes prompt tokens as fast as with caching disabled ((A4) of Appendix~\ref{app:proofs}), which our engines do to within a few percent at the stationary turn mix and to within 12\% over the windows we measured. We therefore take the lower end of $\mathcal W$ to be $\lmiss$, estimated by $\lref$, and write windows as $(1,x)\,\lref$; if $\muinf<\lmiss$, $\mathcal W$ is wider, and every rate in $(\lmiss,\min(\mu^*,\ledge))$ is still in it. We do not read $\muinf$ off the measured closed curves, which are not monotone beyond the peak (at 300k, 0.68~$\lref$ at $n=48$ and 1.51~$\lref$ at $n=300$ in completed sessions); we attribute both features to their synchronized start, which no longer run has tested (\S\ref{app:calib}).

\begin{proposition}[cap identity; elementary]\label{prop:cap}
Let a FIFO gate admit at most $c$ sessions, each holding its slot from admission to the end of its last turn, tool calls included. In the closed system with $n$ sessions the admitted sessions form a closed system of $\min(n,c)$ sessions, so $\mu_{\mathrm{cap}}(n)=\mu(\min(n,c))$.
\end{proposition}
\begin{proof}
When an admitted session finishes, the gate admits its head at once and the replacement joins the gate.
\end{proof}

\begin{corollary}[cap rescue; quasi-static]\label{cor:rescue}
Let $\mu$ be unimodal as in \S\ref{sec:fluid}, let $\muinf<\lambda<\mu^*$, and let $n_1<n_2$ be the stable and the unstable fixed point of Eq.~\eqref{eq:fluid}. Under Eq.~\eqref{eq:fluid} with the departure rate $\mu_{\mathrm{cap}}$ of Proposition~\ref{prop:cap}, a FIFO cap $c$ engaged when more than $c$ sessions are present (the congested regime included) empties its gate if and only if $n_1<c<n_2$, that is, $\mu(c)>\lambda$. Once the gate is empty the cap no longer binds, and removing it leaves the population in the basin $[0,n_2)$ of $n_1$, where it stays.
\end{corollary}
The proof is at the end of this subsection. That the good state persists after the cap is removed is hysteresis of the system, which the rescue tests of \S\ref{sec:rescue} check on the engine; it differs from the cache hysteresis that Lu et al.\ report in simulation, a lower hit rate after a collapse at a comparable working set~\citep{lu2026cachetrap}, and from the dead band of a controller that pauses and resumes admission at two thresholds. A cap $c>n^*$ raises the congested level to $\mu(c)$ but leaves the window $(\mu(c),\mu^*)$ open. A burst of $b$ sessions that arrives when $n_1<c$ sessions are present fills the $c-n_1$ free slots at once, so the gate empties after about $(n_1+b-c)/(\mu(c)-\lambda)$. On the engine the gates took 3--52\% longer than this to empty at 300k, from 4\% less to 13\% more with the native pool, and 1.05--1.63 times as long on vLLM. The registrations' feasibility check used the cruder $b/(\mu(c)-\lambda)$. The proof below also covers a folding curve and comments on the stochastic counterpart.

\paragraph{The fluid statements of \S\ref{sec:fluid}.} With $\mu$ continuous and piecewise linear, Eq.~\eqref{eq:fluid} has a unique solution from every $x_0\ge0$; it is monotone in time, and it converges to a fixed point or diverges. Let $0<\lambda<\mu^*$. Then $\mu(n)=\lambda$ has exactly one solution $n_1$ in $(0,n^*)$, because $\mu$ increases strictly from 0 to $\mu^*$ there, and, since $\lambda\ne\muinf$, at most one in $(n^*,\infty)$, because there $\mu$ is non-increasing and strictly decreasing wherever $\mu>\muinf$. If $\lambda<\muinf$, then $\mu\ge\muinf>\lambda$ beyond $n^*$, so $\mu<\lambda$ on $[0,n_1)$ and $\mu>\lambda$ on $(n_1,\infty)$, and every solution converges to $n_1$. If $\muinf<\lambda$, the second solution $n_2$ exists, because $\mu(n^*)>\lambda$ and $\mu\to\muinf<\lambda$; then $\mu>\lambda$ on $(n_1,n_2)$ and $\mu<\lambda$ on $(n_2,\infty)$, so $n_2$ is unstable, and for $x_0>n_2$, $\dot n\ge\lambda-\mu(x_0)>0$ along the solution and $n\to\infty$. If $\lambda>\mu^*$, then $\mu<\lambda$ everywhere and every solution diverges. A solution from $x_0>n_1$ converges to $n_1$ if and only if $\mu>\lambda$ on all of $(n_1,x_0]$: if so, it decreases and stays in $(n_1,x_0]$, where $n_1$ is the only possible limit; if $\mu(x)\le\lambda$ for some $x\in(n_1,x_0]$, it never falls below $x$, where the drift is non-negative. For $x_0>n^*$, $\mu>\lambda$ on $(n_1,n^*]$ and the minimum of $\mu$ on $[n^*,x_0]$ is $\mu(x_0)$, so the condition is $\lambda<\mu(x_0)$, and $\mu(x_0)$ does not increase and tends to $\muinf$ as $x_0\to\infty$. Without the monotone tail the same argument gives the condition $\lambda<\min_{[n^*,x_0]}\mu$, which tends to $\inf_{n\ge n^*}\mu(n)$, and a point beyond $n^*$ where $\mu$ crosses $\lambda$ from below is a stable congested fixed point. Under the assumptions of Proposition~\ref{prop:A} this infimum is at most $\liminf_{n\to\infty}\mu(n)\le\lmiss$, and it equals $\lmiss$ if, in addition, $\mu(n)\ge\lmiss$ for every $n\ge n^*$.

\paragraph{Proof of Corollary~\ref{cor:rescue}.} While $N\ge c$ the admitted sessions form a closed system of $c$ sessions (Proposition~\ref{prop:cap}), so $\dot N=\lambda-\mu(c)$ and the gate, $N-c$, shrinks if and only if $\mu(c)>\lambda$. By unimodality $\mu>\lambda$ exactly on $(n_1,n_2)$, so the gate empties, from any initial population $N_0>c$, if and only if $n_1<c<n_2$; if $c\le n_1$ or $c\ge n_2$, then $\mu(c)\le\lambda$ and it never empties. Once $N<c$ the cap does not bind, $\dot N=\lambda-\mu(N)$ with $N<c<n_2$, and by the statements above $N$ stays in $[0,n_2)$ and converges to $n_1<c$, so the cap never binds again and removing it changes nothing. This is the standard closed-system equivalence of a system with a bounded multiprogramming level~\citep{denning1976optimal}. If the closed curve itself folds, $\mu$ is multivalued on a band of populations and we take $n_2$ on the good branch; a cap $c\ge n_b$ engaged inside the trap keeps the all-miss state, which exists at every $n\ge n_b$, so $n_1<c<\min(n_2,n_b)$ is necessary, and we have not shown that it is sufficient. The stochastic counterpart (positive recurrence for $\lambda<\mu(c)$) should follow from standard arguments; we have not written them out.

\subsection{The registered paired tests}\label{app:e2tests}

\paragraph{Decision rule and registration.}\label{sec:prereg}
Each run of a registered test lasts 8700\,s and is judged over $[t_b+4500,\,8700]$\,s; in the rule, $\lmiss$ is the cell's $\lref$. A BURST run is TRAPPED if $\lambda>\lmiss$ and (a) its uncached prefill token rate is within $\pm15\%$ of the cell's E0a all-miss rate (11,243 tokens/s at 300k) and (b) the slope of its backlog exceeds $0.5(\lambda-\lmiss)$, the slope being the least-squares slope of 10\,s bin means of the sessions in the system (arrived and not ended, including any held at the gate) over the window. A run is RECOVERED if over the last 300\,s its hit fraction exceeds 0.8 and its outstanding backlog (requests in flight plus sessions held at the gate) is below 20; NOBURST is good if RECOVERED. A run that is neither is DRAINING if its slope is negative and INDETERMINATE otherwise; the labels take precedence in the order trapped, recovered, draining, indeterminate. A rate PASSES if, in at least two of three seeds, BURST is trapped, NOBURST good and CAPPED recovered in that same seed. Criterion (a) uses tokens because a trapped engine with a growing backlog completes sessions at a rate other than $\lmiss$ (faster for our workload, slower with a 20-step budget, \S\ref{sec:mirror}) while it prefills at the all-miss token rate (Theorem~\ref{thm:premium}, \S\ref{sec:validation}); the paragraphs below give the evidence.

The registration was filled by formula from committed copies of the Stage~C curve, the E0a calibration and the E0b bracket: the window $(1,\,2.0)\,\lref$, its upper edge E0b's largest all-stable multiple; the rates one third and two thirds of the way across it, rounded to 0.05 (1.35 and 1.65 $\lref$), each of which must lie strictly between 1.1 times the window's lower edge and 0.9 times its upper edge, here $(1.1,\,1.8)\,\lref$, or the fill returns INCOMPLETE; and the cap, the smallest $n$ with $\mu(n)\ge0.95\mu^*$, which gives $\ncap=20$. We call a test pre-registered when its registration file was committed and pushed to GitHub before its runs, with GitHub's push log of our development repository as the record (not an entry in a third-party registry; pre-registered gates have been used in ML-systems measurement before~\citep{suram2026cacheable}); that repository is private, so readers see the log only as our export of it in \fpath{provenance/} (Appendix~\ref{app:prov}), which cannot be checked against GitHub independently. We call a test registered when a FILLED registration file (Table~\ref{tab:ledger}) binds its runs and was committed before its jobs were submitted; neither label says whether a prediction for the test was blind (\S\ref{sec:frozen}). The E2 and native tests are pre-registered, and so is every later registered test but one, although the text mostly calls these simply registered: the rescue tests R1 and R4 (\S\ref{sec:rescue}), the basin tests BASIN-1 and BASIN-2 (\S\ref{sec:basin}) and the TOOLSLOW and SHED tests, whose registrations reached GitHub 22 to 131 minutes before their jobs were submitted, and the 600k H200 and Llama-3.1-8B paired tests, whose registrations reached it 63 and 47\,s before theirs (\S\ref{sec:h200llama}). The exception is vLLM: its registration reached GitHub after its jobs were submitted, when two of them had started, so that test is registered, not pre-registered. On the H200 with the 300k pool the rule could not place its test points, so that cell has no registered test (\S\ref{sec:h200}). Appendix~\ref{app:prereg} gives the fill, its enforcement by the job script and the push time of every registration, and Appendix~\ref{app:timeline} every prediction round with its time record.

\paragraph{E2.} One arm per 8700\,s job, each on a fresh server; burst of 200 fresh sessions within 60\,s at $t_b=1200$\,s for BURST and CAPPED; CAPPED gate $\ncap=20$; evaluation window $[t_b+4500,\,8700]$\,s. The registered rule (Table~\ref{tab:rule}) was applied by the analysis tool frozen at the registration commit, with the same parameter overrides that the jobs themselves used. Native E2 used the same protocol with the native registration's rates, reference rates and gate $\ncap=64$. Twelve of its 18 runs started after the cluster checkout moved from the registration commit to \texttt{a78ce07}, which adds only the vLLM registration, its input file and two notes files (the research log and the status notes), with no code change, and its verdict ran on that commit; two of its 18 servers sized the pool one token smaller, at 955,756. The vLLM paired test uses the same protocol with the vLLM registration and $\ncap=20$. It uses the same seeds, rates and reference rate as the exploratory vLLM BURST runs, so its BURST arm repeats their workloads rather than replicating them independently.

\paragraph{Why the rule uses tokens.} Criterion (a) uses tokens because a trapped engine with a growing backlog serves mostly young sessions, whose turns are short, so at the all-miss token rate it completes more turns per second than the stationary all-miss mix, and, for our workload, sessions faster than $\lmiss$ (\S\ref{sec:validation}; slower with a 20-step budget, \S\ref{sec:mirror}): on traps simulated in our earlier simulator (\S\ref{sec:sim}), a session-rate test fired on at most 1 of 29 at 2 $\lmiss$ and none of 39 at 2.5, while the token test fired on all 84 traps at 1.5--2.5 $\lmiss$.

\begin{table}[htbp]
\centering\footnotesize
\caption{The registered decision rule.}\label{tab:rule}
\begin{tabular}{@{}l l@{}}
\toprule
Rate estimator & uncached prefill tokens/s\\
Reference rate & E0a all-miss, 11,243 tokens/s\\
 & \quad (native: 11,282; vLLM: 10,349)\\
$\lmiss$ in the rule & the cell's $\lref$\\
Rate tolerance & $\pm0.15$\\
Backlog slope & least squares over 10\,s bin means of\\
 & \quad sessions in the system, gate included\\
Minimum window & 600\,s\\
Trapped & $\lambda>\lmiss$, rate within tolerance,\\
 & \quad and slope $>0.5\,(\lambda-\lmiss)$\\
Recovered & hit $>0.8$ and outstanding $<20$, last 300\,s\\
Draining & slope $<0$\\
Indeterminate & none of the above\\
Precedence & trapped, recovered, draining, indeterminate\\
Hit metric & turns with cached/prompt $>0.5$\\
Pass & $\ge2/3$ of $\ge3$ seeds with all three\\
 & \quad arm conditions met in the seed\\
\bottomrule
\end{tabular}
\end{table}

\paragraph{Backlog definitions.} The figures and Table~\ref{tab:e2} report the backlog as sessions in the system (arrived minus ended), which includes sessions asleep in tool calls (8--12 in the good state at 300k, 16--34 with the native pool). The registered RECOVERED test uses the outstanding backlog (requests in flight plus sessions held at the gate). The simulator's predictions count waiting requests plus the gate queue. The rescue rule's outstanding work also counts sessions paused by the gate (Fig.~\ref{fig:rescue}).

\begin{table*}[t]
\centering\footnotesize
\caption{The three registered tests: the two pre-registered SGLang tests, E2 (300k-token pool, cap 20, $\lref=0.0715$\sps) and native E2 (955,757-token pool, cap 64, $\lref=0.0718$\sps), and the registered vLLM test (300k-token pool, cap 20, $\lref=0.0657$\sps). States for seeds 0, 1, 2: T trapped, I indeterminate, D draining, G good, R recovered. Hit fraction, uncached prefill rate in multiples of the all-miss rate, mean backlog (sessions in the system) and its slope (sessions/s) over the evaluation window; ranges over seeds; good-arm slopes in absolute value. Slope thresholds: 0.0125, 0.0232, 0.0538 and 0.1077 at 1.35, 1.65, 2.5 and 4.0 $\lref$ on SGLang, 0.0115 and 0.0213 at 1.35 and 1.65 $\lref$ on vLLM. Rates are nominal; the base arrivals realized in the window, which the three arms of a seed share, were 1.45, 1.27, 1.42; 1.73, 1.61, 1.75; 2.73, 2.48, 2.58; 4.28, 3.89, 4.09; and on vLLM 1.45, 1.24, 1.41 and 1.76, 1.56, 1.74 $\lref$ for seeds 0, 1, 2.}\label{tab:e2}
\setlength{\tabcolsep}{3pt}
\begin{tabular}{@{}c l c c c c c@{}}
\toprule
$\lambda/\lref$ & arm & states & hit & tokens & backlog & slope\\
\midrule
\multicolumn{7}{@{}l}{\emph{300k-token pool (E2)}}\\
1.35 & BURST   & T I T & 0.000 & 1.009--1.025 & 200--232 & 0.0036--0.0222\\
     & NOBURST & G G G & 0.930--0.934 & 0.08--0.09 & 7.9--9.7 & ${<}0.001$\\
     & CAPPED  & R R R & 0.930--0.934 & 0.08--0.09 & 7.9--9.7 & ${<}0.001$\\
\addlinespace
1.65 & BURST   & T T T & 0.000 & 1.025--1.029 & 283--332 & 0.0277--0.0362\\
     & NOBURST & G G G & 0.921--0.934 & 0.10--0.13 & 10.9--11.8 & ${<}0.002$\\
     & CAPPED  & R R R & 0.929--0.934 & 0.10--0.11 & 10.9--11.7 & ${<}0.002$\\
\midrule
\multicolumn{7}{@{}l}{\emph{native pool (native E2)}}\\
2.5  & BURST   & T T T & 0.000 & 1.074--1.078 & 557--601 & 0.079--0.085\\
     & NOBURST & G G G & 0.929--0.935 & 0.15--0.16 & 15.6--18.5 & ${<}0.003$\\
     & CAPPED  & R R R & 0.929--0.935 & 0.15--0.16 & 15.6--18.5 & ${<}0.003$\\
\addlinespace
4.0  & BURST   & T T T & 0.000 & 1.106--1.124 & 1098--1139 & 0.146--0.164\\
     & NOBURST & G G G & 0.929--0.935 & 0.24--0.27 & 28.2--33.6 & ${<}0.004$\\
     & CAPPED  & R R R & 0.929--0.935 & 0.24--0.27 & 28.1--33.5 & ${<}0.004$\\
\midrule
\multicolumn{7}{@{}l}{\emph{vLLM 0.32.0, 300k-token pool (registered, not pre-registered)}}\\
1.35 & BURST   & T D T & 0.000 & 0.995--1.012 & 201--222 & $-0.00005$ to 0.0170\\
     & NOBURST & G G G & 0.930--0.931 & 0.07--0.09 & 7.0--8.4 & ${<}0.001$\\
     & CAPPED  & R R R & 0.930--0.931 & 0.07--0.09 & 7.0--8.4 & ${<}0.001$\\
\addlinespace
1.65 & BURST   & T I T & 0.000 & 0.948--1.018 & 281--330 & 0.0183--0.0341\\
     & NOBURST & G G G & 0.932--0.933 & 0.10--0.11 & 9.3--10.8 & ${<}0.001$\\
     & CAPPED  & R R R & 0.932--0.933 & 0.10--0.11 & 9.3--10.9 & ${<}0.001$\\
\bottomrule
\end{tabular}
\end{table*}

The registered analysis of E2, run on all 18 runs with the code frozen at the registration commit, passes at both rates (Table~\ref{tab:e2}, top); by the definition of \S\ref{sec:prereg} the test is pre-registered. Every BURST run had a hit fraction of 0.000 over the evaluation window, an uncached prefill rate of 1.009--1.029 times the all-miss rate and a mean backlog of 200--332 sessions, and in Fig.~\ref{fig:e2} each is at zero hits from the 100\,s window starting at 1300\,s to the end of the run. One (1.35 $\lref$, seed 1) is stuck in the same way, but its backlog was flat over the window, so the rule calls it indeterminate: its seed delivered 5.8\% fewer arrivals than nominal in the window, and the trapped engine completed sessions about as fast as they arrived. The pass at 1.35 $\lref$ therefore rests on 2 of 3 seeds, the minimum the rule allows. Every NOBURST and CAPPED run is good (hit fractions 0.921--0.934, backlogs of 8--12 sessions, most asleep in tool calls); in the window CAPPED cannot be told apart from NOBURST, because the gate had drained the burst 1547--2693\,s after it. Outside the registered criteria, the CAPPED arm's hit fraction dips while the gate drains the burst: its lowest 100\,s window is 0.48--0.77 in four of six runs, against 0.82--0.89 for NOBURST. Counting each seed once, since both rates use the same three seeds with the same burst and nested arrivals, a post-hoc one-sided sign test of BURST's hit fraction against NOBURST's gives $p=1/8$, the smallest value three seeds allow. The trapped runs still completed 1.13--1.35 $\lref$, above $\lmiss$, because their young backlog has short turns; only the uncached prefill rate sits at the all-miss level, which is why the registered rule uses tokens (Theorem~\ref{thm:premium}, \S\ref{sec:validation}). So at loads that the engine carried from empty, 200 sessions arriving within a minute, an order of magnitude more than the 8--12 present in the good state, moved it into an all-miss state that it did not leave within 7500\,s, and a cap of 20 sessions, within 5\% of the measured peak of $\mu(n)$, prevented that, at the price of admission delay that the registered criteria do not measure (349--543 sessions per CAPPED run waited at the gate, 405--471\,s on average). Good NOBURST arms are the expected outcome even for a metastable good state: by the heuristic H2 of \S\ref{sec:fp-heur}, a run at these rates leaves it on its own within 8700\,s with probability about 0.5\% at 1.35 and 5\% at 1.65 $\lref$. 

\begin{figure*}[t]
\centering
\includegraphics[width=\textwidth]{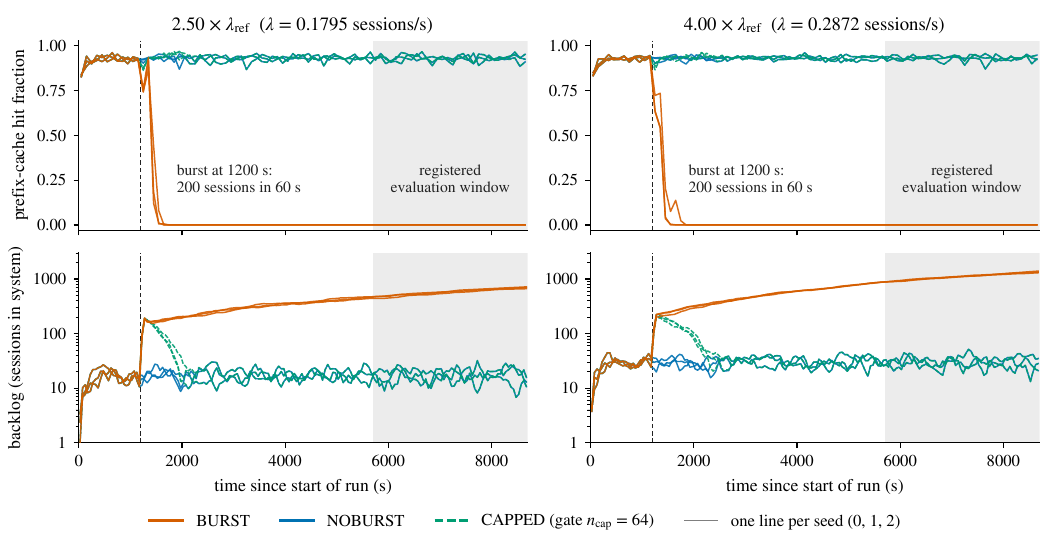}
\caption{The registered test with the native pool (native E2) at 2.5 $\lref$ (left) and 4.0 $\lref$ (right), $\lref=0.0718$\sps; layout as in Fig.~\ref{fig:e2}, with the CAPPED gate at $\ncap=64$ and the backlog axis extended to 3000.}\label{fig:e2native}
\end{figure*}

\paragraph{Native pool.}\label{sec:e2native}
A second registration, filled by the same formulas from the native curve, E0a calibration and E0b bracket, gives the window $(1,\,5.5)\,\lref$, the rates 2.5 and 4.0 $\lref$ and $\ncap=64$, flagged ambiguous because $\mu(80)$ is within 0.35\% of $\mu(64)$; it reached GitHub before the first native E2 job was submitted. The registered analysis passes at both rates, and every run is in the predicted state (Table~\ref{tab:e2}): all six BURST runs were trapped, all six NOBURST runs were good and all six CAPPED runs recovered (Fig.~\ref{fig:e2native}). Pooled post hoc over the two pre-registered SGLang tests (a descriptive count: they share three seeds' session sequences), NOBURST was good and CAPPED recovered in 12 of 12 runs each, and BURST was at zero hits in 12 of 12 and trapped by the rule in 11 of 12.

\paragraph{A second engine: vLLM.}\label{sec:vllm}
vLLM 0.32.0 ran with the SGLang cell's settings where they exist (FCFS, prefix caching on, chunked prefill of 8192 tokens, at most 256 running requests, a 300,000-token pool). Its all-miss capacity is 8\% below SGLang's ($\lref=0.0657\pm0.0030$\sps); its closed curve has SGLang's shape with an earlier, steeper cliff (peak 0.219\sps{} at $n=20$; hit fraction crossing 0.5 near $n=29.1$); and its from-empty edge lies between 2.0 and 2.5 times its $\lref$ (Fig.~\ref{fig:curve}, right). A registration filled from these by the SGLang formulas gives the window $(1,\,2.0)\,\lref$, the rates 1.35 and 1.65 $\lref$ and $\ncap=20$. It reached GitHub only after the 18 jobs were submitted, when two had started and none had ended, so the test is registered, not pre-registered; and its BURST arm reuses the seeds, and so the arrivals, of six earlier exploratory vLLM BURST runs, all of which had been at zero hits after the same burst, so it does not replicate them independently. At 18:21 CDT the memory request of the 15 jobs still pending (17166124--38) was lowered from 128G to 48G so that they could be scheduled; no code, argument or registration changed, and no job had ended. The registered analysis passes at both rates (Table~\ref{tab:e2}, bottom): every BURST run had zero hits over the evaluation window, and NOBURST stayed good and CAPPED recovered in all six runs each; the two BURST runs that the rule does not call trapped (seed 1, the seed with the fewest arrivals) were stuck at the all-miss token rate but failed its slope test. B3v had predicted a pass at both rates, and the trapped runs' token rates lay at or below its range (0.948--1.018 against 0.99--1.07). The first-principles model, pushed at 02:05:30 CDT on 2026-10-31, three minutes before the registered verdict was computed and, by the log of the agent that operated the test (\fpath{notes/LOG.md}), before any of its run directories was opened, put the good-state hit fractions at 0.932 and the gate drains at 1449 and 1725\,s, short of the measured 1521--2027 and 1848--2735\,s, as in three of the four SGLang cells; the theory's own blindness declaration does not cover these runs. Of the mechanism checks frozen before any run had ended, those on the gap between a session's turns (\S\ref{sec:mechanism}), on a full pool rewritten at the all-miss rate and on an engine-side latency rule held in all 18 runs, but the online detector, a follow-up hit fraction below 0.2 over 60\,s, alarmed 100--120\,s after every burst and also in two CAPPED runs while their gates drained, so the frozen decision rule is not met and we do not claim the detector (details below). The bands of these checks were set on the SGLang runs and on the six exploratory vLLM BURST runs, whose arrivals the BURST arm repeats, so only the NOBURST and CAPPED checks are out of sample. Across the three registered tests, BURST was at zero hits in 18 of 18 runs and trapped by the rule in 15, NOBURST was good in 18 of 18 and CAPPED recovered in 18 of 18; the tests share seeds, so these counts are descriptive.

Mechanism predictions for this test were frozen and pushed at 18:20:13 CDT on 2026-10-30, before any of its 18 runs had ended (\fpath{configs/r11_mechanism_predict.yaml} and \fpath{notes/PREDICTIONS_r11_mechanism.md}, commit \texttt{f1338e3}; byte copies under \fpath{r11_mechanism/frozen/} in the test's data directory). They were: (a) in NOBURST and CAPPED, a median gap between a session's turns below 0.05 times the pool's turnover time $R_M=M/\text{churn}$ (\S\ref{sec:mechanism}) and a follow-up hit fraction above 0.95, and in BURST a ratio above 2 and a hit fraction below 0.05; (b) in BURST, a mean KV usage of at least 0.90 and an uncached prompt-token rate within 15\% of the all-miss rate, that is, a full pool rewritten at the all-miss rate; (c) an online detector, a follow-up hit fraction below 0.2 over 60\,s, alarming within 300\,s of the burst in every BURST run and never after 600\,s in a good arm; (d) an engine rule, a mean time to first token above 30\,s over 100\,s, behaving the same way. The frozen decision rule required (b) and (c) in every BURST run and (a) and (c) in all good-arm runs but at most one. (a), (b) and (d) held in all 18 runs: the gap ratio was 6.44--8.78 in BURST and 0.014--0.022 elsewhere; in BURST the KV usage was 0.920--0.937 and the uncached rate 0.871--0.955 times the all-miss rate; and the engine rule alarmed 170--210\,s after the burst in BURST and never in a good arm. (c) failed. The detector alarmed 100--120\,s after the burst in every BURST run, but also in two CAPPED runs, seed 2 at both rates, 880 and 820\,s after the burst, while the gate was draining with 20 sessions admitted and 123--143 waiting; the follow-up hit fraction dipped to 0.06--0.17 for 20--40\,s and was above 0.8 again 110--160\,s later. The frozen decision rule is therefore not met. The detector uses CONCUR's hit-rate threshold (0.2) without its KV-usage condition~\citep{chen2026concur}; we do not claim it, and a deployed alarm on this rule would have fired twice without a trap. The bands of (a)--(d) were chosen on the SGLang runs and on the exploratory vLLM runs, whose arrivals the BURST arm repeats, so only the NOBURST and CAPPED checks are out of sample.

\paragraph{Paired tests on the H200 and on Llama-3.1-8B.}\label{sec:h200llama} Both registrations reached GitHub before their jobs were submitted (by 63 and 47\,s; neither is tagged), so both tests are pre-registered. On the H200 with the 600k pool (jobs 17171020--37) the registered analysis passes at both rates: every BURST run stayed at zero hits with 1.04--1.11 times the all-miss token rate and a growing backlog, trapped by the slope rule in 2 of 3 seeds at 1.5 $\lref$ (seed 1 indeterminate, its slope 1.3\% below the threshold) and 3 of 3 at 2.0 $\lref$, while NOBURST stayed good and CAPPED (cap 48) recovered in 6 of 6 runs each. pred2's trapped backlogs (458 and 867 sessions) lay within 15\% of all six runs (485--509 and 873--913), but its token ratios lay 0.038--0.049 below the measured ones in 5 of 6, outside the $\pm0.03$ tolerance. On Llama-3.1-8B (17170409--26) every BURST run stayed at zero hits at 1.01--1.05 times the all-miss token rate, every NOBURST run was good and every CAPPED run (cap 20) recovered, but the slope rule called only 1 of 3 BURST runs trapped at 1.35 $\lref$ (one indeterminate, one draining) and 2 of 3 at 1.65 $\lref$: the registered analysis passes at 1.65 $\lref$ and fails at 1.35. That all six BURST runs were stuck is a descriptive reading, not the registered criterion. The simulator had predicted a pass at both rates, so its pass at 1.35 $\lref$ missed. The registration-conditional predictions put the trapped backlog within 10\% of all six runs (191 and 266 sessions, against 196--210 and 267--288), the NOBURST hit fraction within 0.007 of every run's (0.932) and the trapped token ratio (1.004, 1.013) within $\pm0.03$ in 3 of 6 runs; in the other 3 it lay 0.031--0.046 below, outside the tolerance.

\paragraph{A faster GPU narrows the window.}\label{sec:h200}
The H200 cell repeats the 300k SGLang cell on one H200 NVL with SGLang's default attention backend for that GPU (FlashAttention~3; the A100 cells used FlashInfer), so the comparison is of GPU and default backend together. Against the A100, the all-miss rate rises 2.71 times ($\lref=0.194$\sps, turns-based), the peak of the closed curve only 1.63 times ($\mu^*=0.367$\sps{} at $n=32$, turns-based; 1.65 times in completed sessions) and the from-empty edge, the largest all-stable nominal multiple of each GPU's seed-0 grid, 1.76 times: the engine stayed stable from empty at 1.1, 1.2 and 1.3 $\lref$ and was unstable at 1.4, 1.55 and 1.75 $\lref$. All four reseeds (two at each of 1.3 and 1.4 $\lref$, added post hoc) were stable, so 1.3 $\lref$ was stable in 3 of 3 seeds and 1.4 in 2 of 3 (seed 0 unstable). The seed-0 grid received 7--8\% more arrivals than nominal, and, as on the A100, stability over all ten runs is monotone in the realized rate, with the edge between 1.496 and 1.503 $\lref$; the two 1.4 reseeds received 1.429 and 1.496. That bracket is only four sessions wide and the 1.503 run ended just above the queue threshold (50.4 against 50), so the data neither show a sharp edge nor need a stochastic band to explain them. The window shrinks from $(1,\,2.0)\,\lref$ on the A100 (nominal, on the seed-0 grid) to $(1,\,1.3)\,\lref$, too narrow for our registration rule: it places the test points at 1.10 and 1.20 $\lref$, outside the band $(1.1,\,1.17)$ that keeps them at least 10\% from the window's ends, so no registration, paired test or decisive claim exists for this GPU at this pool. The cliff moves out: the hit fraction crosses 0.5 near $n=38.9$, against 32.2 on the A100, and past it falls gradually, to 0.46 at $n=40$ and 0.16 at 52, reaching zero only between 52 and 96. Memory and tool time do not speed up with the GPU, so a faster GPU raises the all-miss capacity more than the peak, as the speed law of Appendix~\ref{app:entry} predicts for a uniform speed-up. With the pool doubled to 600k tokens on the same GPU (a separate, registered cell), the peak ratio $\mu^*/\lref$ rose 1.57-fold, from 1.89 to 2.98 (turns-based), and the from-empty edge moved from 1.3 to 2.5 $\lref$. The window $(1,\,2.5)\,\lref$ is then wide enough for the registration rule, which placed the test at 1.5 and 2.0 $\lref$ with cap 48, as the frozen model had predicted. That pre-registered paired test passes at both rates: every BURST run stayed at zero hits with a growing backlog (trapped by the rule in 2 of 3 seeds at 1.5 $\lref$ and 3 of 3 at 2.0), every NOBURST run was good and every CAPPED run recovered (\S\ref{sec:h200llama}). \S\ref{sec:frozen} scores the model's frozen predictions for both H200 cells. KAIROS shows qualitatively that a slower clock tips a fixed arrival rate into thrashing~\citep{yuan2026kairos}, and VAMP locates KV capacity boundaries on H100 and H200 GPUs~\citep{kim2026vamp}; what we add is the measurement in units of $\lmiss$ and the opposite movement of the onset and the peak ratio. In Lu et al.'s simulations a larger and faster GPU (H200 profiles with about twice the cache and decode memory) raises the first trapped load in absolute terms, and across their GPU and host-tier sweep it stays at 0.46--0.70 of the steady-state overload rate~\citep{lu2026cachetrap}; the two statements differ in normalization and in memory (our 300k H200 cell keeps the pool fixed), and both say that the vulnerability persists on the H200.

\subsection{Dose response, tool slowdown and arrival stop}\label{app:basin}

\paragraph{The basin of the good state.}\label{sec:basin}
The paired tests use one burst size, 200 sessions, which in our earlier simulator trapped every work-conserving policy. Two registered BURST-only tests at 1.65 $\lref$ on the 300k pool vary it, with the same seeds: BASIN-1 with 25 and 100 sessions and BASIN-2 with 40, 50 and 60. BASIN-1 passed: a burst of 100 sessions left the engine stuck (hit fraction 0.000 over the evaluation window) and trapped by the registered slope rule in 3 of 3 seeds, and a burst of 25 left it good (RECOVERED, hit fraction 0.92--0.93, within 0.003 of the no-burst run with the same arrivals) in 3 of 3. The good state of the real engine therefore has a finite basin, whose boundary lies between 25 and 100 sessions in every seed, as in the calibrated simulator, where 25 sessions never trapped and 50 trapped about half the time. BASIN-2 tests the first-principles model's critical fresh burst of 53.4 sessions, from a heuristic of \S\ref{sec:fp-heur} that counts a burst session at its peak footprint. Its registered rule held: a burst of 40 sessions left the engine good (RECOVERED, hit fraction 0.925--0.934, within 0.004 of the no-burst run with the same arrivals) and one of 60 left it stuck and trapped, in 3 of 3 seeds each, while 50 sessions, which the rule does not use, trapped seed 1 only. Over the seven sizes from 0 to 200 each seed switches from good to stuck once, between 50 and 60 sessions in seeds 0 and 2 and between 40 and 50 in seed 1: the boundary lies between 40 and 60 sessions in every seed but not at the same size, so 53.4 lies inside two of the three seeds' edge intervals. BASIN-1's and BASIN-2's registrations reached GitHub 68 and 131 minutes before their jobs were submitted; \S\ref{sec:frozen} scores the frozen predictions.

\paragraph{The basin.} Two registered BURST-only tests at 1.65 $\lref$ on the 300k pool vary the burst size: BASIN-1 with 25 and 100 sessions and BASIN-2 with 40, 50 and 60. The theory predicts whether a run is stuck, with a hit fraction below 0.2 over the evaluation window; it does not model the slope test of the registered trap rule, which is co-primary only for BASIN-1's 100-session cell. The model's critical fresh burst is 53.4 sessions, and its per-seed probabilities of a stuck run are 0.04, 0.14, 0.55 and 1.0 for 25, 40, 50 and 60 or more sessions; the stationary-composition estimate, 21.3 sessions, would have 40 stuck. BASIN-2's frozen rule calls the boundary bracketed if 40 stays good and 60 is stuck, each in at least two of three seeds. The second prediction set (pred2, pushed at 04:20:20 CDT on 2026-10-31; the nine BASIN-2 jobs were submitted at 04:21) conditions these probabilities on each seed's burst, rebuilt with the client's own plan generator. Seed 0's first 40 burst sessions are long and seed 2's short, so the probabilities of a stuck run at 40, 50 and 60 sessions become 0.78, 0.86 and 1.00 for seed 0, 0.36, 1.00 and 1.00 for seed 1, and 0.09, 0.22 and 0.81 for seed 2; the rule then holds with probability 0.66 (0.23--0.88 over the ways of mapping a plan to a burst size), against 0.94 with every burst at its mean composition. The critical burst and the stationary-composition estimate are both defined at $n_2$, where the quasi-static model is least reliable and a burst's fate depends on its composition as well as on $\mu(n)$ (\S\ref{sec:fp-weak}), so BASIN-2 tests the heuristic H3, not a consequence of the propositions.

\paragraph{The basin: BASIN-1 outcome.} The paragraph on the basin above reports the outcome, and \S\ref{sec:frozen} scores it. The 100-session runs were stuck (hit fraction 0.000 over $[5700,8700)$\,s) and also trapped by the co-primary slope rule, in 3 of 3 seeds. BASIN-1 does not test the critical burst of 53.4 sessions; BASIN-2 does. The registered analysis ran at \texttt{19c58e9}, which contains BASIN-2's rule (\texttt{621d88f}) and an analysis-only change (one vote per seed; only runs bound to their registration count), both made after BASIN-1's tag; the code at the tag (\texttt{9b67c26}) gives the same decision and per-run values.

\paragraph{The basin: BASIN-2 outcome.} The nine runs (17169893--901, submitted at 04:21:40--41 from a clone at the registration tag, started 10:10--12:41 and pulled at 15:42) are bound to their registrations, and the registered analysis (data in Table~\ref{tab:prov}) calls the boundary bracketed: with 40 sessions every seed was good (RECOVERED; hit fraction 0.934, 0.934 and 0.925 over $[5700,8700)$\,s), with 60 every seed was stuck (hit 0.000) and trapped, and with 50, which the rule does not use, seed 1 was stuck and trapped while seeds 0 and 2 stayed good (0.933 and 0.924). The analysis ran at \texttt{4dd21e0}, which includes \texttt{19c58e9}, an analysis-only change pushed at 03:12:59, after the registration tag and before submission, that counts one vote per seed and only runs bound to their registration; the code at the tag gives the same bracket, curve and per-run values. Each seed's arrivals are common random numbers across sizes, and along 0, 25, 40, 50, 60, 100 and 200 sessions each seed switches from good to stuck once; the burst of $n$ uses the first $n$ arrival times and session specifications of the 200-session draw, so the chain is nested as sets, not by session identity. Descriptively, the stuck runs' hit fraction stayed below 0.2 from the 1400 or 1500\,s window on; in the good runs the lowest 100\,s window was 0.70--0.83 and the next one above 0.8; and every burst session had finished by 1680\,s in the good runs and by 4719\,s in the stuck ones, before the evaluation window, so the stuck runs stay at zero hits on the base sessions alone, which the no-burst runs serve at 0.92--0.93. Against pred2 (pushed at 04:20:20, 80\,s before the submission): the rule held (probability 0.66); the primary per-seed probabilities of a stuck run for seed 0, 0.78 at 40 and 0.86 at 50, were on the wrong side for both of its good runs, and the other seven were on the more likely side; over the nine runs their Brier score is 0.174, against 0.096 for the frozen average-burst layer, which ignores the plans, and 0.259 for the uncalibrated mapping (declared scores; the frozen composition-averaged column, which the declaration does not name, scores 0.111); and the secondary deterministic level crossing on the exact plans called all nine runs. pred2's declared claim that seed 0 is the most trap-prone at 40 sessions was not borne out: no seed was stuck at 40, and seed 1, the only one stuck at 50, has the lowest edge. The critical burst of 53.4 lies inside the edge intervals of seeds 0 and 2, $(50,60]$, and outside seed 1's, $(40,50]$, while the 50\% points of the committor and of the stuck probability (48.4--49.5) lie inside seed 1's only, so no single size fits all three seeds. The round-7 simulator trapped 6 of 10 runs at 50 sessions; the engine trapped 1 of 3. With three seeds per size, 0 of 3 and 3 of 3 have 95\% intervals $[0,0.71]$ and $[0.29,1]$.

\paragraph{Causes.} TOOLSLOW lengthens every tool call that starts in $[1200,2100)$\,s by $\delta=10$ or 30\,s in the NOBURST arm at 1.35 $\lref$, a trigger that involves neither a burst nor a lowered clock. Its registered rule expects the 10\,s slowdown to leave the engine good and the 30\,s slowdown to leave it stuck, each in at least two of three seeds. In the frozen model (pred2) the slowdown acts through capacity and population: for $\delta$ above 13.7\,s the slowed curve's peak lies below $\lambda=0.0966$\sps, so it has no good state (at $\delta=30$\,s the peak is the all-miss level $\mu_b=0.0718$\sps); at $\delta=30$\,s the population reaches 46.9 sessions by 2100\,s, above $n_2=32.2$, while at $\delta=10$\,s it reaches 21.5, near the slowed curve's good state (21.8 sessions, against 8.5 without the slowdown). A pass would show that a temporary increase in tool latency, with no burst, can trigger the trap; it would not separate the longer gap from the capacity loss and the larger population that come with it. LPM-INF disables stock LPM's 128-request fallback (\S\ref{sec:grid}) with a job-local hook that raises the limit to $10^9$ and changes nothing else; the BURST arm at 1.65 $\lref$ is repeated on seeds 0--2 with byte-identical arrival plans. The simulator predicted (pushed at 21:02:46 CDT on 2026-10-30, 65 minutes before the jobs were submitted) that LPM then avoids the all-miss collapse on the two seeds on which stock LPM collapsed, in a partial-hit state that the registered rule still calls trapped.

\paragraph{SHED and TOOLSLOW: outcome.} The ten jobs (17169902--11) were submitted at 04:21:42--43 CDT on 2026-10-31, after the registrations (pushed at 02:28:09) and after pred2 (04:20:20), so both tests are pre-registered and pred2's predictions for them are blind in time. Every run is bound to its registration and paired with its committed 1.35 $\lref$ control of the same seed; the registered analysis (\fpath{analysis/r14_cells.py} at \texttt{db24efa}, an analysis-only fix made before any job) is in \fpath{analysis/agentload_analysis_r14/20261101-154909/} in \fpath{results/nova/toolslow_shed_1716990x_pulled_20261101-154800/}. STUCK is a hit fraction below 0.2 over $[5700,8700)$\,s; GOOD is RECOVERED, or a hit fraction above 0.8 with a backlog slope of at most $0.5(\lambda-\lref)$. SHED passed: with arrivals stopped for 600\,s both runs stayed stuck (hit fraction 0.000; seed 0 trapped, seed 1 indeterminate by the slope rule, with 219 and 168 requests outstanding at the end), and with 3600\,s both were good (0.933 and 0.930, 1--2 outstanding); the same-seed BURST controls stayed stuck. The stops refused 59 and 56, and 355 and 378, sessions, equal to pred2's protocol check, which the arrival plans fix; the R1 cap at the same rate refused none. TOOLSLOW did not meet its rule: with $\delta=10$\,s all three runs were good (hit fractions 0.930--0.934, as their NOBURST controls; 923--1034 tool calls slowed), and with $\delta=30$\,s seeds 0 and 1 were good (0.931, 0.930) and seed 2 trapped (0.000, slope 0.0233\sps, 127 outstanding at the end; 664--770 calls slowed). pred2 had predicted good at 10\,s with probability 0.71 per seed and stuck at 30\,s with probability 1.00, the rule with 0.80, and ranked seed 0 the most trap-prone at 30\,s (49 sessions at 2100\,s, against 37 and 39); seed 0 recovered. Either the slowed engine kept a good state that the fluid layer says it lacks, or the population it reached did not keep the engine stuck once the slowdown ended; these runs do not separate the two. The hit fractions of all ten runs were recomputed from the request logs and agree; the slopes and slope labels were not recomputed independently.

\subsection{The rescue tests}\label{app:rescuetests}\label{sec:rescue}

\paragraph{Rescue (R1, R4).} One rescue run per 8700\,s job, each on a fresh server, from a frozen clone at the registration tag; the client's arrival plan is the BURST plan of the same seed and rate, byte for byte, so until $t_{\mathrm{on}}$ the rescue run replays the BURST run's arrival plan (a fresh run on a fresh server, not the same run). At $t_{\mathrm{on}}$ the gate starts to bind with cap $c$ by attrition and pause: a session whose tool call ends while more than $c$ sessions are admitted gives up its slot and joins a resume queue, which is served before the queue of new arrivals, each in FIFO order; a paused session's KV is not protected while it waits. The cap is removed at the first 10\,s tick at or after $t_{\mathrm{on}}+600$\,s at which both queues have been empty for 300\,s; there is no forced release. The registered per-run rule uses the follow-up hit fraction, the share of a session's turns after its first, attributed by completion time, with more than half of the prompt cached: (a) VALID if the follow-up hit fraction over $[t_{\mathrm{on}}-300,\,t_{\mathrm{on}})$ is below 0.2 and more than 50 requests or sessions are outstanding at $t_{\mathrm{on}}$ (requests in flight plus sessions waiting at the gate or paused), on the registered hardware; (b) RELEASED by 7200\,s; (c) POST-GOOD over $[t_{\mathrm{rel}}+300,\,8700)$: every 300\,s window, tiled from the start with a short remainder covered by the window ending at 8700\,s, has a follow-up hit fraction of at least 0.5, the whole interval above 0.8, an uncached token rate below 0.5 times the all-miss reference, and the registered RECOVERED test holds over the last 300\,s; a window with no completed follow-up fails (c) without being a relapse. RESCUED is (a), (b) and (c); RELAPSE is any such window below 0.2. A same-seed BURST control, from the paired test, is STUCK if its follow-up hit fraction over $[t_{\mathrm{on}},\,8700)$ is below 0.2 and it is not RECOVERED over the last 300\,s. A cell passes if at least 2 of its 3 seeds are RESCUED and every same-seed control is STUCK; R1 is refuted if 2 of its 6 runs relapse or are not released by 7200\,s, and R4 if 2 of its 3 relapse; a refutation takes precedence over a pass, and a failed session in any counted run withdraws the claim that nothing was shed. The pairing requires common random numbers with the committed BURST and NOBURST runs (the BURST plan byte-identical, the NOBURST plan the rescue plan without the burst sessions), equal resolved rates, references, hardware and burst time. Each job ran only a mechanical check of the gate (\texttt{analysis.rescue check-run}: engage within 10\,s of $t_{\mathrm{on}}$, release on the 10\,s grid after the 300\,s hold, no request errors), which passed in all ten jobs; no hit fraction of a rescue run was read before all runs of its cell had ended. The controls ran on 2026-10-24 to 26 as part of the paired tests, not alongside the rescue runs. A smoke job under its own registration checked the gate's mechanics before the registered runs and is never decisive.

\begin{table}[htbp]
\centering\footnotesize
\caption{The registered rescue tests R1 and R4 (seeds 0, 1, 2); ranges over seeds. After release: follow-up hit fraction and mean
outstanding work over $[t_{\mathrm{rel}}+300,\,8700)$\,s for the rescue run and, over the same interval, its same-seed controls.
Delay: session-hours of pauses and admission waits at the client. Rows below the release time are descriptive, computed after the
registered verdict.}\label{tab:rescue}
\setlength{\tabcolsep}{3pt}
\begin{tabular}{@{}>{\raggedright\arraybackslash}p{2.3cm} c c c@{}}
\toprule
 & R1, 300k & R1, 300k & R4, native\\
 & 1.35 $\lref$ & 1.65 $\lref$ & 4.0 $\lref$\\
\midrule
cap; engaged at & 20; 1800\,s & 20; 1800\,s & 64; 2100\,s\\
rescued (relapses) & 3/3 (0) & 3/3 (0) & 3/3 (0)\\
release $t_{\mathrm{rel}}$ (s) & 3980--4550 & 4430--4920 & 5980--6140\\
\addlinespace
\multicolumn{4}{@{}l}{\emph{after release: follow-up hit fraction}}\\
\quad rescue & 0.999--1.000 & 0.994--1.000 & 1.000\\
\quad NOBURST & 0.999--1.000 & 0.991--1.000 & 1.000\\
\quad BURST & 0.000 & 0.000 & 0.000\\
\multicolumn{4}{@{}l}{\emph{after release: mean outstanding work}}\\
\quad rescue & 1.9--2.2 & 2.5--2.8 & 8.8--11.8\\
\quad NOBURST & 1.9--2.2 & 2.5--3.1 & 8.8--11.8\\
\quad BURST & 195--219 & 272--317 & 1133--1178\\
\addlinespace
sessions paused & 121--130 & 138--143 & 220--251\\
delay (session-h) & 38--65 & 50--78 & 222--230\\
\bottomrule
\end{tabular}
\end{table}

\paragraph{The rescue tests in detail.}\label{app:rescue}
Table~\ref{tab:rescueruns} gives the registered runs one by one. In every run the cap was engaged within 0.003\,s of $t_{\mathrm{on}}$, every pause was matched by a resume, no session was paused or gated at the end, and the in-job mechanical check passed; the smoke job ended with 142 sessions gated, 20 in the engine and none failed. Hits came back within three minutes of the engage and draining the backlog took the rest of the capped phase: in the 60\,s bins of Fig.~\ref{fig:rescue} the follow-up hit fraction fell again to minima of 0.13--0.60 per run (the last bin below 0.8 began 780--1620\,s after the engage), and over 300\,s bins it was 0.56--1.00 during R1's capped phase, and in R4 it dipped to 0.47--0.57 in the 300\,s bin from 3000\,s (0.56--0.78 in the bin before), while the native backlog in the engine, at the gate and paused peaked near 430, before it reached 1.00 once the queue drained. The registered RECOVERED test, evaluated every 10\,s, first held 540--700\,s (R1) and 300--350\,s (R4) before the release, while the cap was still on, and the cap last held a session back 301--310\,s before each release, the hold the release rule requires. Each paused session waited 386--569\,s on average and the longest total pause of one session was 817--1240\,s; averaged over all sessions of a run, pauses took 30--63\,s. Every session of the cohort that arrived in $[1200,\,3000)$\,s completed in every rescue run, with a median duration of 630--909\,s (300k) and 1138--1146\,s (native), against 62--73\,s without the burst and 793--1099\,s in the uncapped controls, where 2--24\% of that cohort had not finished at 8700\,s. After release the uncached token rate was 0.08--0.12 (300k) and 0.23--0.27 (native) times the all-miss rate, and every 300\,s window had a follow-up hit fraction of at least 0.938. Against the simulator's exact-plan predictions, the engine paused 0--13 more sessions, its client queue held up to 1.5 times as many sessions with the native pool (387--397 against 258--342), its delay there was 1.5--3 times as large, and its completed shares equal the predicted ones to four decimals in 6 of 9 runs. The persistence, cost and trace tables behind \S\ref{sec:rescue} and Fig.~\ref{fig:rescue} are in \fpath{results/rescue_paper_data/20261031-071938/} in the data directory, written after the verdicts by \fpath{figures/rescue_paper_data.py}.

\begin{table}[htbp]
\centering\footnotesize
\caption{The registered rescue runs (R1: 300k pool, cap 20 from 1800\,s; R4: native pool, cap 64 from 2100\,s). Outstanding at $t_{\mathrm{on}}$: requests in flight plus sessions held at the client. Sim.: the release time the simulator predicted on the run's exact arrival plan (tag \texttt{r11-rescue-pred}). Held: the largest number of sessions waiting at the gate or paused. After: the largest number of sessions in the system after release. Delay: session-hours of pauses and admission waits.}\label{tab:rescueruns}
\setlength{\tabcolsep}{2pt}
\begin{tabular}{@{}c c c c c c c c c@{}}
\toprule
$\lambda/\lref$ & seed & outst.\ at $t_{\mathrm{on}}$ & $t_{\mathrm{rel}}$ (s) & sim.\ (s) & paused & held & after & delay (h)\\
\midrule
1.35 & 0 & 146 & 4550 & 3960 & 124 & 134 & 16 & 53.1\\
     & 1 & 148 & 3980 & 3470 & 130 & 146 & 17 & 38.4\\
     & 2 & 139 & 4400 & 3940 & 121 & 148 & 20 & 65.4\\
\addlinespace
1.65 & 0 & 172 & 4920 & 4900 & 141 & 158 & 23 & 77.8\\
     & 1 & 171 & 4430 & 4440 & 143 & 159 & 21 & 50.1\\
     & 2 & 143 & 4570 & 4280 & 138 & 145 & 27 & 60.8\\
\addlinespace
4.0 & 0 & 324 & 6000 & 4250 & 251 & 397 & 55 & 222.0\\
    & 1 & 322 & 5980 & 5180 & 240 & 397 & 43 & 229.6\\
    & 2 & 298 & 6140 & 4410 & 220 & 387 & 49 & 225.0\\
\bottomrule
\end{tabular}
\end{table}

\paragraph{Rescue and shedding.} Two registered tests engage a cap inside an established trap and release it once the gate has been empty for 300\,s: R1 on the 300k pool (cap 20 at 1800\,s, at 1.35 and 1.65 $\lref$) and R4 with the native pool (cap 64 at 2100\,s, at 4.0 $\lref$). The simulator and the model both predict rescue. At the fluid rate of the paused sessions, whose next turns miss, the model predicts release at 3830, 4360 and 5040\,s ($\pm15\%$; with H1, 3920, 4480 and 5250\,s); it predicts release before the 7200\,s deadline only if those sessions complete at no less than 32, 42 and 51\% of that rate. The gate pauses sessions only at tool-call boundaries and defers work rather than dropping it. Pausing whole programs at tool boundaries is ThunderAgent's and NVIDIA Dynamo's mechanism~\citep{kang2026thunderagent,dhanani2026dynamo}; CONCUR cuts its window of active agents multiplicatively when it detects thrashing (hit rate below 0.2 with KV usage above 0.5) to exit that regime, and MISA-T contracts its cap temporarily on a confirmed hit-rate decline, both in closed or bounded populations~\citep{chen2026concur,hong2026misat}; KAIROS bounds aggregate context and, when bursty arrivals have already broken the bound, defers new agents and boosts the clock, but reports no recovery from an established trap~\citep{yuan2026kairos}; inhibiting new traffic until a bistable channel returns to its lower state was proposed by Carleial and Hellman~\citep{carleial1975bistable}, and shedding load is the usual response in production~\citep{huang2022metastable}. Lu et al.\ restore service on hardware by terminating the burst sessions at the end of the burst, when the trap is incipient, prevent it with their controller active before the surge, rescue an established trap in simulation by activating their controller late (and leaving it armed: its admission releases itself after a cool-down, but no reported run removes the controller and tests persistence), and report one hardware run in which CONCUR recovered slowly after its hit rate fell~\citep{lu2026cachetrap}. The new parts here are a registered, replicated rescue of an established trap (follow-up hit fraction 0.000--0.002 before the cap; one hit among 570 follow-ups in one run) on the engine at the same offered load without shedding, a measured test that the good state persists after the cap is released, and a rescue condition and release times frozen beforehand. The SHED test measures the alternative: arrivals stop for $\tau=600$ or 3600\,s from 1800\,s, refusing about $\lambda\tau=58$ or 348 sessions, against none for the cap.

\paragraph{Rescue: outcome.} \S\ref{sec:frozen} scores the outcome; the per-run values follow (Table~\ref{tab:rescueruns}). The simulator's release times on the exact plans were 3960, 3470 and 3940\,s at 1.35 $\lref$ and 4900, 4440 and 4280\,s at 1.65, against the engine's 4550, 3980 and 4400\,s and 4920, 4430 and 4570\,s, and 4250, 5180 and 4410\,s for R4, against 6000, 5980 and 6140\,s; all nine releases lie inside its 20-seed bands. Against the model's 3830 and 4360\,s, the R1 releases were $+18.8$, $+3.9$ and $+14.9\%$ and $+12.8$, $+1.6$ and $+4.8\%$ late; against 5040\,s, the R4 releases were $+19.0$, $+18.7$ and $+21.8\%$ late (against the H1 value, 5250\,s: $+14.3$, $+13.9$ and $+17.0\%$). The engine's first phase in R4 was slower than the fluid rate but far from collapsed. The smoke run ended before the model's push, but the model made no prediction for it, and by their record the agents that derived the model had no Nova access and opened no rescue output before the push.

\subsection{Other orderings (exploratory)}\label{app:grid}

\paragraph{Other orderings (exploratory).}\label{sec:grid}
Three other engine cells ran the BURST arm at the registered rates and seeds, with no control arms (details below). Disabling SGLang's waiting-prefix refresh made no difference for FCFS. Stock LPM partly helps: from the 200-session burst it recovered in 4 of 6 runs at 1.35 $\lref$ but in 1 of 6 at 1.65. On our reading of its source it stops re-sorting its queue when more than 128 requests wait, and in each run that ended trapped the last hit came when the queue last stood at or below 128. With that fallback disabled by a job-local hook (LPM-INF, exploratory, with its decision rule and the simulator's prediction frozen beforehand), LPM did not collapse on the two arrival streams on which stock LPM had collapsed in 4 of 4 runs (token hit fraction, the share of prompt tokens served from cache, pooled by completion time over the evaluation window: 0.59 and 0.42 against 0.000; unlike the turn hit fraction it has no 0.932 ceiling), but no run recovered by the registered rule; with one rate and two informative seeds this is directional evidence, and it does not separate LPM's ordering from the LRU touch of its prefix matching.

LPM partly helps. From the 200-session burst, at 1.35 $\lref$ it recovered in 4 of 6 runs and was draining in a fifth (hit fractions over the evaluation window; RECOVERED is judged on the last 300\,s, where every recovered run was at 0.92--0.94; with the refresh: recovered at 0.71 and 0.81, draining at 0.51; without: recovered at 0.80 and 0.93, trapped once). At 1.65 $\lref$ it was trapped in 4 of 6, indeterminate in one (hit 0.44) and recovered in one (0.82). The paths behind these window means differ from FCFS's, which had no hits from 100\,s after the burst in all twelve FCFS runs (a post-hoc reading in 100\,s windows). Each LPM run that ended trapped kept some hits until 1500--1900\,s after the burst (3900\,s in one run) and then had none; each run that recovered first exceeded a hit fraction of 0.8 3100--6500\,s after the burst and stayed above it to the end, so that had the runs ended 2000\,s earlier, only two of the five would have met the RECOVERED rule. This is a finite-burst result for the BURST arm alone, not an estimate of LPM's congested service rate $\muinf$; with mixed outcomes in every cell, three seeds do not bracket a threshold. Post hoc, the timing matches SGLang's 128-request fallback: in the five runs that recovered the engine's waiting queue never exceeded 128 after the burst, and in each of the five that ended trapped the last hit fell in the 100\,s window in which the queue last stood at or below 128. This is a correlation in ten runs, not an isolation; the paragraphs below give the details and the other candidate causes.

\paragraph{Stock LPM and its fallback.} In deep congestion stock LPM stops re-sorting its queue, so, on our reading of its source, Proposition~\ref{prop:A} applies to it (Appendix~\ref{app:proofs}); consistent with this, Stage~C's LPM run at $n=300$ had zero hits with a queue above 128, and round 5 of our earlier simulator, which modelled the 128-request fallback, found no LPM advantage in congestion. In the same earlier, uncalibrated simulator, pure LPM without the fallback has a congested service rate of about 1.17 times that simulator's own $\lmiss$ and recovered from 300 cold sessions only up to about 1.1 times it; these are deep-congestion quantities and do not compare directly with recovery from a finite burst. We do not claim that LPM fails, and we have not isolated why it recovered. Post hoc, the timing matches the 128-request fallback: in the five runs that recovered, the engine's waiting queue (5\,s samples) never exceeded 128 after the burst (at most 116--126), and in each of the five that ended trapped the last hit fell in the 100\,s window in which the queue last stood at or below 128. This is a correlation in ten runs, not an isolation. An exploratory intervention followed (LPM-INF, below), which switches off LPM's ordering fallback and the LRU touch of its prefix matching together. Other candidates are partial prefix hits and LPM's own prefix matching, which moves the cached part of every waiting request's prefix to the recent end of the LRU order on each scheduling attempt; the simulator omits both. Under LPM the waiting-prefix refresh is skipped, because that matching already touches the waiting prefixes, so the two LPM configurations differ only while more than 128 requests wait and LPM has fallen back to FCFS. Seed 0 at 1.35 $\lref$ recovered with the refresh and was trapped without it; since the refresh made no difference under FCFS, single runs near LPM's threshold appear sensitive to small differences.

\paragraph{LPM without its fallback (LPM-INF, exploratory).} To test whether the fallback causes stock LPM's collapse, a job-local hook raised SGLang's 128-request limit to $10^9$ and changed nothing else, and the BURST arm reran at 1.65 $\lref$ on seeds 0--2 with byte-identical arrival plans (jobs 17168743--45); the exploratory registration, the endpoints with their decision rule and the simulator's predictions were frozen under tag \texttt{r11-lpminf-pred} before the jobs were submitted. The controls are the grid's stock-LPM runs of 2026-10-24 and 25, partly on other nodes, with the same arrival plans. On seeds 0 and 1, where stock LPM had collapsed in all four runs (with and without the refresh), LPM without the fallback did not collapse: its token hit fraction over the evaluation window was 0.59 and 0.42, against 0.000, and its waiting queue peaked at 165 and 184 requests, against 287--345. This is the outcome that the rule frozen for the test reads as consistent with the fallback causing stock LPM's collapse. On seed 2 the queue never exceeded 128 (at most 120), so the hook never acted and the run does not enter the rule. It is not a fix: no run recovered by the registered rule (draining, indeterminate and draining on seeds 0, 1 and 2), and on seed 1 the engine still prefilled 0.92 times the all-miss token rate with 194 sessions unfinished at 8700\,s. The blind simulator predicted the direction, no collapse on any seed, but not the level (token hit fractions of 0.08--0.10) or the registered state, trapped in its prediction and in none of the three runs; it was flagged in advance as pessimistic about LPM, because it has no partial prefix hits. With the fallback off, LPM keeps both its ordering and the LRU touch that its prefix matching gives waiting requests, so the test does not say which of the two prevents the collapse. Replicate noise in this partial-hit regime is large: on seed 2, three LPM runs that were in effect one configuration, since none of their queues passed 128, gave token hit fractions of 0.44, 0.56 and 0.82. With one rate and two informative seeds this is directional evidence.

\subsection{The entry model: a first-principles throughput curve}\label{app:entry}

\begin{figure*}[t]\centering
\includegraphics[width=\textwidth]{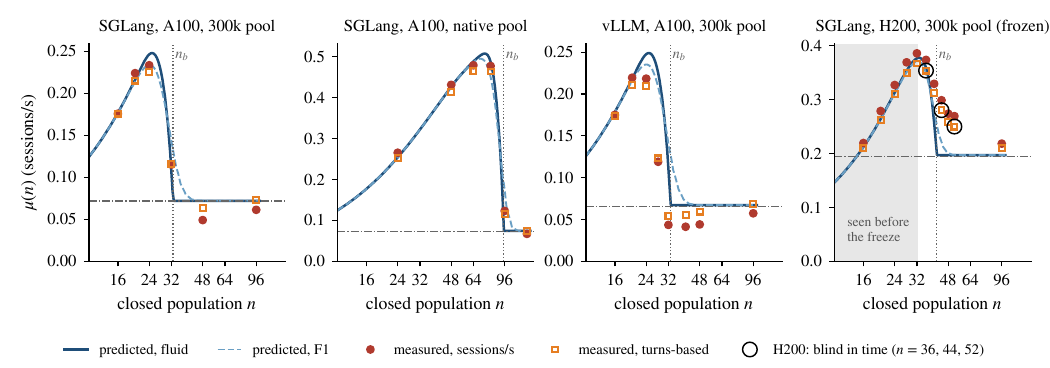}
\caption{Predicted thrashing curves $\mu(n)$ against the engine. Lines: the first-principles model (solid, fluid; dashed, with
the demand-fluctuation layer H1, labelled F1 in the legend), from the pool, the workload and a cost model fitted to server-log timing only; vLLM and the
H200 also get one speed factor fitted to their all-miss token rate. Points: measured $\mu(n)$, sessions/s (filled) and
turns-based (open). Dotted: the predicted all-miss onset $n_b$ (Proposition~\ref{prop:nb}); dash-dotted: the measured $\lref$. No
measured $\mu(n)$ point enters any prediction. The first three panels are in sample (the agents that derived the model knew
every point). The H200 panel is out of sample: predicted under tag \texttt{r13-theory-pred} before those agents read it; the runs at
$n=36$, 44 and 52 (ringed) are blind in time, those at $n\le32$ (shaded) had been seen by the cell's operating agent and the first author, the others are
blind by access only; no blind simulator prediction was made for this cell.}\label{fig:fpmu}
\end{figure*}

Model~F describes the trap once it exists. Whether a load can be trapped at all is set by the closed-population throughput
curve $\mu(n)$ (Fig.~\ref{fig:fpmu}), the session completion rate with $n$ sessions present: Denning's thrashing
curve~\citep{denning1968thrashing}. It rises with batching, peaks at $\mu^*$, and collapses to the all-miss level once
admitting waiting turns evicts the idle KV of sessions in tool calls. With open arrivals the classical quasi-static
approximation $\dot n=\lambda-\mu(n)$~\citep{courtois1975decomposability,kleinrock1975aloha} makes the good state metastable at
every load in the window $\mathcal W=(\lmiss,\min(\mu^*,\ledge))$, where $\ledge$ is the measured from-empty edge
(\S\ref{sec:fluid}); Lu et al.\ write the same population balance with an empirical hit curve~\citep{lu2026cachetrap}. We use it
for the window and the cap condition only, never for the trap.

\paragraph{The curve from first principles.} We predict $\mu(n)$ from the pool size $M$, the workload, the tool-time law and an
iteration-cost model fitted only to SGLang's per-batch timing lines, with no parameter fitted to any $\mu(n)$, hit or trap
measurement. Two couplings produce the cliff. Under prefill priority, misses starve decode, so running requests
stay longer and reserve more of the pool, $K_{\mathrm{run}}(\nu,h)$; and idle KV, evicted least recently used first at a
characteristic idle age $R$ (Che's approximation~\citep{che2002hierarchical}), gives hits $h=F_Z(R)$ only if it survives the tool
time $Z$. The pool balance
\[
M=K_{\mathrm{run}}(\nu,h)+\frac{\nu}{E[T]}\,\mathcal D(R),\qquad h=F_Z(R),
\]
with $\nu$ the turn rate, $\mathcal D(R)$ the idle KV per unit session throughput, Little's law for $n$ and a no-slack fluid
closure, reduces to the characteristic-time equation of Yao et al.~\citep{yao2026multiturn} when nothing runs or queues. With the
couplings it gives a curve with a rise, a peak, a cliff and an all-miss branch, on which the queue, not the tool, sets the reuse gap.

\paragraph{The all-miss onset.} Let $\nu_b$ be the turn rate at which running reservations fill the pool with every turn
missing, $\mu_b=\nu_b/E[T]$ (the model's $\lmiss$), $\bar z'$ the tool time per turn and $E_S$ the service-weighted KV of a running
turn. Proposition~\ref{prop:nb} shows that the all-miss branch exists exactly for
\[
n\ge n_b=\nu_b\bar z'+M/E_S ,
\]
the sessions in tool calls at the all-miss rate plus one pool's worth of running sessions, and that if every GPU cost is divided by
$s$, then $n_b(s)=n_b(1)+(s-1)\nu_b(1)\bar z'$: a faster GPU moves the onset out.

The pool part $M/E_S$ is a capacity ratio, the form of MISA-T's session cap~\citep{hong2026misat}, of AgentKV's collapse
predictor~\citep{anon2026agentkv} and of Lu et al.'s empirical cliff $N_c\approx1.9\,C/\bar c$~\citep{lu2026cachetrap}. The
tool-call term, which grows with GPU speed, is the difference. In colocated serving the cache and the running reservations share
one pool, so the onset of the all-miss branch is exactly where running reservations fill it; the cache-first versus decode-first
distinction of PD-disaggregated serving collapses into one condition. A second result, Proposition~\ref{prop:scaling}, shows that
GPU speed and tool time enter the peak ratio $\mu^*/\lmiss$ only through their product, which it lowers: memory and tools do not
speed up with the GPU, so a faster GPU lowers the peak ratio, the window's upper end when the from-empty edge does not bind. On an H200 the all-miss capacity rose 2.71-fold and the peak only
1.63-fold, and the window shrank from $(1,2.0)$ to $(1,1.3)\,\lref$ (\S\ref{sec:h200}).

\paragraph{Out-of-sample record.} The model's predictions were frozen and pushed, under tags except the untagged Llama-3.1-8B registration-conditional
predictions, before the agents that derived it read the results (Table~\ref{tab:frozensum}, misses included). They held for the rescue verdicts, the
arrival stop, both basin tests' rules (BASIN-1's blind by access only), a doubled A100 pool, the 600k H200 registration and the H200's from-empty edge in nominal
multiples. They missed the H200's softer tail past the peak (at 600k also the 0.5-hit crossing, 4.4 sessions late), a halved
pool's peak height and edge, 4 of the 9 rescue release times, one seed's basin probabilities, the trapped token rates of two
paired tests, Llama-3.1-8B's from-empty edge (one grid step high, a near miss) and a 30\,s tool slowdown that it said would
trap every seed (it trapped 1 of 3; lengthened from 900 to 1800\,s in a pre-registered test designed after that miss, it trapped 3
of 3, as predicted). On TOOLSLOW-D15, a 15\,s slowdown at 1.65~$\lref$ chosen so that the model and the simulator disagree, its
fluid verdict (TRAPS on every seed) missed: one run of three trapped, as the simulator predicted seed by seed, and the model's
per-seed layer, with the right modal calls, gave that outcome a probability of only 0.41. The 0.5-hit crossings of the five cells
predicted before the runs that set them were read (one of them partly seen) lie within 0.1--4.4 sessions of the measured ones (Table~\ref{tab:lu}).

\paragraph{Model.}\label{sec:fp-model}
Let $w_k=P(T\ge k)/E[T]$ be the turn mix, $E_k=L_k+d$ the KV of a session after turn $k$, $\bar E=\sum_k w_kE_k=10{,}569$ tokens, $A_1=\sum_k P(T\ge k+1)E_k=144{,}960$ the KV a session carries into tool calls over its life, $E[E_T]=11{,}045$ its final KV, $\bar z=E[Z]$, and $\bar z'=\bar z(1-1/E[T])$ its tool time per turn. The tool time $Z$ has distribution $F_Z$ and density $f$. \emph{Server.} At turn completion rate $\nu$ (turns/s) and follow-up hit probability $h$, turn $k$ needs prefill GPU time $t_k(h)=t^m_k-h\,\mathbf 1\{k\ge2\}(t^m_k-t^h_k)$, with $t^m_k,t^h_k$ the miss and hit prefill times of the cost model (a hit prefills $\Delta+1=551$ tokens). SGLang runs prefill and decode in separate iterations with prefill priority, so prefill takes a fraction $p=\nu\sum_kw_kt_k(h)$ of the GPU and the number $R_D$ of decoding requests solves $R_D(1-p)=\nu(d-1)\tau_D(R_D)$, where $\tau_D$ is the cost model's decode iteration time. A turn occupies the engine for $S_k=t_k(h)+S_D$ with $S_D=R_D/\nu$ and reserves its KV $E_k$ throughout, so running requests hold $K_{\mathrm{run}}(\nu,h)=\nu\sum_kw_kE_kS_k$ tokens. \emph{Memory.} Idle KV (sessions in tool calls, and the KV of finished sessions, which the engine cannot know is dead) is evicted least recently used first.
\begin{assumption}[Che~\citep{che2002hierarchical,fricker2012versatile}]
In stationarity an idle item is evicted at a deterministic idle age $R$.
\end{assumption}
\noindent By $H=\lambda G$ the idle KV is then $\mu\mathcal D(R)$ with $\mathcal D(R)=A_1E[\min(Z,R)]+E[E_T]R$, and the pool is full, so
\begin{equation}\label{eq:balance}
\begin{gathered}
M=K_{\mathrm{run}}(\nu,h)+\frac{\nu}{E[T]}\mathcal D(R),\qquad h=F_Z(R),\\
n=\nu\,(\bar S+\bar z')+Q,\qquad \mu=\nu/E[T],\\
Q\ge0,\quad K_{\mathrm{run}}\le M,\quad Q(M-K_{\mathrm{run}})=0,
\end{gathered}
\end{equation}
Here $\bar S=\sum_kw_kS_k$, $Q$ is the stationary fluid population waiting for admission, and $Q=\nu W_Q$ with mean admission wait $W_Q$. The last line is an additional no-slack fluid closure, assumption (S5) of Appendix~\ref{app:entry}, not a consequence of whole-request FCFS admission (S4). Under this closure a positive queue forces running reservations to fill the pool, hence $R=0$ and $h=0$; if $R>0$, then $Q=0$ and Little's law reduces to the queue-free equation. Whole requests can instead leave packing slack while the head request waits, a finite-request effect that this model omits. With $K_{\mathrm{run}}=0$ and no queue, Eq.~\eqref{eq:balance} is the characteristic-time equation of Yao et al.~\citep{yao2026multiturn} term for term. Three additions make the hit rate feed back on itself. Running reservations depend on $h$, because under prefill priority misses starve decode. Throughput is set by a server and a closed population, not by an exogenous arrival rate. And there is a memory-gated all-miss branch. On every partially hitting state the reuse gap is still the exogenous tool time; only on the all-miss branch does the queue set it. The queue-free cliff therefore comes from hit-dependent running reservations, with exogenous gaps. The additional no-slack closure defines the memory-gated all-miss branch; positive $Q$ lengthens the reuse gap, but the turnover diagnostic below is not a separate proof that queueing causes eviction.

\paragraph{Results.}\label{sec:fp-results}
\begin{proposition}[stationary states and the all-miss onset]\label{prop:nb}
Let $\nu_b$ solve $K_{\mathrm{run}}(\nu,0)=M$, $\mu_b=\nu_b/E[T]$, and $E_S=\sum_kw_kE_kS_k/\sum_kw_kS_k$ at $(\nu_b,0)$.
(a) Each $R>0$ gives exactly one queue-free stationary state; under the no-slack closure (S5), call the memory-gated states ($K_{\mathrm{run}}=M$, hence $R=0$, $h=0$ and $\nu=\nu_b$) the all-miss states; every state with a positive queue is an all-miss state; $\mu(n)=\mu_b$ for all large $n$, so $\mu_b$ is the model's $\lmiss$ and $\muinf$ (the model's counterpart of Proposition~\ref{prop:A} of Appendix~\ref{app:proofs} under its assumption (A4), that the congested engine computes prompt tokens as fast as with caching disabled).
(b) All-miss states exist exactly for $n\ge n_b$, where
\begin{equation}\label{eq:nb}
n_b=\nu_b\,\bar z'+M/E_S,
\end{equation}
the sessions in tool calls at the all-miss rate plus one pool's worth of running sessions; at population $n\ge n_b$ the all-miss state has a queue wait of $(n-n_b)/\nu_b$ seconds.
(c) If every GPU cost is divided by $s$ (a uniformly faster GPU), then $\nu_b\mapsto s\nu_b$, $E_S$ is unchanged, and $n_b(s)=n_b(1)+(s-1)\,\nu_b(1)\,\bar z'$.
\end{proposition}
\noindent If $F_Z$ vanishes on some $(0,r_0)$, as with a minimum tool latency, queue-free states with $h=0$ and $\nu<\nu_b$ also exist below $n_b$; we do not call them all-miss states, and for the laws discussed here (exponential, mixtures of exponentials, lognormal), which have $F_Z(R)>0$ for every $R>0$, there are none. At 300k, $n_b=4.94+27.76=32.70$ (native pool: 94.86; vLLM: 32.83). Parts (a) and (c) and Proposition~\ref{prop:scaling}(b)--(c) rest on Lemmas~\ref{lem:server} and~\ref{lem:rho}, which are for the separate-iteration (SGLang) server; for vLLM's mixed-iteration model the identity of part (b) uses the same no-slack closure and a well-defined $\nu_b$ with a nonbinding chunk budget (Appendix~\ref{app:entry}). The pool part $M/E_S$ has the form of MISA-T's cap, a reserved block budget over a conservative session length~\citep{hong2026misat}, and of AgentKV's collapse predictor (from its abstract)~\citep{anon2026agentkv}. Here $E_S$ weights each turn's KV by its service time at the all-miss rate (about 10.8k tokens at 300k), so $M/E_S=27.8$ lies just below $\Nlb=M/\Elb\approx28.5$ of \S\ref{sec:theory-curve} and $M/\bar E=28.4$; all three are capacity ratios of this form. The proposed distinction from those capacity ratios is the tool-call term and the identification of $n_b$ as the onset of the all-miss branch under the no-slack fluid closure. On the two SGLang cells the measured 0.5-hit crossing (32.2 and 93.4) lies just below $n_b$; on vLLM it lies at 29.1, nearer the pool part (28.2) than $n_b$. Part (c) says that a uniformly faster GPU moves the onset \emph{outward}: on an H200 at the same pool ($s=2.742$, fitted to its all-miss token rate) $n_b=41.3$. On the H200 the measured crossing moved out, from 32.2 to about 38.9, though past it the engine lost its hits more gradually than the model (\S\ref{sec:h200}, \S\ref{sec:frozen}). With the A100 pool halved and doubled, the measured crossings, 16.5 and 60.1, lay within 0.4 sessions of the frozen fluid predictions and 2.9 and 4.0 sessions beyond $M/E_S$, passing, blind in time, the test of the tool-call term declared with the second prediction set of \S\ref{sec:frozen} (a crossing more than halfway from $M/E_S$ to $n_b$), at 150k by 0.6 sessions. On Llama-3.1-8B it failed by 0.24 sessions and on the H200 at 300k it passed, but for those two cells it was declared after their curves had been read (Table~\ref{tab:frozensum}).

\begin{proposition}[speed and tool time enter only through $s\bar z$]\label{prop:scaling}
Write $\omega_b=\mu^*/\mu_b$, where $\mu^*$ is the supremum of the throughputs of all stationary states of this fluid model, including when its population curve folds. Let $Z=\bar zY$ with a fixed nonnegative $Y$ having a density and $E[Y]=1$, and let $\bar z>0$. (a) $\mu_b(\mathcal C/s)=s\,\mu_b(\mathcal C)$ (the model's $\lmiss$, Proposition~\ref{prop:nb}(a)) and $\omega_b(\mathcal C/s,\bar z)=\omega_b(\mathcal C,s\bar z)$ exactly, for cost model $\mathcal C$. (b) $\omega_b$ is non-increasing in $s\bar z$, strictly while $\omega_b>1$. (c) $\omega_b\to1$ as $s\bar z\to\infty$.
\end{proposition}
\noindent A uniformly faster GPU therefore lowers the peak ratio, the upper end of the window when the from-empty edge does not bind: the good state is limited by memory and tool time, neither of which speeds up. The same holds for slower tools. On the H200 the measured peak ratio against $\lref$ was 1.89 (turns-based), against 3.15 on the A100 and 1.92 predicted (\S\ref{sec:frozen}; the H200 is not a uniform speed-up: it also changes the attention backend, and its 64-to-32 all-miss token-rate ratio is 1.046 against 1.019 under uniform scaling, \S\ref{sec:fp-weak}); this comparison is not blind to the first author, who had seen the H200 values for $n\le32$ before the freeze. The pool enters only through the model's numerical curve; \S\ref{sec:frozen} tests it with the A100 pool halved and doubled and the H200 pool doubled.

\begin{proposition}[cycle accounting identity]\label{prop:trapid}
Consider a tagged session's completion-to-completion interval $I$ of length $g>0$ in the all-miss round-robin approximation. Suppose this interval accounts for one complete turn of each session $i$ in a set $\mathcal P$, with total KV writes $V_I=\sum_{i\in\mathcal P}(L_{k_i}+d)$ and no net in-flight token carry-over at its endpoints. Define the churn on this same interval as $V_I/g$ and $R_M=M/(V_I/g)$. Then
\[
\frac{g}{R_M}=\frac{V_I}{M}
=\frac{\sum_{i\in\mathcal P}(L_{k_i}+d)}{M},
\]
independently of the GPU cost model. A periodic cycle with fixed completion phases and the stated write-volume accounting gives the same identity for every session.
\end{proposition}
\noindent This is an accounting identity conditional on the cycle assumptions, not an eviction theorem or a proof that the all-miss regime sustains itself. One completion per session in an arbitrarily defined cycle does not by itself make every completion-to-completion gap equal to that cycle's duration; likewise, turns crossing the endpoints can make their total written tokens differ from the completed-turn sum. The engine meets the cycle assumptions only approximately, with arrivals, departures, overlapping service and a growing backlog. On the three 300k BURST runs at 1.35 $\lref$ the measured backlog and prompts give 7.24, 6.43 and 6.81 against measured ratios of 6.92, 6.35 and 6.79, within 5\%. Those window-level comparisons combine a median gap with average churn, so the proposition does not make them exact. The proposition says nothing about the good state; there we approximate the median gap by the median tool time plus the mean service time, so that gap$/R_M\approx(\bar z\ln 2+\bar S)\,\text{churn}/M$. Section~\ref{sec:mechanism} compares both with the engine. The underlying cycle argument is EfficientAgent's: in a backlogged pool the other agents' contexts are referenced between two turns of one agent, giving a reuse-distance scale of the pool's total context (its Eq.~5, an approximation for a host KV tier in a closed pool)~\citep{shao2026efficientagent}. Proposition~\ref{prop:trapid} states the corresponding time-domain accounting relation under explicit cycle assumptions.

\paragraph{Heuristic layers (nothing fitted).}\label{sec:fp-heur}
Five heuristic layers, RR and H1--H4, with nothing fitted, extend the model where the propositions do not reach; they are defined in brief below. (The frozen files and the code call H1--H4 F1--F4; we rename them here to keep them apart from the assumptions F1--F6 of model~F, \S\ref{sec:trap}.) \emph{RR}, the trapped layer, replaces Eq.~\eqref{eq:fluid} in the trap by a fluid FCFS round robin in which every follow-up misses and every present session completes one turn per cycle; the model column of Table~\ref{tab:gapR} comes from it. \emph{H1} averages the curve over the KV-demand fluctuation of $n$ sessions. \emph{H2} predicts the from-empty edge from a one-dimensional diffusion of the demand. \emph{H3} gives the probability that a burst leaves a run stuck; burst sessions are fresh, so a burst counts by its peak footprint (4,331 tokens per session, at turn 12), not by its number. \emph{H4} gives the release time of a cap engaged inside the trap.

\paragraph{The sustaining variable on the engines.}\label{sec:mechanism}
At the same arrival rate the good and the trapped state differ in how long a session's KV lies idle between turns compared with how long the pool keeps idle KV. Let the gap of a follow-up turn be the time from the completion of its session's previous turn to its own completion, the tool time plus the turn's latency, and let $R_M=M/\text{churn}$, where churn is the rate at which new KV is written (uncached prompt tokens plus completion tokens per second). Here $R_M$ is a pool-turnover scale computed from average writes, not a pathwise upper bound on eviction age and not the characteristic time $R$ of Assumption MF. An average churn rate does not control the writes during a particular idle interval. Even for a constant rate, identifying gross writes with LRU reuse distance requires additional assumptions about distinct content, reclamation, locking and recency updates. Denning's reuse-window criterion~\citep{denning1968working} and Che's characteristic-time approximation~\citep{che2002hierarchical} motivate this diagnostic, but do not make an idle time exceeding $R_M$ a sufficient miss condition here. Moreover, the measured gap includes the turn's own service, and SGLang's refresh of waiting prefixes (\S\ref{app:harness}) resets the age of some of them. We therefore treat gap$/R_M$ as a descriptive proxy. Table~\ref{tab:gapR} gives its median in every arm of the three registered tests, computed by the script frozen for the vLLM mechanism checks. In every good-state run the ratio is 0.010--0.028: the gap is 4.9--5.9\,s, mostly tool time, and $R_M$ is 188--505\,s. In every BURST run it is 4.1--9.7: $R_M$ falls to 26--78\,s because the trapped engine rewrites the pool at the all-miss rate, and the gap rises to 166--516\,s, of which the mean tool time is 4.8--5.1\,s and the rest is the turn's latency, mostly time in the queue (a turn's own service takes seconds). Turn latency, not tool time, carries the gap past $R_M$. The trapped value is close to the identity of Proposition~\ref{prop:trapid}, which is exact for the specified common-interval cycle accounting, rather than for arbitrary window averages: in that approximation gap$/R_M$ equals the complete-turn KV volume divided by $M$, and the round-robin layer RR, with nothing fitted to any trapped measurement (it uses the arrival rate, the burst and the server-log cost model), gives values within 4\% of the measured means (Table~\ref{tab:gapR}); this agreement is in sample, since the agents that derived the model knew the measured trapped values of the SGLang cells and of the exploratory vLLM runs, whose arrivals the registered vLLM arm repeats. These are correlations across two states at one load; in the registered TOOLSLOW test of \S\ref{sec:frozen}, a temporary increase in tool latency with no burst, 30\,s more per tool call for 900\,s, trapped the engine in 1 of 3 seeds and 10\,s more in none, so a trigger without a burst can trap the engine, but its registered rule (2 of 3) was not met, and the test does not separate the gap from the capacity loss that the slowdown also causes.

\begin{table}[t]
\centering\footnotesize
\caption{The sustaining variable: median gap between a session's turns over the pool's turnover time $R_M=M/\text{churn}$, over the evaluation window, ranges over seeds (good arms: NOBURST and CAPPED). Gap: median, seconds. Model: Proposition~\ref{prop:trapid} applied to the turn mix and backlog that the round-robin layer RR of the first-principles model predicts (\S\ref{sec:fp-heur}; tag \texttt{r13-theory-pred}; for vLLM validated on the exploratory runs, whose arrivals the registered BURST arm repeats). In sample: the SGLang BURST values and the exploratory vLLM values (6.51--6.72 and 8.05--8.63) were known to the agents that derived the model; the registered vLLM values shown had not been opened at the freeze, but the theory's blindness declaration does not cover them. Computed by \fpath{analysis/r11/loopvars_frozen.py} (commit \texttt{f1338e3}), unchanged.}\label{tab:gapR}
\setlength{\tabcolsep}{2pt}
\begin{tabular}{@{}l c c c c c c@{}}
\toprule
 & & \multicolumn{2}{c}{good arms} & \multicolumn{3}{c}{BURST}\\
test & $\lambda/\lref$ & gap$/R_M$ & $R_M$ (s) & gap$/R_M$ & gap (s) & model\\
\midrule
SGLang 300k & 1.35 & 0.016--0.020 & 255--307 & 6.35--6.92 & 166--178 & 6.93\\
            & 1.65 & 0.022--0.028 & 188--230 & 8.18--9.67 & 210--249 & 8.98\\
SGLang native & 2.5 & 0.010--0.012 & 446--505 & 4.14--4.57 & 322--354 & 4.45\\
            & 4.0 & 0.018--0.022 & 272--310 & 6.68--6.96 & 504--516 & 7.06\\
vLLM 300k   & 1.35 & 0.014--0.017 & 286--337 & 6.44--6.79 & 183--196 & 6.64\\
            & 1.65 & 0.019--0.022 & 229--264 & 8.16--8.78 & 231--261 & 8.52\\
\bottomrule
\end{tabular}
\end{table}

\paragraph{Weaknesses.}\label{sec:fp-weak}
Every proposition is about the deterministic model; we have no limit theorem that takes the engine, or the simulator, to it. Its onset and positive-queue classification additionally use the no-slack closure (S5), which omits whole-request packing slack. Assumption MF is a characteristic-time approximation motivated by concentration with many small objects, whereas our pools hold few: one session's mean footprint is 3.5\% of a 300k pool, and the mean live demand of the predicted (H1) peak population is 85\% of the pool on the A100 at 300k and on vLLM, 99\% at 150k and 114\% on the H200 at 300k, where it already exceeds the pool. The frozen notes named this lack of concentration, worst on the H200, as the likeliest failure there and expected it to bias $\mu^*$ upward; the H200 peak was matched at 300k and within tolerance at 600k (7.3\% high), but past the peak the measured hit fraction stays above the prediction at both pools, the opposite of the overstatement of hits near the peak that a deterministic $R$ implies for these exponential tool times, and we have not traced this miss to a specific assumption (\S\ref{sec:frozen}). At 150k the H1 peak was 11.5\% high, in the expected direction, and H2 put the from-empty edge two grid steps too low. The H200 predictions divide every A100 cost by one factor; uniform scaling predicts a 64-to-32-session ratio of all-miss token rates of 1.019, against 1.046 measured, and a variant that fits both gives the low end of the frozen band. On vLLM the cliff comes earlier and steeper than predicted (crossing at 29.1 against 31.8; $\mu(28)$ 0.123 against 0.225\sps, turns-based), and its plateau below $\lmiss$ is not reproduced. The heuristics have their own misses: H2 was pessimistic about the 300k from-empty edge on fresh seeds, having been chosen knowing the seed-0 bracket, whose unstable run had received 2.73 $\lref$; H2 put vLLM's from-empty edge one grid step high ($[2.5,3.0)$ against the measured $[2.0,2.5)$; in sample); and H4, like the simulator, drains the native-pool gate too fast (R4 released 19--22\% late). Cell-wise unimodality and the absence of folds in the predicted closed curves have numerical, not globally certified, support; RR and H1--H4 are heuristics; choosing this model over a second derivation, and H2 over another edge criterion, were selections made on the three measured cells; and the cost model was fitted to the server logs of the 300k Stage~C runs, so that comparison is out of sample in its target but not in its data source. The basin predictions rest on quantities defined at $n_2$. The curve $\mu(n)$ fixes $n_2-n_1$ (21.3 sessions), the critical burst under Eq.~\eqref{eq:fluid} for sessions with the stationary turn mix, which would have 40 stuck; but a burst of fresh sessions changes the composition, which $N$ alone does not track, and near $n_2$ the quasi-static model is least reliable. So $\mu(n)$ does not determine the critical fresh burst (53.4 sessions), and BASIN-2 tests the heuristic H3, not a consequence of the propositions.

\paragraph{Cost model and server.} The prefill GPU time of $x$ new tokens on a cached prefix of $y$ tokens is $t_P(x,y)=a\max(1,x/B_c)+\max(f_0,bx)+c_p\,x\,(y+(x+1)/2)$, with $a=9.24$\,ms, $f_0=1.26$\,ms, $b=62.6\,\mu$s/token, $c_p=2.24$\,ns per causal pair and chunk $B_c=8192$ (\fpath{calibration/round6_server_log_fit/calibration.yaml}). For a miss, $t^m_k=t_P(L_k,0)$; for a hit, $t^h_k=t_P(551,L_k-551)$; turn~1 always misses. The mean prefill time per turn is $\bar t_P(h)=\sum_kw_kt_k(h)$, so $p=\nu\bar t_P(h)$ and $\partial_h\bar t_P=-\beta_P$, with $\beta_P=\sum_{k\ge2}w_k(t^m_k-t^h_k)$; we also write $\beta_E=\sum_{k\ge2}w_kE_k(t^m_k-t^h_k)$ and $\xi=\nu\,\partial_\nu S_D$. A decode iteration over $R_D$ requests takes $\tau_D(R_D)=a+\max(f_0,bR_D)+b_{\mathrm{att}}R_D$, with $b_{\mathrm{att}}=c_d\bar y_D/8192$, $c_d=0.31$\,ms and $\bar y_D$ the mean decode context. We assume: (S1) a turn's prefill latency is its own prefill GPU time, with no queueing behind other prefills; (S2) the GPU is busy whenever a request runs; (S3) a running request reserves $E_k$ from admission to completion, with no retraction; (S4) admission is first-come-first-served, work-conserving and memory-gated: a request waits only if admitting it would make the running reservations exceed $M$ (idle KV is always evictable), and the engine's limit of 256 running requests and, on SGLang, the prefill chunk budget never bind. For vLLM, prefill and decode share iterations: $S_D=(d-1)/r_{\mathrm{it}}$, with iteration rate $r_{\mathrm{it}}=(1-\nu c_{\mathrm{turn}})/a$, where $c_{\mathrm{turn}}=b\bar x+c_p\bar\pi+(b+b_{\mathrm{att}})(d-1)$ is the token-proportional GPU time of a turn ($\bar x$ and $\bar\pi$ are the mean prefilled tokens and causal pairs per turn), as long as an iteration's tokens fit its chunk budget. Lemmas~\ref{lem:server} and~\ref{lem:rho}, and the results that use them, Proposition~\ref{prop:nb}(a) and (c), and Proposition~\ref{prop:scaling}(b)--(c), are for the separate-iteration (SGLang) server; for vLLM, $n_b$ follows from the identity of Proposition~\ref{prop:nb}(b) whenever the mixed server's $\nu_b$ exists, that is, while the chunk budget does not bind. The queue classification and the identification of $n_b$ as an onset also use the additional fluid closure (S5) below.

\paragraph{No-slack fluid admission.} In addition to (S1)--(S4), we impose (S5): in a stationary fluid state, the queued population $Q\ge0$ and running reservation $K_{\mathrm{run}}\le M$ satisfy $Q(M-K_{\mathrm{run}})=0$. The population balance is $n=\nu(\bar S+\bar z')+Q$, with $Q=\nu W_Q$. This is an idealized divisible-fluid, no-packing-slack closure; it does not follow from (S4) for indivisible requests and is not an assertion about either engine. For example, a pool of 10 units can have 6 reserved and a head request requiring 5 waiting, even though 4 units remain. More generally, at an atomic admission blockage with head reservation $E_{\mathrm{head}}$, (S4) gives only $0\le M-K_{\mathrm{run}}<E_{\mathrm{head}}$, not equality to zero. We retain (S5) to specify the model whose onset and stationary branches are analyzed here; Appendix~\ref{app:proofs}'s request-level FCFS congestion bound does not require it.

\begin{lemma}[fluid server]\label{lem:server}
Fix $h$, and let $\nu_{\mathrm{cap}}(h)=1/(\bar t_P(h)+(d-1)(b+b_{\mathrm{att}}))$. For $0<\nu<\nu_{\mathrm{cap}}(h)$ the decode equation has exactly one positive root; for $\nu\ge\nu_{\mathrm{cap}}(h)$ it has none. $S_D$ is continuous, strictly increasing in $\nu$, non-increasing in $h$, and $S_D\to\infty$ as $\nu\uparrow\nu_{\mathrm{cap}}(h)$. At $\nu=0$ the decode root is $R_D=0$ and $S_D$ is defined by its continuous limit $(d-1)\tau_D(0)$. For $\nu>0$, $K_{\mathrm{run}}(\nu,h)$ increases strictly in $\nu$ from $0$ to $\infty$ and decreases strictly in $h$.
\end{lemma}
\begin{proof}
For $\nu>0$, if $p=\nu\bar t_P(h)\ge1$, the left side of the original decode equation $R_D(1-p)=\nu(d-1)\tau_D(R_D)$ is nonpositive while its right side is positive, so no positive root exists. For $p<1$, define $\psi(R)=R/\tau_D(R)$; it is continuous and strictly increasing, since $\tau_D$ is affine with positive intercept on each side of the kink $f_0/b$. Moreover $\psi(0)=0$ and $\psi(\infty)=1/(b+b_{\mathrm{att}})$. The decode equation reads $\psi(R_D)=\nu(d-1)/(1-p)$, which has a unique root if and only if the right side is below $1/(b+b_{\mathrm{att}})$, that is, $\nu<\nu_{\mathrm{cap}}(h)$. The right side increases in $\nu$ and does not increase in $h$, because $t^h_k\le t^m_k$; so does $R_D$, which diverges at $\nu_{\mathrm{cap}}$. Then $S_D=R_D/\nu=(d-1)\tau_D(R_D)/(1-p)$ is a product of positive factors with the same monotonicity. $K_{\mathrm{run}}=\nu\sum_kw_kE_k(t_k(h)+S_D)$ inherits it; strict decrease in $h$ follows from $\partial_hK_{\mathrm{run}}=-\nu(\beta_E+\bar E\chi\xi)<0$ (Lemma~\ref{lem:rho}).
\end{proof}

\begin{lemma}[decode sensitivity]\label{lem:rho}
Away from the kink, $-\partial_hS_D=\chi\,\nu\partial_\nu S_D$, with $\chi=\beta_P/(\bar t_P(h)+(d-1)\tau_D'(R_D))$ and $\nu\partial_\nu S_D=R_D(\bar t_P+(d-1)\tau_D')/J>0$, where $J=(1-p)-\nu(d-1)\tau_D'>0$.
\end{lemma}
\begin{proof}
Write $G(R_D,\nu,h)=R_D(1-\nu\bar t_P(h))-\nu(d-1)\tau_D(R_D)$. Then $\partial_{R_D}G=J$; at the root, $J=(1-p)\alpha_D/\tau_D>0$, with $\alpha_D>0$ the intercept of the affine piece of $\tau_D$. Next, $\partial_\nu G=-R_D\bar t_P-(d-1)\tau_D$ and $\partial_hG=\nu R_D\beta_P$, since $\partial_h\bar t_P=-\beta_P$. Hence $\partial_hS_D=\partial_hR_D/\nu=-R_D\beta_P/J$ and $\nu\partial_\nu S_D=\partial_\nu R_D-R_D/\nu=[R_D\bar t_P+(d-1)\tau_D-JR_D/\nu]/J$. Substituting $JR_D/\nu=R_D/\nu-R_D\bar t_P-R_D(d-1)\tau_D'$ and $R_D/\nu=(d-1)\tau_D/(1-p)$ gives $(d-1)\tau_D-R_D/\nu=-pR_D/\nu=-\bar t_PR_D$, so the numerator equals $R_D(\bar t_P+(d-1)\tau_D')$.
\end{proof}

\begin{proof}[Proof of Proposition~\ref{prop:nb}]
(a) Fix $R>0$ and $h=F_Z(R)$. The function $\nu\mapsto K_{\mathrm{run}}(\nu,h)+\nu\mathcal D(R)/E[T]-M$ is continuous and strictly increasing (Lemma~\ref{lem:server}). It equals $-M$ at $\nu=0$ and is unbounded at $\nu_{\mathrm{cap}}$, so it has exactly one root $\nu(R)$, which is continuous in $R$. By (S5), a state with $Q>0$ has $K_{\mathrm{run}}=M$, so $\mu\mathcal D=0$, hence $R=0$, $h=0$ and $\nu=\nu_b$. From $\nu\mathcal D(R)/E[T]\le M$ and $\mathcal D(R)\ge E[E_T]R$ we get $\nu(R)\to0$ as $R\to\infty$. As $R\downarrow0$, $\nu(R)\to\nu_b$ by continuity, since $\nu_{\mathrm{cap}}(0)$ is excluded. As $R\to\infty$, $h\to1$, the prefill times remain bounded and $S_D\to(d-1)\tau_D(0)$, so $n(R)=\nu(R)(\bar S+\bar z')\to0$. Thus $n(R)$ is bounded on $(0,\infty)$, and beyond its supremum and $n_b$ only the all-miss state remains.
(b) $K_{\mathrm{run}}(\nu_b,0)=\nu_b\sum_kw_kE_kS_k=M$ gives $\nu_b\bar S=M/E_S$; Little's law with no queue gives $n_b=\nu_b\bar S+\nu_b\bar z'$. For $n\ge n_b$ the remaining population is $Q=n-n_b$, which is nonnegative exactly on this range. Its mean admission wait is $W_Q=Q/\nu_b=(n-n_b)/\nu_b$ seconds; these values satisfy (S5) and the population balance. No all-miss state with $K_{\mathrm{run}}=M$ is possible below $n_b$.
(c) Divide every cost by $s$ and time by $s$: every service time scales by $1/s$ and every rate by $s$, and the decode equation and $p$ are invariant. So $(\nu,h)$ solves $K_{\mathrm{run}}=M$ for costs $\mathcal C$ if and only if $(s\nu,h)$ does for $\mathcal C/s$. Hence $\nu_b(s)=s\nu_b(1)$; $E_S$, a ratio of $S$-weighted sums, is invariant; substitute into (b).
\end{proof}

\begin{proof}[Proof of Proposition~\ref{prop:scaling}]
(a) Rescaling time by $s$ maps a stationary state $(n,\nu,h,R)$ of $(\mathcal C/s,\bar z)$ to $(n,\nu/s,h,sR)$ of $(\mathcal C,s\bar z)$. Each equation of Eq.~\eqref{eq:balance} is invariant: Little's law and $K_{\mathrm{run}}$ are rate$\times$time, $E[\min(\bar zY,R)]=s^{-1}E[\min(s\bar zY,sR)]$, and $h$ compares times. Both $\mu^*$ and $\mu_b$ scale by $s$, so their ratio depends on $s\bar z$ only.
(b) Parametrize the queue-free states by $R\in(0,\infty)$, with $h=F_Y(R/\bar z)$. The balance reads $K_{\mathrm{run}}(\nu,F_Y(R/\bar z))+\nu(A_1E[\min(\bar zY,R)]+E[E_T]R)/E[T]=M$. Its left side increases strictly in $\nu$ (Lemma~\ref{lem:server}) and does not decrease in $\bar z$, since $h$ falls, $K_{\mathrm{run}}$ decreases in $h$ and $E[\min(\bar zY,R)]$ rises; where $h>0$ it increases strictly in $\bar z$, because then $E[\min(\bar zY,R)]$ does. So $\nu(R;\bar z)$ does not increase in $\bar z$, while $\nu_b$ (the equation at $h=0$) does not involve $\bar z$, and taking the supremum over $R$, $\omega_b$ is non-increasing in $\bar z$. For strictness, let $\bar z_1<\bar z_2$ and $\omega_b>1$ at $\bar z_2$. Since $Z$ has a density, $\nu(R;\bar z)$ is continuous in $R$, and by the proof of Proposition~\ref{prop:nb}(a) it tends to $\nu_b$ as $R\downarrow0$ and to 0 as $R\to\infty$. So $\sup_R\nu(R;\bar z_2)>\nu_b$ is attained at some $R_2\in(0,\infty)$, where $h>0$, because a state with $h=0$ and $R>0$ has $K_{\mathrm{run}}(\nu,0)<M$ and so $\nu<\nu_b$. Hence $\sup_R\nu(R;\bar z_2)=\nu(R_2;\bar z_2)<\nu(R_2;\bar z_1)\le\sup_R\nu(R;\bar z_1)$. This holds for any $Y$ with a density, bounded or not and whether or not $F_Y$ has flat parts. Apply (a) for $s$.
(c) Fix $h_0\in(0,1)$ and let $\widehat\nu(h_0)$ be the unique root of $K_{\mathrm{run}}(\nu,h_0)=M$. On any queue-free state with $h\le h_0$, monotonicity gives $K_{\mathrm{run}}(\nu,h_0)\le K_{\mathrm{run}}(\nu,h)\le M$, hence $\nu\le\widehat\nu(h_0)$. Since $F_Y(0)=0$ and $F_Y$ is continuous, choose $r_0>0$ with $F_Y(r_0)<h_0$. A state with $h>h_0$ has $R>\bar z r_0$, so the memory balance and $\mathcal D(R)\ge E[E_T]R$ imply
\[
\nu\le\frac{ME[T]}{E[E_T]\,\bar z r_0}.
\]
Consequently
\[
1\le\omega_b(\bar z)\le
\max\left\{\frac{\widehat\nu(h_0)}{\nu_b},
\frac{ME[T]}{\nu_b E[E_T]\,\bar z r_0}\right\}.
\]
First let $\bar z\to\infty$ and then $h_0\downarrow0$. Continuity of the root gives $\widehat\nu(h_0)\to\nu_b$, proving the limit without assuming that $F_Y$ is invertible.
\end{proof}

\begin{proof}[Proof of Proposition~\ref{prop:trapid}]
By definition the interval's average churn is $V_I/g$, so $R_M=Mg/V_I$ and $g/R_M=V_I/M$. Substituting the assumed complete-turn accounting gives the final sum. The cost model affects $g$ but cancels from the ratio. No claim about an individual prefix's eviction age is used or follows.
\end{proof}

\paragraph{Not modelled.} SGLang's retraction of running requests under memory pressure, which the theory notes list as a candidate cause of the 1--3 point deficit in the trapped token rate, although none of our 73 SGLang server logs records one (Appendix~\ref{app:proofs}); prefill queueing (S1); the limit of 256 running requests, which (S4) assumes never binds (at the all-miss state of Proposition~\ref{prop:nb}, $\nu_b\bar S=M/E_S$ requests run, so this needs $M<256E_S\approx2.8$ million tokens, which every cell satisfies); and the reset of the release hold when sessions arrive just after the gate empties.

\subsection{Frozen predictions and their outcomes}\label{sec:frozen}

The registered tests of \S\ref{app:e2tests} check rules written before them. For the later tests, predictions were also frozen: by the calibrated simulator (\S\ref{sec:sim}) and by the first-principles model of Appendix~\ref{app:entry}, under tags \texttt{r13-theory-pred} (pushed at 02:05:30 CDT on 2026-10-31) and \texttt{r13-theory-pred2} (pred2 below, pushed at 04:20:20), whose new drivers call the model's frozen functions unchanged, with primaries, estimators, tolerances and refutation rules fixed in advance (\fpath{notes/THEORY_r13_scoring.md}). Each was committed under its tag before any of its results was read by the agents that made it (the Llama-3.1-8B registration-conditional predictions were pushed untagged, at 07:52 CDT, after their jobs were submitted but before any started); what each prediction is blind to differs and is stated with it. Table~\ref{tab:frozensum} summarizes the outcomes in hand, and the prediction files (\fpath{notes/PREDICTIONS_*.md}, with the scoring rules in \fpath{notes/THEORY_r13_scoring.md}) give every prediction in full. Every outcome is reported against these rules, misses included.

\begin{table*}[!tp]
\centering\footnotesize
\caption{Frozen predictions and the outcomes in this version. ``Simulator'': the calibrated simulator of \S\ref{sec:sim}; ``theory'': the model of Appendix~\ref{app:entry}. Blind in time: pushed before the runs were submitted or started; by access only: the runs had ended, or were in progress, at the push, and were first read after it by anyone on the project. The prediction files \fpath{notes/PREDICTIONS_*.md} give every prediction with its scoring rule.}\label{tab:frozensum}
\setlength{\tabcolsep}{3pt}
\begin{tabular}{@{}>{\raggedright\arraybackslash}p{3.0cm} >{\raggedright\arraybackslash}p{7.4cm} >{\raggedright\arraybackslash}p{6.5cm}@{}}
\toprule
Test & Frozen prediction & Outcome\\
\midrule
H200, 300k pool (a measurement, not registered) & Theory: peak ratio 1.92, 0.5-hit crossing $37.3\pm3$, all-miss onset 41.3, from-empty edge 1.39 $\lref$. Peak not blind to the first author; $n=36$, 44, 52 blind in time; the other runs blind by access only & Peak ratio 1.89, crossing $\approx38.9$, edge in $[1.3,1.4)\,\lref$ nominal (a hit; H2 about 7\% pessimistic in realized rates); no refutation rule fires. Tail softer than predicted from $n=40$, so the onset (41.3) came early\\
\addlinespace[0.3pt]
H200, 600k pool (registered) & pred2: peak ratio 3.20, crossing $66.6\pm3$, edge 2.57 $\lref$, registration at 1.5 and 2.0 $\lref$ with cap 48; there, trapped backlog 458 and 867, token ratio 1.054 and 1.064. Blind in time & Peak ratio 2.98, nominal edge in $[2.5,3.0)$ and registration as predicted; crossing $\approx71.0$, a miss (tail softer, as at 300k). Paired test passes at both rates (BURST trapped 2 and 3 of 3, at zero hits 6 of 6); backlogs within 15\%, token ratios outside $\pm0.03$ in 5 of 6\\
\addlinespace[0.3pt]
Pool sweep, A100, 150k and 600k & Theory: peak ratio 1.87 and 5.47, crossing 16.9 and 60.0, edge 1.09 and 4.53 $\lref$. Blind in time & 600k: all within tolerance. 150k: cliff within tolerance; peak ratio (1.68; predicted 11.5\% high) and edge ($[1.2,1.3)$, two steps above) missed\\
\addlinespace[0.3pt]
Llama-3.1-8B, 300k (registered) & Theory, a priori: peak ratio 3.37, crossing $30.9\pm3$, edge 2.55 $\lref$ (2.46 times its a priori $\lmiss$). pred2, declared after the curve was read: tool-call term if the crossing exceeds 29.89; the nearest of $M/\Elb$, $M/E_S$ and the frozen crossing wins. Simulator: peak at $n=20$; primary mode: edge 2.5, rates 1.5, 2.0 $\lref$; a pass at both registered rates. Closed runs by access only, from-empty runs blind in time & Peak ratio 3.28 at $n=20$, crossing 29.65: within tolerance, but 0.24 below 29.89 (a fail); the nearest-predictor rule picks $M/\Elb$ (28.44), by 0.06 on the listed crossing. Edge in $[2.0,2.5)\,\lref$, so the theory's edge is one grid step high (a near miss). Simulator: peak and cap right; edge and rates one step high. Paired test passes at 1.65 $\lref$ (trapped 2 of 3) and fails at 1.35 (1 of 3; all six BURST runs at zero hits), so the simulator's pass at 1.35 missed\\
\addlinespace[0.3pt]
Fresh-seed E0b, 300k (exploratory) & Simulator: stable in 6 of 6 (blind in time). Theory (H2, not declared for this test; pred2's table pushed after the runs were read): unstable with probability 0.69 per run at 2.5 $\lref$ & Stable in 6 of 6: the seed-0 bracket is not replicated (the rule needed 2.5 $\lref$ unstable in 2 seeds). H2 pessimistic (all three stable at 2.5: 0.03). Later same-plan replicates (exploratory) reproduced seed 0's collapse and seed 4's stability, 3 of 3 each: H2 does not describe same-plan replicates (not a refutation of its cell-level escape probability)\\
\addlinespace[0.3pt]
BASIN-1, 1.65 $\lref$ (registered) & Simulator: pass (blind in time). Theory: this outcome with probability 0.879 (blind by access only) & PASS: 100 sessions trapped and 25 absorbed, 3 of 3 seeds each; simulator right in 6 of 6 runs\\
\addlinespace[0.3pt]
BASIN-2, 1.65 $\lref$ (registered) & Theory: stuck with probability 0.14, 0.55, 1.0 at 40, 50, 60 (critical burst 53.4). pred2, per seed (primary): the rule holds with probability 0.66; secondary, a deterministic crossing per seed. Blind in time & Bracketed: 40 good, 60 stuck, 3 of 3 each; 50 stuck in 1 of 3; 53.4 inside two of three per-seed edges; pred2's primary wrong for seed 0 at 40 and 50 (stuck with probability 0.78, 0.86; both good), right side in the other 7; secondary right in 9 of 9\\
\addlinespace[0.3pt]
Rescue R1, 300k (registered) & Simulator: pass, per-seed release times. Theory: rescued, release at 3830 and 4360\,s ($\pm15\%$). Both blind in time & PASS, 6 of 6 rescued, no relapse. Releases within 600\,s of the simulator's in 6 of 6 and within $\pm15\%$ of the theory's in 5 of 6\\
\addlinespace[0.3pt]
Rescue R4, native pool (registered) & Simulator: pass, per-seed release times. Theory: rescued, release at 5040\,s ($\pm15\%$). Both blind in time & PASS, 3 of 3 rescued, no relapse. Releases 800--1750\,s after the simulator's and 19--22\% after the theory's, outside its $\pm15\%$ (a miss, or a refutation by the headline table; see text)\\
\addlinespace[0.3pt]
LPM-INF, 1.65 $\lref$ (exploratory) & Simulator: no collapse on the seeds on which stock LPM collapsed (blind in time) & No collapse (token hit fraction 0.59 and 0.42); level missed (0.08--0.10 predicted; no partial hits, as flagged); no recovery\\
\addlinespace[0.3pt]
SHED, 1.35 $\lref$ (registered) & pred2: arrivals stopped from 1800\,s for 600\,s do not drain the trap (1.00 per seed), for 3600\,s drain it (0.998); rule 0.997. Blind in time & PASS: 600\,s stuck, 3600\,s good, 2 of 2 each; 56--59 and 355--378 sessions refused (the cap: none)\\
\addlinespace[0.3pt]
TOOLSLOW, 1.35 $\lref$, no burst (registered) & pred2: tool calls 10\,s longer for 900\,s leave the engine good (0.71 per seed), 30\,s longer trap it (1.00); rule 0.80. Blind in time & Rule not met: 10\,s good in 3 of 3, but 30\,s stuck in 1 of 3; pred2 missed\\
\bottomrule
\end{tabular}
\end{table*}

\begin{table*}[!tp]
\centering\footnotesize
\caption{Frozen predictions of rounds 15 and 16 (2026-11-05, after Lu et al.'s paper appeared) and their outcomes, with the conventions of Table~\ref{tab:frozensum}; times in Table~\ref{tab:timeline-r15}.}\label{tab:frozensum15}
\setlength{\tabcolsep}{3pt}
\begin{tabular}{@{}>{\raggedright\arraybackslash}p{3.0cm} >{\raggedright\arraybackslash}p{7.4cm} >{\raggedright\arraybackslash}p{6.5cm}@{}}
\toprule
Test & Frozen prediction & Outcome\\
\midrule
TOOLSLOW-W, 1.35 $\lref$, no burst, 30\,s longer for 1800\,s (registered, round 15) & Simulator, both modes: stuck in 3 of 3 (TRAPS); pred2: TRAPS, stuck with probability 1.00 per seed; trap-state law: S1--S5 of \texttt{icml-trap-pred}. Blind in time; designed after TOOLSLOW and after reading simulator outputs & TRAPS, 3 of 3; every model right; pooled S1--S4 pass; outstanding requests at 3300\,s off the reference accuracy by 8 and 6 in seeds 0 and 2 (Appendix~\ref{app:blind})\\
\addlinespace[0.3pt]
TOOLSLOW-D15, 1.65 $\lref$, no burst, 15\,s longer for 900\,s (registered, round 15) & Simulator, both modes: stuck, good, good (NO-TRAP); pred2 fluid verdict: TRAPS; pred2 per seed: modal stuck, good, good, $P(\text{TRAPS})=0.58$. Blind in time & NO-TRAP: seed 0 stuck, seeds 1 and 2 good; the simulator right in 3 of 3 runs, pred2's fluid verdict wrong, its per-seed layer modal-right with $P(\text{NO-TRAP})=0.41$; the stuck run passes S1--S4, and S5's closeness is missed by 0.0004\,/s (\S\ref{sec:validation})\\
\addlinespace[0.3pt]
CUT, BURST at 1.65 $\lref$, arrivals cut at 2400\,s to 1.15 (CUT-A) or 0.85 $\lref$ (CUT-B) (registered, round 15) & Model~F (\texttt{icml-cut-pred}), from each seed's exact plan: all six stuck; backlog slopes over $[3000,8700)$\,s of $+0.0039$, $-0.0026$, $+0.0028$ (CUT-A) and $-0.0095$ to $-0.0131$\,/s (CUT-B). Blind in time; the rule was registered after the predictions were read & CUT-A: PASS, theory-favoured 3 of 3, late sign 2 of 3 as predicted; the rule rejects only the naive and quasi-static readings (a constant rate at the cut fits as well). CUT-B: PASS, theory-favoured 3 of 3, $S<0$ in 3 of 3, still stuck at 8700\,s (\S\ref{sec:control})\\
\addlinespace[0.3pt]
REPLICATE-EDGE, 2.5 $\lref$ from empty, 300k, the plans of seeds 0 and 4 (exploratory, round 15) & Every deterministic predictor (the simulator in both modes, pred2's level crossing, the arrival, turn and token counts): each plan reproduces its original; H2 read as independent draws: reproducible with probability 0.0097. Blind in time; the plans were chosen knowing the originals & Reproducible: seed 0 collapsed in 3 of 3 (at 1600, 1600 and 1800\,s), seed 4 stable in 3 of 3; H2 does not describe same-plan replicates (\S\ref{sec:limits})\\
\addlinespace[0.3pt]
MIRROR, BURST at 1.65 (decisive) and 1.35 $\lref$, step-capped workload, at most 20 turns (registered, round 16) & Model~F (\texttt{r16-mirror-reg}), from each seed's exact plan, at 1.65 $\lref$: completions 0.05--0.09 $\lref$ over $[2400,3600)$\,s, a wave of the burst cohort at 5135--5195\,s, 0.49--0.60 $\lref$ over $[6600,8700)$\,s, backlog slopes nearer model~F's than the quasi-static ones. Blind in time; the workload was chosen after these predictions had been computed & PASS at 1.65 and 1.35 $\lref$, RENEWAL in 3 of 3 seeds each; all six stuck at zero hits, NOBURST good 3 of 3; at 1.65 $\lref$: 0.06--0.09 $\lref$ over $[2400,3600)$\,s, the wave at 4945--5089\,s, 0.50--0.61 $\lref$ over $[6600,8700)$\,s, below the measured all-miss capacity (1.062 $\lref$); the calibrated wave (W5) missed in 1 of 6 (\S\ref{sec:mirror})\\
\bottomrule
\end{tabular}
\end{table*}

\paragraph{H200, 300k pool.} The model's only H200 inputs were that GPU's E0a measurements, through one speed factor, $s=2.742$, fitted to its all-miss token rate (Proposition~\ref{prop:nb}(c)). At 300k its peak ratio, crossing and from-empty edge held (Table~\ref{tab:frozensum}; the edge on the original seed-0 runs in nominal multiples, while in realized rates, 1.496--1.503 $\lref$, H2 was about 7\% pessimistic), no refutation rule fires, and from $n=40$ the engine keeps more hits than the model (\S\ref{sec:fp-weak}). The agent that operated the 300k H200 cell showed its closed-curve values for $n\le32$, which include the predicted peak, to the first author before the freeze, though not to the agents that derived the model, so the peak ratio is not blind to the first author; the runs at $n=36$, 44 and 52 are blind in time, and the other closed runs and the original from-empty runs, which had ended before the freeze, blind by access only.

\paragraph{Pools and the window law.} The pool sweep repeated Stage~C and E0b on the A100 with the pool halved (150k tokens) and doubled (600k), with no parameter refitted; its runs started after both prediction sets were pushed. At 600k every scored quantity is within tolerance. At 150k the model located the cliff (crossing 16.5 against 16.9) but overstated the peak ratio by 11.5\%, in the direction the frozen notes expected from Assumption MF, and H2 put the edge two grid steps low; no refutation rule fires. The declared log-likelihoods of the six from-empty outcomes favour H2 over the second (level-crossing) edge criterion at 150k ($-2.14$ against $-2.95$) and on the 300k H200 ($-1.35$ against $-2.82$), but not at 600k ($-0.93$ against $-0.54$). On the fresh 300k seeds, for which the frozen 300k file declares no prediction, its level-crossing probabilities also did better than H2 ($-1.47$ against $-4.01$ for all six runs stable); pred2's comparison, declared after those runs were read, scores the deterministic level crossing on each seed's exact plan, which called all six stable. At both pools the frozen crossing, which counts the sessions in tool calls, was nearer than the capacity ratio $M/E_S$ (2.9 and 4.0 sessions early) and the exogenous-gap closures, in which a session's reuse gap is its tool time (7--46 sessions late); on Llama-3.1-8B $M/\Elb$ won that rule narrowly (below). On the H200 with 600k (pred2) the peak ratio, peak position, edge and registration came as predicted, but the crossing came 4.4 sessions late, a miss. Over six cells (Table~\ref{tab:windowlaw}) the measured window shrinks with GPU speed and grows with the pool, consistent with the model's numerical curves. Proposition~\ref{prop:scaling} proves monotonicity of the peak ratio, not of the from-empty edge that sets these measured windows. Doubling the pool raised the peak ratio 1.81-fold on the A100 and 1.57-fold on the H200, against 1.67 for both; H2's edge lies inside the measured nominal bracket in five of the six cells (two of the three blind in time).

\paragraph{A second model.} With Llama-3.1-8B-Instruct on the same GPU, engine and pool, $\lref$ is 17\% lower (0.0596\sps) and the curve has Qwen's shape (Table~\ref{tab:frozensum}), so the registration rule gives Qwen's rates and cap. The model's a priori prediction, pushed while the closed runs were in progress (blind by access only), came within tolerance on the peak ($+6.7\%$), peak ratio, peak position and crossing, and its edge was a near miss. The crossing, 29.65, failed pred2's test of the tool-call term, by 0.24 sessions, and pred2's nearest-predictor rule picks $M/\Elb$ (28.44, 1.21 sessions away) over the a priori crossing it lists (30.92, 1.27 away); read at the measured speed factor, the frozen crossing would win narrowly. Both rules were declared with pred2 after this curve had been read.

\paragraph{Basin.} For BASIN-1 the simulator, pushed 68 minutes before the jobs were submitted, predicted a pass and matched all six per-run calls, with hit fractions within 0.006 and backlog slopes within 0.0005\sps{} of the engine's. The model gave the observed outcome the probability $1.000\times0.958^3=0.879$, blind by access only: every run had passed its burst before the push, and the 100-session runs had ended. For BASIN-2, pred2, conditioning on each seed's burst, gave the registered bracketing rule probability 0.66, blind in time. It held. The per-seed probabilities, pred2's primary prediction, were on the more likely side in 7 of the 9 runs but gave seed 0 a stuck run with probability 0.78 at 40 and 0.86 at 50 sessions, and both runs stayed good (Brier score 0.17, against 0.10 for the frozen average-burst probabilities); its secondary prediction, a deterministic level crossing on each seed's exact arrivals, called all nine runs. The critical burst of 53.4 lies inside the edge interval of seeds 0 and 2, $(50,60]$, not of seed 1, $(40,50]$.

\paragraph{Rescue.} Both prediction sets were pushed before the first rescue run started (the simulator's at 21:30:40 CDT on 2026-10-30, the model's at 02:05:30; first start 02:23:44 on 2026-10-31), so all nine runs are blind in time to both. The simulator's pass and no-relapse calls all held, and its release times were early in R4 (Table~\ref{tab:frozensum}). The model predicted rescue in every cell, almost no relapse, and releases at 3830, 4360 and 5040\,s ($\pm15\%$): R1 released within the tolerance in 5 of 6 runs, all late ($+1.6\%$ to $+18.8\%$), and R4 outside it in 3 of 3, 19--22\% late. The scoring declaration calls a release outside $\pm15\%$ a miss and makes only NOT RESCUED refute, while the headline table frozen with it lists the release time as falsifying; under that table the R4 release prediction, and the R1 prediction at 1.35 $\lref$ seed 0, are refuted. R4's engage time was moved from 2400 to 2100\,s after the simulator had predicted 13 of 20 rescued at 2400\,s, and the agents that froze the model's rescue predictions knew the simulator's.

\paragraph{Earlier prediction rounds.} The calibrated simulator's first three prediction rounds (rounds 6, 8 and 9 of \S\ref{sec:sim}: B1--B4, B1n--B4n, B1v--B4v) predicted the registered paired tests and the curves and edges behind them (\fpath{notes/PREDICTIONS.md}, \fpath{notes/PREDICTIONS_native.md}, \fpath{notes/PREDICTIONS_vllm.md}). Every predicted pass held (B3n's at its own test points, 6--16\% above the registered rates), with trapped token rates and good-arm hit fractions close to the engine's, except that vLLM's trapped token rates lay at or below B3v's range (0.948--1.018 against 0.99--1.07). B4 and B4v got the peak's level within 9\% and 2\% but not the cliff (B4's came early, mostly because it was predicted for a smaller pool, and B4v's late); B2 and B2v got the from-empty edge in nominal multiples, and the primary mode of B2n put the native edge at least one grid step too high. Appendix~\ref{app:timeline} gives what each round is blind to.

\paragraph{Arrival stop and tool slowdown.} Two registered tests at 1.35 $\lref$ on the 300k pool, submitted after pred2's push (blind in time), probe another exit from the trap and another trigger. SHED stops arrivals from 1800\,s, 600\,s after a burst of 200, for $\tau=600$ or 3600\,s: with $\tau=600$\,s both runs stayed stuck (hit fraction 0.000 over the evaluation window) and with $\tau=3600$\,s both were good (0.930 and 0.933), as pred2 predicted (rule probability 0.997; its shortest rescuing stop, 1795\,s, lies between the two). The stops refused 56--59 and 355--378 sessions; the cap of \S\ref{sec:rescue} refused none. TOOLSLOW lengthens by $\delta$ every tool call that starts in $[1200,2100)$\,s, with no burst. With $\delta=10$\,s all three runs stayed good, but with $\delta=30$\,s only seed 2 was trapped while seeds 0 and 1 recovered, so the registered rule (stuck in at least 2 of 3) was not met. pred2 had predicted a trap in every seed at 30\,s (probability 1.00 per seed; rule 0.80), and its per-seed walk put seed 0, which recovered, highest (49 sessions at 2100\,s, against 37 and 39): a miss of the model. A 30\,s slowdown of every tool call for 900\,s, with no burst, can trap the engine, but did so in one of three seeds.

\begin{table}[t]
\centering\footnotesize
\caption{The window across pool and GPU. Peak ratio $\mu^*/\lref$, measured (turns-based; completed sessions in parentheses) against the frozen H1 prediction; from-empty edge, measured bracket (nominal multiples of $\lref$ on seed-0 grids: the native cell's own, otherwise the 300k value of the same GPU, as in Table~\ref{tab:frozensum}) against the frozen H2 edge (at 300k three fresh seeds were stable at 2.5, where seed 0 was not, so the bracket is not replicated). Status: b, blind in time; i, in sample (H2 was chosen knowing these two brackets); s/a, peak seen by the first author before the freeze, edge blind by access only. Predictions: tag \texttt{r13-theory-pred}; H200 at 600k, pred2.}\label{tab:windowlaw}
\setlength{\tabcolsep}{3pt}
\begin{tabular}{@{}l c c c c c@{}}
\toprule
cell & status & $\mu^*/\lref$ & H1 & edge & H2\\
\midrule
A100, 150k & b & 1.68 (1.65) & 1.87 & $[1.2,1.3)$ & 1.09\\
A100, 300k & i & 3.15 (3.26) & 3.27 & $[2.0,2.5)$ & 2.34\\
A100, 600k & b & 5.71 (5.98) & 5.47 & $[4.5,5.0)$ & 4.53\\
A100, native & i & 6.48 (6.67) & 6.89 & $[5.5,6.5)$ & 6.26\\
H200, 300k & s/a & 1.89 (1.99) & 1.92 & $[1.3,1.4)$ & 1.39\\
H200, 600k & b & 2.98 (3.05) & 3.20 & $[2.5,3.0)$ & 2.57\\
\bottomrule
\end{tabular}
\end{table}

\subsection{The registrations}\label{app:prereg}

The registration is \fpath{configs/agentload_prereg_2026-10-24.yaml}, status FILLED, committed in \texttt{d279b7b} at 15:09:12 CDT on 2026-10-24 and pushed to GitHub at 15:09:54, before the E2 jobs were submitted at 15:13:39, SHA-256 \texttt{2a6d680a9615bba962a5673d6db6b3e4\allowbreak ca5102bf778ca974cea46147f5cc0ac6}; the job script writes this SHA-256 into every run's \fpath{meta.json}. Its fields:

\begin{itemize}
\item \textbf{Cell and inputs.} Engine cell \texttt{sglang-fcfs-r1} (SGLang FCFS, waiting-prefix refresh on); committed copies of the Stage~C curve, the E0a calibration and the E0b bracket under \fpath{prereg/inputs/2026-10-23/}; protocol 8700\,s, burst of 200 within 60\,s at 1200\,s.
\item \textbf{Measured.} $\mu(n)$ in sessions/s: 0.1756 ($n=16$), 0.2239 (20), 0.2333 (24), 0.1150 (32), 0.0489 (48), 0.0611 (96), 0.1078 (300). $\mu^*=0.2333$; $\ncap=20$ by the smallest-within-5\% rule (argmax 24; flagged ambiguous). $\lmiss=0.07167$ (sessions) and 0.07153 (turns-based, $=\lref$); all-miss rate 11,243 tokens/s; E0b largest all-stable multiple 2.0, first unstable 2.5, monotone (arrival seed 0 only; the fresh seeds of \S\ref{app:calib} came later); hardware A100 80GB PCIe.
\item \textbf{Predicted boundaries.} Recovery threshold $\lref$ (not $\mu(300)$, which a synchronized start biases upward). Window $(0.0715,\,0.1431)$\sps{} $=(1.0,\,2.0)\,\lref$, upper edge from E0b. CAPPED drain must fit within 80\% of the 7200\,s available after the burst (5760\,s), which holds for $\lambda$ up to 2.64 $\lref$; predicted drain 1571\,s at 1.35 and 1889\,s at 1.65.
\item \textbf{Rates.} 1.35 and 1.65 $\lref$ by the one-third/two-thirds rule, rounded to 0.05, inside the margin band $(1.1,\,1.8)$; seeds 0, 1, 2. On 2026-10-22 the runbook proposed the rate rule and a cap of 20, with keeping 24 as the alternative (\texttt{288b801}, 13:13 CDT); both rules were adopted in the project status notes at 13:50 (\texttt{4435c51}) and reached GitHub at 17:45, before E0b was submitted on 2026-10-23; the registration's own text dates them 2026-10-23.
\item \textbf{Predictions.} Inside the window: BURST trapped, NOBURST good through the run horizon, CAPPED recovered if the drain fits. Below $\lref$: BURST not trapped. Above the window: NOBURST may escape, so no such rate was registered.
\item \textbf{Kill conditions.} NOBURST not good in at least one third of the seeds at a registered rate; BURST not trapped in two thirds of the seeds at every registered rate; CAPPED not recovering while BURST traps and the drain fits; a window that is empty or narrower than the noise.
\item \textbf{Warnings recorded at filling.} $|\mu(300)/\lmiss-1|=0.507$ (a check only), and the ambiguous cap.
\end{itemize}

\noindent The verdict is in \fpath{results/agentload_analysis_burst2/20261025-144012/} inside the E2 data directory: the registered analysis passes at 1.35 and 1.65 $\lref$ (the tool's labels, assigned by commit order, are pre-registered and decisive; by the definition of \S\ref{sec:prereg} the test is pre-registered). The three grid cells have their own derived registrations, \fpath{configs/agentload_prereg_2026-10-24_grid_<cell>.yaml} for the cells \texttt{sglang-fcfs-r0}, \texttt{sglang-lpm-r1} and \texttt{sglang-lpm-r0}, with status EXPLORATORY; the analysis never calls them decisive.

\paragraph{The vLLM registration.} \fpath{configs/agentload_prereg_2026-10-30_vllm.yaml}, status FILLED, SHA-256 \texttt{8f2d5b44d6365795fedde8bf673d51e5\allowbreak 4bf8a685a14d563c86de44492bb95f84}; the same template, formulas, trap rule, kill conditions and protocol as the E2 registration. Cell \texttt{vllm-fcfs}; inputs: the vLLM Stage~C curve and E0b bracket under \fpath{prereg/inputs/2026-10-30-vllm/}, byte-identical to the analyses in the vLLM Stage~C and E0b data directories, and the vLLM E0a calibration above. Measured: $\mu(n)$ 0.1750 ($n=16$), 0.2194 (20), 0.2183 (24), 0.1189 (28), 0.0433 (32), 0.0411 (40), 0.0439 (48), 0.0572 (96), 0.1050 (300); $\mu^*=0.2194$; $\ncap=20$ by the smallest-within-5\% rule (also the argmax; flagged ambiguous, $n=24$ within 0.5\%); $\lref=0.06566$ (turns-based); all-miss rate 10,349 tokens/s; vLLM E0b largest all-stable multiple 2.0, first unstable 2.5, monotone. Window $(0.0657,\,0.1313)$\sps{} $=(1,\,2.0)\,\lref$, upper edge from vLLM E0b, below $\mu^*/\lref=3.34$. Rates 1.35 and 1.65 $\lref$ (0.0886 and 0.1083\sps) by the same rule, inside the margin band $(1.1,\,1.8)$; this rule and the cap rule were recorded in the project status notes (\texttt{836063a}, pushed 12:16 CDT on 2026-10-30) after the vLLM Stage~C and E0b jobs were submitted and before any of them had ended. Predicted drains 1529\,s and 1800\,s; the drain fits up to 2.81 $\lref$. Warnings recorded at filling: $|\mu(300)/\lmiss-1|=0.599$ and the ambiguous cap. The file was filled from five of the six vLLM E0b runs while the 3.0 $\lref$ run was still running; with 2.5 $\lref$ unstable that run could not change the edge, and it was unstable too, so the six-run bracket equals the registered one. The file was committed in \texttt{5a6fbcd} at 16:23:59 CDT, 37\,s before the 18 jobs of the paired test (17166121--38) were submitted at 16:24:36 (the research log gives 16:30), and the jobs run from the cluster's copy of the repository at that commit; Slurm's record of the submission names the file but not its hash. It reached GitHub at 16:26:35, 1\,min 59\,s after the submission, when two of the jobs had started (at 16:25:42 and 16:26:17) and none could have ended. This registration therefore does not meet the definition of \S\ref{sec:prereg}: it was committed before the jobs were submitted, but pushed after the first two had started, so we call this test registered, not pre-registered. As for E2, the job script writes the file's SHA-256 into every run's \fpath{meta.json}. The verdict tool nevertheless prints the labels pre-registered and decisive, which it assigns by commit order.

\paragraph{Ledger of registrations.} Table~\ref{tab:ledger} lists every registration file we filled, with its status. Commit times are the committer's; push times come from GitHub's push log (\fpath{provenance/push_times_2026-11-01.tsv}, built from the export of 2026-11-01, \fpath{provenance/push_times_2026-11-05.tsv} for the round-15 registrations of 2026-11-05, \fpath{provenance/push_times_2026-11-05-r16.tsv} for MIRROR's, or the archived log itself for the grid cells). Each is the first branch update that contains the commit; the BASIN-1 push also carried \texttt{9b67c26}, the commit of tag \texttt{r11-basin-reg}.

\begin{table*}[tbp]
\centering\footnotesize
\caption{Every registration, in order. FILLED registrations bind a test's cell, rates, cap, references, hardware and protocol, and the job script refuses to run a test without one; EXPLORATORY ones are derived from a FILLED parent and are never called decisive. Times CDT, except the five rows of 2026-11-05 (CST), which were registered after Lu et al.'s paper appeared.}\label{tab:ledger}
\begin{tabular}{@{}>{\raggedright\arraybackslash}p{5.0cm} l l >{\raggedright\arraybackslash}p{2.2cm} >{\raggedright\arraybackslash}p{4.0cm}@{}}
\toprule
File (\fpath{configs/agentload_prereg_*}) & status & committed & on GitHub & runs and outcome\\
\midrule
\fpath{2026-10-24.yaml} (E2) & FILLED & \texttt{d279b7b} 10-24 15:09:12 & 15:09:54 & 18 runs; pass (\S\ref{app:e2tests})\\
\fpath{2026-10-24_grid_sglang-*.yaml} (3 grid cells) & EXPLORATORY & \texttt{499c0df} 10-24 23:30:55 & 23:31:32 & 18 BURST runs (\S\ref{sec:grid})\\
\fpath{2026-10-25_native.yaml} (native E2) & FILLED & \texttt{b8c19ec} 10-25 17:31:16 & 17:31:56 & 18 runs; pass (\S\ref{sec:e2native})\\
\fpath{2026-10-24_grid_vllm-fcfs.yaml} & EXPLORATORY & \texttt{a78ce07} 10-25 18:06:18 & 18:06:33 & 6 BURST runs (\S\ref{sec:vllm})\\
\fpath{2026-10-30_vllm.yaml} (vLLM E2) & FILLED & \texttt{5a6fbcd} 10-30 16:23:59 & 16:26:35, after two jobs started & 18 runs; pass; registered, not pre-registered\\
\fpath{2026-10-30_b100.yaml}, \fpath{_b25.yaml} (BASIN-1) & FILLED & \texttt{9128995} 10-30 19:51:20 & 20:04:30 & 6 BURST runs; pass (\S\ref{app:basin})\\
\fpath{2026-10-30_grid_sglang-lpminf.yaml} & EXPLORATORY & \texttt{5064580} 10-30 21:02:19 & 21:02:46 & 3 BURST runs; no collapse on the two informative seeds, no recovery (\S\ref{app:grid})\\
\fpath{2026-10-31_rescue.yaml} (R1), \fpath{_rescue_native.yaml} (R4), \fpath{_rescue_smoke.yaml} & FILLED & \texttt{f0bc3ad} 10-30 22:10:47 & 22:39:02 & 9 runs, submitted 23:01; R1 and R4 pass (\S\ref{sec:rescue}); the smoke job is never decisive\\
\fpath{2026-10-31_b40.yaml}, \fpath{_b50.yaml}, \fpath{_b60.yaml} (BASIN-2) & FILLED & \texttt{621d88f} 10-31 01:58:38 & 02:10:48 & 9 runs, submitted 04:21; bracketed: 40 good, 60 stuck, 3 of 3 each (\S\ref{sec:basin})\\
\fpath{2026-10-31_toolslow_d10.yaml}, \fpath{_d30.yaml} & FILLED & \texttt{b006291} 10-31 02:27:40 & 02:28:09 & 6 runs, submitted 04:21; rule not met: 10\,s good in 3 of 3, 30\,s stuck in 1 of 3 (\S\ref{sec:frozen})\\
\fpath{2026-10-31_shed_t600.yaml}, \fpath{_t3600.yaml} & FILLED & \texttt{b006291} 10-31 02:27:40 & 02:28:09 & 4 runs, submitted 04:21; pass: 600\,s stuck, 3600\,s good, 2 of 2 each (\S\ref{sec:frozen})\\
H200 cell, 300k pool & none & & & window $(1,\,1.3)\,\lref$ too narrow: the rule's rates 1.10 and 1.20 fall outside its margin band $(1.1,\,1.17)$, and the fill returns INCOMPLETE; no test\\
\fpath{2026-10-30_llama31.yaml} (Llama-3.1-8B E2) & FILLED & \texttt{1559f8e} 10-31 07:49:13 & 07:49:15 & 18 runs, submitted 07:50:02 (17170409--26); window $(1,\,2.0)\,\lref$, 1.35 and 1.65 $\lref$, cap 20; pass at 1.65 $\lref$, fail at 1.35 (\S\ref{sec:h200llama})\\
\fpath{2026-10-31_h200_p600k.yaml} (H200 600k E2) & FILLED & \texttt{4d1d431} 10-31 09:26:38 & 09:26:41 & 18 runs, submitted 09:27:44 (17171020--37); window $(1,\,2.5)\,\lref$, 1.5 and 2.0 $\lref$, cap 48; pass at both rates (\S\ref{sec:h200llama})\\
\fpath{2026-11-05_toolslow_d30w1800.yaml} (TOOLSLOW-W) & FILLED & \texttt{e05a262} 11-05 03:17:48 & 03:18:36 & 3 runs, submitted 03:21:38 (17256674--76); TRAPS, stuck 3 of 3 (\S\ref{sec:validation})\\
\fpath{2026-11-05_toolslow_d15_165.yaml} (TOOLSLOW-D15) & FILLED & \texttt{e05a262} 11-05 03:17:48 & 03:18:36 & 3 runs, submitted 03:57:14 (17256873--75); NO-TRAP: seed 0 stuck, seeds 1 and 2 good (\S\ref{sec:validation})\\
\fpath{configs/r15_edge_replicate.yaml} (REPLICATE-EDGE) & EXPLORATORY & \texttt{e05a262} 11-05 03:17:48 & 03:18:36 & 6 runs in two jobs, submitted 03:21:38 (17256677--78); both originals reproduced, 3 of 3 each (\S\ref{sec:limits})\\
\fpath{2026-11-05_cut_a115.yaml}, \fpath{_cut_b085.yaml} (CUT) & FILLED & \texttt{11b0e4e} 11-05 03:54:28 & 03:54:49 & 6 runs, submitted 03:57:14 (17256867--72); CUT-A and CUT-B pass (\S\ref{sec:control})\\
\fpath{2026-11-05_mirror_k20.yaml} (MIRROR) & FILLED & \texttt{bc81ee7} 11-05 14:07:08 & 14:07:35 & 9 runs and one E0a, submitted 14:50:52--53 (17264055--64; the E0a, 17264061, failed its time check and was resubmitted as 17264093 at 14:53:08); pass at 1.65 and 1.35 $\lref$, RENEWAL 3 of 3 each (\S\ref{sec:mirror})\\
\bottomrule
\end{tabular}
\end{table*}

\noindent Superseded before registration: a native rescue engaged at 2400\,s, whose simulator prediction (13 of 20 rescued) led to the engage time of 2100\,s that R4 registers (\fpath{notes/PREDICTIONS_r11_rescue.md}). The SHA-256 of every tagged file is in its tag message (\texttt{r11-basin-reg}, \texttt{r11-lpminf-pred}, \texttt{r11-rescue-reg}, \texttt{r14-basin2-reg}, \texttt{r14-extra-reg}) and in \fpath{provenance/registration-proofs/}; the Llama-3.1-8B and 600k H200 registrations are not tagged, and the job script writes their SHA-256 into each of their runs' \fpath{meta.json}.

\subsection{The calibrated simulator, prediction rounds and timeline}\label{app:timeline}

\paragraph{The calibrated simulator.}\label{sec:sim}
Our discrete-event simulator models a continuous-batching engine whose iteration cost has five parameters fitted only to SGLang's per-batch timing lines in server logs, never to $\mu(n)$; it failed its declared level test, missing one Stage~C cell by 42\%, but reproduced the curve's shape. LLM coding agents used it to make blind predictions for the registered tests, each in its own copy of the repository and instructed not to read the results they predicted; every round was committed and pushed under a tag before its results were read by its authors, and what each round is blind to differs (this subsection gives the model, the validation, the rounds and their time record). Round predictions are labelled B$k$: B1 the all-miss capacity, B2 the from-empty edge, B3 the paired test, B4 the closed curve, with suffix n for the native pool and v for vLLM (\fpath{notes/PREDICTIONS.md}, \fpath{notes/PREDICTIONS_native.md}, \fpath{notes/PREDICTIONS_vllm.md}). Rounds are numbered in project order: these three prediction rounds are rounds 6, 8 and 9, round 7 was a robustness study in the simulator, and rounds 11 to 14 made the predictions and registrations of \S\ref{sec:frozen}. Our earlier simulator (rounds 1--5), which is not calibrated to the engine and uses assumed costs (all-miss capacity 0.0474\sps), gave the earlier-simulator results cited here, in units of its own $\lmiss$. Each experiment cell was also run by its own Claude Code agent, which submitted its jobs and kept a log in \fpath{notes/LOG.md} (the cell's operating agent), and a further, coordinating agent dispatched the cells and drafted the paper; the first-principles model of Appendix~\ref{app:entry} was derived and frozen by agents in the same way. In this paper ``session'' always means a workload session.

\paragraph{Validation.} The declared level test, simulated $\mu$ within 15\% of the real value in each of 8 Stage~C cells, failed in both modes: mixed batching missed one cell, FCFS $n=48$, by $+42\%$ (0.070 against 0.049\sps), separate batching two. The shape was reproduced: the hit fraction crosses 0.5 at $n=32.5$ against 32.2, the plateau ($n=16$--24) agrees within 9.2\% and $n=300$ within 2\%, and $\mu(300)/\mu^*$ is 0.50 against 0.46. The single real run at $n=48$ dipped below the simulator's all-miss level; its turns-based estimate, 0.063\sps, is within 15\%. Keeping a finished session's KV as evictable cache, a labelled sensitivity arm, left mixed batching the closer mode (0.118 against 0.146) but lowered the plateau ($-15.5\%$ at $n=24$; 5 of 8 cells within 15\%).

\begin{table}[p]
\centering\footnotesize
\caption{Order of events, 2026-10-21 to 10-26 (times CDT). Commits are those of the development repository (availability statement); commit and tag times are recorded by the committer, push times by GitHub's activity log, job times by Slurm.}\label{tab:timeline}
\begin{tabular}{@{}p{1.75cm} >{\raggedright\arraybackslash}p{5.6cm}@{}}
\toprule
When & Event\\
\midrule
10-21 & Stage 0 passes (17060165); Stage C submitted 15:28 (17062103--05)\\
10-22 12:54 & Stage C results pulled\\
10-22 13:13 & runbook proposes the rate rule and a cap of 20 (\texttt{288b801})\\
10-22 13:14 & E0a submitted (17072634--37); the last job ends 16:34\\
10-22 13:50 & rate rule and cap adopted in the status notes; simulator cost model fitted to Stage 0/C server logs (\texttt{4435c51})\\
10-22 17:03 & Stage 0 and C data committed (\texttt{d0235a7})\\
10-22 17:45 & first push to GitHub, with the three commits above\\
10-23 14:32 & E0a results committed (\texttt{a76f5e5})\\
10-23 14:42 & E0b submitted (17084750--55); jobs end by 16:09\\
10-24 03:13 & validation of the calibrated simulator committed (\texttt{813cfc1})\\
10-24 03:23 & prediction spec and code committed before any prediction run (\texttt{630ab9e}); on GitHub 03:24\\
10-24 03:49 & predictions B1--B4 committed (\texttt{64bc4e3})\\
10-24 04:05 & independent verification recorded (\texttt{877b936})\\
10-24 14:52 & tag \texttt{r6-predictions-frozen} (tagger time)\\
10-24 14:53 & B1--B4 and the verification reach GitHub\\
10-24 15:06 & E0b results pulled\\
10-24 15:09 & E0b data and registration committed in the same second (\texttt{cfbc702}, \texttt{d279b7b}); registration on GitHub 15:09:54\\
10-24 15:13 & E2 submitted (17093387--404)\\
10-24 15:38 to 10-25 02:36 & E2 jobs run\\
10-25 02:28 & native-pool Stage C and E0a start (17096856--58)\\
10-25 14:40 & registered E2 verdict computed (data \texttt{49fbc7b})\\
10-25 14:42 & native Stage C curve analysed on the cluster; native curve and $\lmiss$ in the status notes at 14:47 (\texttt{6793fb7})\\
10-25 14:45 & native E0b (17100938--39) and vLLM E0a (17100940) start\\
10-25 14:49 & native prediction spec committed (\texttt{29728e3})\\
10-25 14:52 & native Stage C and E0a data committed (\texttt{fa265c2})\\
10-25 14:54 & B1n--B4n run on the clean code commit \texttt{0336c91}\\
10-25 15:06 & B1n--B4n committed (\texttt{e3ea48f}); on GitHub 15:08:28\\
10-25 15:08 & tag \texttt{r8-native-predictions-frozen} (tagger time); native rate rule in the status notes (\texttt{6cb43e9})\\
10-25 17:26 & native E0b jobs end; pulled 17:30\\
10-25 17:31 & native E0b data and native registration committed (\texttt{f3f55ab}, \texttt{b8c19ec}); registration on GitHub 17:31:56\\
10-25 17:42 to 10-26 01:11 & native E2 jobs (17102173--90)\\
10-25 18:06 & vLLM calibration and exploratory registration committed (\texttt{65b65e2}, \texttt{a78ce07}), on GitHub 18:06:33; vLLM BURST submitted 18:07 (17102318--23)\\
10-25 22:42 to 10-26 03:24 & vLLM BURST jobs\\
10-26 10:07 & native E2 and vLLM BURST verdicts computed (data \texttt{f116cf3})\\
\bottomrule
\end{tabular}
\end{table}

\begin{table}[htbp]
\centering\footnotesize
\caption{Order of the vLLM events on 2026-10-30 (times CDT, recorded as in Table~\ref{tab:timeline}).}\label{tab:timeline-vllm}
\begin{tabular}{@{}p{1.75cm} >{\raggedright\arraybackslash}p{5.6cm}@{}}
\toprule
When & Event\\
\midrule
12:10 & vLLM Stage C (17162959--61) and E0b (17162962--63) submitted; 17162959--60 start 12:11\\
12:16 & rule for the vLLM cap and rates in the status notes, on GitHub (\texttt{836063a})\\
12:31 & B1v--B4v committed (\texttt{cc01dbf}); tag \texttt{r9-vllm-predictions-frozen} 12:32 (tagger time)\\
12:32 & B1v--B4v reach GitHub (12:32:44)\\
14:36 & first vLLM jobs end (Stage C, 17162959--60)\\
15:35 & vLLM Stage C curve analysed on the cluster\\
16:23 & bracket of the five finished vLLM E0b runs (16:23:19); vLLM registration committed (\texttt{5a6fbcd}, 16:23:59)\\
16:24 & paired test submitted (17166121--38, 16:24:36); first job starts 16:25:42, second 16:26:17\\
16:26 & vLLM registration and Stage C data (\texttt{276c5f0}) reach GitHub (16:26:35)\\
17:17 & vLLM E0b job 17162963, with the 3.0 $\lref$ run, ends\\
17:28 & vLLM E0b data committed and pushed (\texttt{582c2a6})\\
\bottomrule
\end{tabular}
\end{table}

\noindent Commit and tag times are set by the committer. The external record is GitHub's activity log for the development repository, archived in \fpath{provenance/github_activity_2026-10.json} and, refreshed after the last registered job, \fpath{provenance/github_activity_2026-11-01.json}, together with the cluster's accounting records of the cited jobs (\fpath{provenance/sacct_cited_jobs_2026-10-30.txt}, \fpath{provenance/sacct_cited_jobs_2026-11-01.txt}). The development repository is private; the archived log is an unsigned export made by the first author with GitHub's API, and the hash chains in \fpath{provenance/registration-proofs/} show that the files are unchanged, not when they were pushed; \fpath{provenance/push_times_2026-10.tsv} and its refresh \fpath{provenance/push_times_2026-11-01.tsv}, which adds the rounds 11 to 14, list, for each registration and prediction commit, its first arrival on GitHub and the first job it concerns. GitHub records no push time for tags, so the push times given here are those of the first branch update that contains each commit. The branch holding B1--B4 (\texttt{64bc4e3}) and the verification reached GitHub at 14:53:07 CDT on 10-24, and the E2 registration (\texttt{d279b7b}) at 15:09:54, before the E2 jobs were submitted at 15:13:39. On 10-25, B1n--B4n (\texttt{e3ea48f}) arrived at 15:08:28, the native registration (\texttt{b8c19ec}) at 17:31:56 and the exploratory vLLM registration (\texttt{a78ce07}) at 18:06:33. Pull times are from our local records only.

B3 (E2) and B4 (native pool) reached GitHub before their experiments were submitted, so they are blind in time. B2 (E0b) was committed after E0b had finished on the cluster but before its results were pulled. The LLM agents that made the predictions worked in separate copies of the repository at a commit without the E0a or E0b results and were instructed not to contact the cluster and not to open the E0a or E0b result directories or the primary copy of the repository; they could technically have done so. Their tool-call transcripts, which are not in the repository, were checked afterwards by the orchestrating Claude Code session, which found no cluster command and no access to E0b before the predictions reached GitHub. For B2 this instruction-level control, checked after the fact, is the only safeguard. A scan of every line added between the commit the agents started from and the predictions found no use of withheld data, but a text scan cannot prove that nobody read it. B1 (E0a) rests on the same instruction only: E0a was committed before the predictions. The independent verifier, a separate OpenAI Codex session, had read the project status notes, which mention E0a, so it is independent but not blind; it also reused the prediction agents' copy of the repository instead of a fresh one, as its record states. Its evidence is in \fpath{provenance/codex_prediction_verification/}. Two conditions of the verification failed and are disclosed: the strict containment of B3 passes within the B2 all-stable range, and the platform-specific test failures of the full suite.

In the second round, B2n (native E0b) and B3n (native E2) reached GitHub while native E0b was still running, before any of its runs had finished, and before the native registration or any native E2 job existed. B3n is therefore blind in time. B2n reached GitHub 23\,min after the two native E0b jobs were submitted, when both were running, so it is blind in the sense that no result existed, not in submission order. B1n and B4n are not: native Stage C and E0a had finished, and their headline numbers were in the project status notes before the round's specification was written. That specification is shared by all four, so B2n and B3n are blind to the native E0b and E2 outcomes but were specified after the native curve was known, and B3n's cap of 64 is B4n's output. The LLM agent that ran the round worked in a separate copy of the repository at a commit that predates the native data (\texttt{fb3510e}) and, by its own record, read nothing under the real-engine results directories; that record has not been checked independently. No independent verifier reran the second round.

In the third round (\fpath{notes/PREDICTIONS_vllm.md}), an LLM agent in a separate copy of the repository at \texttt{a4bab1d}, which holds the exploratory vLLM BURST data but no vLLM closed curve or edge, ran the calibrated simulator with its token-proportional costs scaled by one factor per batching mode (1.0882 mixed, 1.0806 separate), chosen so that its all-miss token rate equals vLLM's measured 10,349 tokens/s; since the factor is fitted, B1v is not a prediction. By its own record, which has not been checked independently, it read nothing under the real-engine results directories; the two real vLLM numbers it used, $\lmiss$ and the all-miss token rate, were given in its task. The predictions (\texttt{cc01dbf}, committed 12:31:58) reached GitHub at 12:32:44, 21\,min 48\,s after the five vLLM Stage~C and E0b jobs were submitted (Table~\ref{tab:timeline-vllm}); two of these jobs had been running since 12:11 and none ended before 14:36:50. So B2v and B4v are blind in the sense that no result existed when they were pushed, not in submission order; B3v predates the vLLM registration and the paired test. The orchestrating Claude Code session that specified the round knew the outcome of the exploratory vLLM BURST runs, which were in the repository by then, and chose the candidate rates, so the BURST-arm predictions at 1.35 and 1.65 $\lref$ are blind at most on the predicting agent's side. The notes' remark that their branch was never pushed predates the push. No independent verifier reran the third round. The vLLM registration's own timing is in Appendix~\ref{app:prereg}.

Table~\ref{tab:timeline-r11} also gives what each prediction is blind to. The simulator's BASIN-1 and fresh-seed predictions reached GitHub before those jobs were submitted, and its LPM-INF predictions before the LPM-INF jobs, so all three are blind in time. For the theory, no BASIN-1 run is blind in time. The three 25-session runs had started 1.7 to 2.2\,h before the push (6230--7807\,s of 8700\,s elapsed, past the burst and into the window [5700, 8700)\,s), and the three 100-session runs had ended 0.2 to 1.6\,h before it, so for all six the prediction is blind only by access: no one is known to have read them before the push, and the first recorded reading of any BASIN-1 or fresh-seed outcome was at about 04:07, but no file-access record can show this, because the cluster's class storage is mounted without access times. The same holds for the H200 cell's closed runs at $n=40$, 48, 96 and 300 and its original E0b runs. They had ended before the freeze, and they were first read after it, when the curve and the bracket were combined at 02:30. The runs at $n=36$, 44 and 52 were submitted after the freeze and are blind in time. No blind simulator prediction exists for the H200 cell. The second theory set (pred2) reached GitHub at 04:20:20, before the BASIN-2, TOOLSLOW and SHED jobs were submitted (04:21:40 to 04:21:43), so its predictions for them are blind in time. The 600k H200 jobs had been submitted at 02:50 but could not start before 05:00 and first started at 05:27, so its 600k predictions are blind in time by the start criterion of Table~\ref{tab:frozensum}: no result existed at the push. The pool-sweep jobs had been submitted at 22:58 on 10-30, before either set was pushed, but were held until 02:30 and first started at 06:50:13, so both sets are blind in time for them by the same criterion. The model's a priori Llama-3.1-8B predictions (pushed at 02:05:30) and the simulator's (02:14:06) are blind by access only for that cell's closed runs, which ran from 00:50 to 03:31: three had ended at the theory push, three were running and three started after it (from 02:26); by the operator's record none was read before both pushes (the curve was first combined from 03:14, for $n=16$--24, and in full by 03:37). That cell's from-empty runs were submitted after the model's push and 7.5\,min before the simulator's, started from 04:33, and are blind in time for both by the start criterion, as are its paired-test jobs (submitted at 07:50:02, before that push; first start 12:41) for the registration-conditional predictions pushed at 07:52:42 and 07:52:56. pred2's table for the fresh seeds is not blind: it was committed at 04:20:05, after the last fresh-seed run had ended (03:57:56) and after those outcomes were first read (about 04:07); by its record its agents read no result, which is why its own erratum calls it blind by access only. The simulator's rescue predictions (pushed at 21:30:40 on 10-30) and the model's (02:05:30 on 10-31) reached GitHub before the first rescue run started (02:23:44 on 10-31), so all nine rescue runs are blind in time to both; the smoke run ended before the model's push, but the model made no prediction for it, and by their record the agents that derived the model had no Nova access and opened no rescue output before the push.

\begin{table*}[tbp]
\centering\footnotesize
\caption{Order of the round-11 to round-14 events, 2026-10-30 and 10-31 (times CDT; commit and tag times by the committer; submission, start and end times of jobs by Slurm's accounting, run end times by each run's \fpath{meta.json}). Push times marked * were read from GitHub's activity log at the time (recorded in \fpath{notes/PREDICTIONS_r11_mechanism.md}, \fpath{notes/PREDICTIONS_r13_theory.md}, and \fpath{notes/PREDICTIONS_r13_pred2.md}) or on 10-31; each agrees with the export of 2026-11-01 (\fpath{provenance/github_activity_2026-11-01.json}; rows of \fpath{provenance/push_times_2026-11-01.tsv}, except \texttt{f1338e3} and \texttt{d12eef2}, which are in the export only). The last rows run past 02:00 CDT on 11-01, when daylight saving time ended; times after it are CST.}\label{tab:timeline-r11}
\begin{tabular}{@{}p{2.3cm} >{\raggedright\arraybackslash}p{14.4cm}@{}}
\toprule
When & Event\\
\midrule
10-30 18:20 & r11 mechanism predictions for the vLLM test committed (\texttt{f1338e3}, 18:20:11) and pushed (18:20:13*); three of its 18 jobs running, none ended\\
10-30 18:55 & first vLLM paired-test job ends (17166121, 18:55:07)\\
10-30 19:50 & BASIN-1 and fresh-seed predictions (\texttt{c600fd7}); BASIN-1 registrations (\texttt{9128995}, 19:51); tag \texttt{r11-basin-reg} 20:04; all pushed 20:04:30*\\
10-30 21:01 & LPM-INF predictions (\texttt{e2e8474}) and registration (\texttt{5064580}, 21:02); tag \texttt{r11-lpminf-pred} 21:02; pushed 21:02:46*\\
10-30 21:12 & BASIN-1 and fresh-seed jobs submitted (17168500--07, 21:12:45--46)\\
10-30 21:28 & rescue predictions (\texttt{72501f0}); tag \texttt{r11-rescue-pred} 21:30; pushed 21:30:40*\\
10-30 22:07 & LPM-INF jobs submitted (17168743--45, 22:07:40)\\
10-30 22:10 & R1 and R4 registrations (\texttt{f0bc3ad}); tag \texttt{r11-rescue-reg} 22:38; pushed 22:39:02*\\
10-30 22:28 & the H200 operator reads the closed-curve values for $n\le32$ (17166818--22) and shows them to the first author\\
10-30 22:38 to 23:52 & H200 closed runs at $n=40$, 48, 96 and 300 end (17166828--34; not read until 02:30)\\
10-30 22:58 & pool-sweep jobs submitted (17168864--81, 22:58:03); eligible to start from 02:30:32 on 10-31\\
10-30 23:01 & rescue jobs submitted (17168965--74, 23:01:12--13); the smoke job runs 23:24 to 00:08\\
10-31 00:14 to 01:08 & H200 E0b jobs end (17168377--82; not read until 02:30)\\
10-31 00:20 & last vLLM paired-test job ends\\
10-31 00:28 to 01:51 & the three 100-session BASIN-1 runs end (00:27:56, 00:29:27, 01:50:54)\\
10-31 00:50 to 03:31 & Llama-3.1-8B closed runs (17166814--16): three end before the theory push, three start after it (from 02:26); combined from 03:14 ($n=16$--24) to 03:37 (all nine)\\
10-31 01:58 & BASIN-2 registrations (\texttt{621d88f}); tag \texttt{r14-basin2-reg} 02:10; pushed 02:10:48*\\
10-31 02:05 & theory frozen (\texttt{997d2c4}, 02:04:55) and pushed (02:05:30*); tag \texttt{r13-theory-pred}\\
10-31 02:05 to 03:58 & the six fresh-seed runs end (the first at 02:04:59, 31\,s before the theory push)\\
10-31 02:08 & registered vLLM verdict computed on the cluster; pulled 02:09:56, committed 02:22 (\texttt{cacf833})\\
10-31 02:14 & Llama simulator predictions (\texttt{a62aea9}) pushed (02:14:06*); tag \texttt{r12-llama31-predictions-frozen}\\
10-31 02:20 to 02:47 & the three 25-session BASIN-1 runs end\\
10-31 02:23 to 06:57 & the nine registered rescue runs (17168966--74; the first starts at 02:23:44, 18 minutes after the theory push; the last ends at 06:57:06)\\
10-31 02:27 & TOOLSLOW and SHED registrations (\texttt{b006291}); tag \texttt{r14-extra-reg} 02:27:59; pushed 02:28:09*\\
10-31 02:30 & H200 nine-point curve and E0b bracket combined on the cluster: first reading of $n\ge40$ and of the edge\\
10-31 02:33 & H200 closed runs at $n=36$, 44 and 52 submitted (17169612--14, 02:33:06); they end 03:25 to 04:01\\
10-31 02:50 & H200 E0b reseeds (17169668--71, 02:50:20) and 600k jobs (17169672--80, 17169682--87, 02:50:32 to 02:51:20, not eligible to start before 05:00) submitted\\
10-31 03:49 to 04:21 & the three LPM-INF runs end\\
10-31 04:07 & first recorded reading of the BASIN-1 and fresh-seed outcomes (about 04:07)\\
10-31 04:20 & second theory set for BASIN-2, TOOLSLOW, SHED and the H200 600k cell, with the pool-sweep lookups and the rival scoring of the 0.5-hit crossing, committed (\texttt{cfa7712}, 04:20:05) and pushed (04:20:20*); tag \texttt{r13-theory-pred2}\\
10-31 04:21 & BASIN-2, TOOLSLOW and SHED jobs submitted (17169893--911, 04:21:40 to 04:21:43) from clones at the registration tags, by scripts that require the pred2 commit\\
10-31 05:27 & first 600k H200 job starts (17169672--80, 82--87); curve and bracket combined at 09:24\\
10-31 06:50 to 11:04 & pool-sweep runs (17168864--81); pulled at 11:23 and scored against both sets (\texttt{3025139})\\
10-31 07:05 & rescue runs pulled (committed in \texttt{c6ea41e}); registered verdicts computed at 07:17 (\texttt{be50593}, \texttt{454ff7a})\\
10-31 07:49 & Llama-3.1-8B registration (\texttt{1559f8e}, 07:49:13) pushed (07:49:15*); its 18 E2 jobs submitted at 07:50:02 (17170409--26)\\
10-31 07:52 & registration-conditional Llama predictions (\texttt{a36f441}, \texttt{d12eef2}) pushed (07:52:42*, 07:52:56*), before any Llama E2 job started (first start 12:41)\\
10-31 09:26 & 600k H200 registration (\texttt{4d1d431}, 09:26:38) pushed (09:26:41*); its 18 E2 jobs submitted at 09:27:44 (17171020--37)\\
10-31 09:27 to 11-01 00:28 & the 18 600k H200 paired-test runs (17171020--37)\\
10-31 10:10 to 15:11 & the nine BASIN-2 runs (17169893--901); pulled at 15:42 and first read then\\
10-31 12:41 to 23:31 & the 18 Llama-3.1-8B paired-test runs (17170409--26)\\
10-31 21:03 to 11-01 01:00 CST & the six TOOLSLOW and four SHED runs (17169902--11)\\
11-01 15:44 to 15:49 CST & registered analyses over all runs, with the frozen code: the 600k H200 and Llama-3.1-8B verdicts computed on the cluster and pulled (15:45); the TOOLSLOW and SHED runs pulled and analysed locally (\texttt{db24efa}, 15:49)\\
\bottomrule
\end{tabular}
\end{table*}

\paragraph{Round 15.} Table~\ref{tab:timeline-r15} gives the time record of round 15, on 2026-11-05, after Lu et al.'s paper had appeared (2026-11-02) and with it in hand: the trap-state and CUT predictions, the TOOLSLOW-W, TOOLSLOW-D15 and CUT registrations and the exploratory REPLICATE-EDGE rule (times CST, from \fpath{provenance/push_times_2026-11-05.tsv} and \fpath{provenance/sacct_cited_jobs_2026-11-05.txt}). Every registration and prediction commit reached GitHub before the first job of its batch was submitted, by 145\,s to 3\,h\,32\,min; the one later commit, pushed after the TOOLSLOW-W jobs had started but before any run reached its slowdown, added an unregistered column to the scorer and post-registration notes to two prediction files. The same table gives the record of round 16 (MIRROR) that afternoon, from \fpath{provenance/push_times_2026-11-05-r16.tsv} and \fpath{provenance/sacct_cited_jobs_2026-11-05-r16.txt}: its registration reached GitHub 41.5\,min before the first MIRROR job was submitted (Appendix~\ref{app:blind} gives the errata to its prediction note).

\begin{table*}[tbp]
\centering\footnotesize
\caption{Order of the round-15 and round-16 (MIRROR) events, 2026-11-05 (times CST; commit times by the committer, push times from GitHub's activity log, job times by Slurm's accounting, pull and verdict times from the pull records in \fpath{results/nova/r15_*_pulled_*} and \fpath{results/nova/r16_*_pulled_*}).}\label{tab:timeline-r15}
\begin{tabular}{@{}p{2.3cm} >{\raggedright\arraybackslash}p{14.4cm}@{}}
\toprule
When & Event\\
\midrule
11-05 00:24 & trap-state predictions and scoring rules S1--S5 (\texttt{6b47a6c}, 00:24:39); tag \texttt{icml-trap-pred}; pushed 00:24:47\\
11-05 01:09 & the trap-state scorer \fpath{analysis/theory/icml/score_new.py} (\texttt{24b5e13}, 01:09:21) pushed (01:09:25)\\
11-05 02:25 & CUT predictions from each seed's exact plan (\texttt{2f90267}, 02:25:33); tag \texttt{icml-cut-pred}; pushed 02:25:43\\
11-05 03:18 & TOOLSLOW-W and TOOLSLOW-D15 registrations, the simulator and pred2 predictions for them, and the REPLICATE-EDGE rule (\texttt{e05a262}, 03:17:48); tag \texttt{r15-reg}; pushed 03:18:36\\
11-05 03:21 & TOOLSLOW-W and REPLICATE-EDGE jobs submitted (17256674--78, 03:21:38)\\
11-05 03:34 & the scorer's unregistered $\gamma/z$ column (\texttt{fdb80c9}, 03:34:32) pushed (03:34:35), 777\,s after the TOOLSLOW-W submission and 727\,s after its jobs started, before any run had reached its slowdown at 1200\,s of run time and before any output was read; the same commit added post-registration notes, without changing the frozen tags, to \fpath{notes/PREDICTIONS_icml_trap.md} and \fpath{notes/PREDICTIONS_icml_cut.md}\\
11-05 03:54 & CUT registrations (\texttt{11b0e4e}, 03:54:28); tag \texttt{r15-cut-reg}; pushed 03:54:49\\
11-05 03:57 & CUT and TOOLSLOW-D15 jobs submitted (17256867--75, 03:57:14)\\
11-05 05:51 to 06:04 & the five TOOLSLOW-W and REPLICATE-EDGE jobs end; pulled at 06:10\\
11-05 06:38 & CUT-A runs pulled; registered verdict computed at 06:39\\
11-05 06:51 & CUT-B seed~0 pulled\\
11-05 08:23 & CUT-B seeds 1 and 2 and TOOLSLOW-D15 seed~0 pulled; registered verdict over all six CUT runs at 08:25\\
11-05 08:35 & TOOLSLOW-D15 seeds 1 and 2 pulled; registered verdict at 08:36\\
\addlinespace
11-05 14:07 & MIRROR (round 16): registration, frozen predictions and plans (\texttt{bc81ee7}, 14:07:08); tag \texttt{r16-mirror-reg}; on GitHub 14:07:35\\
11-05 14:49 & MIRROR cache-off smoke test submitted (17264039, 14:49:05); the erratum to the prediction note's writing time (\texttt{18ac5e6}, 14:24:49) reached GitHub at 14:49:22\\
11-05 14:50 & MIRROR jobs submitted in the registered order (17264055--64, 14:50:52--53); the E0a, 17264061, failed its pre-run time check (14:51:57) and was resubmitted as 17264093 (14:53:08) with a longer time limit\\
11-05 16:19 & errata to the prediction note's prose (\texttt{d8c49b7}, 16:19:32) pushed (16:19:36), during the runs and, by our own record, before any MIRROR output was read\\
11-05 16:59 & the E0a pulled (it ended at 16:27:41)\\
11-05 17:22 & the nine MIRROR runs pulled (they ended at 17:20:42--17:21:03); registered verdict at 17:31\\
\bottomrule
\end{tabular}
\end{table*}

\paragraph{Slips in the registration record.} The vLLM registration reached GitHub after the first two of its jobs had started (\S\ref{sec:prereg}), and the memory request of its 15 pending jobs was lowered after submission (\S\ref{sec:vllm}). B1n and B4n were specified after the native curve was known, and B2n, B2v and B4v were pushed after their jobs were submitted (Appendix~\ref{app:timeline}). The pool-sweep jobs (for both theory sets), the 600k H200 jobs (for pred2), the Llama-3.1-8B from-empty jobs (submitted at 02:06:31, for the simulator's predictions pushed at 02:14:06; first start 04:33) and the Llama-3.1-8B paired-test jobs (submitted at 07:50:02, for the registration-conditional predictions pushed at 07:52:42 and 07:52:56) were also submitted before the predictions were pushed, but none started until after the push, so we count those predictions as blind in time by the start criterion of Table~\ref{tab:frozensum}. The agent that operated the H200 cell saw its closed-curve values for $n\le32$, which include the predicted peak, before the freeze and showed them to the first author; the agents that derived and froze the model did not see them, so the H200 peak-ratio rule and hit$(32)$ are not blind to the first author, while the runs at $n=36$, 44 and 52 are blind in time and the other runs at $n\ge40$ and the original from-empty runs, which had ended before the freeze, are blind by access only; four H200 E0b reseeds were added after the first H200 E0b had been read (\S\ref{sec:frozen}). No blind simulator prediction was made for the H200 cell. The 100-session runs of BASIN-1 had ended 0.2--1.6\,h before the model's predictions were pushed, and the 25-session runs were then 6230--7807\,s into their 8700\,s; no one is known to have read them before then, but the cluster's file system records no access times, so all six predictions are blind by access only. The H200 runs at $n=36$, 44 and 52 and the fresh-seed from-empty runs ended before pred2 was pushed and were read before it (combined at 04:09; first read at about 04:07), so pred2's tables for them are not blind, nor are its rows for BASIN-1's burst sizes, whose outcomes the agent that operated the basin cell also first read at about 04:07; pred2's agents read no BASIN-1 result, so to them those rows are blind by access only. The freeze record of pred2, \fpath{notes/PREDICTIONS_r13_pred2.md}, says it was pushed before the first job of the H200 runs at $n=36$, 44 and 52 (those jobs were submitted at 02:33 and had ended before the push); its dated erratum (\texttt{bf19385}) corrects this and calls both tables blind by access only, counting only its agents' access. The BASIN-2, TOOLSLOW and SHED jobs were submitted after the push (04:20:20 CDT; jobs submitted 04:21:40--43). pred2's rival scoring of the 0.5-hit crossing, including the test of the tool-call term, was declared after the 300k H200 and Llama-3.1-8B curves had been combined by their operators (the H200 curve at 02:30; the Llama curve from 03:14, in full by 03:37), so for those two cells it is not blind. The model's and the simulator's Llama-3.1-8B predictions were pushed while that cell's closed runs were in progress, so for its curve they are blind by access only; and the Llama and 600k H200 registrations were pushed 47 and 63\,s before their jobs were submitted and are not tagged, nor are the Llama registration-conditional predictions (\texttt{a36f441}, \texttt{d12eef2}). The agents that froze the basin and rescue predictions of the model knew the simulator's burst-size sweep and its registered rescue prediction. The engage time of the native rescue test was chosen after a simulator prediction for an earlier engage time had been seen, before any rescue run. Three stricter changes to the rescue decision rule were accepted after a review of the simulator's predictions and before any rescue run: a refutation takes precedence over a pass, the no-shed part of a pass is withheld if any counted run has failed sessions, and R4, like R1, requires stuck same-seed BURST controls. The bands of the frozen vLLM mechanism checks were chosen on SGLang runs and on exploratory vLLM runs whose arrivals the registered BURST arm repeats, so only its no-burst and capped arms are out of sample for those checks. No registration or prediction is in a third-party registry, and none has a timestamp other than GitHub's push log of the development repository, which is private, so readers see that log only in our export.

\subsection{Provenance}\label{app:prov}

Every number in the paper is listed with its source file and commit or job in \fpath{paper/NUMBERS.md}. Table~\ref{tab:prov} gives the data directory and commit behind each result. The commit hashes cited are those of the development repository; the public repository's history is a redacted copy of it, and \fpath{provenance/registration-proofs/} maps each cited registration and prediction commit to its public counterpart (availability statement). The 103 commits made before 17:45 CDT on 2026-10-22, including \texttt{288b801}, \texttt{4435c51} and \texttt{d0235a7} cited here, predate a history rewrite on that day that removed the AI co-author trailers from their commit messages and changed nothing else: trees, authors, dates and parent relations are unchanged. \fpath{provenance/commit-map.tsv} maps their old hashes, which notes and run records written before the rewrite cite, to the current ones; the repository the old hashes belong to is not on GitHub, so they can be checked only against this map. Their earliest server record is the repository's first push to GitHub, at 17:45:50 CDT on 2026-10-22, with which 97 of them arrived, these three among them, before E0b was submitted on 2026-10-23; one more (\texttt{fc870c2}) arrived 2\,s later, and five are reachable only from archive tags, which have no push record.

\paragraph{Repository, redactions and timing evidence.} Its history is a redacted copy of the authors' development repository: passages unrelated to the paper were removed from four notes files and one clause from each of two commit messages, five local tool-configuration files were dropped from every commit (all listed in \fpath{provenance/REDACTIONS.md}), and no code, experiment configuration, registration, result or paper file changed in it. A final release commit adds licences and README text and replaces allocation names in some notes, job-script comments and tests (\fpath{provenance/}). Its commit hashes therefore differ from those cited in this paper, which are the development repository's. The directory \fpath{provenance/} maps every cited hash to its public counterpart and holds the timing evidence: GitHub's push log of the development repository (\fpath{github_activity_2026-10.json}, refreshed in \fpath{github_activity_2026-11-01.json}), the external record of when each registration and prediction reached GitHub (\S\ref{sec:prereg}, Appendices~\ref{app:prereg} and~\ref{app:timeline}); for every registration and frozen prediction, the original commit, tree and tag objects, the push-log event that first carried them, and a script that recomputes the hash chains and checks that each registration file in the public repository is byte-identical to the registered one (\fpath{provenance/registration-proofs/}); for the registrations and predictions pushed after 2026-10-30, the refreshed push log (\fpath{github_activity_2026-11-01.json}) and the first push of each with the first job it concerns (\fpath{push_times_2026-11-01.tsv}); for round 15 (2026-11-05, CST), \fpath{github_activity_2026-11-05.json}, \fpath{github_tag_refs_2026-11-05.json}, \fpath{push_times_2026-11-05.tsv} and the accounting records of its 14 jobs (\fpath{sacct_cited_jobs_2026-11-05.txt}, and in UTC \fpath{sacct_cited_jobs_2026-11-05_utc.txt}; re-exported after the last job had ended); for round 16 (MIRROR, the same day), \fpath{github_activity_2026-11-05-r16.json}, \fpath{github_tag_refs_2026-11-05-r16.json}, \fpath{push_times_2026-11-05-r16.tsv} and the accounting records of its 12 jobs (\fpath{sacct_cited_jobs_2026-11-05-r16.txt}, and in UTC \fpath{sacct_cited_jobs_2026-11-05-r16_utc.txt}; exported after the last job had ended); the cluster's accounting records of all 258 cited jobs (\fpath{sacct_cited_jobs_2026-11-01.txt}, and in UTC \fpath{sacct_cited_jobs_2026-11-01_utc.txt}, exported after the last job had ended; the earlier \fpath{sacct_cited_jobs_2026-10-30.txt} covers the jobs through the vLLM paired test); and the evidence of the independent verifier of the first prediction round. Simulator outputs are not stored, apart from their summaries; each can be regenerated from the command, commit and configuration recorded with the run (\fpath{meta.json}, \fpath{config.yaml}). The derived tables in \fpath{paper/derived/} name each engine run by its path on the cluster, \fpath{results/agentload_*/<timestamp>}; the run is stored under the pull directory of its jobs in \fpath{results/nova/}, with the same final path components. The agents' tool-call transcripts (\S\ref{sec:sim}, Appendix~\ref{app:timeline}) are not in the repository.

\begin{table*}[tbp]
\centering\footnotesize
\caption{Data behind each result (paths relative to the repository root at tag \texttt{trap-arxiv-v1}).}\label{tab:prov}
\begin{tabular}{@{}>{\raggedright\arraybackslash}p{2.0cm} >{\raggedright\arraybackslash}p{14.7cm}@{}}
\toprule
Result & Directory (commit)\\
\midrule
Stage 0, C & \fpath{results/nova/s0_smoke_17060165_pulled_20261021-152347/}, \fpath{results/nova/stageC_1706210x_pulled_20261022-125454/} (\texttt{d0235a7})\\
E0a & \fpath{results/nova/e0a_1707263x_pulled_20261023-142037/} (\texttt{a76f5e5})\\
E0b & \fpath{results/nova/e0b_1708475x_pulled_20261024-150639/} (\texttt{cfbc702})\\
E2 & \fpath{results/nova/e2_170933xx_pulled_20261025-144558/} (\texttt{49fbc7b})\\
native, grid, vLLM & \fpath{results/nova/r7_native_grid_vllm_pulled_20261025-145126/} (\texttt{fa265c2})\\
native E0b & \fpath{results/nova/e0b_native_1710093x_pulled_20261025-173038/} (\texttt{f3f55ab})\\
vLLM E0a & \fpath{results/nova/e0a_vllm_17100940_pulled_20261025-180814/} (\texttt{b273241})\\
native E2, vLLM BURST & \fpath{results/nova/e2_native_vllm_burst_pulled_20261026-101256/} (\texttt{f116cf3})\\
vLLM C & \fpath{results/nova/stageC_vllm_1716295x_pulled_20261030-153621/} (\texttt{276c5f0})\\
vLLM E0b & \fpath{results/nova/e0b_vllm_1716296x_pulled_20261030-172810/} (\texttt{582c2a6})\\
validation & \fpath{results/r6_validate_stagec/20261024-025222/} (\texttt{813cfc1})\\
B1--B4 & \fpath{results/r6_predict_report/20261024-034707/}, \fpath{notes/PREDICTIONS.md} (\texttt{64bc4e3})\\
B1n--B4n & \fpath{results/r8_native_report/20261025-150146/}, \fpath{notes/PREDICTIONS_native.md} (\texttt{e3ea48f})\\
B1v--B4v & \fpath{results/r9_vllm_report/20261030-122723/}, \fpath{notes/PREDICTIONS_vllm.md} (\texttt{cc01dbf})\\
vLLM E2 & \fpath{results/nova/e2_vllm_1716612x_pulled_20261031-020956/} (\texttt{cacf833}); mechanism checks in its \fpath{r11_mechanism/} and \fpath{derived/}\\
entry model (App.~\ref{app:entry}) & \fpath{results/theory/merged/}, \fpath{notes/PREDICTIONS_r13_theory.md}, \fpath{notes/THEORY_r13_scoring.md}, \fpath{configs/r13_theory_predict.yaml} (\texttt{997d2c4}, tag \texttt{r13-theory-pred}); \fpath{results/theory/paper/}, \fpath{figures/theory/} (\texttt{5271ff6})\\
H200 & \fpath{results/nova/h200_wave1_pulled_20261031-023416/} and \fpath{prereg/inputs/2026-10-31-h200/} (\texttt{d108056}); \fpath{results/nova/h200_stageC_extra_1716961x_pulled_20261031-040933/} (\texttt{748900b}); \fpath{results/nova/h200_e0b_reseeds_1716966x_pulled_20261031-050308/} (\texttt{4e27364}); 600k pool \fpath{results/nova/h200p600k_stageC_e0b_171696xx_pulled_20261031-092822/} (\texttt{8b548c0}), registration \fpath{configs/agentload_prereg_2026-10-31_h200_p600k.yaml} (\texttt{4d1d431})\\
BASIN-1, E0F & \fpath{results/nova/b1_e0f_1716850x_pulled_20261031-040517/} (\texttt{4dd21e0})\\
BASIN-2 & \fpath{results/nova/b2_1716989x_pulled_20261031-154500/}; scoring \fpath{experiments/r14_b2_score.py}, \fpath{experiments/r14_b2_derived.py} (\texttt{eb08a5d})\\
pool sweep & \fpath{results/nova/pool_sweep_1716886x_pulled_20261031-112342/}, scored by \fpath{experiments/r13_pool_score.py} into its \fpath{derived/} (\texttt{3025139})\\
Llama-3.1-8B & \fpath{results/nova/llama31_wave1_pulled_20261031-033750/} (\texttt{c4c9d50}); \fpath{results/nova/llama31_e0b_registration_pulled_20261031-075313/} (\texttt{38f0712}); conditional predictions \fpath{results/theory/pred2/conditional/} (\texttt{a36f441}, \texttt{d12eef2})\\
LPM-INF & \fpath{results/nova/lpminf_1716874x_pulled_20261031-042548/}, \fpath{results/r11_lpminf_eval/20261031-eval/} (\texttt{fc8b028})\\
rescue R1, R4 & \fpath{results/nova/rescue_1716896x_pulled_20261031-070502/} (\texttt{c6ea41e}); verdicts \fpath{results/agentload_analysis_rescue_r1/20261031-071720/} (\texttt{be50593}) and \fpath{results/agentload_analysis_rescue_r4/20261031-071748/} (\texttt{454ff7a}) in it; descriptive tables \fpath{results/rescue_paper_data/20261031-071938/} in it (\texttt{51c803d}); registrations \fpath{configs/agentload_prereg_2026-10-31_rescue*.yaml} (\texttt{f0bc3ad}); predictions \fpath{results/r11_predict_rescue/20261030-195231/} (\texttt{72501f0})\\
pred2 & \fpath{results/theory/pred2/}, \fpath{configs/r13_theory_pred2.yaml}, \fpath{notes/PREDICTIONS_r13_pred2.md} (\texttt{cfa7712}, tag \texttt{r13-theory-pred2})\\
rounds 11, 12 & \fpath{results/r11_predict_basin/20261030-194539/} (\texttt{c600fd7}), \fpath{results/r11_predict_lpminf/20261030-194911/} (\texttt{e2e8474}), \fpath{results/r11_predict_rescue/20261030-195231/} (\texttt{72501f0}), \fpath{results/r12_llama31_report/20261031-021057/} (\texttt{a62aea9})\\
gap$/R_M$ & \fpath{paper/derived/loopvars_*.csv} (frozen script of \texttt{f1338e3})\\
model~F check & \fpath{results/theory/icml/} (\fpath{trap_check.md}, \fpath{support.json}), \fpath{analysis/theory/icml/}, \fpath{configs/icml_trap_predict.yaml} (tag \texttt{icml-trap-pred}, \texttt{6b47a6c})\\
round 15 & \fpath{results/nova/r15_t1_edge_1725667x_pulled_20261105-061000/}, \fpath{results/nova/r15_cut_a_1725686x_pulled_20261105-063800/}, \fpath{results/nova/r15_cut_b0_17256870_pulled_20261105-065100/}, \fpath{results/nova/r15_cut_b12_1725687x_pulled_20261105-082300/}, \fpath{results/nova/r15_a2_s0_17256873_pulled_20261105-082300/}, \fpath{results/nova/r15_a2_s12_1725687x_pulled_20261105-083500/}; \fpath{analysis/r15_cut.py}; tags \texttt{r15-reg}, \texttt{r15-cut-reg} and \texttt{icml-cut-pred}\\
data-only analyses & \fpath{results/r15_dataonly/}, \fpath{analysis/r15_dataonly/}\\
MIRROR & \fpath{results/nova/r16_mirror_*_pulled_20261105-*/}, \fpath{results/agentload_analysis_r16_mirror/} (\texttt{f641b02}); \fpath{results/theory/r16/}, \fpath{analysis/r16_mirror.py}, \fpath{analysis/theory/icml/mirror_predict.py} (tag \texttt{r16-mirror-reg}, \texttt{bc81ee7}; erratum \texttt{d8c49b7})\\
push log, jobs & \fpath{provenance/}\\
figures & \fpath{figures/paper/}, \fpath{figures/rescue_paper_data.py}, \fpath{figures/theory_icml/} (Fig.~\ref{fig:trapuniv}), \fpath{figures/r16/} (Fig.~\ref{fig:mirror}), \fpath{paper/figs/README.md}\\
derived & \fpath{paper/derived/}, \fpath{figures/paper/derived_numbers.py}\\
\bottomrule
\end{tabular}
\end{table*}

\paragraph{Derived numbers.} Quantities that no committed analysis reports directly are computed from committed run data by \fpath{figures/paper/derived_numbers.py} and written to \fpath{paper/derived/}: the CAPPED drain times; the BURST backlog after the burst and over the window; the E0b hit fractions and realized arrival rates; the arm counts and the post-hoc sign test that counts each seed once; per BURST run, the hit fraction before the burst and the first 100\,s window from which no turn hit; and, per run of every arm, the client-side cost (gate waits, turn latency, unfinished sessions) and the realized arrival and completion rates over the window. They cover E0b, E2, native E0b, native E2, vLLM E0b and the exploratory vLLM BURST runs. Each derived file names a run both by its path on the cluster and by its path from the repository root.

\paragraph{Notes that support the text.} The math check of the model, \fpath{notes/MATHCHECK_thrashing_curve.md}, was done by an LLM agent on 2026-10-21; it predates the present assumptions (A1)--(A4) and Lemma~\ref{lem:waits}, and where it differs from \S\ref{app:qs} and Appendix~\ref{app:proofs}, those hold; the verification of Stage~C (three-source hit check, eviction survival times, prompt continuity) is \fpath{notes/STAGEC_VERIFICATION.md}; the audit of the PEEK, Continuum and CONCUR rules against released code is \fpath{notes/BASELINES_AUDIT.md}; an audit of the citations as of 2026-10-25, the scoop check (Lu et al.~\citep{lu2026cachetrap} appeared on 2026-11-02, after the last registered jobs of the earlier version had run (about 2026-11-01), and this version positions our work against it; \fpath{notes/LIT_CHECK_trap_2026-11-05.md} holds the literature check for model~F) and a first reading of the SGLang and vLLM sources are in \fpath{notes/lit/CITATION_AUDIT_2026-10-25.md} (that reading went twice through a page summariser, a web tool that returns a language model's summary of each file, without verbatim code; the line references of \S\ref{app:harness} and Appendix~\ref{app:proofs} were later read verbatim in the files at the tags), and that audit predates the citations added since, among them Yuan et al.~\citep{yuan2026agentic}, which it lists as not cited; the trace statistics are in \fpath{notes/lit/AGENT_TRACES.md}. The theory's derivations and their review are \fpath{notes/THEORY_r13_merged.md}, \fpath{notes/THEORY_r13_fluid_rigorous.md} and \fpath{notes/THEORY_r13_che_endogenous.md}; where they differ from Appendix~\ref{app:entry}, it holds. Where the research log and a committed CSV differ in the last digit (three hit fractions and one backlog of the policy grid), the paper uses the CSV.

\subsection{Proofs of the quasi-static results}\label{app:proofs}

\paragraph{The congestion bound.} Section~\ref{app:qs} states the following proposition in brief.

\begin{proposition}[Request-level FCFS, any online eviction rule]\label{prop:A}
Consider the closed system with $n$ sessions under any online eviction rule, one that does not foresee tool times, and assume that its long-run rates exist and are deterministic (Appendix~\ref{app:proofs}) and that:
\begin{enumerate}[label=(A\arabic*),leftmargin=*]
\item sessions share no cached tokens, each has at most one outstanding request, and evicted KV is discarded, not offloaded to another memory tier; tool times are i.i.d.\ with finite mean, independent of all other randomness, and unknown to the engine until they end;
\item admission is request-level FCFS: no request is admitted for the first time before a request that was already waiting when it joined the queue. More generally, it suffices that the order meets the condition of Appendix~\ref{app:proofs} for some $Q_0$ and $S_0$ independent of $n$, which allows reordering while at most $Q_0$ requests wait and lets a follow-up pass at most $S_0$ requests that were waiting when it joined. At most $B$ requests run at once, and at most $\alpha t+\beta$ admissions, re-admissions after a preemption included, occur in any interval of length $t$, with $B$, $\alpha$ and $\beta$ independent of $n$;
\item in congestion the engine computes prompt tokens no faster than with caching disabled: $\limsup_{n\to\infty}u_n\le C$, where $u_n$ is the long-run rate of prompt tokens that are computed rather than served from cache when their request is first admitted (each prompt token is counted once, so tokens recomputed after a preemption are not), and $C$ is the limit of that rate as $n\to\infty$ for the closed system with caching disabled.
\end{enumerate}
Then the long-run rates of cache-served prompt tokens and of hit turns tend to 0 as $n\to\infty$, so does the hit fraction if $\liminf_{n\to\infty}\mu(n)>0$, and $\limsup_{n\to\infty}\mu(n)\le\lmiss$. If also (A4) $u_n\to C$, that is, the congested engine stays saturated and computes prompt tokens as fast as with caching disabled, then $\mu(n)\to\lmiss$, so the hit fraction tends to 0.
\end{proposition}
\noindent The proof below rests on Lemma~\ref{lem:waits}. The idea is that a follow-up can reuse a cached token only if the token stays in the pool while the follow-up waits, which bounds reuse by $M$ over the wait whatever the eviction rule, and that request-level FCFS makes short waits rare as $n$ grows, because the engine cannot foresee tool times.

\paragraph{Model and conventions.} In the closed system of \S\ref{sec:notation}, $n$ sessions are present and a finished session is replaced at once by a fresh one, whose first request joins the queue. At time 0 every session is fresh, so no tool call is in progress, as in the synchronized start of our closed runs. A request joins the queue when its session sends it, waits, and is admitted; it leaves the queue only by admission. An admitted request runs until it completes or, if the engine preempts it, rejoins the queue. At a common instant joins are processed before admissions, as when an engine receives requests before each scheduling pass, and requests that join at the same instant enter in some order; the requests waiting when request $i$ joins are those in the queue just before it joins and those that join at the same instant ahead of it. When a request that is not its session's last completes, the session starts a tool call, and when the call ends the session's next request, a follow-up, joins the queue. By (A1) evicted KV is discarded, not offloaded. A follow-up's wait runs from its joining the queue to its first admission, and its reuse, and whether it hits, are counted at that admission. SGLang logged no retraction in any of the 73 server logs of our SGLang runs. When a running request cannot get a block for its next tokens, vLLM preempts the most recently admitted running request and returns it to the head of its queue; its closed runs logged 12 to 72 preemptions per run from $n=32$ on (72 at $n=300$, in a run that completed 5,395 requests), and its cache-disabled reference 9 or 10 per 32-session run. Since $u_n$ and $C$ count each prompt token once, at its request's first admission, recomputation after a preemption enters neither, and the proofs below go through with re-admissions counted among the admissions that (A2) bounds. By (A1) each session is at every moment in exactly one of three states: its request waits, its request runs, or it is in a tool call. Tool times have law $F_Z$ and mean $\bar z<\infty$. A long-run rate is $\lim_{t\to\infty}A(t)/t$ for a count or sum $A(t)$ over $[0,t]$; we assume that these limits exist almost surely and are deterministic, as for a positive Harris recurrent Markov process from any initial state, and that $\lmiss$ and $C$, the limits of $\mu(n)$ and $u_n$ as $n\to\infty$ with caching disabled, exist. Every $A(t)/t$ used below is bounded for fixed $n$ and $t\ge1$, since at most $\alpha t+\beta$ requests are admitted by time $t$, so by bounded convergence each rate also equals $\lim_{t\to\infty}E[A(t)]/t$, and bounds on expectations carry over to rates.

\begin{lemma}[short waits are rare]\label{lem:waits}
Assume (A1) and (A2), with $Q_0$ and $S_0$ as in (A2) (both 0 for request-level FCFS). Let $\gamma_n(K)$ be the long-run rate of follow-ups that wait less than $K$, and let $\kappa=\lfloor\alpha+\beta\rfloor+B$, $y=\kappa(1+\bar z)$ and $m=n-B-\alpha K-\beta-Q_0-S_0$. If $m>\kappa$, then
\[
\gamma_n(K)\le\frac{y+y^2+\kappa y}{m-\kappa},
\]
so $\gamma_n(K)=O(1/n)$ for every fixed $K>0$.
\end{lemma}

\noindent The general ordering condition of (A2), the only property of the order that the proof uses, is this: if more than $Q_0$ requests wait at every moment from a follow-up $i$'s joining the queue to its first admission, counting at a moment with admissions the requests that wait just before them, then all but at most $S_0$ of the requests that were waiting when $i$ joined are admitted, for the first time or again, no later than that admission. Request-level FCFS meets it with $Q_0=S_0=0$, since $i$ itself waits.

\begin{proof}
For a follow-up $i$ write $q_i$ for the time it joins the queue, $D_i$ for its wait and $Q_i$ for the number of requests waiting when it joins, and let $X(t)$ be the number of sessions in tool calls at time $t$.

\emph{A short wait needs a short queue.} Let $D_i<K$. At most $\alpha K+\beta$ admissions occur in $[q_i,q_i+D_i]$, $i$'s own included, so fewer than $\alpha K+\beta$ occur before $i$'s. If $Q_i\ge\alpha K+\beta+Q_0+S_0$, then at every moment of $[q_i,q_i+D_i]$, counted as above, more than $Q_0$ of the $Q_i$ requests still wait, because requests leave the queue only by admission; so at least $Q_i-S_0$ of them are admitted no later than $i$, and $Q_i-S_0+1\le\alpha K+\beta$, a contradiction. Hence $Q_i<\alpha K+\beta+Q_0+S_0$.

\emph{A short queue needs almost every session in a tool call.} Just before $q_i$ each of the other $n-1$ sessions is waiting (at most $Q_i$ of them), running or in a tool call, and $i$'s own session is in a tool call or, if that call has length 0, running; at most $B$ sessions run. So $X(q_i-)\ge n-Q_i-B>m$ whenever $D_i<K$.

\emph{From the instants at which tool calls end to fixed times.} Follow-ups join the queue when tool calls end, which happens more often when many sessions are in tool calls, so an average over these instants is not a time average; we avoid it as follows. A request that completes in an interval $(t_0,t_0+1]$ was admitted, for the last time, in that interval or was running at time $t_0$, so at most $\kappa$ requests complete, and at most $\kappa$ tool calls start, in any such interval. Fix an integer $j\ge0$. A follow-up that joins the queue in $(j,j+1]$ ends a tool call that was in progress at $j$ or started in the interval, so at most $X(j)+\kappa$ follow-ups join in it, and $X(q-)\le X(j)+\kappa$ for every $q\in(j,j+1]$. By the previous step, the number of follow-ups that join in $(j,j+1]$ and wait less than $K$ is at most $g(X(j))$, where $g(x)=(x+\kappa)\,\mathbf 1\{x>m-\kappa\}$.

\emph{A bound on $X(t)$ that does not depend on $n$.} Fix $t$ and split $(-\infty,t]$ into the intervals $I_k=(t-k-1,\,t-k]$, $k\ge0$; no tool call starts before time 0, and at most $\kappa$ start in each $I_k$. Let $(Z_{kl})_{k\ge0,\,1\le l\le\kappa}$ be i.i.d.\ with law $F_Z$, independent of everything else. Give the $l$-th call that starts in $I_k$ the duration $Z_{kl}$, and calls that start after $t$ fresh durations. This does not change the law of the system: by (A1) each call's duration is independent of all other randomness, and the system's evolution up to any time depends on a call's duration only through whether the call has ended by then; the index $(k,l)$ of a call is fixed when the call starts, so the variable it receives has not been used and is independent of the past at that moment, and the two constructions agree in law by induction over successive events. A call that started in $I_k$ and is still in progress at $t$ has a duration greater than $k$, so in this construction
\[
X(t)\le\hat X=\sum_{k\ge0}\sum_{l=1}^{\kappa}\mathbf 1\{Z_{kl}>k\},
\]
whose law depends on neither $t$ nor $n$. $\hat X$ is a sum of independent indicators with $E[\hat X]=\kappa\sum_{k\ge0}P(Z>k)\le\kappa(1+\bar z)=y$, so $\mathrm{Var}\,\hat X\le E[\hat X]$ and $E[\hat X^2]\le y+y^2$.

\emph{Conclusion.} Since $g$ is non-decreasing and $\mathbf 1\{\hat X>m-\kappa\}\le\hat X/(m-\kappa)$,
\[
E[g(X(j))]\le E[g(\hat X)]\le\frac{E[\hat X^2]+\kappa E[\hat X]}{m-\kappa}\le\frac{y+y^2+\kappa y}{m-\kappa}.
\]
Summing over the unit intervals of $(0,t]$, for integer $t$, bounds the expected number of follow-ups that join by time $t$ and wait less than $K$ by $t$ times this quantity; dividing by $t$ and letting $t\to\infty$ gives the claim.
\end{proof}

\noindent The bound is of order $1/n$, but its constants grow with the admission rate and the mean tool time, so it is a statement about the limit, not about the populations we ran.

\paragraph{Proof idea.} By (A1), the $r_i$ prompt tokens that follow-up $i$ reuses at its first admission are its session's own and stay in the pool from the end of the session's previous turn until that admission, an interval no shorter than its queue wait $D_i$, and the token-time that these intervals claim does not overlap. Resident token-time accrues at rate at most $M$, so follow-ups that wait at least $K$ reuse tokens at rate at most $M/K$, whatever the eviction rule. By (A2) a follow-up that waits less than $K$ joined behind fewer than $\alpha K+\beta+Q_0+S_0$ requests, so just before it joined more than $n-B-\alpha K-\beta-Q_0-S_0$ sessions were in tool calls; Lemma~\ref{lem:waits} (Appendix~\ref{app:proofs}), which uses that the engine does not see tool times, shows that such follow-ups occur at a rate of order $1/n$, and each reuses at most $L_{\max}=40{,}350$ tokens. Letting $n\to\infty$ and then $K\to\infty$, the cache-served token rate tends to 0, and since a hit reuses more than $L_0/2$ tokens, so does the rate of hits. Counted once, at its request's first admission, each of a session's $W$ prompt tokens is either served from cache or computed, so $u_n$ is $\mu(n)E[W]$ less the cache-served token rate, and with caching disabled the same count gives $C=\lmiss E[W]$; (A3) gives the bound and (A4) the limit. The same count gives turns at the rate $\mu(n)E[T]$, so the hit fraction tends to 0 whenever $\mu(n)$ stays bounded away from 0.

\begin{proof}[Proof of Proposition~\ref{prop:A}]
\emph{Reuse needs waiting token-time.} Let follow-up $i$ of session $\sigma$ join the queue at $q_i$, wait $D_i$, and reuse $r_i$ prompt tokens from cache at its first admission. By (A1) these tokens are $\sigma$'s own, computed by its earlier requests, and evicted KV is discarded, so each of them was in the pool throughout $[q_i,q_i+D_i]$, during which $\sigma$ has no running request to recompute it. These pairs of token and time are distinct for different requests: sessions share no tokens, and the waits of one session's requests do not overlap. At most $M$ tokens are resident at any time, so for every $t$, $\sum_i r_iD_i\le Mt$ over the follow-ups first admitted by $t$. This step uses neither the order nor the eviction rule, so it holds for clairvoyant rules too.

\emph{Reuse vanishes.} Fix $K>0$. By time $t$ the follow-ups that waited at least $K$ reused at most $Mt/K$ tokens, and each of the others at most $L_{\max}=40{,}350$, the longest prompt. So the long-run rate $\rho_n$ of cache-served prompt tokens satisfies $\rho_n\le M/K+L_{\max}\gamma_n(K)$, and Lemma~\ref{lem:waits}, with the order's $Q_0$ and $S_0$, gives $\limsup_{n\to\infty}\rho_n\le M/K$ for every $K$, so $\rho_n\to0$. A hit reuses more than half its prompt, hence more than $L_0/2$ tokens, so hits occur at a rate of at most $2\rho_n/L_0\to0$.

\emph{Work.} A session with $T$ turns sends $W(T)$ prompt tokens (\S\ref{sec:notation}), and each, counted once at its request's first admission, is either served from cache or computed. The closed system has $n$ slots, each holding one session at a time and a fresh one after it finishes, so each slot holds a sequence of sessions with i.i.d.\ turn counts, and by the strong law of large numbers completed sessions bring prompt tokens at the long-run rate $\mu(n)E[W]$ and turns at the rate $\mu(n)E[T]$; sessions in progress hold at most $nW(60)$ tokens and $60n$ turns and do not affect rates. Hence $u_n=\mu(n)E[W]-\rho_n$. With caching disabled nothing is reused, so the same count gives $C=\lmiss E[W]$. So $\mu(n)=(u_n+\rho_n)/E[W]$, and (A3) gives $\limsup_{n\to\infty}\mu(n)\le C/E[W]=\lmiss$. The hit fraction is the rate of hits over the rate of turns, so it is at most $2\rho_n/(L_0\,\mu(n)E[T])$ and tends to 0 whenever $\liminf_{n\to\infty}\mu(n)>0$. Under (A4), $\mu(n)\to\lmiss>0$.
\end{proof}

\paragraph{Scope of Proposition~\ref{prop:A}.} The proof's last step, from vanishing reuse to $\lmiss$, is the bound that MISA-T notes at prefill saturation, a request rate of at most the prefill throughput over the mean uncached prompt tokens per request~\citep{hong2026misat}; what the proposition adds is that request-level FCFS makes reuse vanish in congestion whatever the online eviction rule. (A2) describes an admission pass that stops at the first request that does not fit, as in our simulators and, on our reading of their sources, in SGLang and vLLM under their default FCFS policies (Appendix~\ref{app:proofs}); vLLM also preempts running requests and returns them to the head of its queue, which (A2) allows, because its ordering clause concerns first admissions; with asynchronous scheduling it also lets a follow-up pass a bounded number of just-preempted requests, so it meets the general condition of (A2) with $Q_0=0$ and a bounded $S_0$ (Appendix~\ref{app:proofs}). A diverging mean cycle time would not be enough: Lemma~\ref{lem:waits} uses bounded admissions and i.i.d.\ tool times to bound the tail of the number of sessions in tool calls independently of $n$, which makes it rare that almost all sessions are in tool calls when one returns. The lemma's bound is of order $1/n$ but has large constants, so the proposition describes the limit, not the populations we ran. The lemma needs FCFS order only while more than $Q_0$ requests wait, so, on our reading of its source, the proposition also covers SGLang's stock LPM, which stops re-sorting its queue whenever more than 128 requests wait, so that new requests queue behind all waiting ones ($Q_0=128$). It does not cover program-level FCFS, which puts an old session's follow-up ahead of newer sessions' requests however long the queue: in our earlier simulator its hit fraction stays near 0.03 from $n=64$ to 300, while FCFS has none. A pure resident-first order, such as LPM without the fallback, violates (A2) outright: a returning session whose KV is still resident is admitted ahead of the queue.

(A1) holds for our random-token sessions, apart from rare one-token coincidences (5 of the 19,924 first turns in E2 had one cached token), but not for agents that share a system prompt: one resident copy then serves every session. We expect the argument to apply to the private suffixes, with $\lmiss$ and $C$ measured with the shared prefix cached, but have not shown this. (A1) also rules out offloading evicted KV to host memory, which would let a follow-up reuse tokens that held no pool space while it waited. (A3) and (A4) hold on our engines only approximately: with zero hits, closed runs and the registered BURST runs computed prompt tokens at 0.95--1.12 times the cache-disabled rate $C$, and at 1.03 times it near the stationary turn mix, so under (A4) the proposition puts $\muinf$ about 2--5\% above $\lref$, within the standard error of $\lref$ (Appendix~\ref{app:proofs} gives the measurements).

The bound holds for every online eviction rule, including protection of queued prefixes and TTL pins that keep request-level FCFS order, where (A4) can fail; in our earlier simulator both still hit 10--23\% of turns at $n=300$, so a finite-$n$ hit fraction does not by itself separate covered orders from uncovered ones. Belady's offline rule~\citep{belady1966study} foresees tool times and is outside (A1): the token-time step of the proof covers it, Lemma~\ref{lem:waits} does not.

\paragraph{Checks of (A3) and (A4) on our engines.} (A3) and (A4) treat the engine as a server of prompt tokens, as E0a suggests (\S\ref{sec:edge}). They concern long-run rates of the closed system, in which every order serves turns in the stationary mix, since each completed session contributes all its turns; our measurement windows did not have that mix, and the token rate varied between them, so we can check (A3) and (A4) only approximately. We estimate $C$ by E0a's uncached token rate at 32 sessions, 11,243 tokens/s. With zero hits, closed runs computed prompt tokens at 1.03~$C$ at $n=96$, whose served prompts were within 3\% of the stationary mean, on SGLang and on vLLM (against vLLM's own $C$), at 0.99~$C$ at $n=48$ (prompts 12--14\% longer) and at 1.10--1.11~$C$ at $n=300$ (28--30\% shorter); the registered SGLang BURST runs computed 1.01--1.03~$C$ at 300k (prompts 9--20\% shorter) and 1.07--1.12 times the 32-session all-miss rate of the native pool (prompts 29--41\% shorter). So on our engine (A3) and (A4) hold only approximately: to within a few percent at the stationary mix, and to within 12\% over the windows we measured. We estimate $\lmiss$ by $\lref$, E0a's turns-based rate at 32 sessions (0.0715\sps); $\lref$ and $C/E[W]$ estimate the same cache-disabled limit. $C/E[W]$ is 0.0728\sps, 1.7\% above $\lref$, because the turns in E0a's window were longer than the stationary mix (10,648 against 10,469 prompt tokens on average). E0a's 64-session check computed 1.7\% more tokens per second than the 32-session runs and 3.2\% more than the same seed at 32 sessions, although its window held longer turns (11,456 prompt tokens on average), so the 32-session reference falls short of saturation by about 2--3\%, perhaps more. With the 64-session rate in place of $C$, $C/E[W]$ is 1.034 $\lref$, and with the zero-hit rate at $n=96$ (11,602 tokens/s) it is 1.050 $\lref$. Under (A4) the proposition therefore puts $\muinf$ about 2--5\% above $\lref$, an offset within the standard error of $\lref$ over seeds (0.0038\sps, 5\%); on vLLM the three estimates are 1.020, 1.038 and 1.050 $\lref$ ($\lref=0.0657$\sps).

\paragraph{Engine sources of \S\ref{app:harness}.} The SGLang rules of \S\ref{app:harness} (default policy and eviction rule, the 128-request LPM fallback, the unlocked prefix of a waiting request, and the refresh of waiting prefixes) are in these files under \fpath{python/sglang/srt/} at tag v0.5.22: \fpath{arg_groups/fields/schedule.py}, lines 82--98; \fpath{arg_groups/fields/memory.py}, lines 25--42; \fpath{managers/schedule_policy.py}, lines 133--135, 278--311 and 1591; \fpath{mem_cache/radix_cache.py}, lines 552--580; and \fpath{environ.py}, line 594. vLLM 0.32.0 schedules FCFS, its default, or by priority (\fpath{vllm/config/scheduler.py}, lines 21 and 159), and a freed cached block keeps its hash until it is reallocated (\fpath{vllm/v1/core/block_pool.py}, lines 666--690 and 774--803; at tag v0.32.0). We first read these sources through a web tool that returns a language model's summary of each file (Appendix~\ref{app:prov}); the line references here and above are to the files at the tags, read verbatim.

\subsection{Threats to validity}\label{app:threats}\label{sec:limits-old}

\paragraph{Scope of the registered evidence.} All registered SGLang evidence is for SGLang 0.5.22 and, but for two paired tests, Qwen2.5-7B-Instruct and one A100, at two pool sizes: the 300k-token pool limit and the 955,757-token pool SGLang sizes itself, with which the cliff moved out and the registered tests passed too, so the pool limit did not create the cliff. The two exceptions, Llama-3.1-8B on the A100 at 300k and Qwen2.5-7B on one H200 with a 600k pool, have three seeds per rate, and the first passed at only one of its two rates. vLLM 0.32.0 ran only with the 300k-token pool, and its paired test is registered, not pre-registered, with a BURST arm that repeats the workloads of earlier exploratory runs (\S\ref{sec:vllm}). The 300k H200 cell is a measurement, not a registered test, and the H200 cells compare GPU and default attention backend together. The rescue tests engage the registered cap at a fixed time, not on an online signal, with three seeds per cell; they test persistence once the cap no longer binds, with no perturbation after release, and their cost and persistence comparisons are descriptive. The shedding and tool-slowdown tests ran at one rate on the 300k pool, with two and three seeds per setting. Larger models, parallelism, offloading and GPUs other than the A100 and the H200 are untested.

\paragraph{Synthetic sessions and clients.} Sessions are synthetic (random tokens, fixed tool outputs and output lengths, geometric turn counts, exponential 5\,s tools) and share no prefixes, which removes the case PEEK's scheduler targets. Tool times in real agent traces (TraceLab~\citep{zhu2026tracelab}; mini-SWE-agent~\citep{sweagent2026miniswe,yang2024sweagent} on SWE-bench Verified~\citep{jimenez2024swebench,openai2024swebenchverified}; statistics in \fpath{notes/lit/AGENT_TRACES.md}) have medians 15--33 times shorter than ours and coefficients of variation of 4.3--4.7 against 1, and outputs per call 2.7--5.9 times our 100 tokens. With a shared system prompt the ratio of all-miss to all-hit prompt work, about 16, would barely change, but the hit fraction of a trapped engine would stay near the shared share, so hit-based signals would have to count private-suffix reuse. Our clients never retry or abandon a request; retries would add load in the trapped state and abandonment would remove it, so either could change whether the engine recovers.

\paragraph{Few runs, finite horizons, nominal rates.} The closed curves and the from-empty sweeps have one run per point, so the peak's location is within noise and single runs near the cliff land on either side of it. The registered tests, the grid, the vLLM runs and the rescue tests share the same three seeds, so their runs are not independent across rates, pools, engines or tests. All rates are nominal; each from-empty sweep used a single arrival seed, and at 300k the seed-0 bracket was not replicated on fresh seeds, although stability is monotone in the realized rate, with the edge between 2.665 and 2.726 $\lref$; on the H200 at 300k it is likewise monotone over ten runs, with the edge between 1.496 and 1.503 $\lref$ (the 1.503 run ended just above the queue threshold). ``Does not leave'' means within 7500\,s on the engine, and the 3000\,s horizon of the from-empty runs cannot see slow escapes. The registered rule loses margin at both ends of the window: the slope test near $\lmiss$, and the token test at high multiples. Uncontrolled factors remain, among them the choice of the native cap and the coincidence of the cliff with a change of job and node in both SGLang curves.

\paragraph{Models and simulators.} The fluid model is used for the window, the recovery threshold and drain times, and through the heuristics H2--H4 for the from-empty edge, burst sizes and release times; it is not quantitative in congestion and is least reliable near the cliff, and the frozen SHED and TOOLSLOW predictions used it there; the first held and the second did not. Proposition~\ref{prop:A} is a limit statement under stated assumptions, including private prefixes, no offloading, i.i.d.\ tool times unknown to the engine and an engine that in congestion computes prompt tokens no faster than with caching disabled, which our engine meets only to within a few percent at the stationary turn mix and within 12\% over the windows we measured; that recovery without a cap needs a rate below $\lmiss$ is proved only within model~F (Corollary~\ref{cor:trapthr}) and tested on the engine by SHED's full stop and the CUT runs (a cut above $\lmiss$, CUT-A, left all three traps in place within the run, and a cut below it, CUT-B, shrank their backlogs only slowly), pure LPM's congested rate is not covered by Proposition~\ref{prop:A} (about 1.17~$\lmiss$ in our earlier simulator, \S\ref{sec:grid}), and the capped system's stochastic stability is not proved. The first-principles model's propositions hold for a deterministic fluid model, with no limit theorem to the engine. It misses the H200's tail past the peak at both pools (at 600k also the 0.5-hit crossing, 4.4 sessions late, and the trapped token rate of the paired test, 0.038--0.049 low in 5 of 6 runs), vLLM's earlier cliff and, with a 150k-token pool, the peak height and the from-empty edge. It predicted that a 30\,s slowdown of every tool call for 900\,s, with no burst, would trap the engine in every seed; it trapped 1 of 3. Its fluid verdict for TOOLSLOW-D15, a 15\,s slowdown for 900\,s at 1.65~$\lref$, was again a trap in every seed; one run of three trapped, as the calibrated simulator predicted seed by seed. On Llama-3.1-8B its edge was one grid step high (a near miss), its trapped token rate in the paired test was 0.031--0.046 low in 3 of 6 runs, and its crossing failed a later-declared test of the tool-call term. In BASIN-2 its per-seed probabilities favoured the wrong state for seed 0 at 40 and 50 sessions. Its release-time predictions were early, with observed releases more than 15\% later in 4 of 9 runs (all three with the native pool), and its derivation and edge criterion were selected on the measured cells (\S\ref{sec:fp-weak}). The calibrated simulator failed its level test at one cell, predicted a pass of the Llama-3.1-8B paired test at 1.35 $\lref$, which failed, omits partial prefixes and the waiting-prefix refresh, frees a finished session's KV at once and understated LPM.

\paragraph{The registration record.} Appendix~\ref{app:timeline} lists the slips. The vLLM registration reached GitHub after two of its jobs had started, and some prediction rounds were pushed after their jobs were submitted, or specified after related data were known. The closed-curve values of the 300k H200 cell for $n\le32$, which include the predicted peak, were seen by its operating agent and the first author before the freeze, though not by the agents that derived the model. Some predictions are blind by access only: the model's for BASIN-1 and the other 300k H200 runs that had ended before the freeze, and the model's and the simulator's for the Llama-3.1-8B curve. The 600k H200 and Llama-3.1-8B registrations and the Llama-3.1-8B registration-conditional predictions are not tagged, and R4's engage time, three stricter changes to the rescue rule and the model's rescue predictions were chosen or frozen with knowledge of the simulator's predictions. No registration or prediction is in a third-party registry, and none has a timestamp other than GitHub's push log of a private repository, which readers can see only in our export.

\section{Lu et al.'s concurrent work and other systems in detail}\label{app:lu}

\paragraph{Concurrent work.} Lu et al.~\citep{lu2026cachetrap} study the same phenomenon on SGLang with prefill--decode
disaggregation and a host KV tier, serving a hybrid mixture-of-experts model (Qwen3.6-35B-A3B) on Terminal-Bench and SWE-bench Pro
trajectories. On hardware, in single runs, as their run accounting implies, a ten-minute surge cut goodput by about 88\% over the following hour,
terminating the burst sessions at the end of the burst restored service, and in a run that let them finish the base population kept
growing after the last burst session left. Their fluid model, $\dot N=\lambda-X(N)/K$ with an empirical hit curve $h(W/C)$ of the
working set over the cache, gives the stable and unstable equilibria, the vulnerable window between the low-reuse departure rate
$D_{\mathrm{low}}$ and the peak, the statement that arrivals below $D_{\mathrm{low}}$ let the population drain, a resource condition
for cache-first vulnerability, and a drain-time estimate. In simulation they map vulnerability, show that changing post-surge
arrivals alone reverses recovery, and report cache hysteresis (a lower hit rate after collapse at a comparable working set). Their
controller, Basin\footnote{Their Basin controller is unrelated to our BASIN-1 and BASIN-2 tests, which probe the basin of
attraction of the good state.}, combines fixed session priority, admission by predicted peak context, decode-memory lending and
shedding; on H200 hardware an arrival-time prototype of it achieves 1.20--1.70 times Continuum's goodput, and the predictor-driven
design is evaluated in simulation. Against this, we add an exactly analysed idealization of the trapped state with its tests
(\S\ref{sec:trap}--\ref{sec:validation}), an engine test of the reading that the throughput measured in a trap exceeds the drain
threshold (Corollary~\ref{cor:trapthr}, proved for our workload in model~F, our idealization of an FCFS engine on which every turn misses;
CUT-A and
CUT-B are consistent with it; their systems keep some reuse), a first-principles entry model with an
out-of-sample record (Appendix~\ref{app:entry}), the registered and replicated rescue of an established trap at the same offered
load without shedding, with the control removed and persistence tested (\S\ref{sec:control}), and registered matched-arm evidence on colocated engines,
including a second engine (\S\ref{sec:stampede}). They have real traces, disaggregated serving, a larger model, SLO goodput and a full
mitigation system, which we do not.

\paragraph{The cut and the rescue in their terms.} Lu et al.\ show in
simulation that lowering post-surge arrivals reverses the outcome, and observe that the throughput measured in the collapsed state
sets the scale of the cut~\citep{lu2026cachetrap}. Corollary~\ref{cor:trapthr}, proved in model~F, our idealization of an FCFS engine on which
every turn misses, sharpens this: there the cut must go below $\lmiss$, which for our workload lies below the trapped throughput $D(\lambda)$
(Theorem~\ref{thm:premium}). Their
GPU-only simulations point the same way: the trap persisted at a cut to 0.2 of their reference rate (12 sessions/min) while the
trapped departures they measured were 0.20--0.31 of it. Their
systems keep some reuse in the trap, though, and with a host tier one seed in three recovered at cuts above the trapped departures
they measured, so the corollary does not describe them.
Lu et
al.\ restore service on hardware by terminating the burst sessions at the end of the burst, when the trap is incipient, prevent it
with their controller active before the surge, rescue an established trap in simulation by activating their controller late (and
leaving it armed: its admission releases itself after a cool-down, but no reported run removes the controller and tests
persistence), and report one hardware run in which CONCUR's hit rate fell
to 0.33 before its concurrency window contracted and recovered slowly, with the controller left on and no release
test~\citep{lu2026cachetrap}.

\begin{table}[t]\centering\footnotesize
\caption{The 0.5-hit crossing (sessions): measured; Lu et al.'s hit curve $h(W/C)$, read post hoc with $C=M$,
$\bar c=\bar E=10{,}569$ tokens and $h$ fixed by the measured A100 300k crossing ($W/C=1.133$), so that it predicts
$1.133\,M/\bar E$ on every engine and GPU; our frozen crossing $n_{1/2}$ (fluid; nothing fitted to any curve point; the H200 and
vLLM cells use one speed factor fitted to their all-miss token rate); and the capacity ratio $M/E_S$. Status of our prediction:
b, blind in time; a, blind by access only ($^\ast$the crossing lies between $n=36$, blind in time, and 40; the cell's operating agent and the first
author had seen $n\le32$); i, in sample. Lu et al.'s own constant $N_c\approx1.9\,C/\bar c$, measured on a disaggregated H100 deployment,
would put every crossing 1.4--1.9 times further out, so we fix $h$ on one of our cells; they place their cliff where the hit rate
falls to about 0.3, and we compare 0.5-hit crossings. With its constant fitted to all eight
crossings instead ($W/C=1.098$, in sample), the $h(W/C)$ reading errs by 3.7 sessions on average, against 1.4 for
$n_{1/2}$.}\label{tab:lu}
\setlength{\tabcolsep}{3pt}
\begin{tabular}{@{}lccccc@{}}\toprule
cell & status & meas. & $h(W/C)$ & $n_{1/2}$ & $M/E_S$\\\midrule
A100, 150k & b & 16.5 & 16.1 & 16.9 & 13.6\\
A100, 300k & i & 32.2 & (32.2) & 31.5 & 27.8\\
A100, 600k & b & 60.1 & 64.3 & 60.0 & 56.1\\
A100, native & i & 93.4 & 102.5 & 93.7 & 89.8\\
H200, 300k & a$^\ast$ & 38.9 & 32.2 & 37.3 & 27.8\\
H200, 600k & b & 71.0 & 64.3 & 66.6 & 56.1\\
vLLM, 300k & i & 29.1 & 32.2 & 31.8 & 28.2\\
Llama, 300k & a & 29.65 & 32.2 & 30.9 & 27.8\\\midrule
mean error, b and a (5) & & & 4.1 & 1.5 & 7.0\\
mean error, all but 300k (7) & & & 4.7 & 1.5 & 5.6\\\bottomrule
\end{tabular}
\end{table}

\paragraph{Against Lu et al.'s empirical model.} Lu et al.'s fluid model takes its hit rate $h(W/C)$, a function of the working set
$W=N\bar c$ over the prefix-cache capacity $C$, from measurements of the deployed disaggregated system, and uses it to classify
deployments and explain mitigations rather than to predict the hit rate~\citep{lu2026cachetrap}. Read post hoc on our colocated
cells, with $C=M$, $\bar c=\bar E$ and one $h$ fixed by the A100 300k crossing, it puts the 0.5-hit crossing in proportion to the pool
and, with that curve carried across GPUs, at the same population on the H200. On the five cells whose crossings we predicted
before the runs that set them were read (three blind in time, two blind by access only, one of them partly seen), it errs by 4.1 sessions on average, against 1.5 for our
frozen crossings (Table~\ref{tab:lu}). The measured crossing moved out on the H200, by 6.7 sessions at 300k and 10.9 at 600k, against
5.8 and 6.6 from the tool-call term; refitting $h$ on the H200's own 300k cell still puts its 600k crossing 6.7 sessions out. Their
vulnerability condition and drain estimate are stated for separate prefill and decode instances and for resident-first ordering
with healthy reuse, so we do not treat them as predictions of theirs; the D4 paragraph below applies them post hoc.

\paragraph{Lu et al.'s model on our cells (D4).} Every reading below is ours, post hoc, of a model built for disaggregated serving
whose hit curve is empirical and system-specific. It was computed by one agent and recomputed by a second with its own code, as
the analyses of Appendix~\ref{app:dataonly}; the second computation also judged which readings are fair.
\begin{itemize}[leftmargin=*,nosep]
\item \emph{Onset} (Table~\ref{tab:lu}). With $h$ fixed by the A100 300k crossing, the reading errs by 4.66 sessions on average over
the seven other cells and 4.11 over the five predicted before the runs that set them were read, against 1.53 and 1.55 for our frozen crossings.
The comparison survives every charitable variant tried: the constant fitted to all eight cells (3.74 against 1.42), fitted on the four
A100 pools only (3.57 against 1.53 over seven cells), left out one cell at a time (4.72), and crossings of the token hit fraction
instead of the turn hit fraction (4.05 against 1.53 on the five cells). A hit curve that depends only on $W/C$ cannot move the onset
with the GPU; refitted per GPU it still misses the 600k H200 by 6.7 sessions (ours: 4.4 sessions low). Over the four A100 pools the measured
crossing is affine in the pool, $2.97+1.003\,M/\bar E$ (intercept standard error 0.71; 95\% interval $-0.1$ to 6.0 with two degrees of
freedom), consistent with the tool-call term (4.7--5.1 sessions) net of $n_{1/2}$'s lead over $n_b$, but not established by four points.
\item \emph{Vulnerability condition.} Their Eq.~2 compares a dedicated prefix cache with decode memory. In one pool the two are the
same memory, and it reduces to $f\lesssim1/2$ (1/1.9 with their measured cliff constant) for the fraction $f$ of active sessions
in service, which here moves along the curve:
0.19--0.53 at our model's fluid peak, 0.38--0.90 at the 0.5-hit crossing and 0.67--0.95 at $n_b$, so the verdict depends on where $f$ is
read. All eight cells are vulnerable, so they cannot test the condition's specificity. Their own classifier, which compares
throughput bounds, flags all 17 cache-first cases among 36 simulated deployments (28 of 36 correct), and in simulation they find
that a larger, faster GPU raises the first trapped load in absolute terms. We do not score the condition.
\item \emph{Drain.} Their Eq.~4, $S/(D(N^*)-\lambda)$, is stated for resident-first ordering near the peak with healthy reuse, and its
queue holds sessions awaiting their first call; most of a rescue backlog is admitted sessions paused at tool boundaries, whose next
turn misses. Applied to R1 and R4 with the measured peak rate and backlog (release $=t_{\mathrm{on}}+(n_{\mathrm{on}}-c)/(\mu^*-\lambda)+300$\,s),
it is 19--39\% early: the engine released 24--64\% later. Our model's own single-phase variant, in the same frozen file but not a
declared prediction, is 12--34\% early (released 14--51\% later); the frozen two-phase drain, in which every paused session's next
turn misses, was 2--18\% early (released 2--22\% later). The gain comes from the two-phase structure, not from a different peak.
\item \emph{Window.} The ratio of peak ratios between pools needs their prefill and decode bounds and a smooth minimum whose form
they do not give for colocated serving; we omit that comparison.
\end{itemize}
Sources: measured crossings from the closed-population runs (\fpath{paper/NUMBERS.md}); frozen predictions in
\fpath{results/theory/merged/*.json} and \fpath{results/theory/pred2/h200_600k.json}; rescue releases in the committed
\fpath{rescue_runs.csv}; Lu et al.'s equations from their Section~5. The full record is
\fpath{results/r15_dataonly/verify_d4/verify_D4.md}.

\paragraph{Other systems.} CacheWise evaluates coding agents with sessions started together~\citep{tiwari2026cachewise}; XPerf, a benchmark, sees hit rates fall toward saturation in fixed-rate runs at increasing rates~\citep{wang2026xperf}. Other systems manage idle KV across tool calls or workflow steps by offloading, prefetching or retention~\citep{abhyankar2024infercept,pan2025kvflow,xia2026mori,bian2025tokencake,zhang2026cachescout,liu2026askthetool,biswas2026sutradhara}, schedule agent programs~\citep{luo2026agentix,lin2024parrot}, cache conversation state hierarchically~\citep{gao2024cachedattention,yu2025pensieve,qin2025mooncake} (Mooncake's scheduler also rejects requests early under overload), admit, place or reserve capacity for agent sessions by tool readiness, KV residency or likely returns~\citep{li2026tao,jiang2026coda,meng2026talaria}, or make locality-aware scheduling fair~\citep{cao2025dlpm}. Eviction-order work~\citep{pan2026rrevict,wu2026randomization,liu2026fancy,pan2026scalesim} changes which idle state is evicted; RR-Evict assumes that the active trajectories fit in the cache, ScaleSim prefetches and evicts each agent's state by the estimated order of the agents' next requests, and Liu et al.\ replay traces without feedback from queueing to arrivals, as do Arya et al.\ for production Microsoft M365 agent traces, where LRU with a pooled budget of a tenth of the busiest hour's no-eviction demand makes 63\% of Deep Research calls recompute~\citep{arya2026residency}. Other close neighbours: Metronome calls a collapse in real-time speech-model serving metastable, in the sense that identical runs collapse or survive near a threshold, but there each session's KV stays pinned for the whole conversation until the pinned KV exhausts the pool, a different mechanism from eviction and recomputation of reusable prefixes~\citep{meng2026metronome}; AgentServeSim simulates agent serving with KV retained across tool gaps and successor turns released by simulated completions, so call timing depends on the policy, and validates it against vLLM in 20 paired cells, without a metastability analysis~\citep{rajib2026agentservesim}; and ServeTwin simulates multi-turn agent workloads while tracking KV-cache evolution and reports that a pool 1.4 times larger lifts the hit rate from 22\% to 93\% at a fixed number of users~\citep{park2026servetwin}.

\end{document}